\RequirePackage{fix-cm}
\documentclass{svjour3}                     % onecolumn (standard format)
\smartqed  % flush right qed marks, e.g. at end of proof
\usepackage{graphicx}
\usepackage{color}
\usepackage[dvipsnames]{xcolor}
\usepackage{cite}
\usepackage[sort&compress, square]{natbib}
\usepackage{har2nat}
\usepackage{amssymb}
\usepackage{amsfonts}
\setcitestyle{square}
\usepackage{amsmath}
 \journalname{Space Science Review}

\newcommand{\grl}{    {Geophys. Res. Lett.}}
\newcommand{\jgr}{    {J. Geophys. Res.}}
\newcommand{\ssr}{    {Space Sci. Rev.}}
\newcommand{\planss}{    {Plan. Sp. Sci.}}

\newcommand{\nat}{ {Nature}}
\newcommand{\pre}{ {Phys. Rev. E}}
\newcommand{\prl}{ {Phys. Rev. Lett}}

\newcommand{\red}{\textcolor{red}}
\def\H{{\cal H}}
\def\Q{{\cal Q}}
\def\h{{\cal H}}
\def\B{{\cal B}}
\def\E{{\cal E}}
\def\e{{\epsilon}}
\def\M{{\cal M}}
\def\N{{\cal N}}
\def\mP{{\cal P}}
\def\P{{\cal P}}

\def\Nr{{ n_r}}
\def\C{{\cal C}}
\def\mF{{\cal L}}
\def\a{{\cal A}}
\def\W{{\cal W}}
\def\D{{\cal D}}
\def\Y{{\cal Y}}
\def\R{{\cal R}}
\def\area{{\cal S}}
\def\flux{{\cal F}}
\def\Uw{{\cal U}_{w}}
\begin{document}

%\tableofcontents

\title{Nonlinear resonant interactions of radiation belt electrons with intense whistler-mode waves.
}
\subtitle{}

\titlerunning{Nonlinear resonant wave-particle interactions}        % if too long for running head

%\author*[1]{\sur{Angelopoulos Vassilis}}\email{vassilis@ucla.edu}


%\affil*[1]{\orgdiv{Earth, Planetary, and Space Sciences}, \orgname{University of California, Los Angeles}, \orgaddress{\street{595 Charles E Young Dr E}, \city{Los Angeles}, \postcode{90095}, \state{CA}, \country{USA}}}

%\affil[2]{\orgdiv{Department}, \orgname{Organization}, \orgaddress{\street{Street}, \city{City}, \postcode{10587}, \state{State}, \country{Country}}}

%Albert?

\author{A. V. Artemyev\textsuperscript{1}  \and D. Mourenas\textsuperscript{2,3} \and X.-J. Zhang\textsuperscript{4,1}    \and O. Agapitov\textsuperscript{5}   \and A. I. Neishtadt\textsuperscript{6} \and D. L. Vainchtein\textsuperscript{7}  \and A. A. Vasiliev\textsuperscript{8} \and X. Zhang\textsuperscript{1} \and Q. Ma\textsuperscript{1} \and J. Bortnik\textsuperscript{1}, V. V. Krasnoselskikh\textsuperscript{9}  }


\institute{\at  1 University of California, Los Angeles, Los Angeles, CA 90095, USA\\
              \email{aartemyev@igpp.ucla.edu},
\at 2   CEA, DAM, DIF, Arpajon, France,
\at 3   Laboratoire Matière en Conditions Extrêmes, Université Paris-Saclay, CEA, Bruyères-le-Châtel, France
\at 4	University of Texas at Dallas, Richardson, TX 75080,
\at 5	Space Sciences Laboratory, University of California, 94720, Berkeley, USA,
\at 6	Department of Mathematical Sciences, Loughborough University, Loughborough, LE11 3TU, United Kingdom,
\at 7	Nyheim Plasma Institute, Drexel University, Camden, New Jersey 08103, USA,
\at 8	Harbour.Space University, Carrer de Rosa Sensat 9-11, 08005 Barcelona Spain
\at 9   LPC2E, CNRS-University of Orléans-CNES, 45071, Orléans, France}

\date{Received: date / Accepted: date}

\maketitle

\begin{abstract}
The dynamics of the Earth's outer radiation belt, filled by energetic electron fluxes, is largely controlled by electron resonant interactions with electromagnetic whistler-mode waves. The most coherent and intense waves resonantly interact with electrons nonlinearly, and the observable effects of such nonlinear interactions cannot be described within the frame of classical quasi-linear models. This paper provides an overview of the current stage of the theory of nonlinear resonant interactions and discusses different possible approaches for incorporating these nonlinear interactions into global radiation belt simulations. We focused on observational properties of whistler-mode waves and theoretical aspects of electron nonlinear resonant interactions between such waves and energetic electrons.
\subclass{MSC code1 \and MSC code2 \and more}
\end{abstract}

\section{Introduction}
\begin{figure}
\centering
\includegraphics[width=1\textwidth]{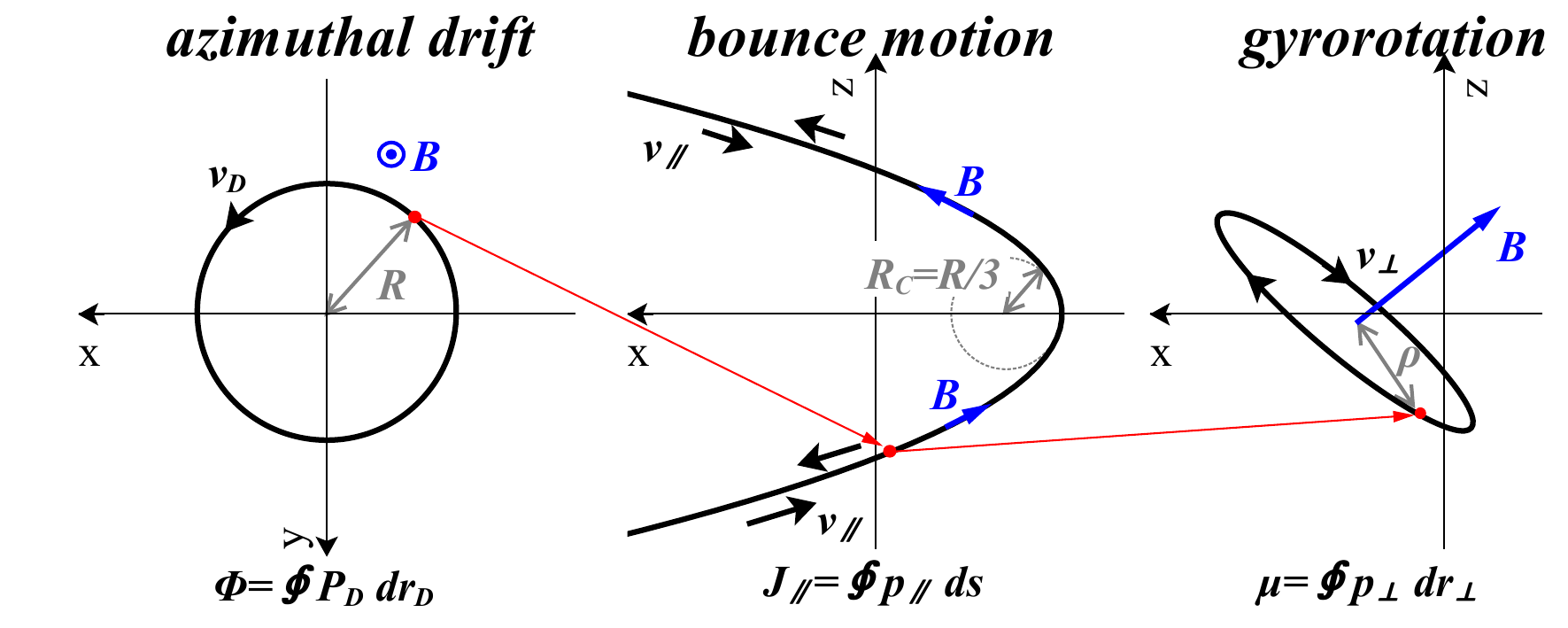}
\caption{Schematic view of electron motion in dipole magnetic field. Three types of periodical motions are shown (from left to right): azimuthal drift around the Earth with the invariant $\Phi=\oint{P_Ddr_D}$ where $P_D=eA_\Phi/c+mv_D\approx eA_\Phi/c$ is the azimuthal momentum and $dr_D=R d\Phi$ is the element of azimuthal trajectory; bounce motion along magnetic field lines between mirror points with the invariant $J_\parallel=\oint{p_\parallel ds}$ where $s$ is the field aligned coordinate; gyrorotation around the magnetic field with the invariant $\mu=\oint{p_\perp dr_\perp}$ where $(p_\parallel, p_\perp)$ are parallel and perpendicular momentum components. The direction of the magnetic field is shown in blue. In dipole field $R=R_EL$ ($R_E$ is the Earth radius and $L$ is the $L$-shell parameter), the magnetic field inhomogeneity scale along magnetic field lines is the curvature radius $R_C=R/3$, the electron gyroradius $\rho\ll R$.}
\label{fig01}
\end{figure}

The outer radiation belt is a near-Earth magnetospheric region filled with energetic electrons reaching relativistic and even ultra-relativistic energies \cite{Horne05Nature,Horne07:NatPhys,Baker14:Nature,Baker16,Allison&Shprits20}. The potentially significant damaging effects of relativistic electron fluxes for the many satellites on orbit continuously drives investigation, modelling, and forecasting of the radiation belt dynamics \cite{Horne13,Baker18:ssr}. Although the magnetic field in the outer radiation belt is dominated by the quite stable dipole field of the Earth, energetic electron fluxes in this region may vary by several orders of magnitude. Wave-particle resonant interaction is the main driver of such flux variations: the radial drift of energetic electrons is provided by drift resonance with ultra-low-frequency (ULF) waves, whereas bounce and cyclotron resonances with extremely and very-low-frequency (ELF/VLF) waves are responsible for electron pitch-angle scattering and energization \cite{bookLyons&Williams,bookSchulz&anzerotti74,Tverskoy69,bookTrakhtengerts&Rycroft08}. Without wave-particle resonant interactions, since electrons are magnetized by the strong dipolar geomagnetic field, three adiabatic invariants are conserved during electron motion, corresponding to three types of electron periodical motions: the magnetic moment, $\mu$, corresponds to the gyrorotation, the bounce invariant, $J_\parallel$, corresponds to bounce oscillations along field lines, and the third invariant, $\Phi$, corresponds to the azimuthal drift motion around the Earth (see schematic \ref{fig01} and \cite{bookSchulz&anzerotti74,Tverskoy69}). The conservation of these invariants can fix the electron energy $\gamma=E/m_{e}c^2$ and equatorial pitch-angle $\alpha_{eq}$, and the $L$-shell, the normalized distance between the Earth center and the farthest (equatorial) point of the magnetic field line along which electrons are bouncing at every point along the electron drift orbit. Therefore, any change of electron phase space density should be attributed to destruction of one or more of these invariants.

\begin{figure}
\centering
\includegraphics[width=1\textwidth]{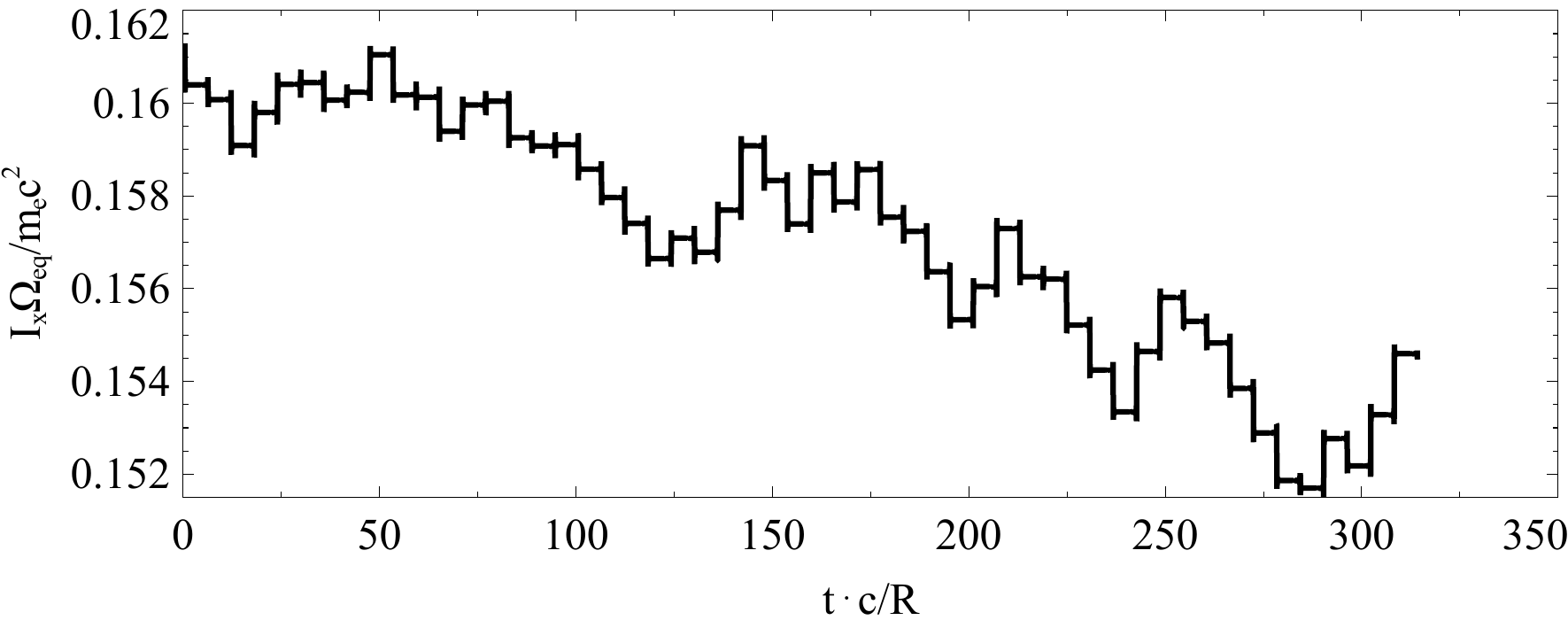}
\caption{Example of electron resonant interactions with a field-aligned whistler-mode wave: effect of electron scattering and magnetic moment violation is shown. Plots show the electron magnetic moment $I_x=m_ec^2(\gamma^2-1)\sin^2\alpha/\Omega_0$ along the particle trajectory during half of an electron bounce period in a dipole magnetic field. System parameters are: whistler-mode wave frequency equal to $0.35$ of equatorial electron gyrofrequency, wave magnetic field amplitude is $10$ pT, the background magnetic field is dipolar \cite{Bell84} with $L-$shell equal to $6$, equatorial plasma density is given by an empirical model \citep{Sheeley01} and it is constant along magnetic field lines. The magnetic moment is $\mu=eI_x/m_ec$, and time is normalized to $c/R$ with $R=LR_E$, with $R_E$ the Earth radius. The total time of this simulation is $\sim 41$s.}
\label{fig02}
\end{figure}

Adiabatic invariants are conserved exponentially well \cite{Kulsrud57,Lenard59,Dykhne60,Slutskin64,Cohen78,Neishtadt00}, and their destruction requires the action of a force varying in space or time faster than the spatial/temporal scale of the periodic motion corresponding to the specific invariant \citep[e.g.,][]{Roberts69}. For example, destruction of $\Phi$ requires some external force with a temporal scale comparable to the electron azimuthal drift period, whereas violation of $\mu$ requires an external force with a temporal scale comparable to the electron gyroperiod. Such external forces can be Lorentz force of wave electromagnetic fields varying with the corresponding temporal scales. Figure \ref{fig02} shows an example of such invariant violation for the electron resonant interaction with circularly polarized electromagnetic waves. Each resonant interaction lasts for a short interval during which the adiabatic invariant experiences a random jump, whereas the longer time intervals between resonant interactions are characterized by conservation of the adiabatic invariant, when the particle is bouncing or drifting far from the wave region.

Therefore, the primary theoretical problem for evaluation of the radiation belt dynamics is to develop an approach describing the evolution of the electron phase space density $f$ due to multiple resonances with realistic electromagnetic waves. Let us start our introduction to this problem with a brief description of a well developed, and most frequently used, theoretical concept of such evolution -- quasi-linear theory \cite{MacDonald&Walt61,Vedenov62,Drummond&Pines62,Trakhtengerts63,Wentworth63,Kennel&Engelmann66,Kennel&Petschek66,Roberts69}. This theory describes the self-consistent dynamics of the charged particle distribution and of the spectrum of electromagnetic waves: waves are generated by unstable particle populations and scatter these populations, moving them in parameter space toward the equilibrium state. The two main equations of the quasi-linear theory are the Fokker-Planck (diffusion) equation
\begin{eqnarray*}
 \frac{{\partial f_0 }}{{\partial t}} &=& \frac{\partial }{{\partial p_ \bot  }}\left( {D_{ \bot  \bot } \frac{{\partial f_0 }}{{\partial p_ \bot  }} + D_{ \bot \parallel } \frac{{\partial f_0 }}{{\partial p_\parallel  }}} \right) + \frac{\partial }{{\partial p_\parallel  }}\left( {D_{\parallel \parallel } \frac{{\partial f_0 }}{{\partial p_\parallel  }} + D_{\parallel  \bot } \frac{{\partial f_0 }}{{\partial p_ \bot  }}} \right) \\
  &+& \frac{{D_{ \bot \parallel } }}{{p_ \bot  }}\frac{{\partial f_0 }}{{\partial p_\parallel  }} + \frac{{D_{ \bot  \bot } }}{{p_ \bot  }}\frac{{\partial f_0 }}{{\partial p_ \bot  }} = \frac{\partial }{{\partial {\bf p}}}\left( {\hat D_{{ pp}} \frac{{\partial f}}{{\partial {\bf p}}}} \right)
 \end{eqnarray*}
and the wave spectrum equation
\[
\frac{{\partial \Uw }}{{\partial t}} + \left( {\nabla  \cdot {\bf v}_g \Uw } \right) = 2\gamma \Uw
\]
where $\hat D_{pp}$ is the $2\times2$ tensor consisting of diffusion coefficients in momentum space $(p_\parallel, p_\perp)$; and $\parallel, \perp$ are relative to the direction of the background magnetic field; $\Uw$ is the wave energy density, that can be expressed through the wave magnetic field energy $B_k^2$ and the Hermitian part of the dielectric tensor $\hat\varepsilon(\omega)$ \cite{Shklyar09:review}, as
\[
\Uw  = \frac{1}{{16\pi \omega }}\left( {{\bf E} \cdot \frac{{d\omega ^2 \hat \varepsilon (\omega )}}{{d\omega }}{\bf E}} \right) = \frac{1}{{16\pi }}\left( {B_k^2 {\bf a} \cdot \frac{{d\omega \hat \varepsilon (\omega )}}{{d\omega }}{\bf a} + {\bf B}^2 } \right) = B_k^2 Y,
\]
where the vector ${\bf a}$ is given by the relationship between wave electric field vector and wave magnetic field magnitude ${\bf E}_w={\bf a}B_k$ and is determined by the wave dispersion relation, as well as coefficient $Y$. Note that the mostly used quasi-linear equations are written for cyclotron resonance between charged particles and electromagnetic waves and, thus, describe the system averaged over electron gyrophase, i.e., these equations reduce the initially 3D momentum space to 2D $(p_\parallel, p_\perp)$ space. Diffusion coefficients $\hat D_{pp}$ can be derived from the linear perturbation theory, the basic assumption of any quasi-linear model \cite{Vedenov62,Drummond&Pines62,Trakhtengerts63,Kennel&Engelmann66}. Although the initial formulation of quasi-linear theory assumes that $\hat D_{pp}$ should be derived for a self-consistent wave spectrum $B_k^2$, the diffusion equation is often solved only for the most energetic (or relativistic) particle population, which usually does not contribute significantly to the variation of $B_k^2$. Therefore, statistical models of the waves, measured by spacecraft, can then be used instead of numerically evaluating the self-consistent evolution of $B_k^2$ \cite[see examples in][]{Summers05, Summers07:rates, Glauert&Horne05, Ni08}.

The quasi-linear equations describe only diffusion $\hat D_{pp}\propto \B_w^2/B_0^2$ and drift $\hat V_p \nabla_{p} \hat D_{pp}\propto \B_w^2/B_0^2$ with $\B_w^2=\langle B_k^2\rangle$ the wave spectrum intensity. These two processes, diffusion and drift, are the results of integration of the wave Lorentz force along unperturbed particle trajectories and, thus, do not include any nonlinear effects. Nonlinearity consists in a significant role played by the wave Lorentz force in charged particle dynamics, and should manifest itself in a nonlinear dependence of diffusion on wave intensity, $\hat D_{pp}\propto (\B_w/B_0)^\kappa$ with $\kappa\ne 2$, strong drifts $\hat V_p \nabla_{p} \hat D_{pp}\propto (\B_w/B_0)^\kappa$ with $\kappa<2$, and nondiffusive/drift terms in the full kinetic equation. The latter terms are the most difficult to include in the basic numerical models of radiation belt dynamics \cite[see discussion in][]{Furuya08,Artemyev14:grl:fast_transport,Omura15}. To explain a possible generalization of the Fokker-Planck equation suitable for including nonlinear wave-particle interactions, let us start with the general form of Smoluchowski coagulation equation \cite{bookVedenyapin11}
\begin{equation}
\frac{{\partial f_0 }}{{\partial t}} = \int\limits_{ - \infty }^\infty  {\left( {\hat K\left( {\left. {\bf p} \right|{\bf p'}} \right)f_0 \left( {{\bf p'}} \right) - \hat K\left( {\left. {{\bf p'}} \right|{\bf p}} \right)f_0 \left( {\bf p} \right)} \right)d{\bf p'}}
\label{eq:smolukhovskiy}
\end{equation}
describing the evolution of $f({\bf p})$. Here, $\hat K({\bf p}|{\bf p’})$ is the coagulation kernel that describes the rate at which particle positions change in 2D $(p_\parallel, p_\perp)$ space from ${\bf p}$ to ${\bf p’}$. Thus, the first term in Eq. (\ref{eq:smolukhovskiy}) describes the particle flux toward ${\bf p}$ and the second term describes the particle fluxes away from ${\bf p}$.

Let us assume that each resonant interaction slightly changes particle momentum, $\Delta p_{\parallel,\perp}/p\ll 1$, so that we can expand $f({\bf p’})$ as
\[
f_0 \left( {{\bf p'}} \right) = f_0 \left( {{\bf p} + \Delta {\bf p}} \right) = f_0 \left( {\bf p} \right) + \frac{{\partial f_0 }}{{\partial {\bf p}}} \cdot \Delta {\bf p} + \frac{1}{2}\frac{{\partial ^2 f_0 }}{{\partial {\bf p}^2 }}\left( {\Delta {\bf p}} \right)^2
\]
Substituting this expression into Eq. (\ref{eq:smolukhovskiy}), we obtain
\begin{eqnarray*}
 \frac{{\partial f_0 }}{{\partial t}} &=& f_0 \left( {\bf p} \right)\int\limits_{ - \infty }^\infty  {\left( {\hat K\left( {\left. {\bf p} \right|{\bf p'}} \right) - \hat K\left( {\left. {{\bf p'}} \right|{\bf p}} \right)} \right)d{\bf p'}}  + \frac{{\partial f_0 }}{{\partial {\bf p}}}\int\limits_{ - \infty }^\infty  {\hat K\left( {\left. {\bf p} \right|{\bf p'}} \right)\Delta {\bf p}d{\bf p'}}  \\
   &+& \frac{1}{2}\frac{{\partial ^2 f_0 }}{{\partial {\bf p}^2 }}\int\limits_{ - \infty }^\infty  {\hat K\left( {\left. {\bf p} \right|{\bf p'}} \right)\left( {\Delta {\bf p}} \right)^2 d{\bf p'}} = {\bf V}\frac{{\partial f_0 }}{{\partial {\bf p}}} + \hat D_{pp} \frac{{\partial ^2 f_0 }}{{\partial {\bf p}^2 }} = \frac{\partial }{{\partial {\bf p}}}\left( {\hat D_{pp} \frac{{\partial f_0 }}{{\partial {\bf p}}}} \right) \\
 \end{eqnarray*}
where the last equality is provided by the divergence free condition ${\bf V}=\partial \hat D_{pp}/\partial {\bf p}$ \cite{bookVedenyapin11,bookSagdeev88,Lichtenberg&Lieberman83:book}. In this case, the Smoluchowski coagulation equation can be reduced to the Fokker-Planck diffusion equation:
\[
\frac{{\partial f_0 }}{{\partial t}} = \frac{\partial }{{\partial {\bf p}}}\left( {\hat D_{pp} \frac{{\partial f_0 }}{{\partial {\bf p}}}} \right)
\]
This is the limit of the quasi-linear theory. If the resonant interaction is nonlinear, but can still be described by $\Delta p_{\parallel,\perp}/p\ll 1$, the diffusion equation is still usable, although diffusion coefficients should then be derived from test particle models including nonlinear wave field effects \citep[see examples in][]{Karpman&Shkliar77,Inan87,Shklyar21,Allanson22,Frantsuzov23:jpp}.

\begin{figure}
\centering
\includegraphics[width=1\textwidth]{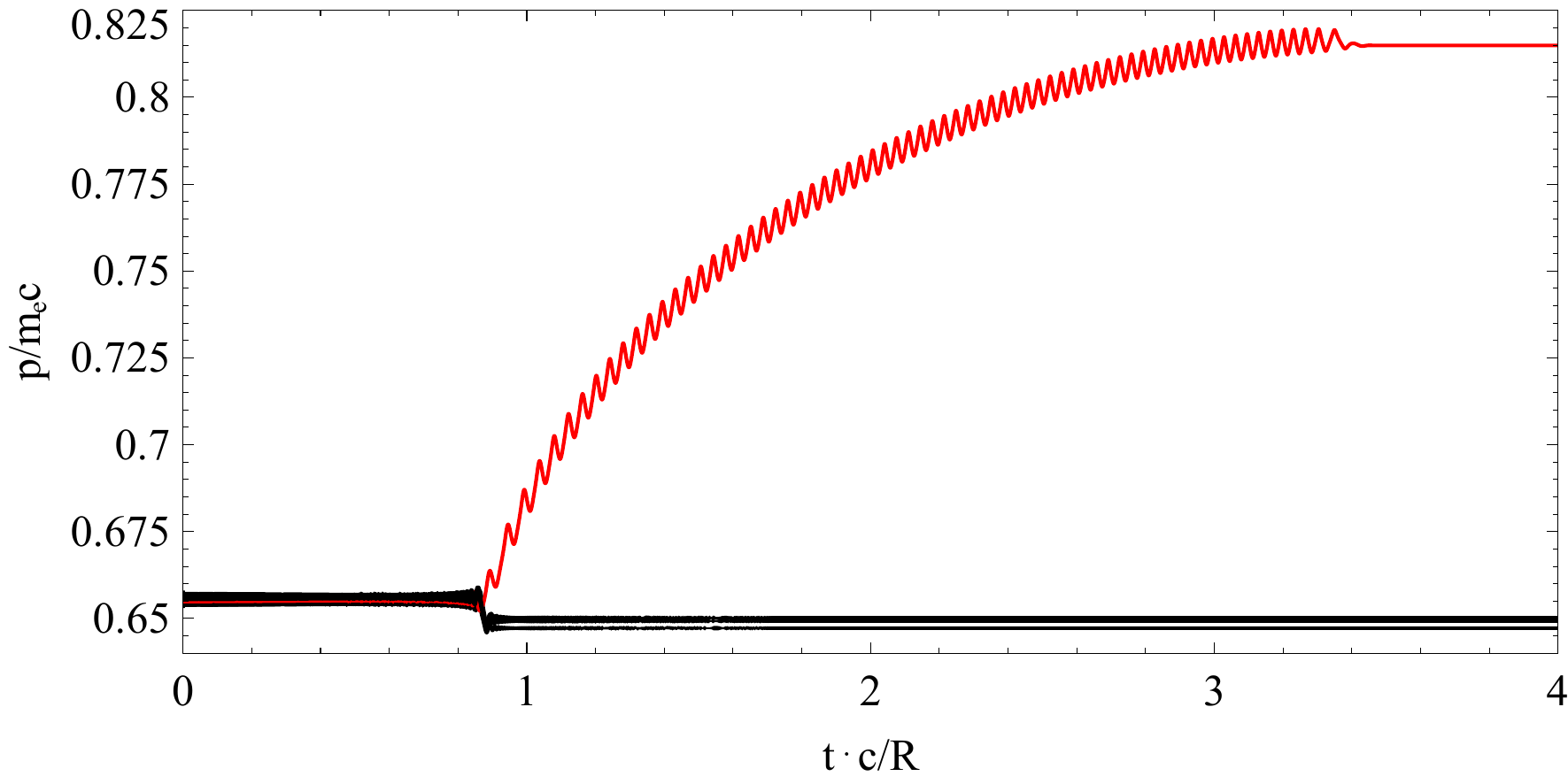}
\caption{Examples of electron resonant interactions with a field-aligned whistler-mode wave: the effect of phase trapping is shown by red color, whereas all black trajectories show the phase bunching effect (small momentum decrease). Plot shows electron momentum $p=m_ec\sqrt{\gamma^2-1}$ for a half of an electron bounce period in a dipole magnetic field. System parameters are the same as in Fig. \ref{fig02}, but wave amplitude is $300$pT. }
\label{fig03}
\end{figure}

The most sophisticated case is when $\hat K$ describe large momentum changes $\Delta p_{\parallel,\perp}/p\sim 1$, and the integral operator $\sim \int \hat K$ cannot be reduced to the differential one. Such situation is common for electron nonlinear resonant interaction with intense whistler-mode waves \citep[e.g.,][]{Demekhov06,Omura07,Bortnik08,Agapitov15:grl:acceleration} and EMIC \citep[e.g.,][]{Albert&Bortnik09,Omura&Zhao12,Grach&Demekhov18:I,Grach&Demekhov20}. Figure \ref{fig03} shows examples of electron trajectories with the large resonant changes of momentum due to the so-called phase trapping effect. A comparison of black (scattered electrons) and red (phase trapped electron) trajectories illustrates the main problem for the description of nonlinear resonant interractions with the differential operators of the Fokker-Planck equation: the momentum change for a single resonant interaction (i.e., during the interval between electron trapping into the resonance and escape from this trapping) is comparable to the initial momentum amplitude. Therefore, the inclusion of such large momentum jumps into the Fokker-Planck equation would either require a significant decrease of the typical time-step of the simulation \citep[such that the time-step of electron distribution evolution is much smaller than the electron bounce period, and phase trapping is modelled as a combination of small consecutive energy changes, see examples in][]{Shklyar81,Foster17}, or the development of a non-differential (integral) operator describing the large energy change on the smallest system time-scale, during an electron bounce period. Therefore, the main challenge for radiation belt models is to construct $\hat K$ or find an approach for taking into account the effects of the integral operator $\sim \int \hat K$ into the diffusive Fokker-Planck equation. This review is devoted to possible solutions to this challenge.

The most direct, and quite effective, approach consists in numerically evaluating the $\hat K$ function. This approach has been applied for test particle simulations of electron interactions with EMIC and electrostatic waves \citep{Zheng19:emic,Artemyev19:cnsns}, but the most elaborate variant of this approach, called the {\it Green function} approach, has been proposed for electron resonant interactions with chorus waves \citep{Furuya08,Omura15}. This is the most developed and advanced approach accounting for multiple properties of resonances: wave frequency drift \citep{Hsieh&Omura17:radio_science}, wave propagation in the form of a train of short wave packets \citep{Kubota&Omura18,Hiraga&Omura20}, wave oblique propagation \citep{Hsieh&Omura17,Hsieh20}, multiple resonances due to wave obliqueness \citep{Hsieh22,Hsieh&Omura23}. The main advance of this approach is that the numerical evaluation of $\hat K$ can be performed for an arbitrary and very realistic wave field model. The main disadvantage is that such numerical evaluation requires a discretization of the momentum space with a sufficient statistics (sufficiently large number) of resonant interactions inside each bin, whereas the lowest bin size is determined by weak diffusive scattering $\Delta p\propto \B_w/B_0$ and the range of $\Delta p$ is determined by the phase trapping property with $\Delta p \propto O(\B_w/B_0)\gg \B_w/B_0$. Thus, a purely numerical evaluation of $\hat K$ may sometimes miss some weak diffusive effects, and this approach should be mostly effective for modeling brief events with not-widely-varying system characteristics (to avoid a recalculation of $\hat K$ for multiple realizations of system parameters).

In this review, we examine the theoretical properties of the $\hat K$ function, and explore different approaches for its analytical evaluation (Sections~\ref{sec:basic}, \ref{sec:nl}, and \ref{sec:1d}). We provide a detailed investigation of $\hat K$ for nonlinear electron interactions with monochromatic intense whistler-mode waves, and provide asymptotic solutions for the kinetic equation including such $\hat K$ (Sections~\ref{sec:1d}). Then, we generalize $\hat K$ for systems with a large wave ensemble, and perform such a generalization via the mapping technique for nonlinear resonant interactions (Section~\ref{sec:mapping}). In Appendix E, we provide several examples of application of this technique for simulations of the observed dynamics of the electron flux. The next natural generalization of the theoretical approach for nonlinear wave-particle interactions consists of the inclusion of the effects of short wave-packets. We discuss the main aspects of this generalization (Section~\ref{sec:packets}) and of the theoretical approaches for inclusion of short wave-packets into the mapping technique (Section~\ref{sec:mapping_short}). Next, we consider an approach allowing the incorporation of nonlinear resonant interactions into existing global numerical models of the radiation belts (Section~\ref{sec:nl&ql}). Finally we discuss several aspects of nonlinear resonant interactions that are not included in this review, but can be important for specific plasma systems (Section~\ref{sec:discussion}). The review also contains Appendix A, with the main equations of the Hamiltonian approach for wave-particle resonant interactions, Appendix B considering a special situation of nonlinear interactions for field-aligned particles, Appendix C describing analytical estimates for electron resonant interaction with short wave-packets, Appendix D describing the problem of the electron phase gain between two resonances, and Appendix E with several examples of observations of nonlinear resonant effects.

\section{Basic properties of electron resonant interactions} \label{sec:basic}
\begin{figure}
\centering
\includegraphics[width=1\textwidth]{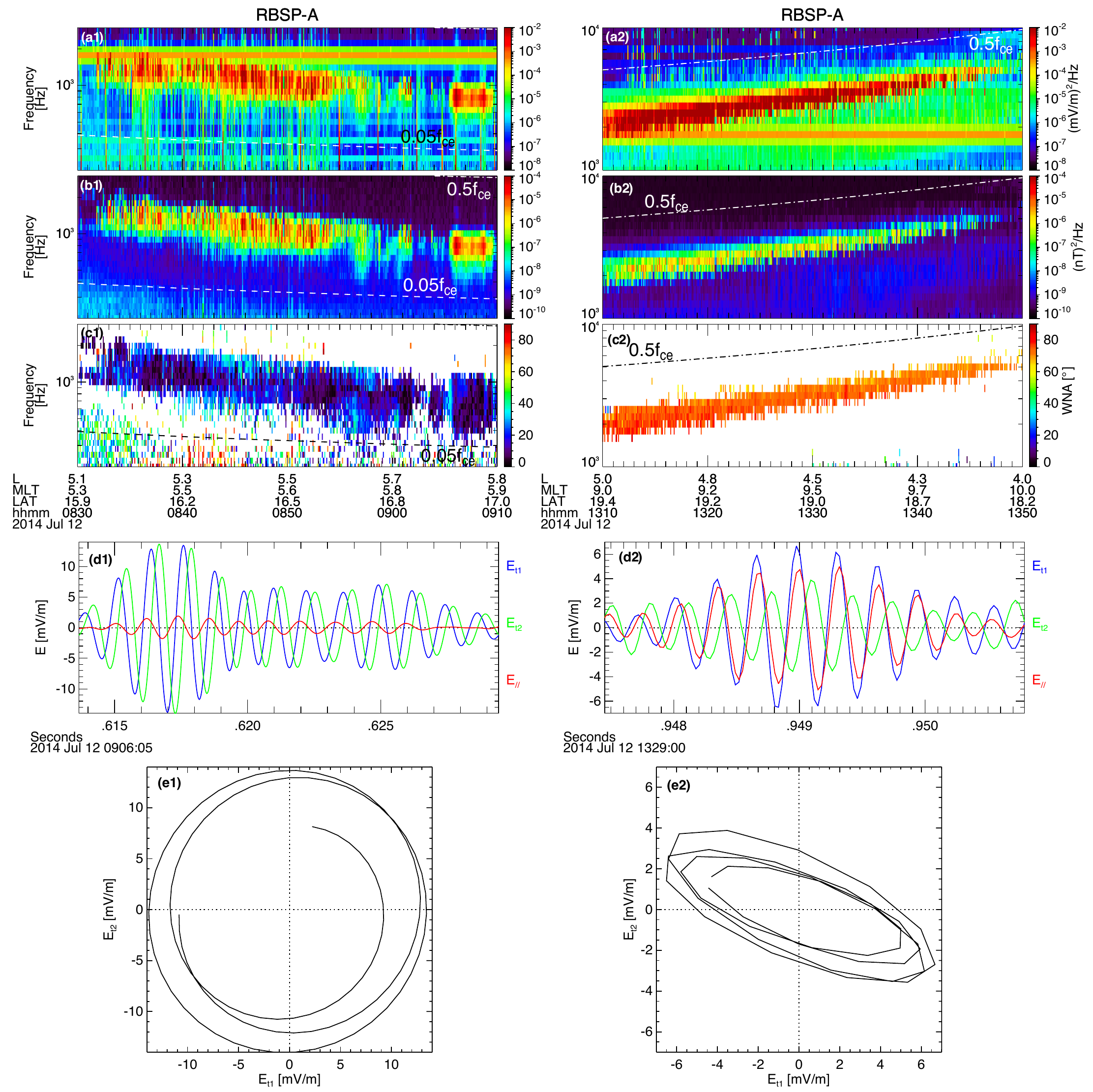}
\caption{Two examples of whistler-mode waves (from top to bottom): (a) wave electric field spectrum, (b) wave magnetic field spectrum, (c) wave normal angle, (d) examples of individual wave-packets (electric field components are shown), (e) examples of electric field polarization (two transverse components are plotted). Left panels show the example of field-aligned intense waves, and right panels show the example of very oblique waves. In panels (a-c) $f_{ce}=\Omega_0/2\pi$ is the electron gyrofrequency. Data are collected by Van Allen Radiation Belt Probes (RBSP) \cite{Mauk13} in the outer radiation belt (see coordinates in $L$-shell, magnetic local time, and latitudes below panels (c)). We use wave field measurements by the Electric and Magnetic Field Instrument Suite and Integrated Science (EMFISIS) on RBSP \cite{Kletzing13}. Details of wave data processing can be found in \cite{Agapitov16:grl,Agapitov15:grl:acceleration}. }
\label{fig04}
\end{figure}

We start with general information about whistler-mode waves observed in the inner magnetosphere, and specifically within the outer radiation belt, outside the plasmasphere. These right-hand circularly polarized electromagnetic waves are mainly generated in the frequency range from $0.1$ to $0.7$ times the equatorial electron gyrofrequency $\Omega_0$ under the form of repetitive rising tones, and have been called chorus waves \cite{Storey53,bookHelliwell65, Burton&Holzer74, Tsurutani&Smith74}. There are two main modes of these waves: electromagnetic mode with nearly parallel propagation relative to the background magnetic field and quasi-electrostatic mode with strongly oblique propagation. Figure \ref{fig04} shows examples of both wave modes. More detailed information about statistics of these two wave modes and their relative occurrence rates can be found in \cite{Meredith12,Li11,Li16:statistics,Agapitov13:jgr,Agapitov18:jgr,Artemyev16:ssr}.

Field-aligned (i.e., propagating parallel to the background magnetic field) whistler-mode waves are generated at the equator by transversely anisotropic electron populations \cite{Sagdeev&Shafranov61,Kennel66}, which are either injected from the plasma sheet \cite{Tao11,Fu14:radiation_belts} or generated by dayside magnetosphere compression \cite{Li15:solarwind}. After an initial linear wave growth \cite{Kennel66}, nonlinear wave growth takes over once the generated wave reaches a threshold amplitude for electron trapping in the inhomogeneous magnetic field, leading to the formation of characteristic rising or falling tone elements \citep[see reviews in][]{Helliwell&Crystal73,Nunn74,Demekhov17, Omura08, Omura13:AGU,Omura21:review,Tao20,Tao21}. The electron azimuthal drift from the injection region to the day side and such day side compression determine the domain of presence of near-equatorial field-aligned whistler-mode waves \cite{Meredith12,Li11,Li13,Agapitov13:jgr}. Propagating away from their equatorial source region, these waves become oblique \cite{Alekhin&Shklyar80,Bell02,Shklyar04,Bortnik06} and experience Landau damping by suprathermal electrons \cite{Bortnik07:landau,Chen13,Watt13:ray_tracing}. Such damping is stronger on the night side due to larger magnetic field line curvature, leading to a confinement of these waves near the equator ($<15^\circ$), whereas on the day side field-aligned and weakly oblique waves may propagate up to middle latitudes of $\sim 30^\circ$ \cite{Agapitov13:jgr,Agapitov18:jgr}. A potentially important sub-population of field-aligned waves consists of ducted waves, which are trapped within plasma density perturbations and can propagate without damping to high latitudes \cite{Laird&Nunn75,Karpman&Kaufman82}. Such waves have been observed in-situ \cite{Streltsov&Bengtson20,Chen21:ducting,Chen21:ducting&plasmapause}, reproduced in numerical simulations \cite{Hanzelka&Santolik19,Ke21:ducts,Streltsov&Goyal21}, and detected by ground-based stations \cite{Collier11,Titova15,Titova17,MartinezCalderon15,MartinezCalderon20,Demekhov17:ducts}. However, the population of ducted whistler-mode waves has not yet been precisely quantified and their occurrence rate in each region is not known \cite[see discussion in][]{Artemyev21:jgr:ducts,Artemyev24:jgr:ELFIN&injection,Zhang23:grl:ELFIN&TEC}.

Very oblique waves observed at high latitudes likely result from the diffraction of initially field-aligned waves during their propagation along the inhomogeneous magnetic field \cite{Agapitov13:jgr,Breuillard12:angeo,Breuillard14,Chen13}. However, additionally to this high-latitude population, there are also near-equatorial very oblique waves \cite{Cully08, Cattell08,Agapitov13:jgr}, which are likely generated by transversely anisotropic electrons in the presence of field-aligned electron streams that reduce the Landau damping \cite{Mourenas15,Artemyev16:ssr, Gao16:obliquewaves,Li16}. Generation of very oblique waves, propagating around the resonance cone angle \cite{bookSazhin93}, require specific distributions of suprathermal ($100$ eV to a few keV) electrons with a plateau in parallel velocity space \cite{Mourenas15, Artemyev16:ssr, Chen19:Gao&beams,Kong21:Gao:beams,Ke22:landau}. Such field-aligned electron streams can be formed either by electrostatic parallel fields often observed around plasma sheet injections \cite[see discussion in][]{Artemyev&Mourenas20:jgr} or by ionosphere outflow \cite[see discussion in][]{Artemyev20:jgr:feedback}. Although both scenarios assume specific conditions for very oblique wave generation, this wave population is quite widespread in observations \cite{Agapitov13:jgr,Li16:statistics} and important for energetic electron flux dynamics \cite{Agapitov15:grl:acceleration, Artemyev13:grl, Artemyev15:natcom, Mourenas14, Hsieh20}. Nevertheless, intense very oblique waves are rarely observed simultaneously with intense field-aligned waves, probably due to Landau damping and nonlinear effects \cite{Agapitov16:grl}.

\begin{figure}
\centering
\includegraphics[width=1\textwidth]{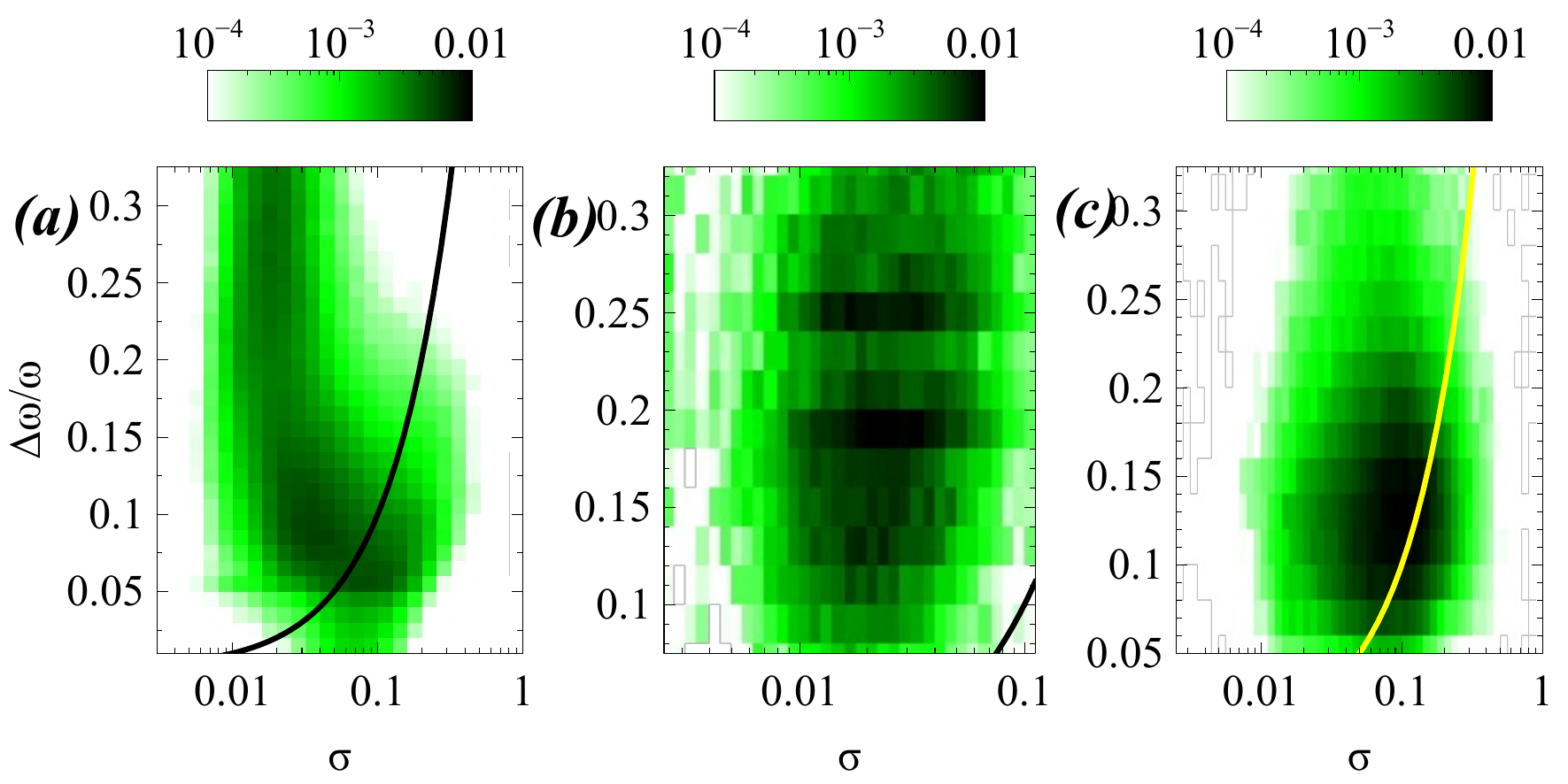}
\caption{Distribution of statistics of field-aligned intense whistler-mode waves in ($\Delta\omega/\omega$, $\sigma$) parametric space where $\Delta\omega$ is the spectrum width, $\sigma  = \sqrt {\B_w /B_0 } \beta _e^{1/4} \left( {\Omega _0 /\omega  - 1} \right)^{-1/4}$ and $\beta_e$ is the electron plasma beta parameter (ratio of electron thermal pressure and background magnetic field pressure). Curves show the $\sigma = \Delta\omega/\omega$ threshold. Panel (a) shows statistics of field-aligned whistler-mode wave packets in the Earth's outer radiation belt \citep[see][]{Zhang19:grl}; panel (b) shows statistics of whistler-mode waves observed in the plasma sheet, around injection regions \cite{Zhang18:whistlers&injections}; panel (c) shows statistics of very oblique whistler-mode waves observed in the outer radiation belt (we use THEMIS \cite{Angelopoulos08:ssr} electric field and magnetic field \cite{Bonnell08,LeContel08} wave measurements and the same criteria for wave identification as \cite{Li16:statistics}).  }
\label{fig05}
\end{figure}

Electron resonant interactions with these two wave modes are quite different \cite{Bell84,Bell86,Solovev&Shkliar86,Shklyar09:review,Artemyev15:natcom,Albert17}. Therefore, nonlinear effects will be considered separately for each wave mode. Both wave modes share an important property – they are coherent and quite narrow band waves, with a high intensity. The importance of this property will become clear if we consider the applicability criteria for quasi-linear theory. This theory is based on the concept of resonance overlap for a wide spectrum of waves \cite{Shapiro&Sagdeev97}. In momentum space, the resonance width is $\Delta p_R\approx \sqrt{eU_wm_e}$ where $U_w  = A_w \sqrt {2I_x \Omega _0 /m_e c^2 } $ for the cyclotron resonance with electromagnetic waves \cite{Karimabadi90:waves} and $U_w  = \varphi$ for Landau resonance with electrostatic waves \cite{Palmadesso72}. The width of the Landau resonance for an electrostatic wave does not depend on electron characteristics and is entirely determined by the wave electrostatic potential, $\varphi$, whereas the width of the cyclotron resonance depends not only on the wave vector potential, $A_w=B_w/k$, but also on electron magnetic moment $I_x$ (i.e., on pitch-angle and energy). Note, however, that Landau resonance with electromagnetic waves is also characterized by a resonant width depending on wave characteristics \cite{Shklyar09:review}. The distance between resonances with two nearby waves in the spectrum would be $\Delta p_\omega   = m_e \left( {\Delta \omega /k} \right) \cdot \left( {1 - p_R /m_e v_g } \right)$, where $\Delta \omega$ is the distance between waves (e.g., spectral width of wave-packet), and $v_g=\partial \omega/\partial k$ is the wave group velocity \cite{Karpman74:ssr}. The overlap condition requires that there are many $\Delta p_R$ within $\Delta p_\omega$, i.e., that the wave spectrum be wide enough ($\Delta \omega$ is large) or that the wave amplitude be weak enough ($U_w$ is small). This condition is generally satisfied for low amplitude whistler-mode waves observed in the near-Earth plasma sheet \citep{Gao22:Luphi_whistlers,Waheed23}, but it is often not satisfied for narrow band intense whistler-mode chorus waves in the plasma injection regions and Earth's outer radiation belt, see Fig. \ref{fig05}. Besides quite small wave spectrum width, $\Delta \omega/\omega$, whistler-mode wave packets have peak amplitudes of about $\in[10^{-2},10^{-3}]B_0$ \citep[e.g.,][]{Cattell11:Wilson,Agapitov14:jgr:acceleration,Zhang19:grl,Tyler19}, that is, a factor $\times 100$ higher than mean wave amplitudes derived from the averaged wave spectra (e.g., compare wave packet statistics in \cite{Zhang18:jgr:intensewaves} or \cite{Zhang19:grl} with time-averaged wave statistics in \cite{Agapitov18:jgr}). Such high intensity wave packets may nonlinearly interact with electrons through cyclotron or Landau resonances.

Therefore, wave-particle resonant interactions should be considered under the assumption that electrons interact resonantly with individual intense waves. The corresponding Hamiltonian for wave particle interactions in dipole field can be written as (see Appendix A and  \citep{Albert93,Albert13:AGU,Vainchtein18:jgr}):
\begin{equation}
H = m_e c^2 \gamma  - eU_w \left( {I_x ,s} \right)\cos \left( {\phi  - \Nr\psi } \right),\quad \gamma  = \sqrt {1 + \left( {\frac{{p_\parallel }}{{m_e c}}} \right)^2  + \frac{{2I_x \Omega _0 \left( s \right)}}{{m_e c^2 }}}
\label{eq:hamiltonian}
\end{equation}
where $\Omega_0(s)=eB_0(s)/m_ec$ is the electron gyrofrequency, $(s,p_\parallel)$ are conjugated field-aligned coordinate and momentum, $(\psi, I_x)$ are conjugated gyrophase and magnetic moment ($I_x=(\gamma^2-1)m_ec^2\sin^2\alpha/\Omega_0$ and $\alpha$ is a local pitch-angle), $\Nr=0,\pm1,\pm2,...$ is the resonance number. The resonance condition $\dot\phi=\Nr\dot\psi$, the wave phase definition $\dot\phi=k_\parallel(s)\dot s-\omega$, and Hamiltonian equations $\dot\psi=\partial H/\partial I_x=\Omega_0/\gamma$, $\dot s = \partial H/\partial p_\parallel=p_\parallel/m_e\gamma$ determine the resonant momentum $p_\parallel=p_R$:
\[
p_R  = m_e \frac{{\gamma \omega  + \Nr \Omega _0 }}{{k_\parallel }}
\]

The function $U_w$ is the generalized wave amplitude, including effects of whistler-mode dispersion (see Appendix A and  \citep{Albert93,Artemyev18:jpp}):
\[
U_w  = A_w \sum\limits_{\Nr =  - \infty }^{ + \infty } {h^{(\Nr)} }
\]
and
\[
 h^{(\Nr)} = \frac{{\rho \Omega _0 }}{{2c\gamma }}\left( {\left( {C_1 - \cos \theta } \right)J_{\Nr - 1}  - \left( {C_1 + \cos \theta } \right)J_{\Nr + 1}} \right)  -\left( {\frac{{p_\parallel }}{{\gamma m_e c}} + C_2 } \right)\sin \theta J_{\Nr}
\]
where $J_{\Nr}$ are Bessel functions with argument $k_ \bot  \rho$, $\rho  = \sqrt {2I_x /m_e \Omega _0 }$, $\theta$ is the wave normal angle (wave number has two components: field-aligned $k_\parallel = k\cos\theta$ and transverse $k_\perp=k\sin\theta$). This equation for $u_w^{(\Nr)}$ has been derived using the relations between wave magnetic and electric field components in a cold plasma \cite{Williams&Lyons74,Tao&Bortnik10} and, thus, the coefficients $C_{1,2}$, given in Appendix A, depend only on wave characteristics. An important property of $C_1$ is that for $\theta=0$ (field-aligned propagation) it is equal to one, whereas $U_w^{(\Nr)}=0$ for all $\Nr$ except $\Nr=-1$, for which we have $h^{(-1)}=\rho \Omega _0/c\gamma=\sqrt{2I_x\Omega_0/m_ec^2}$.

To describe the wave dispersion, $\omega=\omega(k,\theta)$, we again use the cold plasma approximation \cite{bookStix62}:
\[
N^2  = \frac{1}{2}\frac{{(RL - PS)\sin ^2 \theta  + 2PS}}{{S\sin ^2 \theta  + P\cos ^2 \theta }} + \frac{1}{2}\frac{{\sqrt {(RL - PS)^2 \sin ^4 \theta  + 4P^2 D^2 \cos ^2 \theta } }}{{S\sin ^2 \theta  + P\cos ^2 \theta }}
\]
where $N=kc/\omega$ is the refractive index, and Stix coefficients are
\begin{eqnarray*}
 R &=& 1 - \sum\limits_j {\frac{{\Omega _{pj}^2 }}{\omega }\frac{1}{{\omega  + \Omega _{0j} }}} ,\quad L = 1 - \sum\limits_j {\frac{{\Omega _{pj}^2 }}{\omega }\frac{1}{{\omega  - \Omega _{0j} }}}  \\
 S &=& \frac{1}{2}(R + L)  ,\quad D = \frac{1}{2}(R - L) ,\quad P =1-\sum\limits_j\Omega_{pj}^2/\omega^2
\end{eqnarray*}
Here $\Omega_{0j}=e_jB/m_jc$ is the gyrofrequency for particles with a charge $j=e$ for electrons and $j=i$ or $j=p$ for ions or protons, $\Omega_{pj}=\sqrt{4\pi n_0 e_j^2/m_j}$, $n_0$ is the background density ($n_0=n_e=n_i$). Note $e_j=-e$ for electrons and $e_j=e$ for protons.

A simplified form of this dispersion relation for a dense plasma is
\[
\omega ^2  = \frac{{\Omega _{0}^2 \cos ^2 \theta }}{{\left( {1 + (\Omega _{pe} /kc)^2 } \right)^2 }} + \frac{{\nu \Omega _{0}^2 }}{{1 + (\Omega _{pe} /kc)^2 }}
\]
where $\Omega_{pe}(s)\gg \Omega_{0}(s)$ is the electron plasma frequency, $\Omega_{0}=\Omega_{0e}$, and $\nu = m_e/m_i$, $m_i$ is the mass of the ion mixture (i.e., this is the proton mass for purely proton-electron plasma). For quasi-parallel wave propagation, $\theta \sim 0$, the last term in the simplified dispersion relation can be omitted:
\[
\omega  = \frac{{\Omega _{0} \cos \theta }}{{1 + (\Omega _{pe} /kc)^2 }}
\]

Let us consider the Hamiltonian (\ref{eq:hamiltonian}) under the assumption of constant wave frequency, $\omega=const$. Although this assumption is not applicable for the most intense whistler-mode chorus waves \citep{Omura08,Omura21:review,Tao21}, it is quite useful to describe the basic properties of wave-particle interactions. In this case, the time dependence is included only as a linear term $\omega t$ in the wave phase $\phi$ and, therefore, we may remove this dependence (to obtain a conservative Hamiltonian) by changing the variables: from $(\psi, I_x)$ to $(\zeta, I)$ where $\zeta=\Nr\psi+\phi$. The corresponding generating function is
\[
W = \left( {\phi  - \Nr\psi } \right)I + sp +\psi \tilde I_x
\]
and the new Hamiltonian is
\begin{eqnarray}
    \H_I &=&  - \omega I + m_e c^2 \gamma  - eU_w \left( {\tilde I_x  - \Nr I,s} \right)\cos \zeta \nonumber \\
    \label{eq:hamiltonian_const}\\
    \gamma  &=& \sqrt {1 + \left( {\frac{{p  + k_\parallel I}}{{m_e c}}} \right)^2  + \frac{{2\left( {\tilde I_x  -\Nr I} \right)\Omega _0 \left( s \right)}}{{m_e c^2 }}} \nonumber
\end{eqnarray}
where $p=p_\parallel-k_\parallel I$ is the new momentum conjugated to the new coordinate $\tilde s=s$ (keeping the $s$ notation), $I_x=\tilde I_x-\Nr I$, and $I$ is conjugated to the new phase $\zeta$ \citep[see, e.g.,][]{Artemyev18:jpp}.

The Hamiltonian (\ref{eq:hamiltonian_const}) does not depend on time, and thus $\h_I$=const, with the integral of electron motion
\begin{equation}
h=m_ec^2\gamma-\omega I \label{eq:h_const}
\end{equation}
Note that $\tilde I_x$ is a constant, because $\H_I$ does not depend on $\psi$. For the case of Landau resonance, $\Nr=0$, $\tilde I_x=I_x$, i.e., $I_x$ is conserved. Using this conservation law in Eq. (\ref{eq:h_const}) gives $h=\Omega_{eq} I_x=const$. For the first cyclotron resonance, $\Nr=-1$, $\tilde I_x=I_x-I$ and can be set equal to zero, i.e. $I_x=I$. In this case, Eq. (\ref{eq:h_const}) can be written as:
\[
h=m_ec^2\gamma-\omega I_x
\]

\begin{figure}
\centering
\includegraphics[width=1\textwidth]{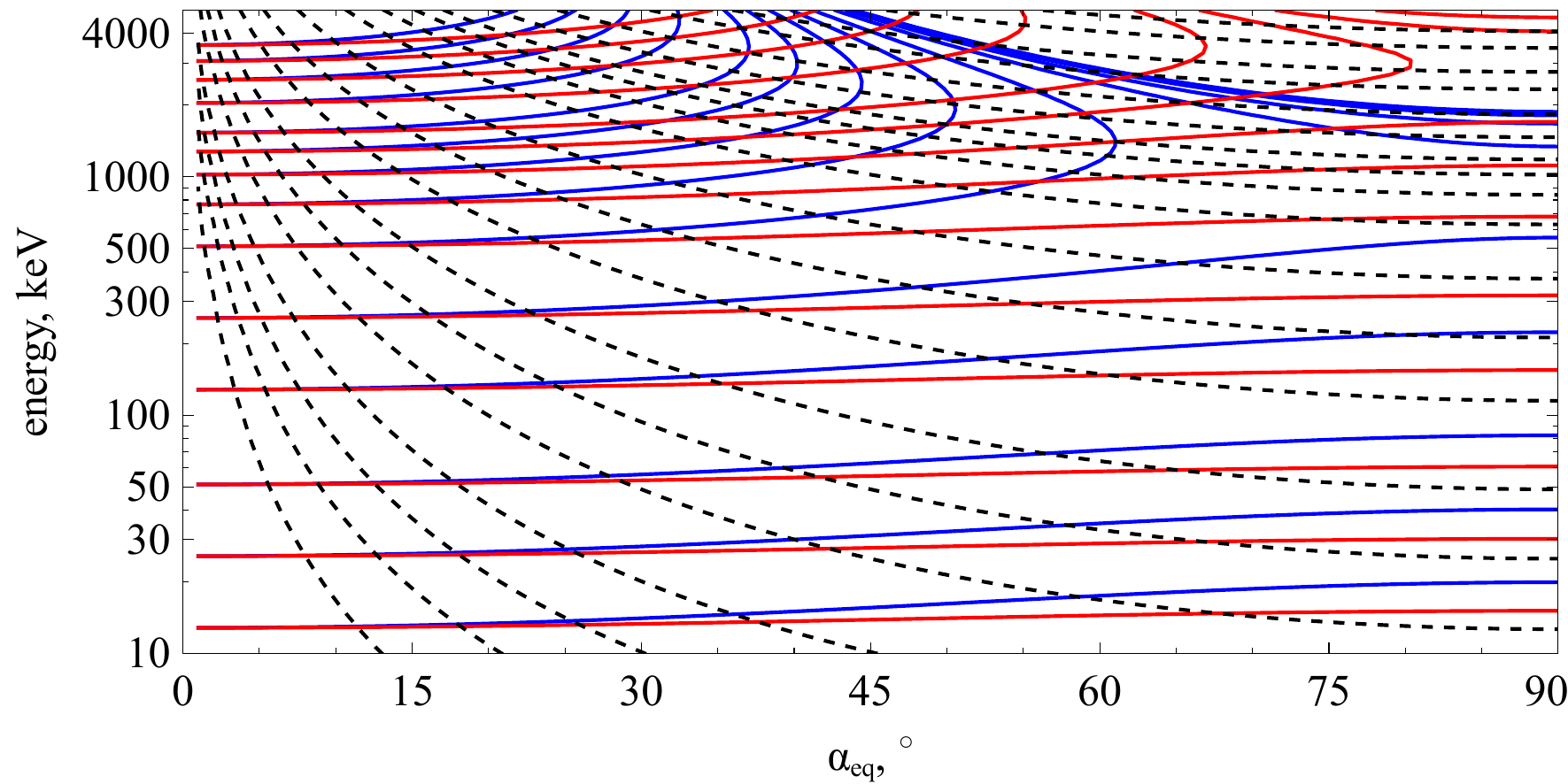}
\caption{Resonance curves given by Eq. (\ref{eq:h_const}) for Landau ($\Nr=0$; dashed black line) and first cyclotron ($\Nr=-1$; solid red and blue lines) resonances. Two wave frequencies are shown: $\omega/\Omega_{eq}=0.15$ (red) and $\omega/\Omega_{eq}=0.35$ (blue). }
\label{fig06}
\end{figure}

The integral given by Eq.(\ref{eq:h_const}) describes trajectories, in the momentum space or in energy, pitch-angle space $(E, \alpha_{eq})$, along which wave-particle interactions are transporting electrons. Note that without such interactions, the energy $E=m_ec^2(\gamma-1)$ and equatorial pitch-angle $\alpha_{eq}={\rm asin}\left(2I_x\Omega_{eq}/(\gamma^2-1)\right)$ would be constants of motion. Figure \ref{fig06} shows these trajectories (also called {\it resonance curves}; e.g., \cite{bookWalker93}) for the Landau ($\Nr=0$) and first cyclotron ($\Nr=-1$) resonances of electrons with whistler-mode waves. The main feature of the cyclotron resonance is that the resonance curves are almost parallel to the pitch-angle axis for small pitch-angles, and show energy increase with pitch-angle increase for larger pitch-angles. Thus, electrons transported to smaller pitch-angles (and finally into the loss-cone, thereby precipitating into the atmosphere) lose their energy, but around the loss-cone (at small pitch-angles) this transport occurs almost without energy change. The main feature of the Landau resonance is the large energy increase of electrons transported along the resonant curves toward smaller pitch-angles, i.e., electron precipitation due to the Landau resonance is accompanied by electron acceleration. In contrast to the cyclotron resonance, electron transport to higher pitch-angles via the Landau resonance is associated with electron energy loss. For the Landau resonance, the shape of resonance curves is dictated by magnetic moment conservation and does not depend on the wave frequency, whereas for the cyclotron resonance a smaller wave frequency means less energy change for the same pitch-angle change. At high energies for cyclotron resonance, the resonance curves show a change of direction: there, energy increase corresponds to pitch-angle decrease. This is the so-called {\it turning acceleration} effect \citep{Omura07}, which occurs when $\gamma\omega/\Omega_0$ is larger than one and $d\gamma/d\alpha$ changes sign for large $\alpha$:
\[
\frac{{d\gamma }}{{d\alpha }} = \frac{{\left( {\omega /\Omega _0 } \right)\left( {\gamma ^2  - 1} \right)\sin \alpha \cos \alpha }}{{1 - \left( {\omega \gamma /\Omega _0 } \right)\sin ^2 \alpha }}
\]
This equation is obtained by differentiation of Eq. (\ref{eq:h_const}) over $\alpha$. Note that $\gamma\omega/\Omega_0=1$ also separates cyclotron resonances with negative resonant momentum (waves and resonant particles are moving in opposite directions) and with positive resonant momentum (waves and resonant particles are moving in the same direction), $p_R=m_e\gamma\omega\left(1-\Omega_0/\gamma\omega\right)/k_\parallel$.

The resonance curves in Fig. \ref{fig06} show that for the same wave characteristics, electrons with different energies and pitch-angles (i.e., with different $h$ integrals of motion) can interact resonantly with a same monochromatic ($\omega=const$) wave. This is due to background magnetic field inhomogeneity: electrons bounce along magnetic field lines and have a different local pitch-angle $\alpha(s)={\rm asin}\left(\sin(\alpha_{eq})\sqrt{\Omega_0(s)/\Omega_{eq}}\right)$ at different magnetic latitudes (coordinate $s$). Accordingly, electrons with a fixed energy will be able to find a specific latitude where their pitch-angles will satisfy the resonance condition with a whistler-mode wave of fixed frequency. There are two important consequences of such magnetic field inhomogeneity: (1) a monochromatic wave may interact resonantly with electrons in a wide energy, pitch-angle range (in contrast to a homogeneous plasma, where only a wide wave spectrum may provide resonances over a wide energy, pitch-angle range), (2) electron bounce motion moves particles into the resonances and moves them out of the resonances, which are therefore of limited size in latitude, and of limited duration. This second effect actually replaces the limitation of the duration of the resonant interaction of electrons with whistler-mode waves that is due to the existence of a broad wave spectrum in classical quasi-linear models \cite[see discussion in][]{Karpman74:ssr,Shklyar81,Shklyar11:angeo,Albert01,Albert10,Allanson22}.

\subsection{Resonant interaction with monochromatic wave}\label{sec:resonance}
Let us now consider the effects of a high wave amplitude on resonant particle dynamics for an arbitrary wave propagation angle. We start with the Hamiltonian (\ref{eq:hamiltonian_const}) and follow the standard procedure for analysis of resonant Hamiltonian systems \citep{Neishtadt05,Neishtadt14:rms,Albert13:AGU,Artemyev18:jpp}. First, let us write the equation for the resonant condition, $\dot\zeta=0$
\[
\dot \zeta  = \frac{{\partial H_I }}{{\partial I}} \approx  - \omega  + m_e\frac{{\Nr\Omega _{0}  + k_\parallel\left( {p + k_\parallel I} \right)}}{m_e\gamma }=0
\]
where we omit the term
\[
\sim\frac{{\partial eU_w }}{{\partial I}}\cos \zeta
\]
that is proportional to the wave amplitude and should not be included into the definition of the resonance (see discussion of exceptions in \cite{Li22:anomalous_resonance} and discussion of the importance of this term in Appendix B).

Equation $\dot\zeta=0$ has a solution $I=I_R$, where
\[
\frac{{k_\parallel I_R }}{{m_e c}} =  - \frac{p}{{m_e c}} - \frac{{\Nr\Omega _{0} }}{{k_\parallel c}} + \frac{1}{{\sqrt {N_\parallel^2  - 1} }}\sqrt {1 + \frac{{2\tilde I_x \Omega _{0} }}{{m_e c^2 }} - \left( {\frac{{\Nr\Omega _{0} }}{{k_\parallel c}}} \right)^2  - 2\frac{{\Nr\Omega _{0} }}{{k_\parallel c}}\frac{p}{{m_e c}}}
\]
where $N_\parallel=k_\parallel c/\omega$.

Expanding Hamiltonian (\ref{eq:hamiltonian_const}) around $I=I_R$, we obtain
\begin{eqnarray*}
 \H_I  &\approx& \Lambda  + \frac{1}{{2M}}\left( {I - I_R } \right)^2  + eU_w \left( {I_R ,s} \right)\cos \zeta \nonumber  \\
 \Lambda  &=&  - \omega I_R  + m_e c^2 \gamma _R ,\quad \gamma _R  = \frac{{N_\parallel }}{{\sqrt {N_\parallel^2  - 1} }}\sqrt {1 + \frac{{2\tilde I_x \Omega _{0} }}{{m_e c^2 }} - \left( {\frac{{\Nr\Omega _{0} }}{{k_\parallel c}}} \right)^2  - 2\frac{{\Nr\Omega _{0} }}{{k_\parallel c}}\frac{p}{{m_e c}}}  \nonumber
 \end{eqnarray*}
In Appendix A, we describe the variable change and the corresponding generating function: $(\zeta, I) \to (\zeta, P_\zeta)$ where $P_\zeta=I-I_R$. After this change and an expansion relative to the resonance, the Hamiltonian $\H_I$ is separated into two parts. The first part
\begin{equation}
\Lambda=-\omega I_R(s,p)+m_ec^2\gamma_R(s,p) \label{eq:hamiltonian_slow}
\end{equation}
describes the dynamics of slow variables, $(s,p)$. The second part
\begin{equation}
\h_\zeta   = \frac{1}{{2M}}P_\zeta ^2  + {\rm A}\zeta  + {\rm B}\cos \zeta
 \label{eq:hamiltonian_zeta}
\end{equation}
describes the dynamics of fast variables $(\zeta, P_\zeta)$, with coefficients depending on slow variables $(p,s)$. Coefficients $M$, ${\rm A}$, and ${\rm B}$ are derived in Appendix A for the general case of arbitrary wave propagation direction. For the first cyclotron resonance with field-aligned waves, these coefficients are
\begin{eqnarray*}
 {\rm A} &=& \left\{ {\Lambda ,I_R } \right\}_{s,p}  \approx  - \frac{{m_e c^2 \D}}{{\gamma _R }}\frac{{N_\parallel^2 }}{{N_\parallel^2  - 1}}\left( {\left( {\frac{{p_{\parallel,R} }}{{m_e c}}} \right)^2 \frac{{\partial \ln N_\parallel }}{{\partial \ln \Omega _0 }} + \frac{{\Omega _0 }}{{\omega N_\parallel }}\frac{{p_{\parallel,R} }}{{m_e c}} - \frac{{\Omega _0 I_{x,R} }}{{m_e c^2 }}} \right) \\
 {\rm B} &=& eU_w \left( {I_R ,s_R } \right),\quad M = \left( {\frac{{\partial ^2 \h_I }}{{\partial I^2 }}} \right)_{I_R }^{ - 1}  = \frac{{m_e c^2 }}{{\omega ^2 }}\frac{{\gamma _R }}{{N_\parallel^2  - 1}}
 \end{eqnarray*}
where $N_\parallel=k_\parallel c/\omega$, $p_{\parallel,R}=(\gamma_R-\Nr\Omega_0/\omega)/N_\parallel$, $\D=c(\partial \ln \Omega_0/\partial s)/N_\parallel\omega\ll 1$ is a dimensionless factor of system inhomogeneity, and $s_R$ and $I_{x,R}$ are coordinate and momentum in the resonance.

\begin{figure}
\centering
\includegraphics[width=1\textwidth]{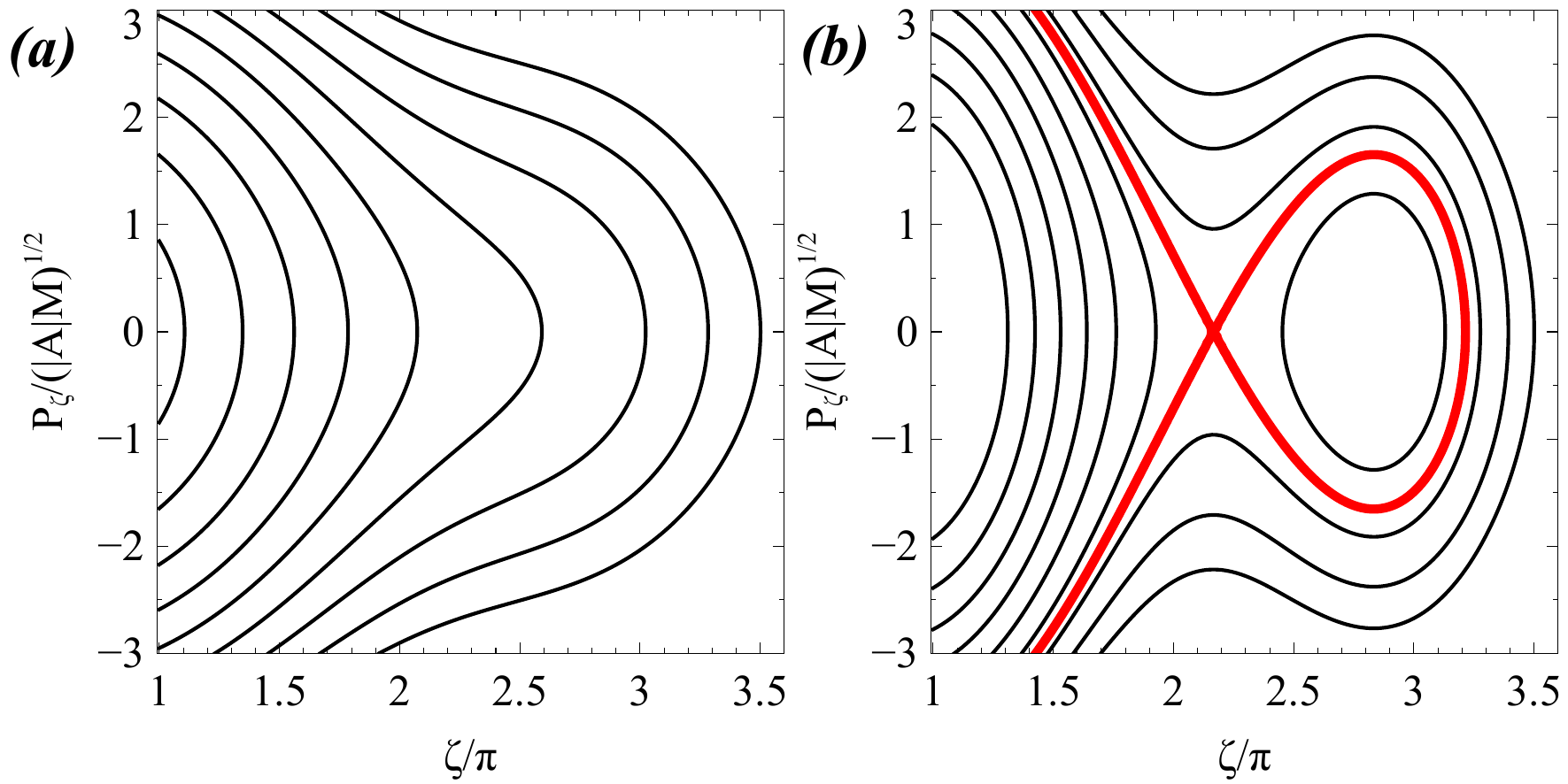}
\caption{Phase portraits for Hamiltonian (\ref{eq:hamiltonian_zeta}) for $a=|{\rm B}/{\rm A}|<1/2$ (a) and $a=|{\rm B}/{\rm A}|=2$ (b).   }
\label{fig07}
\end{figure}

These coefficients determine the character of wave-particle resonant interactions. Figure \ref{fig07} shows the phase portrait for two regimes: $a=|{\rm B}/{\rm A}|<1$ and $a=|{\rm B}/{\rm A}|>1$, where
\[
\frac{\rm B}{\rm A} = -\frac{{eU_w \left( {I_R ,s_R } \right)\gamma _R }}{{m_e c^2 \D}}\frac{{N_\parallel^2  - 1}}{{N_\parallel^2 }}\left( {\left( {\frac{{p_{\parallel,R} }}{{m_e c}}} \right)^2 \frac{{\partial \ln N_\parallel }}{{\partial \ln \Omega _0 }} + \frac{{\Omega _0 }}{{\omega N_\parallel }}\frac{{p_{\parallel,R} }}{{m_e c}} - \frac{{\Omega _0 I_{x,R} }}{{m_e c^2 }}} \right)^{ - 1}
\]
For $a<1$ the phase portrait is filled by trajectories crossing the resonance, $\dot\zeta=P_\zeta/M=0$, only once. These are so-called {\it transient} trajectories, and as particles moving along these trajectories spend a quite limited time around the resonance, the wave field cannot significantly change their orbits. Such regime of wave-particle resonant interactions is principally similar to linear scattering, which is evaluated for unperturbed particle trajectories \cite{Karpman&Shkliar77,Albert01}.

For $a>1$ the phase portrait is divided into two domains: the internal domain is filled by closed trajectories with multiple resonance $\dot\zeta=0$ crossings, whereas the external domain is filled by open trajectories with a single resonance crossing, but particles on these trajectories need to move around the interval domain and thus spend much more time around the resonance in comparison with particles moving along {\it transient} trajectories. Particles inside the internal domain are called {\it phase trapped} and may spend a very long time in the resonance (oscillating around the resonance). Particles in the external domain will be scattered, but this scattering cannot be evaluated under the approximation of unperturbed trajectories. This nonlinear scattering is very different from the linear one, appropriate for {\it transient} trajectories. The entire problematic of nonlinear wave-particle interaction consists in developing an accurate description of such nonlinear scattering and phase trapping effects for a large ensemble of electrons.

The factor $1/a$ is essentially the same parameter as the $S$-parameter used in previous studies of wave-particle resonant interactions by \citet{Helliwell67,Omura08,Omura09} and as the $\tau/\alpha$-parameter used by \citet{Karpman74,Karpman75PS,Shklyar09:review,Shklyar11:angeo}. We can rewrite the expression for $a$ as
\begin{eqnarray*}
 \frac{1}{a} &=& \frac{{m_e c^2 }}{{eB_w \gamma _R }}\sqrt {\frac{{m_e c^2 }}{{2\Omega _0 I_x }}} \frac{k}{\omega }\frac{{N_\parallel }}{{N_\parallel^2  - 1}}\left( {\frac{{\Omega _0 I_x }}{{m_e c^2 }} - \left( {\frac{{p_{\parallel,R} }}{{m_e c}}} \right)^2 \frac{{\partial \ln N_a }}{{\partial \ln \Omega _0 }} - \frac{{\Omega _0 }}{{\omega N_a }}\frac{{p_{\parallel,R} }}{{m_e c}}} \right)c\frac{{\partial \ln \Omega _0 }}{{\partial s}} \\
  &=& \frac{c}{{s'_0 \Omega _w \omega }}\frac{{\partial \Omega _0 }}{{\partial s}}\frac{1}{{2\xi '\delta '}}\left( {\frac{{\gamma _R \omega }}{{\Omega _0 }}\left( {\frac{{V'_{ \bot 0} }}{c}} \right)^2  - 2\gamma _R \frac{\omega }{{\Omega _0 }}\left( {\frac{{V'_R }}{c}} \right)^2 \frac{{\partial \ln N_\parallel }}{{\partial \ln \Omega _0 }} - 2\frac{{V'_R V'_p }}{{c^2 }}} \right) \\
  &=& \frac{c}{{s'_0 \Omega _w \omega }}\frac{{\partial \Omega _0 }}{{\partial s}}\frac{1}{{2\xi '\delta '}}\left( {\frac{{\gamma _R \omega }}{{\Omega _0 }}\left( {\frac{{V'_{ \bot 0} }}{c}} \right)^2  - \left( {2 + \frac{\omega }{{\Omega _0 }}\frac{{\Omega _0  - \gamma _R \omega }}{{\Omega _0  - \omega }}} \right)\frac{{V'_R V'_p }}{{c^2 }}} \right)
 \end{eqnarray*}
where we used notation from \citep{Omura09}:
\begin{eqnarray*}
 s'_0  &=& \frac{{N_\parallel^2  - 1}}{N_\parallel}\frac{{V_{ \bot 0} }}{c},\quad \xi ' = \sqrt {N_\parallel^2  - 1} ,\quad \delta ' = \sqrt {1 - N_\parallel^{ - 2} }  \\
 V'_R  &=& p_{\parallel,R} /m_e \gamma _R ,\quad V'_{ \bot 0}  = c\sqrt {2I_x \Omega _0 /m_e c^2 } /\gamma _R ,\quad V'_p  = \omega /k = c/N_\parallel
 \end{eqnarray*}
and a simplified dispersion relation with constant plasma density (see Appendex A):
\[
\frac{{\partial \ln N_\parallel }}{{\partial \ln \Omega _0 }} =  - \frac{1}{2}\frac{\omega }{{\Omega _0  - \omega }}
\]
In this form, $1/a$ fully coincides with $S$ given for $\partial \omega/\partial t=0$ by Eqs. (10)-(13) from \citet{Omura09}.

\begin{figure}
\centering
\includegraphics[width=1\textwidth]{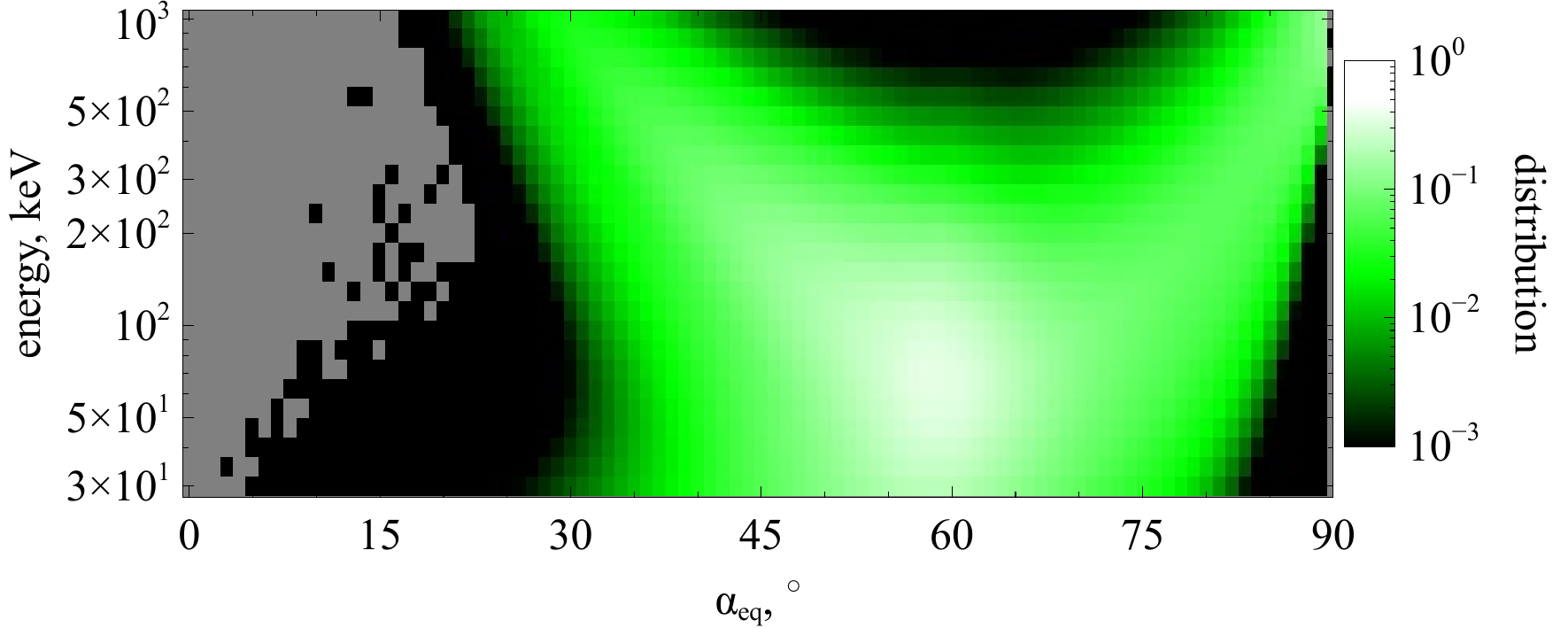}
\caption{Probability distribution of $a>1$ (i.e., of nonlinear wave-particle interaction) for electrons and intense field-aligned whistler-mode waves in the outer radiation belt, for $L$-shell $\in[5,7]$.  Dataset is from \citep{Zhang19:grl}. }
\label{fig08}
\end{figure}

This parameter determines whether the wave field $\sim {\rm B}$ is sufficiently strong to overcome the inhomogeneity of the background magnetic field $\sim {\rm A}$. For $a>1$, the wave-particle resonant interaction is nonlinear, and this nonlinearity both changes the electron diffusion and introduces new effects of phase bunching and trapping. The parameter $a$ depends on background plasma and wave characteristics, but also on electron energy and pitch-angle. Figure \ref{fig08} shows the probability distribution of $a>1$ for field-aligned whistler-mode waves in the outer radiation belt (this probability is displayed in $(E,\alpha_{eq})$ space). Medium to high pitch-angle electrons with not-too-high energies interact resonantly with waves close to the equator (where $\partial \Omega_0/\partial s$ is small), and this increases the probability of nonlinear interaction. Small pitch-angles and/or very high energy electrons interact resonantly with waves at high latitudes (farther from the equator), where the large $\partial \Omega_0/\partial s$ reduces the probability of nonlinear interaction. Note that the equation for $a$ has been derived from Hamiltonian (\ref{eq:hamiltonian_zeta}), whereas this Hamiltonian would not work for very small pitch-angle particles having $I_x\Omega_0\sim eU_w$ (see details in Section \ref{sec:nl_small_alpha} and Appendix B). In this limit of small pitch-angles, another criterion for nonlinear interactions can be derived \cite{Gan24}, which shows the possibility for such interactions.

\subsection{Diffusion by monochromatic waves}
To demonstrate how the inhomogeneity can affect wave-particle interactions, let us derive the electron diffusion rates in the simple case of a field-aligned whistler mode wave. Equations of motion for Hamiltonian (\ref{eq:hamiltonian}) have the form:
\[
\h = mc^2 \gamma  - \sqrt {\frac{{2I_x \Omega _0 }}{{m_ec^2 }}} \frac{{eA_w }}{\gamma }\cos \left( {\phi  + \psi } \right),\quad \gamma  = \sqrt {1 + \left( {\frac{{p_\parallel }}{{m_ec}}} \right)^2  + \frac{{2I_x \Omega _0 (z)}}{{m_ec^2 }}}
\]
and we can use $U_w=\sqrt{2I_x\Omega_{0}/m_ec^2}A_w/\gamma$.

Let us consider the approximation of a small wave amplitude $|{\rm A}|\gg |{\rm B}|$ at the resonance. Then, we can consider unperturbed ($U_w=0$) particle trajectories to determine the phase $\zeta$ variation around the resonance $t=t_R$, $\zeta_R=\zeta(t_R)$. Expanding $\zeta$ around the resonance, we obtain
\[
\zeta  \approx \zeta _R  + \frac{1}{2}\left( {t - t_R } \right)^2 \left. {\ddot \zeta } \right|_{t = t_R }
\]
because $\dot\zeta|_{t=t_{R}}=0$.
For Hamiltonian $\h_{\zeta}$ from Eqs. (\ref{eq:hamiltonian_zeta}) $\ddot \zeta|_{t=t_R}$ can be expressed as
\[
\ddot \zeta  = \frac{{\dot P_\zeta  }}{M} = \frac{{\left\{ {\Lambda ,I_R } \right\}}}{M} + \frac{{U_w }}{M}\cos \zeta \quad
\]
Thus, for $|{\rm A}|\gg|{\rm B}|$ (i.e., $|{\Lambda,I_R}|\gg|eU_w|$) we have
\[
\left. {\ddot \zeta } \right|_{t = t_R } = \frac{\rm A}{M}=\frac{{\left\{ {\Lambda ,I_R } \right\}}}{M}
\]
Therefore, the $\Delta I$ change in the resonance is given by the time integral of $-\partial H_I/\partial \zeta$ and can be written as
\begin{eqnarray}
 \Delta I &=& eU_w \int\limits_{ - \infty }^{ + \infty } {\sin \zeta dt}  = eU_w\int\limits_{ - \infty }^{ + \infty } {\sin \left( {\zeta _R  + \frac{1}{2}\left. {\ddot \zeta } \right|_{t = t_R } \left( {t - t_R } \right)^2 } \right)dt}  \nonumber\\
  &=& 2eU_w\left( {\left. {\frac{1}{2}\ddot \zeta } \right|_{t = t_R } } \right)^{ - 1/2} \left( {\sin \zeta _R \int\limits_0^{ + \infty } {\cos \left( {q^2 } \right)dq}  + \cos \zeta _R \int\limits_0^{ + \infty } {\sin \left( {q^2 } \right)dq} } \right) \nonumber\\
  &=& \sqrt{2\pi}eU_w\left( {\left. {\frac{1}{2}\ddot \zeta } \right|_{t = t_R } } \right)^{ - 1/2}\sin \left( {\zeta _R  + \frac{\pi }{4}} \right) \label{eq:DI_linear}
 \end{eqnarray}
where the $\pm\infty$ limits of integration correspond to the large $\Delta t\omega \gg 1$ approximation for the time $\Delta t$ of motion around the resonance, and we use the {\it Fresnel integral} equation
\[
\int\limits_0^{ + \infty } {\cos \left( {q^2 } \right)dq}  = \int\limits_0^{ + \infty } {\sin \left( {q^2 } \right)dq}  = \sqrt {\frac{\pi }{8}}
\]
The averaged momentum change is zero, $\langle \Delta I \rangle_{\zeta_R}=0$, and we have
\begin{equation}
\left\langle {\left( {\Delta I} \right)^2 } \right\rangle _{\zeta _R }  = \frac{{\pi \left( {eU_w } \right)^2 }}{{\left. {\ddot \zeta } \right|_{t = t_R } }} = \pi \left( {\frac{{m_e c^2 }}{\omega }} \right)^2 \frac{{B_w^2 }}{{B_0^2 }}\frac{{2I_x \Omega _0 }}{{m_e c^2 }}\frac{{\Omega _0^2 }}{{\omega ^2 N_\parallel^4 }}\frac{c}{{\delta v_R }}
\label{eq:DII}
\end{equation}
where $\delta v_R$ is the effective resonance width:
\[
\delta v_R  = c\frac{{\partial \ln \Omega _0 }}{{k_\parallel \partial s}}\left( {\left( {\frac{{p_{\parallel,R} }}{{m_e c}}} \right)^2 \left( {\frac{{\partial \ln N_\parallel }}{{\partial \ln \Omega _0 }} + \frac{1}{2}} \right) + \frac{{\Omega _0 }}{{\omega N_\parallel }}\frac{{p_{\parallel,R} }}{{m_e c}} + \frac{1}{2}\left( {1 - \gamma ^2 } \right)} \right).
\]
In Eq. (\ref{eq:DII}), all variables depending on $s$ should be evaluated at the resonance location.

Taking into account Eq. (\ref{eq:h_const}), the equation for $\left\langle {\left( {\Delta \gamma } \right)^2 } \right\rangle$ can be rewritten as
\[
\left\langle {\left( {\Delta \gamma } \right)^2 } \right\rangle _{\zeta _R }  = \left\langle {\left( {\frac{{\omega \Delta I}}{{mc^2 }}} \right)^2 } \right\rangle _{\zeta _R }  = \pi \frac{{B_w^2 }}{{B_0^2 }}\frac{{2I_x \Omega _0 }}{{m_e c^2 }}\frac{{\Omega _0^2 }}{{\omega ^2 N^4 }}\frac{c}{{\delta v_R }}
\]
The conservation of magnetic moment $I_x=(\gamma^2-1)\sin^2\alpha_{eq}/2\Omega_{0}(0)$ provides an additional relation between energy $mc^2\gamma$ and pitch-angle $\alpha_{eq}$ changes:
\[
\left. {\frac{{\partial \alpha _{eq} }}{{\partial \gamma }}} \right|_{I_x  = const}  = \frac{{\gamma \tan \alpha _{eq} }}{{\gamma ^2  - 1}}
\]
This relation allows the recalculation of $\left\langle {\left( {\Delta \gamma } \right)^2 } \right\rangle _{\zeta _R }$ to $\left\langle {\left( {\Delta \alpha_{eq} } \right)^2 } \right\rangle _{\zeta _R }$.
Therefore, we may characterize wave-particle interactions by diffusion coefficients
\begin{equation}
D_{\gamma \gamma }  = \frac{{\left\langle {\left( {\Delta \gamma } \right)^2 } \right\rangle _{\zeta _R } }}{{2\tau }},\quad D_{\alpha _{eq} \alpha _{eq} }  = \frac{{\left\langle {\left( {\Delta \alpha _{eq} } \right)^2 } \right\rangle _{\zeta _R } }}{{2\tau }} = \left( {\frac{{\gamma \tan \alpha _{eq} }}{{\gamma ^2  - 1}}} \right)^2D_{\gamma \gamma }
\label{eq:Dgg}
\end{equation}
where $\tau=\tau(\gamma,\alpha_{eq})$ is the time between two resonant interactions for a single resonance within half of a bounce period, $\tau=\tau_{bounce}/2$, and Eqs. (\ref{eq:Dgg}) provide the bounce averaged energy and pitch-angle diffusion rates \cite[see also][]{Albert10}.

Note that Eq. (\ref{eq:DI_linear}) provides the $\Delta I$ change for unperturbed particle trajectories (i.e., when particle coordinate and velocity do not depend on $eU_w$) and, thus, $\Delta I \sim \int{eU_w\sin\zeta dt}$ with $\zeta=\zeta_0(t)$ does not depend on $eU_w$. The mean value of such a change is $\langle \Delta I \rangle=0$, whereas its variance is $\langle (\Delta I)^2 \rangle\propto (eU_w)^2$. The next order of particle trajectory perturbations should include terms linearly proportional to $eU_w$, and thus $\zeta=\zeta_0(t)+eU_w\cdot C\sin\left(\zeta_0(t)\right)$ where $C$ is a function of slow coordinates. At this next order, $\Delta I \sim \int{eU_w\sin\zeta dt}$ will contain two terms $\int{eU_w\sin\zeta_0 dt}$ and $\sim \int{eU_w^2\sin^2\zeta dt}$, with a mean value $\langle \Delta I \rangle\propto  (eU_w)^2$ provided by the second term. Thus, to estimate $\langle \Delta I \rangle$ of the same order as $\langle (\Delta I)^2 \rangle\propto (eU_w)^2$, one needs to consider a second order perturbation of particle trajectories. The absence of a mean value $\langle \Delta I \rangle$ for the unperturbed trajectories does not mean that $\langle \Delta I \rangle=0$, but only implies that one has not $\langle \Delta I \rangle \propto eU_w$, but instead $\langle \Delta I \rangle \propto (eU_w)^2$.

\section{Nonlinear resonant characteristics} \label{sec:nl}
\begin{figure}
\centering
\includegraphics[width=1\textwidth]{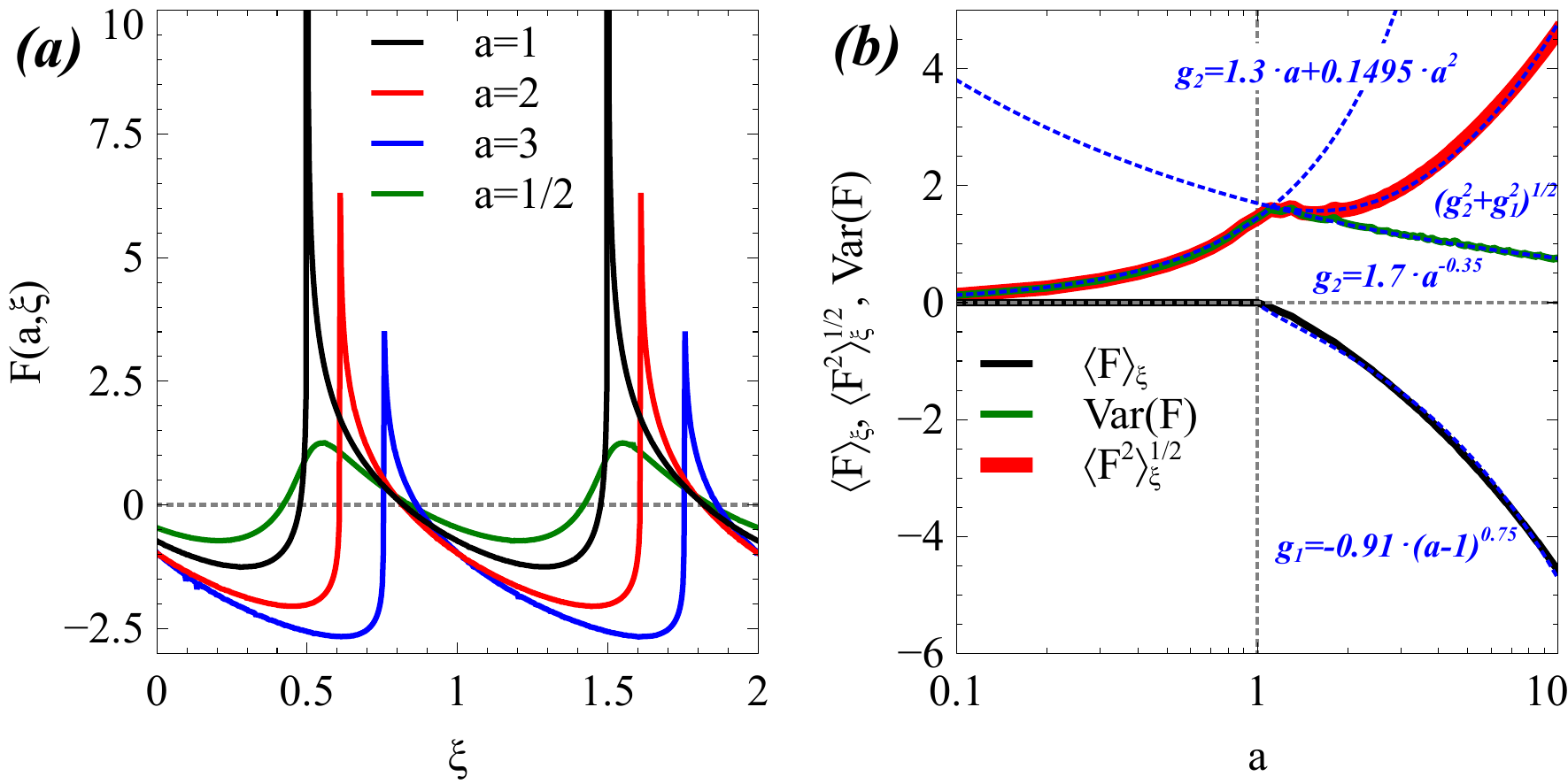}
\caption{Panel (a) shows function $F(\xi,a)$ given by Eq. (\ref{eq:F}). Panel (b) shows functions $\langle F(\xi,a)\rangle_{\xi}$, $\langle F^2(\xi,a)\rangle_{\xi}^{1/2}$, and ${\rm Var}(F)=\sqrt{\langle F(\xi,a)^2\rangle_{\xi}-\langle F(\xi,a)\rangle_{\xi}^2}$. Blue dashed lines present fittings of the displayed functions. }
\label{fig09}
\end{figure}

Let us describe the main characteristics of nonlinear resonant interactions. We start with the resonant energy change experienced by the particles crossing the resonance $\dot\zeta=0$, the so-called transient particles. Using Eq. (\ref{eq:h_const}), we can write $m_ec^2\Delta\gamma = \omega\Delta I$ where $\Delta I$ is given by integration of the Hamiltonian equation for Hamiltonian (\ref{eq:hamiltonian_const}):
\[
 \Delta I =2eU_w \int\limits_{ - \infty }^{t_R } {\sin \zeta dt}
\]
where $t_R$ is the time of resonance crossing ($\dot\zeta=0$). Using Hamiltonian (\ref{eq:hamiltonian_zeta}) for $dt =d\zeta/\dot\zeta=Md\zeta/P_\zeta$, we obtain
\begin{equation}
 \Delta I =2eU_w M\int\limits_{ - \infty }^{\zeta _R } {\frac{{\sin \zeta }}{{P_\zeta  }}d\zeta }  = \int\limits_{ - \infty }^{\zeta _R } {\frac{{eU_w \sqrt {2M} \sin \zeta d\zeta }}{{\sqrt {2\pi\xi{\rm A}   - {\rm A}\zeta  - {\rm B} \cos \zeta } }}}  \nonumber\\
 \label{eq:DeltaI}
\end{equation}
where we introduce $\xi=\h_\zeta/2\pi{\rm A}$, the energy of particles in the phase portrait from Fig. \ref{fig07}. Equation (\ref{eq:DeltaI}) can be written as $\Delta I=\Delta I_0 F(\xi, a)$, where
\[
\Delta I_0 =\sqrt {2M\left| {\rm A} \right|}
\]
and
\begin{equation}
F= \int\limits_{ - \infty }^{\zeta _R } {\frac{{a\sin \zeta d\zeta }}{{\sqrt {2\pi\xi   - \zeta  - a\cos \zeta } }}}   \label{eq:F}
\end{equation}
The function $F$ is shown in Fig. \ref{fig09}(a). It is a periodic function of $\xi$ with the period $1$:
\begin{align*}
    F(a,\xi + 1) &= \int\limits_{-\infty}^{\zeta_R} d\zeta \: \frac{\sqrt{a} \sin\zeta}{\sqrt{2\pi\xi + 2 \pi - \zeta - a\: \cos\zeta}} = \int\limits_{-\infty}^{\zeta_R - 2 \pi} d\tilde{\zeta} \: \frac{\sqrt{a} \sin\tilde{\zeta}}{\sqrt{2\pi\xi - \tilde{\zeta} - a\: \cos\tilde{\zeta}}} = \nonumber \\
    &= \int\limits_{-\infty}^{\tilde{\zeta}_R} d\tilde{\zeta} \: \frac{\sqrt{a} \sin\tilde{\zeta}}{\sqrt{2\pi\xi - \tilde{\zeta} - a\: \cos\tilde{\zeta}}} = F(a,\xi)
\end{align*}
For $a<1$, this periodic function is such that its average $\langle F(a,\xi)\rangle_{\xi\in[0,1)}$ is zero, whereas for $a>1$ the profile of $F$ becomes asymmetric relative to zero and $\langle f(a,\xi)\rangle_{\xi\in[0,1)}$ is finite (see Appendix A and \cite{Neishtadt75}). As function $F$ takes both positive and negative values, resonant electrons can increase and decrease their moment $I$, with $\Delta I \sim F$. The $\xi$-averaged $F$ is always non positive, and this determines the conventional description of electron resonant scattering (for $a>1$ and the first cyclotron resonance this scattering is called {\it phase bunching}, see \cite{Matsumoto&Omura81,Omura&Matsumoto82,Winglee85}) as a process with electron momentum (and energy) decrease \citep[e.g.,][]{Albert01}. However, for individual $\xi$ values (i.e., for specific values of electron resonant energy $\h_\zeta$) $\Delta I$ is positive, and this effect is called {\it positive phase bunching} and has been considered in \citep{Albert22:phase_bunching,Vargas23:pop}. Although such {\it positive phase bunching} may be important for transient electron scattering by very intense whistler-mode wave packets \citep[see discussion in][]{Lundin&Shkliar77,Inan78}, the overall electron ensemble dynamics can be described by $\xi$-averaged system characteristics that do not include {\it positive phase bunching} \citep{Vargas23:pop}.

The energy $\xi$ depends on initial particle gyrophase and can be considered as a random variable with uniform distribution within $[0,1]$ (see Fig. 8 in \citep{Itin00} and Fig. 5 in \citep{Frantsuzov23:jpp}). Therefore, to estimate the actual energy variation of transient particles we shall average the function $F$ over $\xi$ \citep[see also][]{Neishtadt14:rms,Artemyev18:cnsns,Frantsuzov23:jpp}:
\begin{equation}
\langle F\rangle_\xi=  - \frac{1}{\pi }\int\limits_{\zeta _ -  }^{\zeta _ +  } {\sqrt {\zeta _ +   - \zeta  + a\left( {\cos \zeta _ +   - \cos \zeta } \right)} d\zeta } \label{eq:Fa}
\end{equation}
This equation has been derived in \cite{Neishtadt75}, and we repeat the derivations in Appendix A.

Equation (\ref{eq:Fa}) shows that $\langle \Delta I\rangle_\xi =\Delta I_0 \langle F\rangle_\xi$ can be written as
\begin{equation}
\langle \Delta I\rangle_\xi = \sqrt{2M|{\rm A}|}\langle F\rangle_\xi
 \label{eq:DeltaI_average}
\end{equation}
and is equal to $-\area/2\pi$, where $\area$ is the area surrounded by the separatrix in Fig. \ref{fig07}(b):
\begin{equation}
\area  = {\sqrt {8M\left| {\rm A}\right|} }\int\limits_{\zeta _ -  }^{\zeta _ +  } {\sqrt {\zeta _ +   - \zeta  + a\left( {\cos \zeta _ +   - \cos \zeta } \right)} d\zeta } \label{eq:S},
\end{equation}
Here the two $\zeta_\pm$ values are  the coordinate of the saddle point ($\sin\zeta_-=1/a$) and a solution of $\zeta+a\cos\zeta=\zeta_-+a\cos\zeta_-$ equation, different from $\zeta_-$. From the definition of $\area$ it is clear that $\area=0$ for $a\leq 1$. The equality $\langle \Delta I\rangle_\xi=-\area/2\pi$ is an important property of  the Hamiltonian system (\ref{eq:hamiltonian_zeta}) that determines a balance between phase trapping and phase bunching processes \citep[see][]{Solovev&Shkliar86,Itin00,Artemyev16:pop:letter}. A useful asymptotic expression of $\area$ is
\[
\area_{a \gg 1}  \approx \sqrt {8M\left| B \right|} \int\limits_0^{2\pi } {\sqrt {1 - \cos \zeta } d} \zeta  = 2^{5/2} \sqrt {8M\left| B \right|}  = 16\sqrt {M\left| B \right|}
\]
where $\zeta_-\to 0$ and $\zeta_+\to 2\pi$. This asymptote shows that $\area$ scales with wave amplitude as $\area \propto \sqrt{|{\rm B}|}\propto \sqrt{U_w}\propto \sqrt{\B_w/B_0}$.

Figure \ref{fig09}(b) shows functions $\langle F(\xi,a)\rangle_{\xi}$, $\langle F^2(\xi,a)\rangle_{\xi}^{1/2}$, and $\sqrt{\langle F(\xi,a)^2\rangle_{\xi}-\langle F(\xi,a)\rangle_{\xi}^2}$ that describe the mean energy change, variance, and energy dispersion. Comparing Figs. \ref{fig09}(a) and (b), we see that although $F$ can be both positive and negative, the averaged $\langle F(\xi,a)\rangle_{\xi}$ is always positive (or zero, for $a<1$). Thus, individual particles may experience $\Delta I$ increase due to the phase bunching, but the $\xi$-averaged effect of such bunching corresponds to $\Delta I<0$. Such positive phase bunching ($\Delta I>0$) has been explained and investigated in \citet{Albert22:phase_bunching}, whereas \citet{Vargas23:pop} demonstrated that positive bunching does change the overall dynamics of a charged particle ensemble, which can be fully described by the averaged $\langle F(\xi,a)\rangle_{\xi}$ function.

Let us now consider particles experiencing phase trapping. In contrast with transient particles merely crossing the resonance $\dot\zeta=0$, trapped particles cross the separatrix (the curve demarcating open and closed trajectories in Fig. \ref{fig07}(b)), and their characteristic motion is qualitatively changed, i.e., they start oscillating around the resonance $\dot\zeta=0$. To estimate the amount of such particles, we shall compare the variation of the area surrounded by the separatrix, $\area$, and the total phase space flux $\flux$ crossing the resonance:
\[
\flux = {\int\limits_0^{2\pi } {\dot P_\zeta  d\zeta } }=2\pi{\rm A}
\]
This comparison provides the so-called probability of trapping:
\begin{equation}
  \Pi  \approx \frac{{\left\{ {\area,\Lambda } \right\}}}{{\left| \flux \right| + \left\{ {\area,\Lambda } \right\}/2}} = \frac{{2\left\{ {\area,\Lambda } \right\}}}{{4\pi \left| {\left\{ {I_R ,\Lambda } \right\}} \right| + \left\{ {\area,\Lambda } \right\}}} \label{eq:P0}
\end{equation}
where $\dot\area=\{\area, \Lambda\}$ and $\Pi=0$ for area decrease, $\dot\area<0$. If $\area$ changes slowly enough ( $\dot\area < 4\pi |\{I_R ,\Lambda\}|$), which is the case in almost all systems under consideration, we can use the approximation
\begin{equation}
\Pi  \approx \frac{{\left\{ {\area,\Lambda } \right\}}}{{2\pi \left| {\left\{ {I_R ,\Lambda } \right\}} \right|}} \label{eq:P}
\end{equation}
We note that ${I_R,\Lambda}=\dot I_R$, and thus the probability of trapping can be rewritten as
\[
\Pi  \approx \frac{{\dot \area}}{{2\pi \dot I_R }} = \frac{1}{{2\pi }}\left. {\frac{{d\area}}{{dI}}} \right|_{I = I_R }
\]
Taking into account that $\langle\Delta I\rangle_{\xi}=-\area/2\pi$, we obtain
\[
\Pi  \approx  - \frac{{d\left\langle {\Delta I} \right\rangle _\xi  }}{{dI}}
\]
This equation provides a direct relationship between the probability of trapping and the average variation $\langle\Delta I\rangle_{\xi}$ due to nonlinear scattering.

\begin{figure}
\centering
\includegraphics[width=1\textwidth]{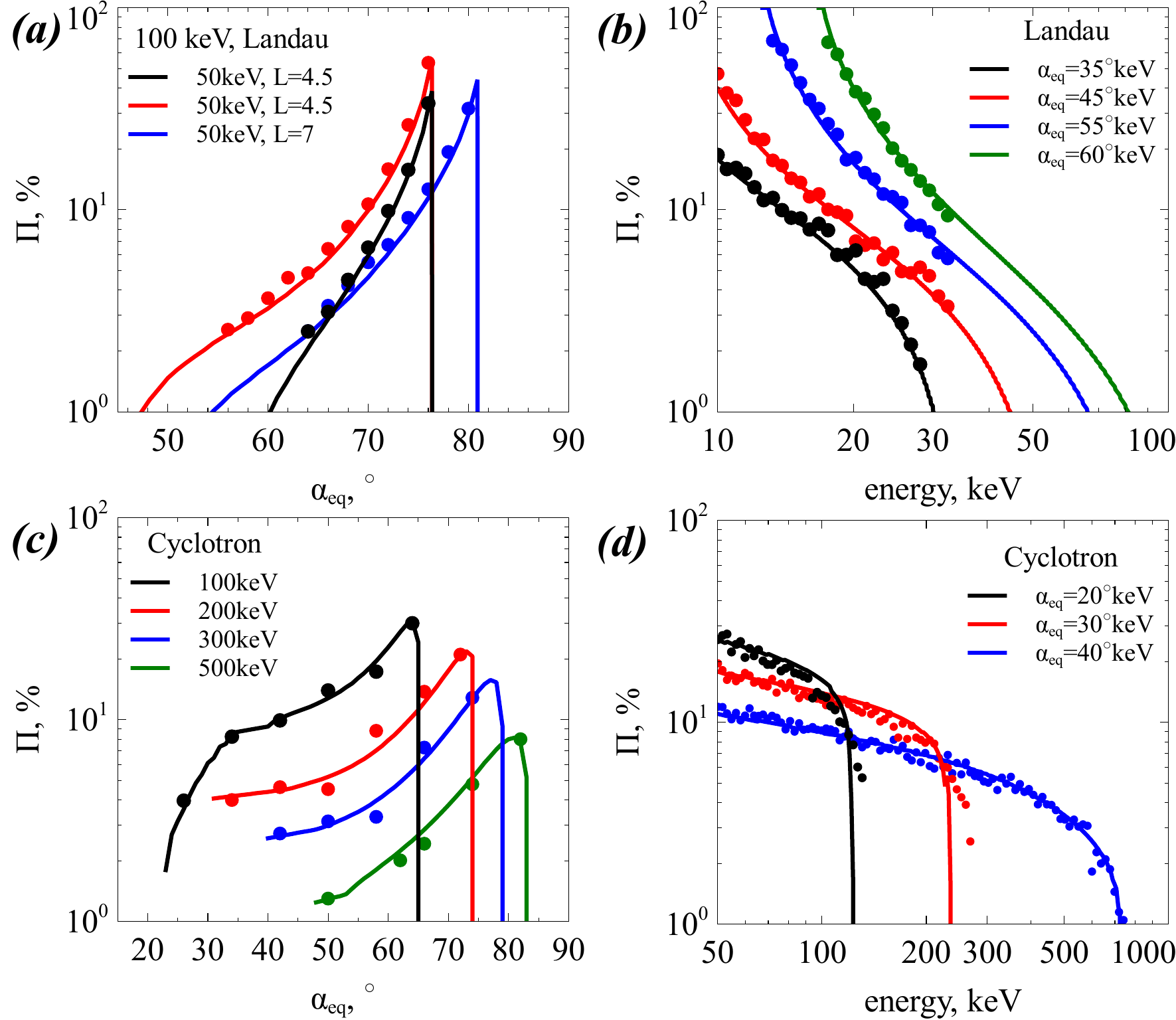}
\caption{Four sets of examples of numerical verification of the equation for the probability of trapping: curves show theoretical results from Eq. (\ref{eq:P}) and circles show results of test-particle simulations (a relative number of particles experiencing the phase trapping for a single resonant interaction).  Panel (a) shows results for Landau resonance from \cite{Artemyev13:pop} (wave amplitudes are the same for black and blue curves, but a factor of $2$ larger for the red curve), panel (b) shows results for Landau resonance from \cite{Artemyev14:grl:fast_transport}, panel (c) shows results for the cyclotron resonance from \cite{Artemyev15:pop:probability}, and panel (d) shows results for the cyclotron resonance from \cite{Vainchtein18:jgr}. Details of wave models and background magnetic field conditions can be found in the corresponding studies.}
\label{fig10}
\end{figure}

The probability of trapping can be understood as the ratio of particles experiencing trapping for a single resonant interaction to the total number of particles crossing the resonance. For fixed system characteristics and $h$ constant, this probability depends only on the initial electron energy. Figure \ref{fig10} shows examples of verification of Eq. (\ref{eq:P}) for the first cyclotron resonance with field-aligned waves and for the Landau resonance with very oblique waves \citep[see more examples in][]{Artemyev12:pop:nondiffusion,Artemyev14:grl:fast_transport,Artemyev15:pop:probability,Vainchtein18:jgr}. The usual scheme for an evaluation of the probability of trapping in numerical test particle simulations includes an integration of trajectories for a large particle ensemble with the same initial energy and pitch-angles, but random initial gyrophases. This ensemble passes through the resonance once and we can count the number of particles trapped into the resonance. The ratio of such trapped particles to the initial total number of particles in the ensemble will provide an estimate of the probability of trapping. Each colored circle in Fig. \ref{fig10} has been obtained via such a scheme, applied to different particle (energy, pitch-angle) and system ($L$-shell, wave amplitude) parameters. This numerically evaluated probability of phase trapping is compared with the analytical formula of $\Pi$, confirming that Eq. (\ref{eq:P}) accurately describes the trapping probability. Note that we can resort to probabilities for the description of trapping, because in any realistic system the initial electron gyrophase is an unknown parameter. Therefore, we can average the system over this gyrophase to reduce its dimensionality, which leads to an inherently probabilistic description of trapping \cite{Neishtadt75,Shklyar81}.

\begin{figure}
\centering
\includegraphics[width=1\textwidth]{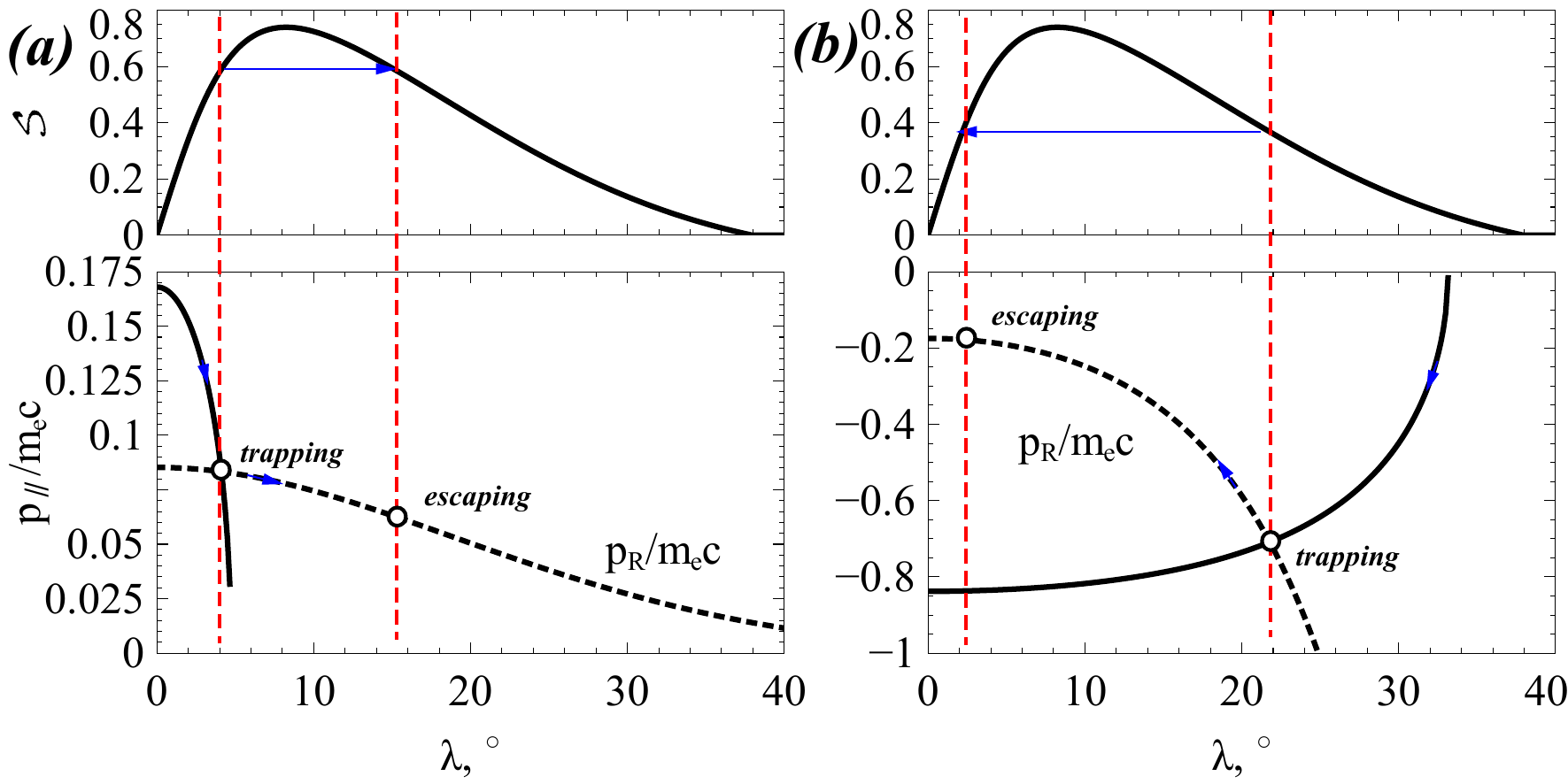}
\caption{Schematic view of electron trapping/de-trapping (escape) for the Landau (a) and cyclotron (b) resonances. Top panels show typical profiles of area $\area(\lambda)$, and bottom panels show unperturbed electron trajectory $p_{\parallel}(\lambda)$ and resonant condition $p_\parallel=p_{R}(\lambda)$. Trapping and escape positions are shown by vertical red lines.}
\label{fig11}
\end{figure}

To evaluate the energy variation of an electron due to phase trapping, let us consider the motion of a trapped electrons. After crossing the separatrix in the phase portrait from Fig. \ref{fig07}(b), electrons start rotating around the resonance along closed trajectories. This rotation occurs with a frequency $\Omega_{tr}\sim \sqrt{{\rm B}/M} \propto (\B_w/B_0)^{1/2}\Omega_{0}$, whereas the phase portrait evolves with the rate of $(s,p)$ change, that is $\sim 2\pi/\tau_b \gg \Omega_0$ (the bounce period being the longest time scale in this system). Note that the resonance condition for whistler-mode waves assumes that $\Omega_0$ is of the same order as $k_\parallel \dot s \sim k_\parallel R/\tau_b$, where $R$ is the spatial scale of the background magnetic field inhomogeneity. On the other hand, the condition for nonlinear resonant interaction, $a>1$, assumes that $k_\parallel^{-1} \partial \ln\Omega_0/\partial s\sim 1/k_\parallel R$ is of the same order as the wave strength $\sim \B_w/B_0$, i.e. $k_\parallel R \sim B_0/\B_w$ and $\Omega_0\tau_b \sim B_0/\B_w \gg 1$. Thus, trapped electron oscillations around the resonance are much faster than the phase portrait evolution, $\Omega_{tr}\tau_b\propto (\B_w/B_0)^{1/2}\Omega_0\tau_b \propto (\B_w/B_0)^{-1/2}\gg 1$. Such a fast periodical motion should introduce an adiabatic invariant $I_\zeta=(2\pi)^{-1}\oint{P_{\zeta}d\zeta}$ \citep{bookLL:mech60}, which is equal to $\area_{tr}/2\pi$ evaluated at the time of trapping. Thus, trapped electrons move within the region surrounded by the separatrix, and at the time of trapping their invariant (area surrounded by their trajectories) is $I_\zeta=\area_{tr}/2\pi$ with an increasing $\area_{tr}$. Electrons will stay trapped until $\area/2\pi $ becomes larger than $I_\zeta$. Accordingly, electrons will escape from the trapping regime when $\area=\area_{tr}$ and $\area$ decreases. Illustration of this trapping/de-trapping dynamics is shown in Fig.  \ref{fig11}.

Figures \ref{fig09}-\ref{fig11} show that all nonlinear resonant effects are described by the $\area(I)$ (or $\area(\gamma)$) function: the energy variation of phase bunched particles is $\Delta \gamma = -\omega\area/2\pi m_ec^2$, the probability of phase trapping is $\Pi\approx (2\pi)^{-1}d\area/d I = (2\pi)^{-1}(d\area/d \gamma)\cdot(\omega/m_ec^2)$, and the energy variation due to trapping is determined by the equation $\area(\gamma)=\area\left(\gamma+\Delta\gamma_{tr}\right)$ \citep[see also][]{Artemyev18:jpp}. Conversely, electron diffusion requires knowing $\langle F^2\rangle_\xi$, which cannot be expressed through the $\area=-2\pi\sqrt{2|{\rm A}|M}\langle F\rangle_\xi$ function, and should be evaluated separately. In the next section, we will use a quite universal property of $\area$ to construct a kinetic equation including the effects of nonlinear resonant interactions.

\subsection{Small pitch-angle limit} \label{sec:nl_small_alpha}
There is one important limitation of system (\ref{eq:hamiltonian_zeta}, \ref{eq:hamiltonian_slow}): the factor $\sim \B_w\sqrt{2I\Omega_0}$ in the wave term is implicitly assumed to be constant within a typical time-scale of resonant interaction $\sim \Omega_{tr}^{-1}$. This assumption is valid as long as the phase variation $\dot \zeta$ is controlled by the $O(\B_w/B_0)$ term, while the term $\sim \B_w\sqrt{2\Omega_0/I}\sin\zeta$ remains a small correction. This assumption is naturally violated for sufficiently small $I$, when the term $\sim \B_w\sqrt{2\Omega_0/I}$ becomes important and the variation rate of $I$ becomes comparable to the $\zeta$ variation rate \citep{Lundin&Shkliar77}. A detailed description of this case can be found in \citep{Albert21,Artemyev21:pop} and in Appendix B, whereas multiple important effects of small $I$ (small $I_x$) on resonant electron motion are described in \citep{Kitahara&Katoh19,Grach&Demekhov20,Gan22}. Here, we only briefly discuss these effects.

\begin{figure}
\centering
\includegraphics[width=1\textwidth]{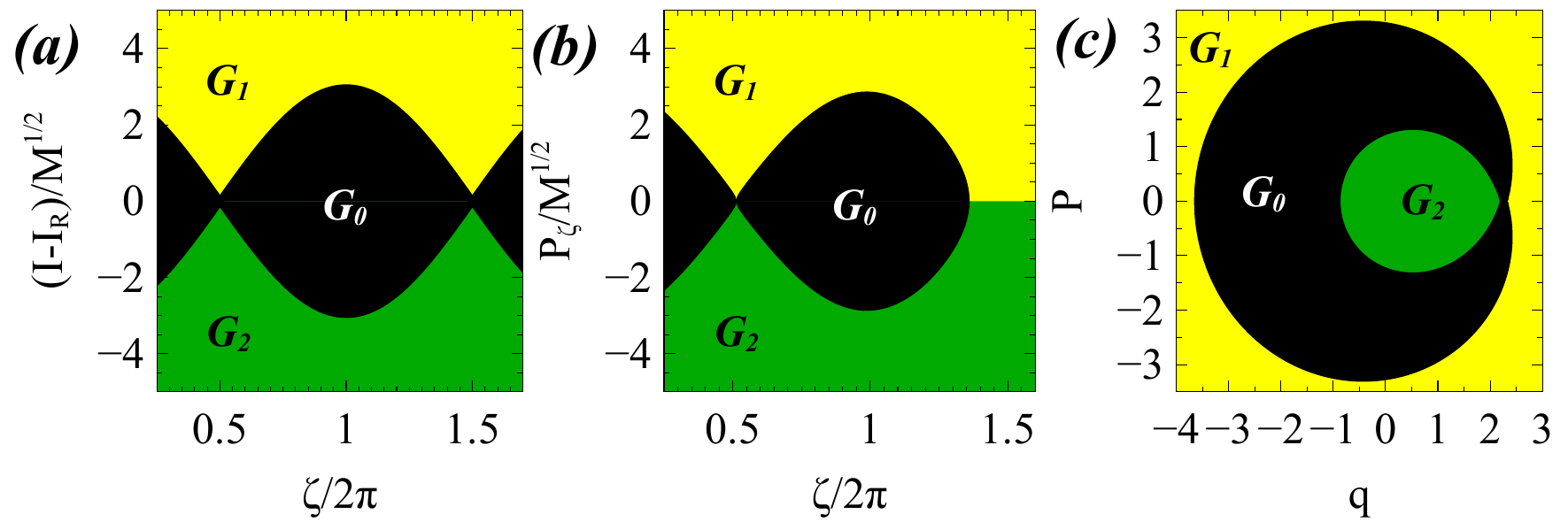}
\caption{Comparison of phase portraits of (a) Hamiltonian (\ref{eq:hamiltonian_const}), (b) Hamiltonian (\ref{eq:hamiltonian_zeta}), and (c) Hamiltonian (\ref{eq:H_XY}). }
\label{fig12}
\end{figure}

Let us consider resonant nonlinear scattering of electrons with small $I_x=I$ and $\Nr=-1$, and expand the Hamiltonian (\ref{eq:hamiltonian_const}) for small $I$:
\[
\tilde{\H}_I  \approx \Lambda _0  + \frac{1}{{2M_0 }}\left( {I - I_R } \right)^2  + \sqrt {\frac{{2I\Omega _0 }}{{m_e c^2 }}} \frac{{e\B_w }}{k}\cos \zeta
\]
where $M_0=\partial^2 \H_I/\partial I^2|_{I=0}$, $\Lambda_0=\H_I|_{I=0}$, and $I_R$ is given in Appendix B. The main difference from the Hamiltonian expanded around the resonant $I_R$ is that wave amplitude in $\tilde{\H}$ depends on $I$. The Hamiltonian $\tilde{\H}_I$ describes fast $\zeta$ motion and slow $(s,p, I)$. An important property of this Hamiltonian is that for $I\sim (\B_w/B_0)^{2/3}$ the dynamics of $I$ becomes as fast as the dynamics of $\zeta$ and, thus, there is no time separation between $\zeta$ and $I$, which become fast variables. In Appendix B we show that the Hamiltonian of these fast variables is
\begin{equation}
\mF  = \frac{1}{2}\left( {\frac{1}{2}P^2  + \frac{1}{2}q^2  - Y_R } \right)^2  + u q \label{eq:H_XY}
\end{equation}
 where $P=(\B_w/B_0)^{-1/3}\sqrt{2I\Omega_0/m_ec^2}\sin\zeta$, $q=-(\B_w/B_0)^{-1/3}\sqrt{2I\Omega_0/m_ec^2}\cos\zeta$, $Y_R = (\B_w/B_0)^{-2/3}I_R\Omega_0/m_ec^2$, and $u$ is a normalization constant. The phase portrait of Hamiltonian (\ref{eq:H_XY}) is shown in Fig. \ref{fig12}(c). Let us compare this phase portrait with phase portraits of the Hamiltonian expanded around $I_R$,
\begin{equation}
\H_I =  - \omega I_{R} +\gamma_{R}  + \frac{1}{{2M}} \left( {I - I_{R} } \right)^2
+ \sqrt {\frac{{2I_{R} \Omega _{0} }}{{m_ec^2 }}} \frac{{e\B_w }}{{k}}\sin \zeta \label{eq:H_Ires}
\end{equation}
The main difference between this Hamiltonian and the Hamiltonian from Eq. (\ref{eq:H_XY}) is that the effective wave amplitude does not depend on $I$. The phase portrait of Hamiltonian (\ref{eq:H_Ires}) with frozen slow variables is shown in Fig. \ref{fig12}(a).

Instead of introducing $(q,P)$ coordinates, it is more convenient to introduce $P_\zeta=I-I_{R}$ and rewrite Hamiltonian (\ref{eq:H_Ires}) into Eq. (\ref{eq:hamiltonian_zeta}). The phase portrait of Hamiltonian (\ref{eq:hamiltonian_zeta}) with frozen slow variables  is shown in Fig. \ref{fig12}(b). This is the classical portrait of the pendulum with torque with three main phase space regions: before resonance $P_\zeta=0$ crossing, particles are in $G_{1}$, and resonance crossing can result in trapping (particles appear in $G_{0}$) or scattering (particles appear in $G_{2}$). Therefore, there is a direct relation between three regions $G_{0,1,2}$ of Hamiltonian of (\ref{eq:H_XY}) and Hamiltonian (\ref{eq:hamiltonian_zeta}).

For the initial system given by Eq. (\ref{eq:hamiltonian}) with $\Nr=-1$ the phase bunching (transition from $G_{1}$ to $G_{2}$) always appears with $I_x$ decrease, and this effect is well seen in the phase portrait (c) of Fig. \ref{fig12}: the area $\area_{1}$ is always larger than the area $\area_{2}$. But when the initial invariant $I_x$ is sufficiently small, particles become trapped within region $G_{0}$ as soon as this region appears during particle motion along their trajectories. In phase portrait (c) of Fig. \ref{fig12} this trapping means that the area surrounded by the particle trajectory $2\pi\oint{Pdq}\sim 2\pi \sqrt{2I\Omega_0/m_ec^2}$ is smaller than $S_{0}$ at the moment when $G_{0}$ appears. The threshold $I_x$ value is $2I_x\Omega_{0}/m_ec^2\sim (\B_w/ B_0)^{2/3}$ (see Appendix B and \cite{Artemyev21:pop,Albert21}).

This is the so-called {\it autoresonance} phenomena of 100\% probability of trapping in resonant systems \citep[see, e.g.,][and references therein]{Fajans&Friedland01, Friedland09, Neishtadt13,Neishtadt75, Sinclair72:MNRAS}. Being trapped into $G_{0}$, particles can both increase or decrease their $I_x$ during the transition from $G_{0}$ to $G_{2}$: the $I_x$ change depends on the ratio of $S_{0}/S_{2}$ at the moment when $S_{0}$ becomes equal to $2\pi\oint{Pdq}$. For sufficiently small $2\pi\oint{Pdq}$ (small $I_x$), the ratio $S_{0}/S_{2}=2\pi\oint{Pdq}/S_{2}$ will be below one, and particles will increase their $I_x$ due to the resonant interaction. Formally, this interaction cannot be called bunching, because particles are trapped into $G_{0}$ from the beginning.

\section{Kinetic equation with nonlinear resonant effects} \label{sec:1d}
In this section, we use the function of $\area(\gamma)$ to derive a kinetic equation for the distribution function of electrons nonlinearly interacting resonantly with whistler-mode waves for a constant $h$ given by Eq. (\ref{eq:h_const}). Let us start with the simplified situation of a constant bounce period $\partial\tau_b/\partial \gamma=0$. We separate the $\hat K$ kernel into two parts: one part describing diffusion and the phase bunching effect $\hat K_{bc}$ and another part describing the trapping effect $\hat K_{tr}$. Using the approximation of small energy change due to bunching, we expand $\hat K_{bc}$ as
\[
\int\limits_0^\infty{\hat K_{bc}(\gamma,\gamma')f(\gamma')d\gamma'}  = \frac{\partial }{{\partial \gamma }}\left( {D_{\gamma \gamma } \frac{{\partial f}}{{\partial \gamma }}} \right) - \frac{\partial }{{\partial \gamma }}\left( {V_\gamma  f} \right)
\]
Note that there are two main contributions to the particle drift term $V_\gamma$: the gradient of the diffusion rate, $V_\gamma \propto \partial D_{\gamma\gamma}/\partial \gamma \propto (\B_w/B_0)^2$, and nonlinear scattering (phase bunching effect), $V_\gamma\propto \area \propto (\B_w/B_0)^{1/2}$. The first term determines the divergence-free condition for the Fokker-Planck diffusion equation and is taken into account in the form of a $\propto D_{\gamma\gamma}$ operator \cite{Lieberman&Lichtenberg73,bookSagdeev88}. Thus, in the equation for $\hat K_{bc}$ we may include only the second term $V_\gamma\propto (\B_w/B_0)^{1/2}$, which is much larger than the first one, because $\B_w/B_0 \ll 1$.

For the trapping kernel, we use the following formulation:
\[
\int\limits_0^\infty  {\hat K_{tr+} \left( {\left. \gamma  \right|\gamma '} \right)f\left( {\gamma '} \right)d\gamma '}  - \int\limits_0^\infty  {\hat K_{tr-} \left( {\left. {\gamma '} \right|\gamma } \right)f\left( \gamma  \right)d\gamma '}
\]
where the second term describes particle transport from the energy of trapping, $\gamma$, and the first term describes  particle transport toward the energy of escape from the resonance, $\gamma$, from the energy of trapping, $\gamma'$, given by $\area(\gamma')=\area(\gamma)$. Substituting $K_{bn}$ and $K_{tr}$ into the Smoluchowski equation (\ref{eq:smolukhovskiy}), we obtain
\begin{eqnarray*}
 \frac{{\partial f}}{{\partial t}} &=& \frac{\partial }{{\partial \gamma }}\left( {D_{\gamma \gamma } \frac{{\partial f}}{{\partial \gamma }}} \right) - \frac{\partial }{{\partial \gamma }}\left( {V_\gamma  f} \right) \\
  &+& \int\limits_0^\infty  {\hat K_{tr+} \left( {\left. \gamma  \right|\gamma '} \right)f\left( {\gamma '} \right)d\gamma '}  - \int\limits_0^\infty  {\hat K_{tr-} \left( {\left. {\gamma '} \right|\gamma } \right)f\left( \gamma  \right)d\gamma '}
 \end{eqnarray*}
The term responsible for particle transport from the energy of trapping, $\gamma$, can be written as
\[
 \int\limits_0^\infty  {\hat K_{tr-} \left( {\left. {\gamma '} \right|\gamma } \right)f\left( \gamma  \right)d\gamma '}  = \int\limits_0^\infty  {\frac{\Pi \left( \gamma  \right)}{\tau_b}\delta \left( {\gamma  - \gamma '} \right)f\left( \gamma  \right)d\gamma '} =\frac{\Pi(\gamma)}{\tau_b}f(\gamma)\left(1-\Theta\right)
\]
where
\begin{equation}
\Theta \left( \gamma  \right){\rm  = }\left\{ {\begin{array}{*{20}c}
   {{\rm 0,}} & {dS/d\gamma  < 0}  \\
   {{\rm 1,}} & {dS/d\gamma  > 0}  \\
\end{array}} \right.\label{eq:theta}
\end{equation}

The term responsible for particle transport from the energy $\gamma'$ into the energy of escape, $\gamma$, can be written as
\begin{eqnarray*}
\int\limits_0^\infty  {\hat K_{tr+} \left( {\left. \gamma  \right|\gamma '} \right)f\left( {\gamma '} \right)d\gamma '}  &=& \int\limits_0^\infty  {\Pi \left( {\gamma '} \right)\delta \left( {\gamma  - {\rm T}\left( {\gamma '} \right)} \right)f\left( {\gamma '} \right)d\gamma '} \\&=&
\frac{\Pi \left( {\gamma^* } \right)}{\tau_b}\left| {\frac{{d{\rm T}\left( {\gamma^* } \right)}}{{d\gamma^* }}} \right|^{ - 1} f\left( {\gamma^* } \right)
\end{eqnarray*}
where ${\rm T}(\gamma^*)=\gamma$ is the solution of the equation $\area(\gamma^*)=\area(\gamma)$. Note that Eq. (\ref{eq:h_const}) provides a linear relation between electron energy $\gamma$ and momentum $I$: $d\gamma/dI=\omega/m_ec^2=const$. This relation allows us to use the equation $d\gamma^*/d\gamma=dI^*/dI$ to write:
\[
\frac{{dT\left( {\gamma ^*} \right)}}{{d\gamma ^*}} = \frac{{d\gamma }}{{d\gamma ^*}} = \frac{{dI}}{{dI^*}} = \frac{{dS\left( {I^*} \right)/dI^*}}{{dS\left( I \right)/dI}} =  - \frac{{\Pi \left( {I^*} \right)/\tau _b }}{{dS\left( I \right)/dI}} =  - \frac{{\Pi \left( {\gamma ^*} \right)/\tau _b }}{{dV_\gamma\left( \gamma  \right)/d\gamma }}
\]

Thus, for $\hat K_{tr+}$ we obtain for $\Theta=1$
\[
\int\limits_0^\infty  {K_{tr + } \left( {\left. \gamma  \right|\gamma '} \right)f\left( {\gamma '} \right)d\gamma '}  = \frac{{dS\left( \gamma  \right)}}{{d\gamma }}f\left( {\gamma ^*} \right)
\]

\begin{figure}
\centering
\includegraphics[width=1\textwidth]{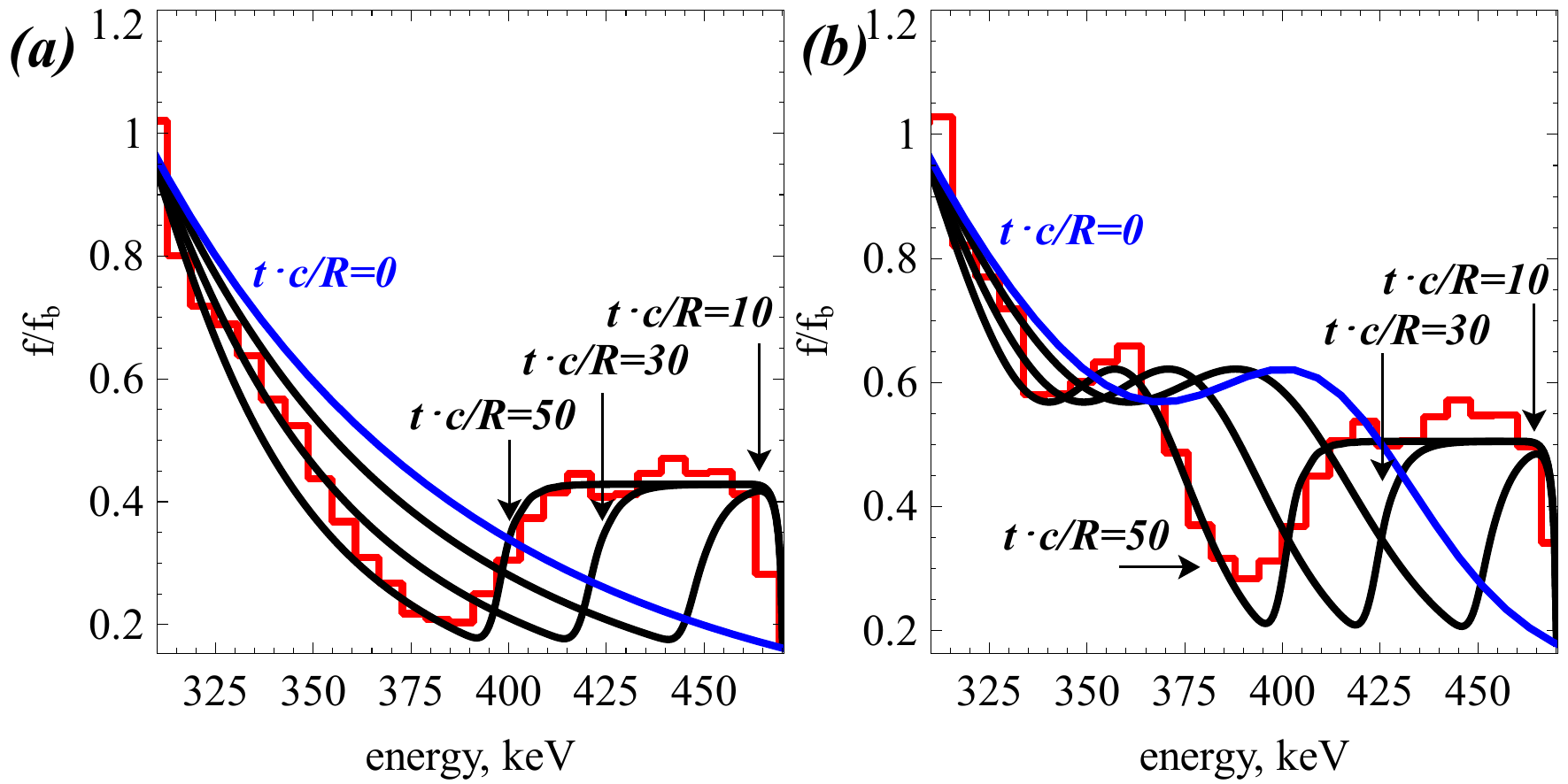}
\caption{Solutions of Eq. (\ref{eq:kinetic1D}) for two initial conditions (shown in blue). The function $f$ is normalized to its initial value at the left boundary, $f_b$. Results are shown at three different times, and the solution for $tc/R=50$ is compared with results of test particle simulations (red) with $10^6$ trajectories. Solutions are obtained for $h/m_ec^2=1.454$ (this value of $h$ corresponds to, e.g., an equatorial pitch-angle $\alpha_{eq}=45^\circ$ for 300 keV electron energy). We use a curvature-free dipole magnetic field \citep{Bell84} at $L$-shell$=4.5$. The whistler-mode wave frequency is $0.35$ times the electron cyclotron frequency at the equator, and the plasma frequency is $4.5$ times the electron cyclotron frequency at the equator. Wave amplitude is $300$ pT. The distribution of the wave amplitude along magnetic field lines is modeled by the empirical function $\tanh((\lambda/\delta\lambda_1)^2)\exp(-(\lambda/\delta\lambda_2)^2)$ with $\delta\lambda_1=2^\circ$, $\delta\lambda_2=20^\circ$. This empirical function fits the typically observed whistler-mode wave intensity distribution \cite{Agapitov13:jgr}.
%For each case we show $\area(\gamma)$ and $D(\gamma)$ profiles
\label{fig13}}
\end{figure}

Therefore, combining all these terms in the equation for $\partial f/\partial t$, we obtain:
\begin{equation}
\frac{{\partial f}}{{\partial t}} = \frac{\partial }{{\partial \gamma }}\left( {D_{\gamma \gamma } \frac{{\partial f}}{{\partial \gamma }}} \right) - \frac{\partial }{{\partial \gamma }}\left( {V_\gamma  f} \right) + \frac{{dV_\gamma\left( \gamma  \right)}}{{d\gamma }}f\left( {\gamma ^*} \right)\Theta - \frac{{\Pi \left( \gamma  \right)}}{{\tau _b }}f\left( \gamma  \right)\left(1-\Theta\right)
\label{eq:kinetic1D}
\end{equation}
Taking into account that $\Pi=(2\pi)^{-1}(d\area/dI)=-dV_\gamma/d\gamma$, we can rewrite Eq.(\ref{eq:kinetic1D}) as:
\[
\frac{{\partial f}}{{\partial t}} =\frac{\partial }{{\partial \gamma }}\left( {D_{\gamma \gamma } \frac{{\partial f}}{{\partial \gamma }}} \right) - V_\gamma \frac{\partial f}{{\partial \gamma }}+ \frac{{dV\left( \gamma  \right)}}{{d\gamma }}\left( f\left( {\gamma ^*} \right) - f\left( {\gamma } \right) \right)\Theta
\]
This is a basic formulation of a kinetic equation describing electron distribution dynamics due to multiple nonlinear resonant interactions. The numerical solution of this equation has been verified by comparisons with test particle simulations in \cite{Artemyev16:pop:letter,Artemyev17:pre,Leoncini18}. Figure \ref{fig13} shows two  examples of such verifications. We use two different initial conditions for the $f(\gamma)$ function and solve Eq. (\ref{eq:kinetic1D}) within the energy range of $\area\ne0$, i.e., we consider electron dynamics only within the energy range of nonlinear resonant interactions, $\gamma \in[\gamma_{\min},\gamma_{\max}]$. The diffusion rate $D_{\gamma\gamma}$ does not necessary vanish at $\gamma_{\min,\max}$ and particles can diffusively move in/out of the energy range of nonlinear resonant interactions. Thus, to verify Eq. (\ref{eq:kinetic1D}) within the range $\gamma \in[\gamma_{\min},\gamma_{\max}]$ we numerically modify $D_{\gamma\gamma}\to D_{\gamma\gamma}\cdot 4\left(\gamma-\gamma_{\min}\right)^2\left(\gamma_{\max}-\gamma\right)^2/\left(\gamma_{\max}-\gamma_{\min} \right)^4$, to guarantee that electrons will not leave the range $\gamma \in[\gamma_{\min},\gamma_{\max}]$ of nonlinear resonant interactions. We also numerically integrate $10^6$ test particle trajectories described by the Hamiltonian (\ref{eq:hamiltonian}).

The solutions shown in Fig. \ref{fig13}(a) start with a power law distribution $f\sim \gamma^{-3}$. Phase trapping forms an accelerated population of electrons at large energies, and with time this population drifts to smaller energies due to the phase bunching. The fine balance of bunching speed and trapping probability prevents an accumulation of electrons in the region where electrons are released from trapping acceleration: the new accelerated population only replaces a previously accelerated population moved to smaller energy, and the electron population at large energy does not increase in magnitude but occupies a larger energy range as time goes on. At the time $50R/c$, corresponding to about fifty bounce periods with a single resonant interaction during each period, we compare the solution of Eq. (\ref{eq:kinetic1D}) with test particle simulation results: the corresponding red and black curves are quite close, demonstrating that the kinetic equation describes well the dynamics of the electron energy distribution.

The solutions shown in Fig. \ref{fig13}(b) start with a distribution having a small maximum at intermediate energies and a plateau. The resonant dynamics of the electron energy distribution includes the same components shown in Fig. \ref{fig13}(a): formation of an accelerated population due to phase trapping and the following drift of this population to smaller energies due to phase bunching. This dynamics is supplemented by the evolution of the maximum initially present at intermediate energies: due to an absence of phase trapping at these energies, this maximum merely drifts toward smaller energies via phase bunching. The enhanced electron population at smaller energies (due to the initial plateau) provides more particles for trapping acceleration, and the accelerated population in Fig. \ref{fig13}(b) has a larger magnitude in comparison with the population shown in Fig. \ref{fig13}(a). The comparison of solution of Eq. (\ref{eq:kinetic1D}) with results of test particle simulations (at time $50R/c$) confirms that the kinetic equation describes correctly the dynamics of the electron energy distribution.

We note that in the case of test particle simulations, we cannot perform the modification of the diffusion coefficient that prevents particle escape from the energy range of nonlinear wave-particle interactions in the numerical solution of Eq. (\ref{eq:kinetic1D}). Thus, to perform fair comparisons between test particle simulations and solutions of Eq. (\ref{eq:kinetic1D}), we rerun each particle escaping $\gamma \in[\gamma_{\min},\gamma_{\max}]$ and save the total number of particles within this energy range. The diffusion $D_{\gamma\gamma}$ changes the electron distribution much slower than nonlinear resonant phase trapping and bunching, and such a weak effect of diffusion may help the comparison of test particle simulation results with solutions of Eq. (\ref{eq:kinetic1D}), despite the different descriptions of diffusive scattering in the frame of these two approaches (due to $D_{\gamma\gamma}$ modification in the kinetic equation).

\subsection{Boundary conditions}
\begin{figure}
\centering
\includegraphics[width=1\textwidth]{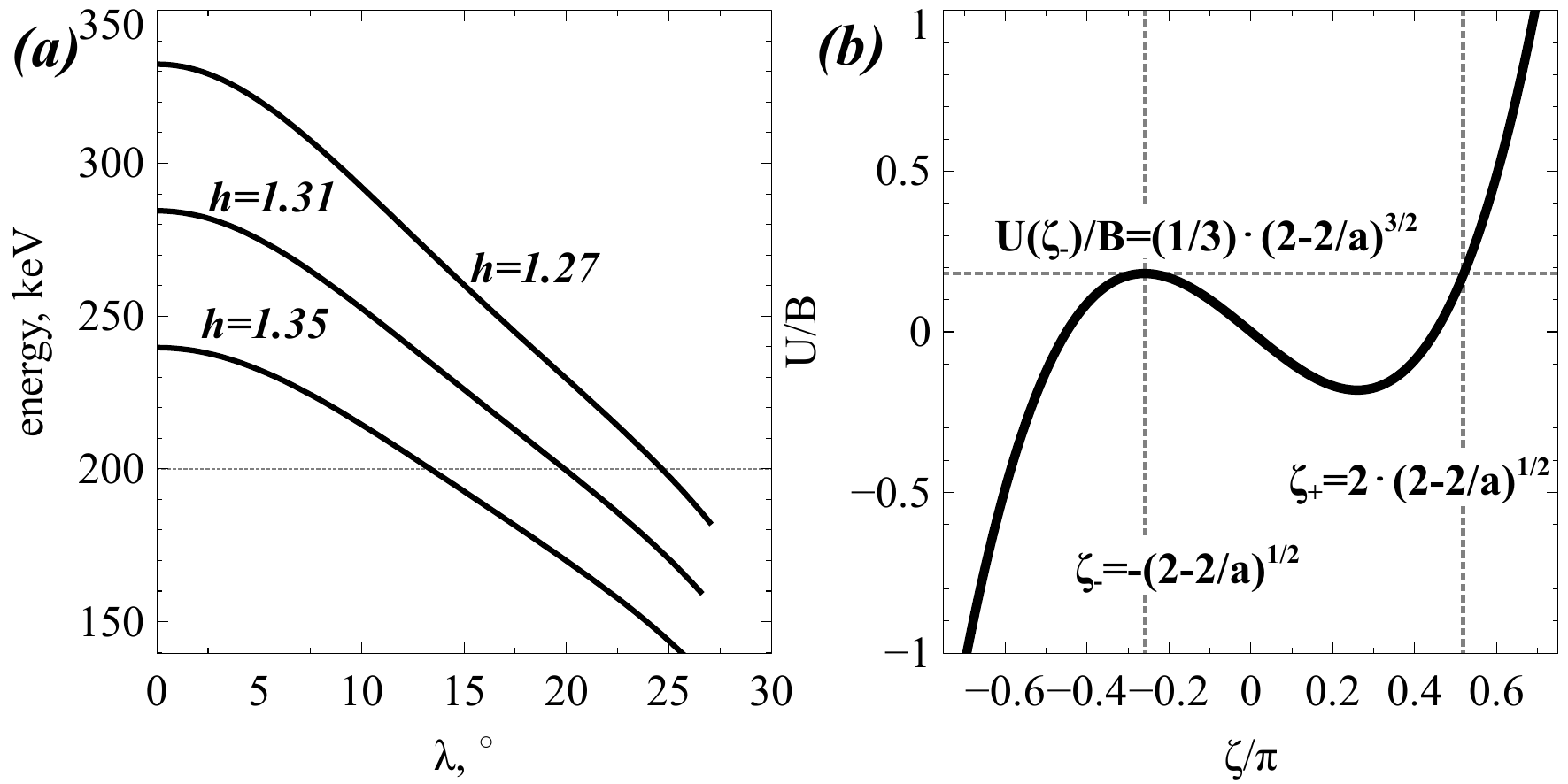}
\caption{(a) Resonant energy $(\gamma -1)m_ec^2$ as a function of magnetic latitude for first cyclotron resonance with field-aligned waves ($\omega/\Omega_{eq}=0.35$, $L=6$) for several $h$ (equatorial pitch-angle $60^\circ$, $45^\circ$, and $30^\circ$ for $200$keV electron) values. (b) Profile of of {\it potential energy} $U/B=\zeta^3/6-\zeta(1-a^{-1})$ for Hamiltonian (\ref{eq:Hamiltonian_a1}). }
\label{fig14}
\end{figure}

The profile $\area(\gamma)$ determines all the main terms of the kinetic equation (\ref{eq:kinetic1D}) and, therefore, the properties of $\area(\gamma)$ are very important for understanding the solution of this equation. The important characteristic of $\area(\gamma)$ is its asymptotic behavior near zero values. Depending on the specific distribution of wave electromagnetic field (the term $\sim B_w$ in Eq. (\ref{eq:hamiltonian})0, there are different numbers of zeros of $\area(\gamma)$, but the simplest case corresponds to the cyclotron resonance with field-aligned whistler-mode waves when the term $\sim \B_w$ varies monotonically along the magnetic latitude. Note that for a conserved invariant $h$ from Eq. (\ref{eq:h_const}), the resonant energy $\gamma$ is a linear function of $I=I_x$ and a monotonic function of magnetic latitude $\lambda$ (or coordinate $s$). Figure \ref{fig14}(a) shows examples of latitudinal profiles of resonant energy for different $h$ values.

Near the equatorial plane, $\area(\gamma)$ should drop to zero because $\B_w\to 0$ in the wave source region, and thus $a \propto \B_w$ is less than one \citep[see][for descriptions of alternative $\B_w$ profiles for whistler-mode waves generated by lightning in the ionosphere and propagating from high latitudes to the equator]{Shklyar17,Shklyar21,Shklyar&Luzhkovskiy23}. At high latitudes, the gradient $\partial\Omega_0/\partial s$ becomes quite large and thus $a\propto \left(\partial\Omega_0/\partial s \right)^{-1}$ becomes less than one. Thus, for sufficiently large $\B_w$ we find that $\area$ is above zero between the equatorial plane and some high-latitude location. Let us consider the behavior of $\area$ around these two zeros, where $a\to 1$. First, we expand Hamiltonian (\ref{eq:hamiltonian_zeta}) as \citep[see][]{Artemyev19:pd,Artemyev21:pre}:
\begin{equation}
 \H_\zeta = \frac{1}{2M}P_\zeta^2  + {\rm B} \cdot \left( {\frac{1}{a}\zeta  + \cos \zeta } \right) \approx \frac{1}{2M}P_\zeta^2  - {\rm B} \cdot \left( {\left( {1-\frac{1}{a}} \right)\zeta  - \frac{1}{6}\zeta ^3 } \right)  \label{eq:Hamiltonian_a1}
\end{equation}
where we expand $\cos\zeta$ around $\zeta=\pi/2$ and shift $\zeta\to\zeta-\pi/2$. The profile of {\it potential energy}\, $U/{\rm B}=-\left(1-a^{-1}\right)\zeta+\zeta^3/6$ for this Hamiltonian is shown in Fig. \ref{fig14}(b): potential energy has a local minimum in the interval $ (\zeta_-, \zeta_+)$. The corresponding area surrounded by the separatrix takes the form
\begin{eqnarray*}
 \area_{a\approx 1} &=& \sqrt {8M|{\rm B}|} \int\limits_{\zeta _ -  }^{\zeta _ +  } {\sqrt {\left( {1  - a^{ - 1}} \right)\left( {\zeta  - \zeta _ -  } \right) - \frac{1}{6}\left( {\zeta ^3  - \zeta _ - ^3 } \right)} d\zeta }  \\
  &=& \sqrt {8M|{\rm B}|} \left( { 1 - a^{ - 1}} \right)^{5/4} \int\limits_{\tilde \zeta _ -  }^{\tilde \zeta _ +  } {\sqrt {\left( {\tilde \zeta  - \tilde \zeta _ -  } \right) - \frac{1}{6}\left( {\tilde \zeta ^3  - \tilde \zeta _ - ^3 } \right)} d\tilde \zeta }  \\
  &=& \sqrt {8M|{\rm B}|} \left( {1  - a^{ - 1}} \right)^{5/4} \frac{{12 \cdot 3^{3/4} }}{5}=\frac{{12 \cdot 3^{3/4} }}{5}\sqrt {8M\frac{|{\rm A}|^{5/2}}{|{\rm B}|^{  3/2}} } \left( {a - 1} \right)^{5/4}
 \end{eqnarray*}
where $\tilde\zeta=\zeta/\left(a^{-1}-1\right)^{1/2}$. This equation provides the asymptotic variation of $\area$ for $a\to 1$, and we can rewrite this asymptote as a function of the resonant energy. Indeed, $a$ is a function of $s$, and for a conserved $h$ given by Eq. (\ref{eq:h_const}) there is a profile of resonant energy $\gamma(s)$, see Fig. \ref{fig14}(a). Therefore, we can write $a=a(\gamma)$ and then expand $a$ around the energy $\gamma_b$ corresponding to $a=1$ and $\area=0$: $a\approx 1+C\cdot(\gamma-\gamma_b)$, where $C=const$. Substituting this expansion into equation $\area_{a\approx 1}$, we obtain
\[
\area_{a\approx1}\propto\left|\gamma-\gamma_b\right|^{5/4}
\]
Note that $\area\ne 0$ within some $\gamma\in[\gamma_{\min},\gamma_{\max}]$ domain, and we should have an asymptotic form $\area\propto |\gamma-\gamma_{\min,\max}|^{5/4}$ at both boundaries. This asymptotic form of $\area$ around $a\approx 1$ shows that $\Pi\propto d\area/d\gamma \propto |\gamma-\gamma_b|^{1/4}\to 0$ at the energy boundary, i.e., both drift $V\propto \area$ and probability of trapping $\Pi$ drop to zero at the boundary.

\begin{figure}
\centering
\includegraphics[width=1\textwidth]{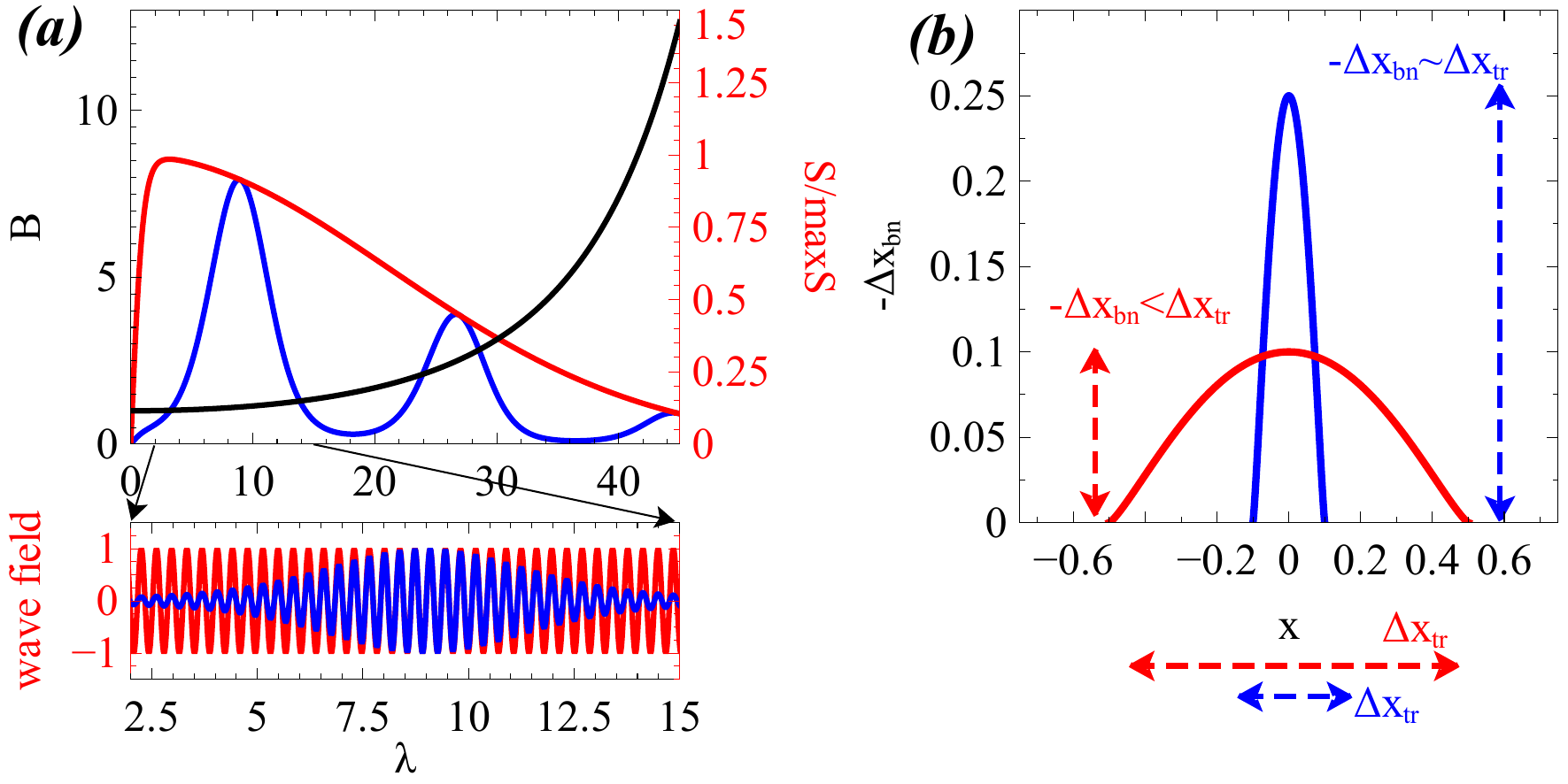}
\caption{Panel (a) shows a schematic view of $\area$ profiles for plane wave (red) and wave-packet (blue). Panel (b) shows the difference of $S(x)/2\pi=-\Delta x_{bn}(x)$ profiles for a long wave-packet having $\Delta x_{tr}\gg |\Delta x_{bn}|$ and for short wave-packets having $\Delta x_{tr}\sim |\Delta x_{bn}|$}
\label{fig15}
\end{figure}

The equation for $\area_{a\approx1}$ determines the $\area$ profile around the zeros, and can be adopted to describe systems with intrinsically small $\area$. The most natural example of such systems is a system with wave-packets. The corresponding schematic is provided in Fig. \ref{fig15}(a). The plane wave $\sim \sin(\zeta)$ is a very simplified approximation of much more complicated and realistic situations, where the wave field often consists of a series of wave-packets \citep{Zhang19:grl,Zhang20:grl:phase}. Each packet has a finite length, and small-scale packets will correspond to a small latitudinal range of $\area\ne 0$, i.e., the $a\approx 1$ regime will be applicable for the entire range of energies with $\area\ne 0$. Then, $\area$ can be approximated as
\[
\area_{small}=S_0\cdot\left(\gamma_{\max}-\gamma\right)^{5/4}\left(\gamma-\gamma_{\min}\right)^{5/4}
\]
where $S_0$ is the magnitude of $\area$ and $\gamma_{\max,\min}$ are boundary values of energies where $a=1$. Using changes of variables, $x=2\gamma/(\gamma_{\max}-\gamma_{\min})-(\gamma_{\max}+\gamma_{\min})/(\gamma_{\max}-\gamma_{\min})$ and $S_0\to S_0\cdot\left((\gamma_{\max}-\gamma_{\min})/2\right)^{5/2}$, we can write
\begin{equation}
\area_{small} = S_0(1-x^2)^{5/4} \label{eq:area_small}
\end{equation}
with the equation for the probability of trapping $\Pi=d\area_{small}/dx=-(5/2)\cdot x S_0 (1-x^2)^{1/4}$. Equation (\ref{eq:area_small}) provides a very convenient and useful model for investigating the properties of systems with nonlinear wave particle resonant interactions. This equation works as long as the area is sufficiently large that $\Delta x_{tr}\gg \Delta x_{bn}\sim S_0$ (see scheme in Fig. \ref{fig15}(b)), and with the decrease of $(\gamma_{\max}-\gamma_{\min})$ (with increase of $S_0$) this equation becomes inapplicable (see \cite{Artemyev21:pre} and Appendix C).

\subsection{Asymptotic solutions}\label{sec:asymptote}
Let us consider the asymptote of the solution of Eq. (\ref{eq:kinetic1D}) for $t\to\infty$ \citep[see details in][]{Artemyev19:pd}. We use the normalized variable $x=2\gamma/(\gamma_{\max}-\gamma_{\min})-(\gamma_{\max}+\gamma_{\min})/(\gamma_{\max}-\gamma_{\min})$ and rewrite this equation as
\begin{equation}
   \frac{{\partial f}}{{\partial t}} = \left\{ {\begin{array}{*{20}c}
   { - V\left( x \right)\frac{{\partial f}}{{\partial x}} + \frac{1}{2}\frac{\partial }{{\partial x}}\left( {D\left( x \right)\frac{\partial }{{\partial x}}} \right),} & {x \le0 }  \\
   { - V\left( x \right)\frac{{\partial f}}{{\partial x}} - \frac{{dV\left( x \right)}}{{dx}}\left( {f - f' } \right) + \frac{1}{2}\frac{\partial }{{\partial x}}\left( {D\left( x \right)\frac{\partial }{{\partial x}}} \right),} & {x \ge 0 }  \\
\end{array}} \right. \label{eq:1D_x}
\end{equation}
where $V(x)=V_0(1-x^2)^{5/4}$ for $x\in[-1,1]$,  $f^*=f(x^*)$ and $x^*=-x$. This equation, as well as Eq. (\ref{eq:kinetic1D}), conserves the total number of particles:
\begin{eqnarray*}
 &&\frac{d}{{dt}}\int\limits_{ - 1}^1 {f\left( {x,t} \right)dx}  = \int\limits_{ - 1}^1 {\frac{{\partial f\left( {x,t} \right)}}{{\partial t}}dx}  = \int\limits_{ - 1}^1 {\frac{1}{2}\frac{\partial }{{\partial x}}\left( {D\frac{{\partial f}}{{\partial x}}} \right)dx}  + \int\limits_{ - 1}^0 {\left( { - \frac{{\partial \left( {Vf} \right)}}{{\partial x}} + f\frac{{\partial V}}{{\partial x}}} \right)dx}  \\
  &+& \int\limits_0^1 {\left( { - \frac{{\partial \left( {Vf} \right)}}{{\partial x}} + f^* \frac{{\partial V}}{{\partial x}}} \right)dx}  = \int\limits_{ - 1}^0 {f\left( {x,t} \right)\frac{{\partial V\left( x \right)}}{{\partial x}}dx}  + \int\limits_0^1 {f^* \left( {y,t} \right)\frac{{\partial V\left( y \right)}}{{\partial y}}dy}
 \end{eqnarray*}
Changing the integration variable in the second integral $y\to x$ and
using $V(x)=V(y)$, $f(x,t)=f'(y,t)$ we get
\[
\frac{d}{{dt}}\int\limits_{ - 1}^1 {f\left( {x,t} \right)dx}  = \int\limits_{ - 1}^0 {f\left( {x,t} \right)\frac{{dV(x)}}{{dx}}dx}  + \int\limits_0^1 {f\left( {x,t} \right)\frac{{dV(x)}}{{dx}}dx}  = 0
\]

To find an asymptotic solution to Eq. (\ref{eq:1D_x}) for $t\to \infty$, we restrict our consideration to the case with $D=0$. This is a reasonable approximation, because diffusion leads to much slower changes in the electron distribution than nonlinear resonant phase bunching and trapping. This approximation also requires that bunching and trapping do not compensate each other, leaving diffusion the main process (see discussion in Appendix C).  Thus, we consider
\[
   \frac{{\partial f}}{{\partial t}} = \left\{ {\begin{array}{*{20}c}
   { - V\left( x \right)\frac{{\partial f}}{{\partial x}},} & {x \le0 }  \\
   { - V\left( x \right)\frac{{\partial f}}{{\partial x}} - \frac{{dV\left( x \right)}}{{dt}}\left( {f - f^* } \right),} & {x \ge 0 }  \\
\end{array}} \right.
\]
We start with the construction of the general solution to this equation. The characteristic curves of this equation are
\begin{eqnarray*}
 \frac{{dt}}{1} &=& \frac{{dx}}{{V\left( x \right)}} = \frac{{df}}{0},\quad x \le 0 \\
 \frac{{dt}}{1} &=& \frac{{dx}}{{V\left( x \right)}} = \frac{{df}}{{ - {\textstyle{{dV} \over {dx}}}\left( {f - f^* } \right)}},\quad x \ge 0
 \end{eqnarray*}
The first equation gives
\[
t - \int\limits_0^x {\frac{{dy}}{{V\left( y \right)}} = C_1 }  = const,
\]
and thus for $x\leq 0$ we have the general solution
\[
f\left( {x,t} \right) = \W\left( {t - \int\limits_0^x {\frac{{dy}}{{V\left( y \right)}}} } \right),\quad x \le 0
\]
where $\W$ is an arbitrary smooth function.

The characteristics for $x\geq 0$ give
\[
t - \int\limits_0^x {\frac{{dy}}{{V\left( y \right)}} = const = C_2 } ,\quad \quad \frac{{df}}{{dx}} =  - \frac{1}{{V\left( x \right)}}\frac{{dV}}{{dx}}\left( {f - f^* } \right),
\]
and thus for $x\geq 0$ we have the general solution
\[
f\left( {x,t} \right) = \frac{1}{{V\left( x \right)}}\int\limits_0^x {\frac{{dV(y)}}{{dy}}f^* \left( {y,t + \int\limits_x^y {\frac{{dV(z)}}{{dz}}} } \right)dy}  + \frac{1}{{V\left( x \right)}}\Y \left( {t - \int\limits_0^x {\frac{{dV(z)}}{{dz}}} } \right),\quad x \ge 0
\]
where $\Y$ is an arbitrary smooth function. This solution differs from one for $x\leq 0$ by the term $\sim (f-f^*)$ where $f^*$ can be considered as a source term. The requirement that the solution to be continuous at $x=0$ gives $\W=V_0\Y$ where $V_0=V(0)$.

We consider a simple symmetric $V(x)$ function, such that if $x'<0$ is the trapping value then $x=-x^*$ is the value of release from the trapping. Thus, we can write (see the more general case with arbitrary $V(x)$ function in \cite{Artemyev19:pd}):
\[
f^* \left( {x,t} \right) = f\left( { - x,t} \right) = \W\left( {t - \int\limits_0^{ - x} {\frac{{dy}}{{V\left( y \right)}}} } \right)
\]
Using this equation and $\W=V_0\Y$, we obtain for the general solution at $x>0$:
\[
f\left( {x,t} \right) = \frac{1}{{V\left( x \right)}}\int\limits_0^x {\frac{{dV\left( y \right)}}{{dy}}\W\left( {t - \int\limits_0^x {\frac{{dz}}{{V\left( z \right)}}}  - \int\limits_y^{ - y} {\frac{{dz}}{{V\left( z \right)}}} } \right)}  + \frac{{V_0 }}{{V\left( x \right)}}\W\left( {t - \int\limits_0^x {\frac{{dy}}{{V\left( y \right)}}} } \right)
\]

Let $f_0(x)$ be an initial distribution function, $f(x,0)=f_0(x)$. For negative arguments of $\W$ we can define $\W$ through  $f_0(x)$:
\[
f_0 \left( x \right) = \W\left( { - \int\limits_0^x {\frac{{dy}}{{V\left( y \right)}}} } \right)
\]
and thus $\W$ is a bounded function for negative values of its argument.
%Taking into account that $V(x)\to 0$ for $x\to 1$,
We define the function $\W$ for positive argument through the initial distribution as a solution of the  equation
\[
f_0 \left( x \right) = \frac{1}{{V\left( x \right)}}\int\limits_0^x {\frac{{dy}}{{V\left( y \right)}}\W\left( { - \int\limits_0^x {\frac{{dz}}{{V\left( z \right)}}}  - \int\limits_y^{ - y} {\frac{{dz}}{{V\left( z \right)}}} } \right)dz}  + \frac{{V_0 }}{{V\left( x \right)}}\W\left( { - \int\limits_0^x {\frac{{dz}}{{V\left( z \right)}}} } \right),
\]
One can show that this equation determines a unique bounded function $W$.

Let us define
\[
\bar \W = \mathop {\lim }\limits_{x \to 1} \int\limits_0^x {\frac{{dV\left( y \right)}}{{dy}}\W\left( { - \int\limits_0^x {\frac{{dz}}{{V\left( z \right)}} - \int\limits_y^{ - y} {\frac{{dz}}{{V\left( z \right)}}} } } \right)dy}
\]
This is a finite limit, because $dV/dy=O\left(|y \mp1|^{1/4}\right)$ if $y\sim \pm 1$ and $dV/dy=O\left(y\right)$ if $y\sim 0$.

One can show that
$\W(arg)\to-\bar \W/V_0$ for $arg\to +\infty$.

Let us consider the limit $\lim _{t \to \infty } f\left( {x,t} \right)$ at a fixed $x\ne \pm 1$. For $x\leq 0$ we can write
\[
f\left( {x,t} \right) = \W\left( {t - \int\limits_0^x {\frac{{dy}}{{V\left( y \right)}}} } \right) \to  - \frac{{\bar \W}}{{V_0 }} = const,\quad {\rm as}\;t \to \infty
\]
Thus, for $x\geq 0$ we can write
\[
f^* \left( {x,t} \right) \to  - \frac{{\bar \W}}{{V_0 }} = const,\quad {\rm as}\;t \to \infty
\]
and
\[
f\left( {x,t} \right) \to \frac{1}{{V\left( x \right)}}\left( { - \frac{{\bar \W}}{{V_0 }}\int\limits_0^x {\frac{{dV\left( y \right)}}{{dy}}dy}  - \bar \W} \right) =  - \frac{{\bar \W}}{{V_0 }},\quad {\rm as}\;t \to \infty
\]
Therefore,
%we have proved that
$f(x,t) \to const$ as $t \to \infty$, and the asymptotic solution to Eq. (\ref{eq:kinetic1D}) is a constant function $f=const$.  This is an important result, because it theoretically proves that the final stage of the evolution of the electron distribution in the presence of multiple nonlinear resonant interactions is identical to the final stage of a diffusive evolution, that is, a plateau with a null gradient along the resonance curve. Figure \ref{fig16} shows a numerical verification of this result. For the numerical solution, we use  $\area(x)$ given by Eq. (\ref{eq:area_small}), but we do not make any assumption of smallness $\gamma_{\max}-\gamma_{\min}$, i.e., the function $\area(x)$ does not necessarily corresponds to the short wave-packet approximation. Numerical results show that the initially localized maximum of $f(x)|_{t=0}$ quickly (over a time scale $\sim (\B_w/B_0)^{-1/2}$) evolves toward $f=const$.
Therefore, there are two main differences between nonlinear and quasi-linear evolution: (1) the formation of new phase space gradients (like beam structures) due to nonlinear resonant interactions in the transient initial phase, (2) the much shorter time-scale of evolution to the final stage in the case of nonlinear interactions. The first difference is important mainly when we compare very quick phenomena associated with resonant wave-particle interactions which include only a few such interactions; a good example is the phenomenon of microburst precipitation \citep[see details in the Section~\ref{sec:discussion} and in][]{Shumko21,Chen20:microbursts,Kang&Bortnik22:microburst,Chen22:microbursts}. The second difference is more important for long-term radiation belt dynamics \citep[see details in the Section~\ref{sec:nl&ql} and in ][]{Artemyev22:jgr:NL&QL}.

\subsection{Effect of a non-constant bounce period}
Equation (\ref{eq:kinetic1D}) can be generalized for the case when the period between resonant interactions depends on energy. For electron resonance with a monochromatic whistler-mode wave, this period $\tau$ is equal to the half of the bounce period along a magnetic field line (in case of waves propagating away from the equatorial source region). Thus, in dipole magnetic field we have
\[
\tau=\frac{\tau_{b}}{2}=
2\int\limits_0^{s_{\max } } {\frac{{m_e \gamma ds}}{{p_\parallel }}}  = \frac{{2\, m_e }}{{c\sqrt {1 - \gamma ^{ - 2} } }}\int\limits_0^{s_{\max } } {\left( {1 - \sin ^2 \alpha _{eq} \frac{{\Omega _{0} \left( s \right)}}{{\Omega _{0} \left( 0 \right)}}} \right)^{ - 1/2} ds}
\]
where $s_{\max}$ is determined by $\sin^2\alpha_{eq}\Omega_{0}(s_{\max})=\Omega(0)$. Thus, for $h=const$ from Eq. (\ref{eq:h_const}) we have $\tau=\tau(\gamma; h)$:
\[
\tau=
 \frac{{2m_e }}{{c\sqrt {1 - \gamma ^{ - 2} } }}\int\limits_0^{s_{\max } } {\left( {1 - \frac{{2\left( {\gamma  - h/m_e c^2 } \right)}}{{\gamma ^2  - 1}}\frac{{\Omega _{0} \left( s \right)}}{\omega }} \right)^{ - 1/2} ds}
\]
The generalized form of Eq. (\ref{eq:kinetic1D}) can be written as \citep[][]{Artemyev17:arXiv,Artemyev21:book}:
\begin{eqnarray}
 \frac{{\partial f}}{{\partial t}}&=& \frac{1}{2}\frac{\partial }{{\partial \gamma }}\left( {D_{\gamma \gamma } \frac{{\partial f}}{{\partial \gamma }}} \right) - \frac{\partial }{{\partial \gamma }}\left( {\frac{{\left\langle {\Delta \gamma } \right\rangle }}{\tau }f} \right) + \frac{f}{\tau }\frac{{d\left\langle {\Delta \gamma } \right\rangle }}{{d\gamma }}\left( {1 - \Theta } \right) + \frac{{d\left\langle {\Delta \gamma } \right\rangle }}{{d\gamma }}\frac{{f^* }}{{\tau^* }}\Theta \nonumber \\
 \label{eq:kinetic1D_tau}\\
 & =& \frac{1}{2}\frac{\partial }{{\partial \gamma }}\left( {D_{\gamma \gamma } \frac{{\partial f}}{{\partial \gamma }}} \right) - V\frac{{\partial f}}{{\partial \gamma }} - \frac{{dV}}{{d\gamma }}\left( {f - f^* \frac{\tau }{{\tau^* }}} \right)\Theta  + V\frac{{d\ln \tau }}{{d\gamma }}\left( {\frac{\tau }{{\tau^* }}f^* \Theta  - f\left( {1 - \Theta } \right)} \right) \nonumber
\end{eqnarray}
where $V=\langle\Delta\gamma\rangle/\tau$, $f^*=f(\gamma^*)$ and $\tau^*=\tau(\gamma^*)$ with $\area(\gamma)=\area(\gamma^*)$, and $\Theta$ is defined by Eq. (\ref{eq:theta}). Note $\langle\Delta \gamma \rangle=\omega\langle\Delta I\rangle_\xi/m_ec^2$ and $\langle\Delta I\rangle_\xi$ is given by Eq. (\ref{eq:DeltaI_average}). We may introduce a new variable $J$ as
\[
\frac{{d\gamma }}{{dJ}} = \frac{{2\pi }}{{\tau \left( \gamma  \right)}}
\]
Using this new variable, we introduce $\tilde{f}(J,t)$, $D_{JJ}$, $V_J$:
\[
f = \frac{{\tilde f\tau }}{{2\pi }},\quad V   = \frac{{2\pi V_J }}{\tau },\quad D_{\gamma \gamma }  = \frac{{4\pi ^2 D_{JJ} }}{{\tau ^2 }}
\]
Then, we can rewrite Eq. (\ref{eq:kinetic1D_tau}) as
\begin{equation}
\frac{{\partial \tilde f}}{{\partial t}} =  - V_J \frac{{\partial \tilde f}}{{\partial J}} + \frac{1}{2}\frac{\partial }{{\partial J}}\left( {D_{JJ} \frac{{\partial \tilde f}}{{\partial J}}} \right) + \frac{{\partial V_J }}{{\partial J}}\left( {f^* - f } \right)\Theta  \label{eq:kinetic1D_J}
\end{equation}
This equation coincides with Eq. (\ref{eq:kinetic1D}), but instead of $\gamma$ (or $I=m_ec^2\gamma/\omega-const$) we should use the variable $J$. The asymptotic solutions of Eqs. (\ref{eq:kinetic1D}) and (\ref{eq:kinetic1D_J}) are the same. Therefore, the system equations are not changed in the more general case with $\tau_b=\tau_b(\gamma)$, but the energy space is modified: instead of having $\gamma$ linearly proportional to $I_x$ through Eq. (\ref{eq:h_const}), we now have $J$ given by $dJ/d\gamma=\tau_b(\gamma)/2\pi$.

\subsection{Simulation Techniques}\label{sec:mapping}
In this section, we briefly review several possible schemes for the numerical simulation of electron dynamics in a system with multiple nonlinear resonances. We focus on the mapping technique, which is first introduced in the case of a single monochromatic wave, and then generalized for a wave ensemble. But we also compare this technique with the well-developed and quite powerful {\it Green function} approach \cite{Furuya08,Omura15} and with an analytically derived version of the generalized Fokker-Planck equation, which relies on a {\it Probabilistic approach}. All these techniques are based on the same equations and physical concepts, and the main (if not only) difference concerns the numerical implementation of these techniques. Note that, although we do not provide results of simulations of realistic (observed by spacecraft) events in this section, Appendix E includes a detailed analysis of two observational events modeled with the mapping technique.

\subsection{Mapping for a single wave}
Kinetic equations (\ref{eq:kinetic1D}) and (\ref{eq:kinetic1D_tau}) describe the dynamics of the electron distribution function in a system with multiple nonlinear resonances. Since the evolution of the distribution consists of information about multiple electron trajectories, instead of solving the kinetic equation we may solve a large set of equations describing each individual electron trajectory. The most detailed Hamiltonian equations (\ref{eq:hamiltonian}) trace all electron coordinates, fast and slow, but kinetic equations  (\ref{eq:kinetic1D},\ref{eq:kinetic1D_tau}) describe only the electron energy evolution (or $I$) for $h=const$. Therefore, the corresponding equation for electron trajectories should also involve only equations for energy, and should not describe electron motion between resonant interactions. The closest analog of such equations describing electron energy change at resonances is the Chirikov map \cite{bookChirikov87} that should give $\gamma_{n+1}=\gamma_n + \Delta\gamma(\gamma_n)$, where $n$ is the number of resonant interactions \citep[see such maps for electron scattering by whistler-mode waves in][]{Benkadda96,Khazanov14}. For kinetic equations (\ref{eq:kinetic1D},\ref{eq:kinetic1D_tau}) this map can be written as \cite{Artemyev20:rcd,Artemyev20:pop}:
\begin{equation}
\begin{array}{l}
 \gamma _{n + 1}  = \gamma _n  + \left\{ {\begin{array}{*{20}c}
   {\Delta \gamma _{trap} \left( {\gamma _n } \right),} & {\xi _n  \in \left[ {0,\Pi \left( {\gamma _n } \right)} \right)}  \\
   {\Delta \gamma _{bunch} \left( {\gamma _n } \right)} & {\xi _n  \in \left( {\Pi \left( {\gamma _n } \right),1} \right]}  \\
\end{array}} \right. \\
 \Pi \left( {\gamma _n } \right) =  - d\Delta \gamma _{bunch} \left( {\gamma _n } \right)/d\gamma  \\
 \end{array}
\label{eq:map}
\end{equation}
where $\xi_n$ is a random variable with a uniform distribution within $[0,1]$. The value of $\xi_n$ is indeed determined by the phase gain between two resonant interactions, and this gain can be considered as a random variable, which is a nontrivial result. A rigorous proof of $\xi_n$ properties can be found in \cite{Gao23:RCD23}. A simplified version of this proof is provided in Appendix D.

\begin{figure}
\centering
\includegraphics[width=1\textwidth]{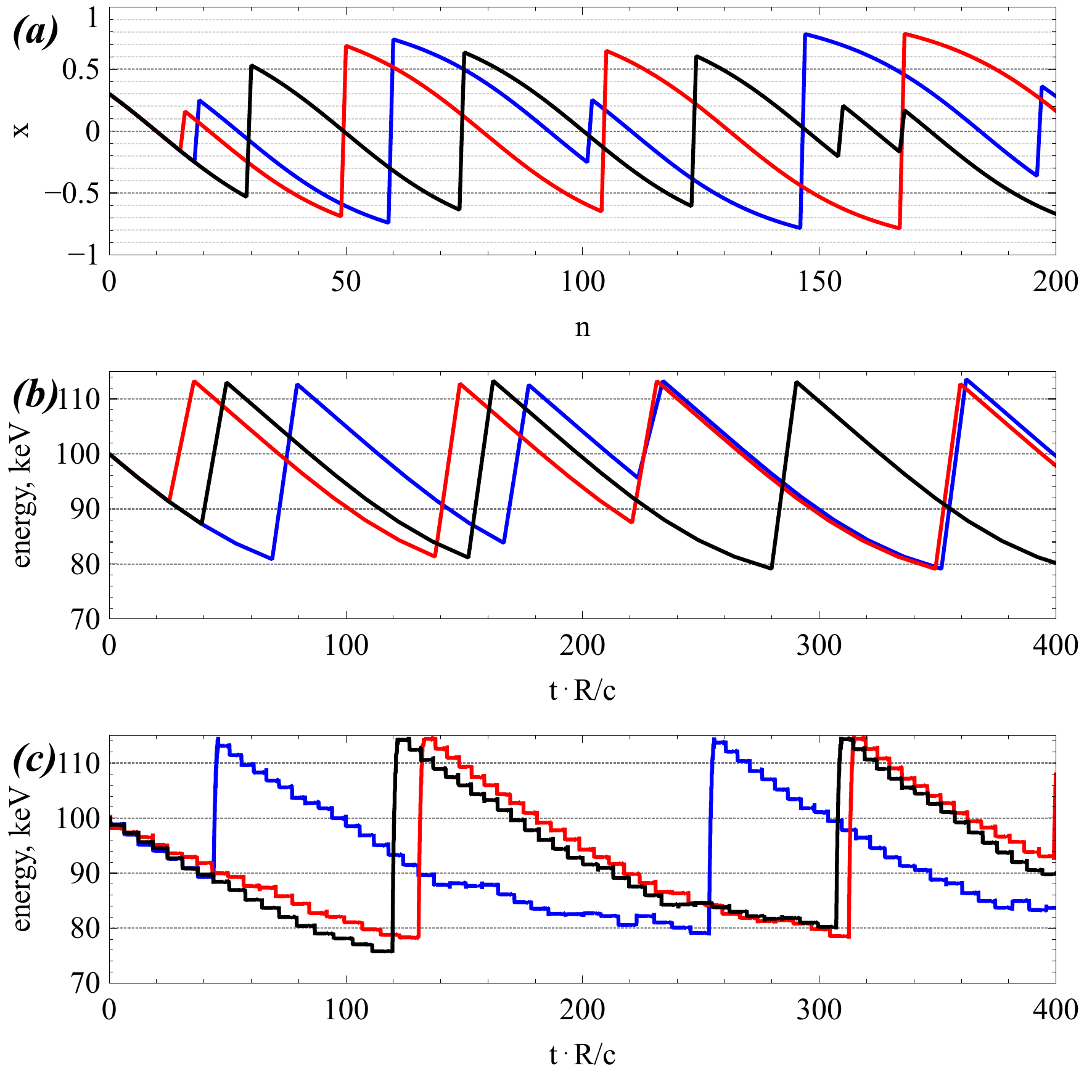}
\caption{Panel (a) shows three trajectories $x_n$ for map (\ref{eq:map_simple}) with $\Delta x_0=10^{-3/2}$. Panel (b) shows three trajectories $m_ec^2(\gamma_n-1)$ for map (\ref{eq:map}). Panel (c) shows three trajectories  obtained by numerical integration of Hamiltonian equations (\ref{eq:hamiltonian}). Results in panels (b,c) are obtained for initial energy $100$ keV, initial pitch-angle $\alpha_{eq}=60^\circ$, wave amplitude $300$ pT, and $L$-shell$=6$. Other systems parameters are the same as in Fig. \ref{fig02}. }
\label{fig16}
\end{figure}

For the simplified model  (\ref{eq:area_small}) this map can be rewritten as
\begin{equation}
\begin{array}{l}
 x_{n + 1}  = x_n  + \left\{ {\begin{array}{*{20}c}
   { - 2x_n ,} & {\xi _n  \in \left[ {0,\Pi \left( {x_n } \right)} \right)}  \\
   { - \Delta x_0  \cdot \left( {1 - x_n^2 } \right)^{5/4} } & {\xi _n  \in \left( {\Pi \left( {x_n } \right),1} \right]}  \\
\end{array}} \right. \\
 \Pi \left( {x_n } \right) =  - \Delta x_0 \frac{5}{2}x_n \left( {1 - x_n^2 } \right)^{1/4}  \\
 \end{array}
\label{eq:map_simple}
\end{equation}
with $\Delta\gamma_{bunch}=-S/2\pi$ and $\Delta x_0=(S_0/2\pi)\propto \sqrt{\B_w/B_0}$.
Figure \ref{fig16}(a) shows several examples of $x_n$ trajectories obtained with the map (\ref{eq:map_simple}). The dynamics of $x$ consists of rare and large jumps (due to phase trapping) with $x$ increase and regular $x$ drift to smaller values (due to phase bunching). This dynamics consists of basic elements, phase trapping and phase bunching, resembling well electron energy changes due to a single resonant interaction, see Fig. \ref{fig03}. The fine balance between trapping and bunching (the probability of trapping $\Pi=(2\pi)^{-1}d\area/dx$, and magnitude of bunching $\Delta x=-\area/2\pi$) results in a quasi-periodical $x$ motion between small ($x\sim -1$) and large ($x\sim 1$) values.

For realistic systems, $\area(\gamma)$ should be derived based on actual wave field and background magnetic field latitudinal profiles. Using such realistic $\area(\gamma)$, we plot $\gamma_n$ trajectories in Fig. \ref{fig16}(b) and compare them with trajectories obtained by direct numerical integration of Hamiltonian equations (\ref{eq:hamiltonian}), which are plotted in Fig. \ref{fig16}(c). This comparison demonstrates that the map (\ref{eq:map}) describes well the main constitutive elements of the dynamics of $\gamma$. Note that the map (\ref{eq:map}) includes a significant randomization factor (the random $\xi$) and, therefore, we cannot expect a one-to-one similarity (at each time) between the $\gamma_n$ profiles obtained from the map and from numerical integration of Hamiltonian equations, even if both trajectories start with the same initial conditions. However, the most important point is that the statistical properties of the $\gamma_n$ dynamics are the same for trajectories obtained by the mapping technique and by direct numerical integration of Hamiltonian equations.

\subsection{Nonlinear resonances with multiple waves}
\begin{figure}
\centering
\includegraphics[width=1\textwidth]{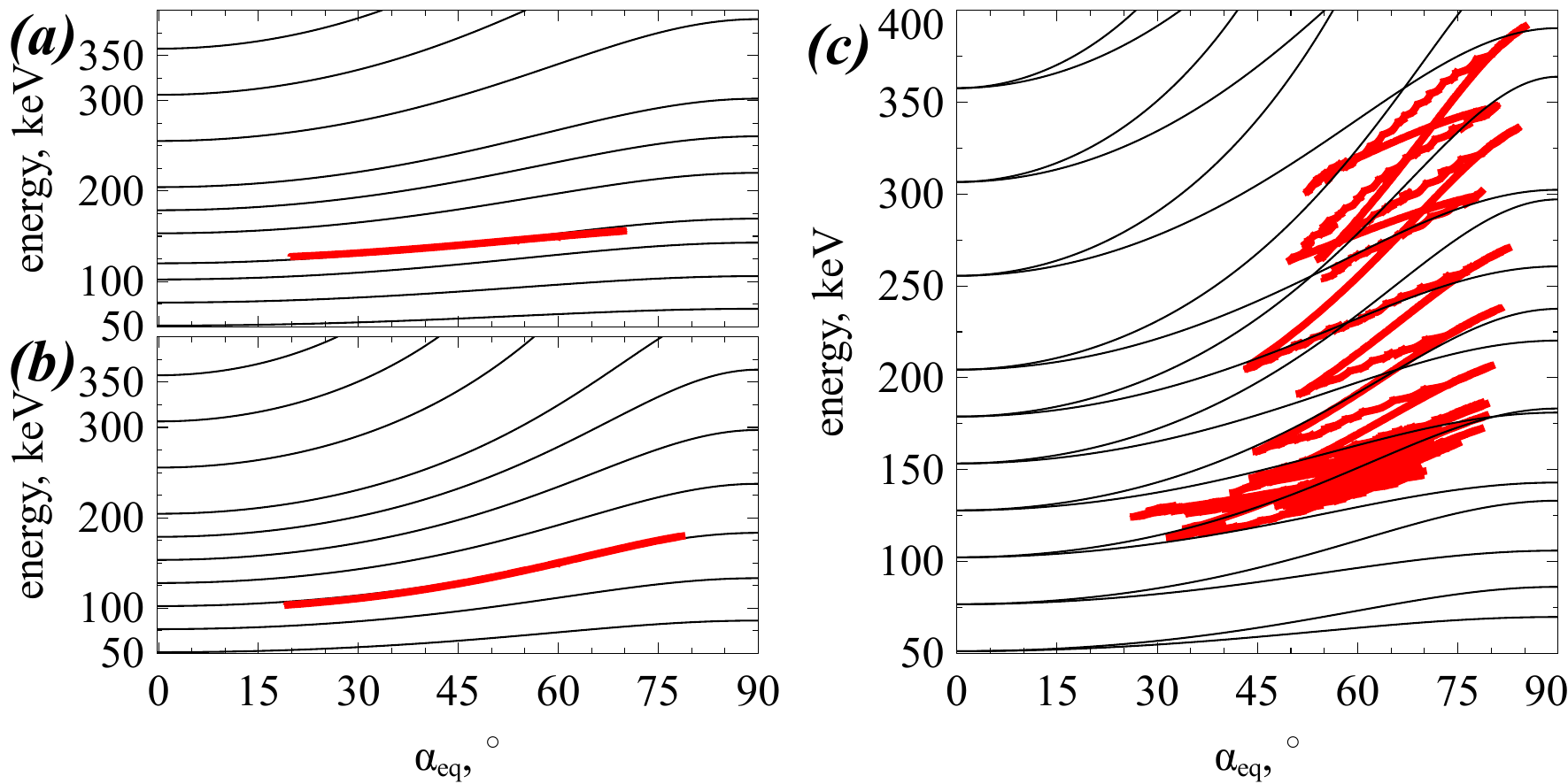}
\caption{Top panels show resonance curves (black) and electron trajectories in the energy/pitch-angle space for Hamiltonian (\ref{eq:hamiltonian}), when the electron interacts only with the first whistler-mode wave (a), only with the second whistler-mode wave (b), or with both whistler-mode waves (c). Wave frequencies are $\omega_1=0.4\Omega_{eq}$, $\omega_2=0.2\Omega_{eq}$, both wave amplitudes are $300$ pT, $L=6$, and other system parameters are from Fig. \ref{fig02}.}
\label{fig17}
\end{figure}

Kinetic equations (\ref{eq:kinetic1D}) and (\ref{eq:kinetic1D_tau}) have been derived for a 1D system with $h=const$, i.e., for a system including only one monochromatic wave. If we consider a more realistic situation where electrons interact resonantly with various waves having different frequencies, we cannot use this 1D approximation, because for each wave frequency we will correspond to a different $h$ given by Eq. (\ref{eq:h_const}). This means that the frequency value determines the shape of resonance curves in the velocity, energy space, and for different frequencies the wave-particle resonance move electrons along different curves. Moreover, $I$ also depends on wave frequency (or wave number $k$), and for each wave frequency we have a corresponding $I$, i.e., in the system with two wave frequencies we have $I_1$ and $I_2$. During the resonant interaction with the first wave, $I_1$ changes but $I_2$ is conserved, and vice versa. Therefore, for each resonance electrons move along the corresponding resonance curve. The conservation of $h$ and of one of the momenta ($I_1$ or $I_2$) leads to a one-dimensional electron dynamics in the energy/pitch-angle space, and if there is only one wave in the system electrons never leave the corresponding single resonance curve. However, electron dynamics becomes 2D in the presence of two waves, when both $I_1$ and $I_2$ change. Figure \ref{fig17} illustrates this effect by showing electron resonant interactions with a single wave and with two waves. The electron moves along resonance curves and jumps between these curves due to $I_{1,2}$ jumps. Accumulation of such jumps between resonance curves ultimately leads a single electron trajectory to cover the entire energy/pitch-angle space \citep[see more details in][]{Artemyev21:jpp}.

As emphasized earlier, to generalize kinetic equations (\ref{eq:kinetic1D}) and (\ref{eq:kinetic1D_tau}) for systems with multiple waves, we need to use a probabilistic approach. We introduce a probability distribution function $\mP(B_w,\omega)$ that determines the wave characteristics $(\omega,B_w)$ during the next resonant interaction. And then we average all operators in these equations over $\mP$. Since such averaging significantly complicates the kinetic equations, we need an alternative approach. Let us discuss three possible methods for modeling the evolution of the electron distribution due to multiple resonances with a wave ensemble. Although we are speaking about a wave ensemble, it is assumed that electrons interact resonantly with only one monochromatic wave at a time, without the wave resonance overlap \citep[see discussion in][]{Tao13,An22:Tao,Gan22}, while the resonant waves may have different properties during different bounce period.

\begin{figure}
\centering
\includegraphics[width=1\textwidth]{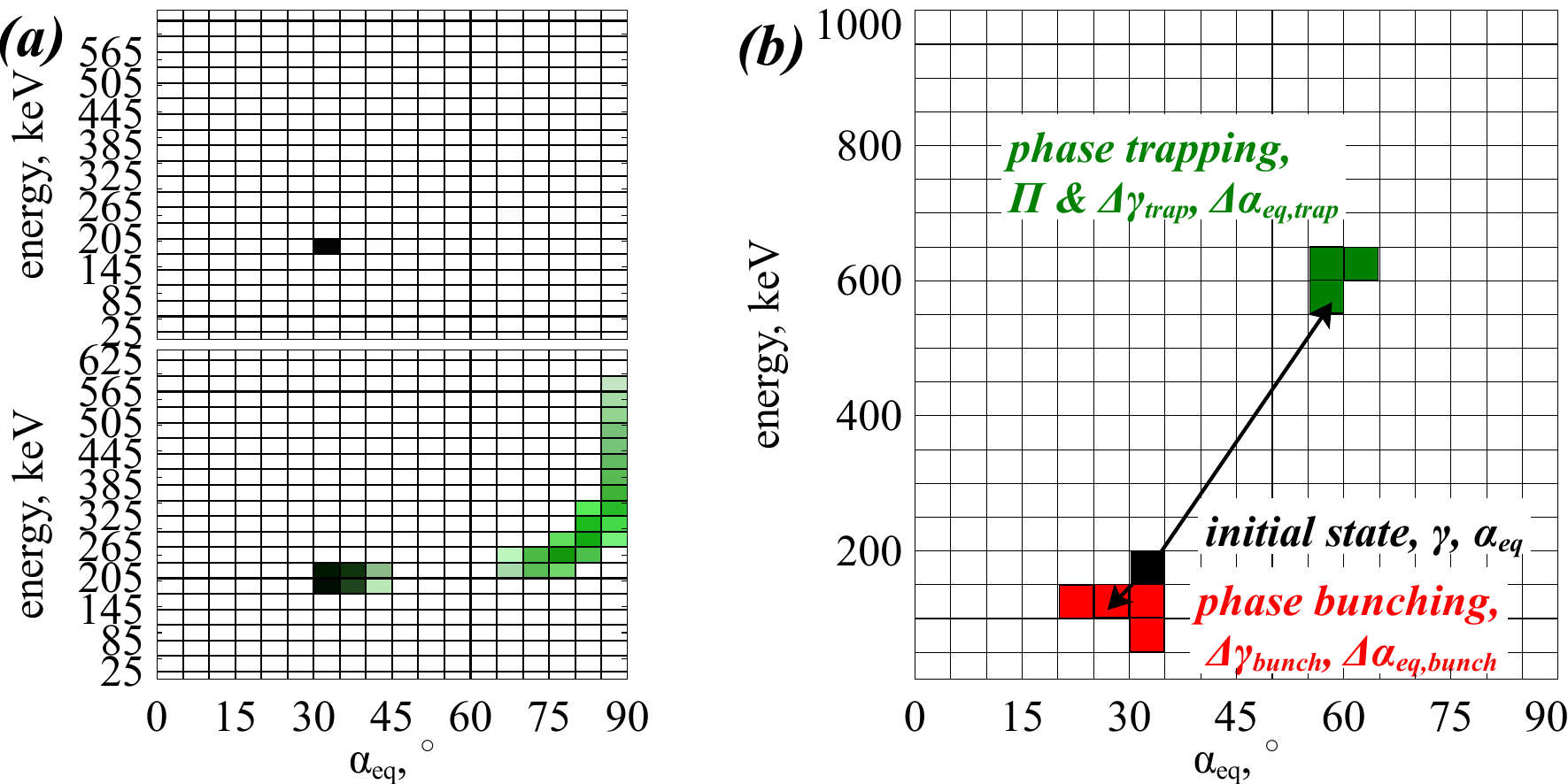}
\caption{(a) Schematic view of the Green function approach. \red{Numerical simulations} are performed to obtain energy, pitch-angle distribution of electrons after resonant interactions (bottom panel) for electrons having the same initial energy, pitch-angle (top panel). Such distributions, obtained for different wave characteristics and for all energy, pitch-angle bins are combined into the Green function $ G\left( {\left. {E,\alpha _{eq} } \right|E',\alpha '_{eq} } \right)$ that describes the probability of $(E',\alpha_{eq}')\to(E,\alpha_{eq})$ transition \citep[see details in][]{Omura15}. (b) Schematic view of the probabilistic approach, where for each energy, pitch-angle bin the analytical model (for a given distribution of wave characteristics) provides the probability $\Pi$ and energy, pitch-angle changes due to phase trapping (green) and due to phase bunching (red) \citep[see details in][]{Vainchtein18:jgr}.  }
\label{fig18}
\end{figure}

\paragraph{Green function approach}
The Green function approach \cite{Furuya08,Omura15} assumes that systems with multiple different waves (or with a single wave with evolving characteristics) can be described using Eq. (\ref{eq:smolukhovskiy}) with a kernel derived from test-particle simulations. Figure \ref{fig18}(a) shows the basic scheme of this approach: test particle simulations provide the probability distribution function of energy and pitch-angle changes, and this function is used to construct the kernel and rewrite the Smoluchowski equation as
\[
\begin{array}{l}
 f_{n + 1} \left( {E,\alpha _{eq} } \right) =
\int\limits_0^\infty  {dE'\int\limits_0^{\pi /2} {f_n \left( {E',\alpha '_{eq} } \right)G\left( {\left. {E,\alpha _{eq} } \right|E',\alpha '_{eq} } \right)d\alpha '_{eq} } }\\
G\left( {\left. {E,\alpha _{eq} } \right|E',\alpha '_{eq} } \right)= \C\left[ {\delta \left( {E - E',\alpha _{eq}  - \alpha '_{eq} } \right)} \right] \\
 \end{array}
\]
where $\partial f/\partial t$ is replaced by a discrete difference $f_{n+1}-f_n$ during \red{each time step of one bounce period}, and $f_n$ is included into the Green function
\[
\int\limits_0^\infty  {dE'\int\limits_0^{\pi /2} {G\left( {\left. {E,\alpha _{eq} } \right|E',\alpha '_{eq} } \right)d\alpha '_{eq} }  = 1}
\]
while $\C$ is the particle scattering operator describing energy and pitch-angle change during \red{a single bounce period}. This operator can be obtained from test particle simulations: a large particle ensemble can be traced across the resonance and the variations of their energy and pitch-angles can be combined to determine the probabilities of all possible transports $(E', \alpha_{eq}')\to(E, \alpha_{eq})$ \cite{Furuya08}. This is quite a powerful approach for quantitatively describing multiple nonlinear resonant interactions affecting the dynamics of energetic electron fluxes. Several important results have been obtained this way, like that description of the formation of relativistic/ultra-relativistic electron populations due to turning acceleration by field-aligned waves \cite{Omura15} or due to the Landau and high-order resonances with oblique waves \cite{Hsieh&Omura17,Hsieh20}, and investigations of rapid scattering and losses of energetic electrons due to a combination of the Landau and cyclotron nonlinear resonances \cite{Hsieh22,Hsieh&Omura23}. These simulations demonstrated that the Green function approach is really promising and, combined with the observed wave distributions, it can potentially replace and supersede standard simulations of radiation belt dynamics based on the Fokker-Planck diffusion equation. The main technical difficulty of this approach is the need to predefine the resolution in energy and pitch-angle for the $\C$ operator that will be determined using test particle simulations. For instance, typical nonlinear wave-particle interactions with intense waves often cover at least three order of magnitude of energies, $[1,1000]$ keV, whereas a simultaneous inclusion of electron scattering by weak waves requires a minimum energy bin size about $10-100$eV \citep[see typical diffusion rate magnitudes in][]{Glauert&Horne05,Horne13:jgr}. Therefore, to accurately incorporate the effects from both intense and weak waves, one would need $10^4 - 10^5$ energy bins and about the same number of pitch-angle bins. Thus filling the corresponding $2D$ matrix for $\C$ is computationally very expensive. Consequently, the Green function approach is useful mostly for describing electron flux dynamics in a system with intense waves (during active geomagnetic conditions), when the energy and pitch-angle bin sizes can remain sufficiently large. The two alternative methods described below, the {\it Probabilistic approach} and the {\it Mapping technique}, would require similarly huge numbers of small energy and pitch-angle bins for an accurate description of electron dynamics in the presence of both intense and weak waves, but as we will see, with potentially different intrinsic accuracy and total CPU time. The main differences of the ({\it Probabilistic approach} and {\it Mapping technique}) from the {\it Green function approach} consists in the analytical evaluation of the basic properties (characteristics) of wave-particle resonant interactions. This improves the accuracy of the evaluation of such characteristics, but reduces the flexibility for including comprehensive details of wave-particle interactions (like wave frequency drift and wave-packet structure). Roughly speaking, the {\it Green function approach} is ideal for a detailed modeling of short-term dynamics of electron fluxes, when peculiarities of wave-field can play the most important role, the {\it Probabilistic approach} is optimal for the inclusion of nonlinear resonant effects into existing numerical schemes of radiation belt dynamics (an even simpler technique for such an inclusion is discussed in Section \ref{sec:nl&ql}), and the {\it Mapping technique} is the most suitable for observation-based modeling of meso-scale events, when wave characteristics are derived from spacecraft observations, and for incorporation of wave-particle resonant interactions into global test-particle simulations (see discussion in \citep{Artemyev22:jgr:DF&ELFIN} and in Section \ref{sec:sde&mapping}).

\paragraph{Probabilistic approach}
An alternative to the Green function approach and a different way of rewriting the Smoluchowski equation was proposed in \citet{Vainchtein18:jgr}. This approach is based on the idea of separating phase trapping and bunching processes and their analytical evaluations. In this case, the discretization of the electron energy, pitch-angle space allows constructing a matrix $\hat R$ of energy, pitch-angle probability jumps. Within this approach, Eq. (\ref{eq:smolukhovskiy}) takes the form
\begin{equation}
\frac{{\partial f_{ij} }}{{\partial t}} =  - \frac{{N_{ij} }}{{\tau _{ij} }}\int\limits_W {\mP\left( W \right)dW}  + \left( {\int\limits_W {\sum\limits_{k,l} {\hat M_{ij}^{kl} \left( W \right)\mP\left( W \right)dW} } } \right)f_{ij} \label{eq:mitia}
\end{equation}
where $f_{i,j}=f(E_i,\alpha_{eq,j})$, $N_{ij}=N(E_i,\alpha_{eq,j})$ is the number of resonant interactions that particles undergo during a single bounce period $\tau_{ij}=\tau(E_i,\alpha_{eq,j})$, $W=(\B_w,\omega)$ are wave characteristics and $\mP(W)$ is the probability distribution function of wave characteristics normalized in such a way that $\int_W\mP(W)dW$ is the ratio of the total time interval of spacecraft wave measurements to the cumulative time interval of observations of intense waves resonating with electrons nonlinearly. The operator $\hat M_{ij}^{kl}(W)$ is a 4D matrix describing the transformation of the 2D matrix $f_{ij}$ of initial energy, pitch-angle to a $f_{kl}$ matrix of energy, pitch-angle after one bounce period.

A schematic view of this probabilistic approach is provided in Fig. \ref{fig18}(b): for each energy, pitch-angle bin and each pair of wave characteristics $(\B_w, \omega)$, we determine the probabilities of particle coming and leaving due to the phase trapping and phase bunching. These probabilities are determined using the analytical model of wave-particle nonlinear interaction and, therefore, do not require large numerical test-particle simulations. Next, the matrices $\hat R(W)$ are averaged over the probability distributions of wave characteristics. The averaged matrix is used in the equation for the electron distribution function. The main advantage of this approach is that it is based on an analytical model of energy changes and trapping probabilities that can be rapidly evaluated over a wide parametric domain with arbitrarily high accuracy. The main disadvantage is that such analytical equations do not allow a simple generalization for more sophisticated and realistic waveforms (see discussion in Section~\ref{sec:packets}). However, even such an idealized approach can still provide important information about systems with nonlinear wave-particle interactions. Specifically, numerical simulations of electron distribution dynamics with realistic and wave distributions, $\mP(\B_w,\omega)$, have shown the importance of appropriately choosing the initial electron distributions \cite{Vainchtein18:jgr}. The evolution of the electron distribution function due to nonlinear wave-particle interactions \red{leads} to a rapid change of the total electron energy, but the energy of a whistler-mode wave is not sufficient in itself to provide such an \red{deceleration/acceleration} \citep[see discussion and estimates in][]{Shklyar11:angeo}. Therefore, the large population of scattered (decelerated) electrons provides the only available energy source for trapped electron acceleration. The corresponding energy balance imposes several constraints on the initial electron distribution function for models describing nonlinear interactions. This initial distribution should be consistent with the conservation of the total energy for solutions of Eq.~(\ref{eq:mitia}). \citet{Vainchtein18:jgr} checked the total particle energy variation for different initial pitch-angle distributions: an initial distribution with transverse anisotropy ($F \propto \sin\alpha_{eq}$) immediately loses energy, whereas a field-aligned anisotropic distribution ($f \propto \cos\alpha_{eq}$) first gains energy before losing it. More specific distributions (butterfly $f \propto \sin2\alpha_{eq}$ sometimes observed in the radiation belts \citep[see statistics in][]{Asnes05} and a distribution with field-aligned beams $f \propto \sin\alpha_{eq}\cos^4\alpha_{eq}$ observed during particle injections \citep[e.g.,][]{Mozer16}) show energy increase during the first stage of the evolution. This suggests that for a $f$ to result in total energy conservation it should be more isotropic (i.e., an intermediate state between transverse anisotropy
$\sim\sin\alpha_{eq}$ and field-aligned anisotropy $\sim \cos\alpha_{eq}$). Moreover, under realistic conditions, particle anisotropy is energy dependent, and can be different for $<100$ keV particles and $\sim 1$ MeV particles. Therefore, further investigations of particle distributions consistent with the assumption of total energy conservation are needed to accurately model nonlinear wave-particle interactions \citep[see, also,][for details on the relation between trapped and scattered electron populations in a system with small variations of wave intensity]{Shklyar17,Shklyar&Luzhkovskiy23}.

\begin{figure}
\centering
\includegraphics[width=1\textwidth]{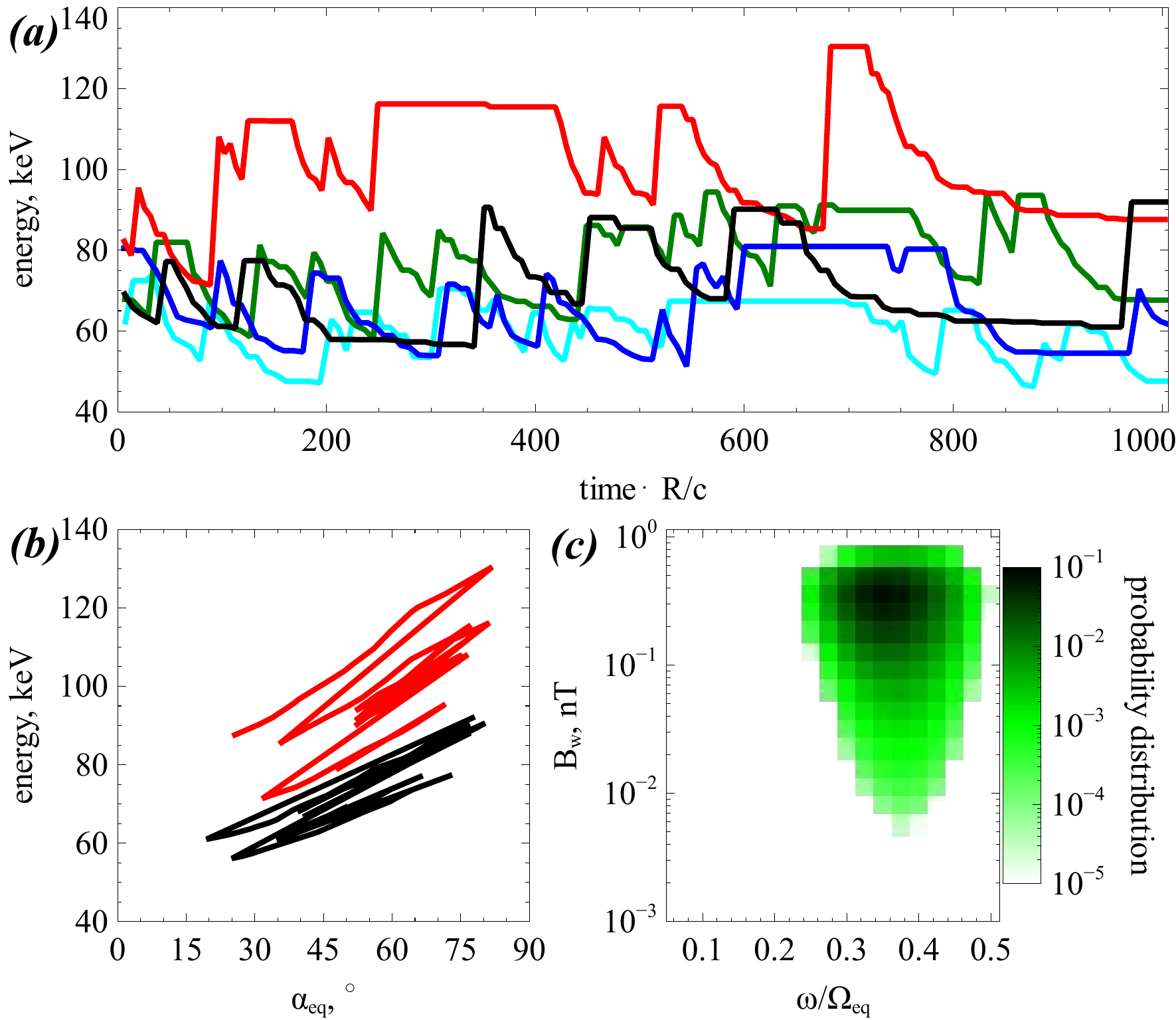}
\caption{Panel (a) shows five examples of electron trajectories (energy vs. time) for cyclotron resonance with waves having the distribution shown in panel (c). All electrons have the same initial energy and pitch-angle. Panel (b) shows two trajectories from panel (a) in the (energy, pitch-angle) space. Background magnetic field and plasma density characteristics are the same as in Fig. \ref{fig02}.}
\label{fig19}
\end{figure}

\paragraph{Mapping technique}
The mapping technique can easily be generalized to multi-wave systems: instead of Eq. (\ref{eq:map}) for a single wave, we should use
\begin{equation}
\gamma _{n + 1}  = \gamma _n  + \left\{ {\begin{array}{*{20}c}
   {\Delta \gamma _{trap} \left( {\eta _n ,\gamma _n } \right),} & {\xi _n  \in \left[ {0,\Pi \left( {\eta _n ,\gamma _n } \right)} \right)}  \\
   {\Delta \gamma _{bunch} \left( {\eta _n ,\gamma _n } \right),} & {\xi _n  \in \left( {\Pi \left( {\eta _n ,\gamma _n } \right),1} \right]}  \\
\end{array}} \right.
\label{eq:map_multi}
\end{equation}
where the index $\eta_n$ is a random number with a probability distribution such that $\eta$ determines the probability of resonant interaction with different waves, which characteristics are given by the distribution $\mP(B_w,\omega)$. Thus, before each resonant interaction (each $\gamma$ change) with a particular particle, we determine from $\mP(B_w,\omega)$ the corresponding wave, and use the corresponding $\area$ to determine energy, pitch-angle change for this particle. Particle pitch-angles should be recalculated with new $\gamma_{n+1}$ and corresponding wave frequency $\omega$ accordingly to Eq. (\ref{eq:h_const}). Figure \ref{fig19}(a) shows several examples of electron trajectories evaluated with the mapping (\ref{eq:map_multi}) for the wave probability distribution $\mP(B_w,\omega)$ from panel (c). This distribution contains waves of different frequencies, and for each frequency the resonant interaction will move electrons along the $h(\omega)=const$ curve (see Eq. (\ref{eq:h_const})). Phase bunching moves electrons along $h=const$ curves with energy and pitch-angle decrease (see the prolonged intervals with energy decrease in panel (a)), whereas phase trapping moves electrons along the same curves, but with an energy increase (see the rare positive energy jumps in panel (b)). For a single frequency system, there would be only one curve $h=const$ for each initial electron energy, pitch-angle, and thus electrons would move along this curve in the energy, pitch-angle space. But the system in Fig. \ref{fig19} contains multiple frequencies and, thus, resonant interactions will provide complex electron dynamics in the 2D energy, pitch-angle space. Figure \ref{fig19}(b) shows two trajectories from panel (a) in the energy, pitch-angle space: drifting from one resonant curve to another one, electrons may increase their energy with a net pitch-angle decrease (see also Fig. \ref{fig17}(c)), which is purely due to resonant interactions with different waves at multiple frequencies. Note that the distribution shown in Fig. \ref{fig19}(c) contains many waves with an amplitude insufficiently large to resonate with a particular electron nonlinearly. Such waves should provide electron diffusive scattering, which is not included into the mapping given by Eq. (\ref{eq:map_multi}). Such diffusive scattering can be incorporated into the mapping scheme using standard (Chirikov-type, see \cite{bookChirikov87}) maps. Examples of such maps can be found in \cite{Vasiliev88,Zaslavskii89:jetp,Khazanov13,Khazanov14}. %In Appendix E we discuss the potential importance of such a diffusive scattering in systems with nonlinear resonant interactions.

The map (\ref{eq:map_multi}) can be used with a 3D wave characteristic distribution $\mP=\mP(B_w,\omega,\theta)$, but then, we need to use a different Eq. (\ref{eq:h_const}) for a different resonance number. Generally two main resonances, Landau resonance with very oblique waves ($\Nr=0$) and the first cyclotron resonance with quasi-parallel waves ($\Nr=1$), are included \cite{Artemyev21:jpp}. The difference is in relation of $I$ from Eq.  (\ref{eq:h_const}) and the electron magnetic moment $I_x$: for $\Nr=0$ Eq. (\ref{eq:h_const}) becomes $I_x=const$, whereas for $\Nr=1$ Eq. (\ref{eq:h_const}) becomes $m_ec^2\gamma-\omega I_x=const$.

\begin{figure}
\centering
\includegraphics[width=1\textwidth]{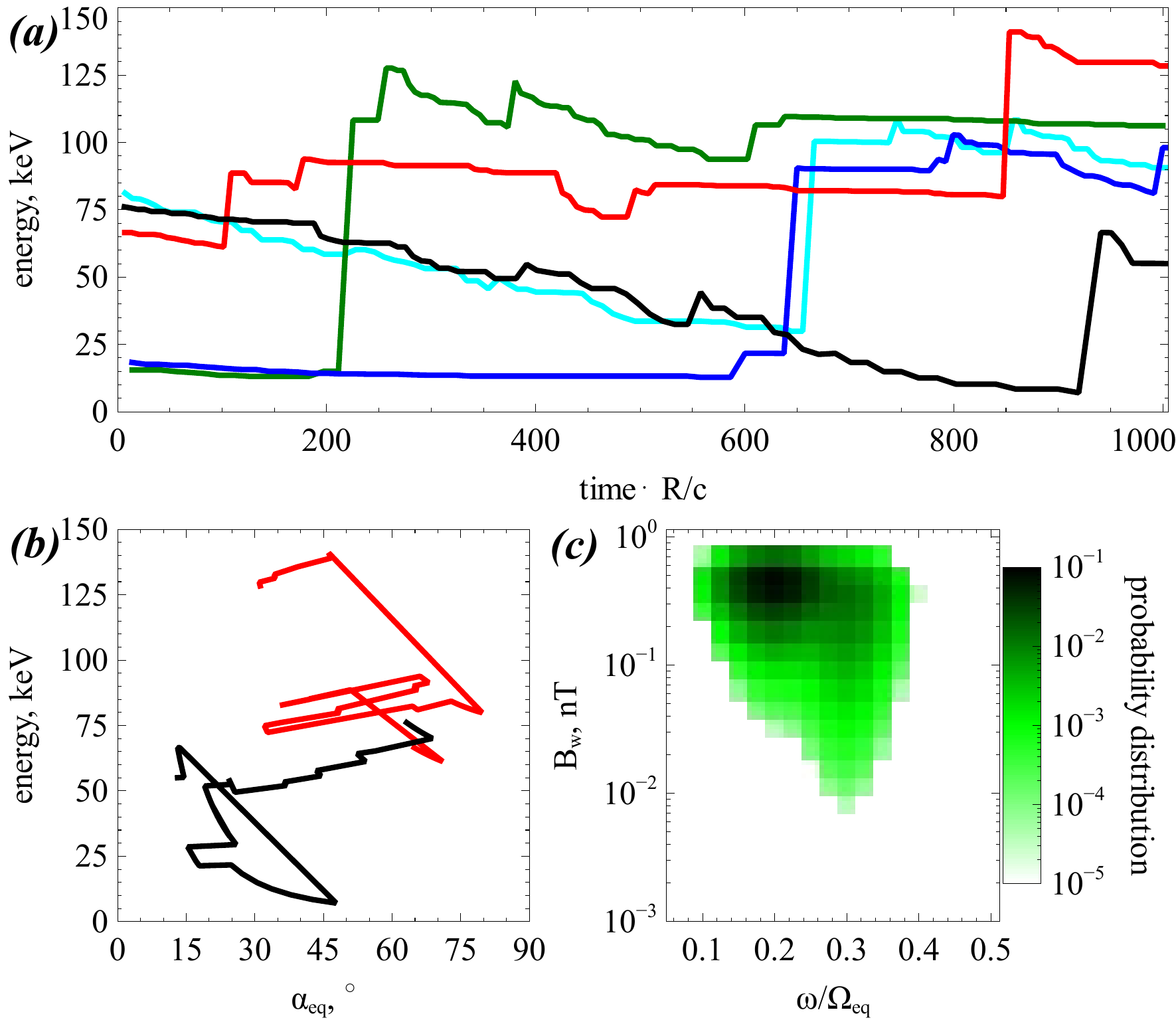}
\caption{Panel (a) shows five examples of electron trajectories (energy vs. time) for Landau and cyclotron resonances with waves having the distribution shown in panel (c). All electrons have the same initial energy and pitch-angle. Panel (b) shows two trajectories from panel (a) in the (energy, pitch-angle) space. Background magnetic field and plasma density characteristics are the same as in Fig. \ref{fig02}.}
\label{fig20}
\end{figure}

Using a similar approach with a probability distribution for the waves, we may generalize the map (\ref{eq:map_multi}) for a system with both types of whistler-mode waves: quasi-parallel and very oblique. In this case, before each resonant interaction, we determine the wave type, which in turn determines the type of resonance -- Landau or cyclotron. Figure \ref{fig20}(a) shows examples of electron trajectories in the case where the probability for electrons to meet very oblique and quasi-parallel waves is the same. Electron energy dynamics resembles results obtained for a system including only cyclotron resonance (see Fig.\ref{fig19}(a)): prolonged periods of energy decrease due to the phase bunching (due to both Landau and cyclotron resonances) are intermittently interrupted by rare positive jumps of energy due to phase trapping (related to both Landau and cyclotron resonances). The main difference with the cyclotron-resonance-only system (including only parallel waves) from Fig. \ref{fig19} is that energy increases/decreases in Landau resonance correspond to pitch-angle decreases/increases, such that Landau phase trapping, in particular, results in simultaneous acceleration and pitch-angle decrease. Together, nonlinear wave-particle interactions through cyclotron and Landau resonances with parallel and oblique waves provide both significant electron acceleration and efficient transport toward the loss-cone (see Fig.\ref{fig20}(b) and compare with Fig.\ref{fig19}(b)) and, therefore, lead to an enhanced precipitation of energetic (accelerated) electrons compared with a system including only cyclotron resonance with parallel waves \citep[see more details of this effect in][]{Mourenas16,Hsieh22}.

Such mapping simulations of individual electron trajectories (energy as a function of the resonant interaction number or time) can be combined to comprehensively describe the full dynamics of the electron distribution function \citep{Artemyev21:jpp,Artemyev22:jgr:Landau&ELFIN,Zhang22:natcom}. Therefore, the mapping technique may provide the same level of system description as the probabilistic approach and Green function approach.

\section{Effects of short wave-packets}\label{sec:packets}
\begin{figure}
\centering
\includegraphics[width=1\textwidth]{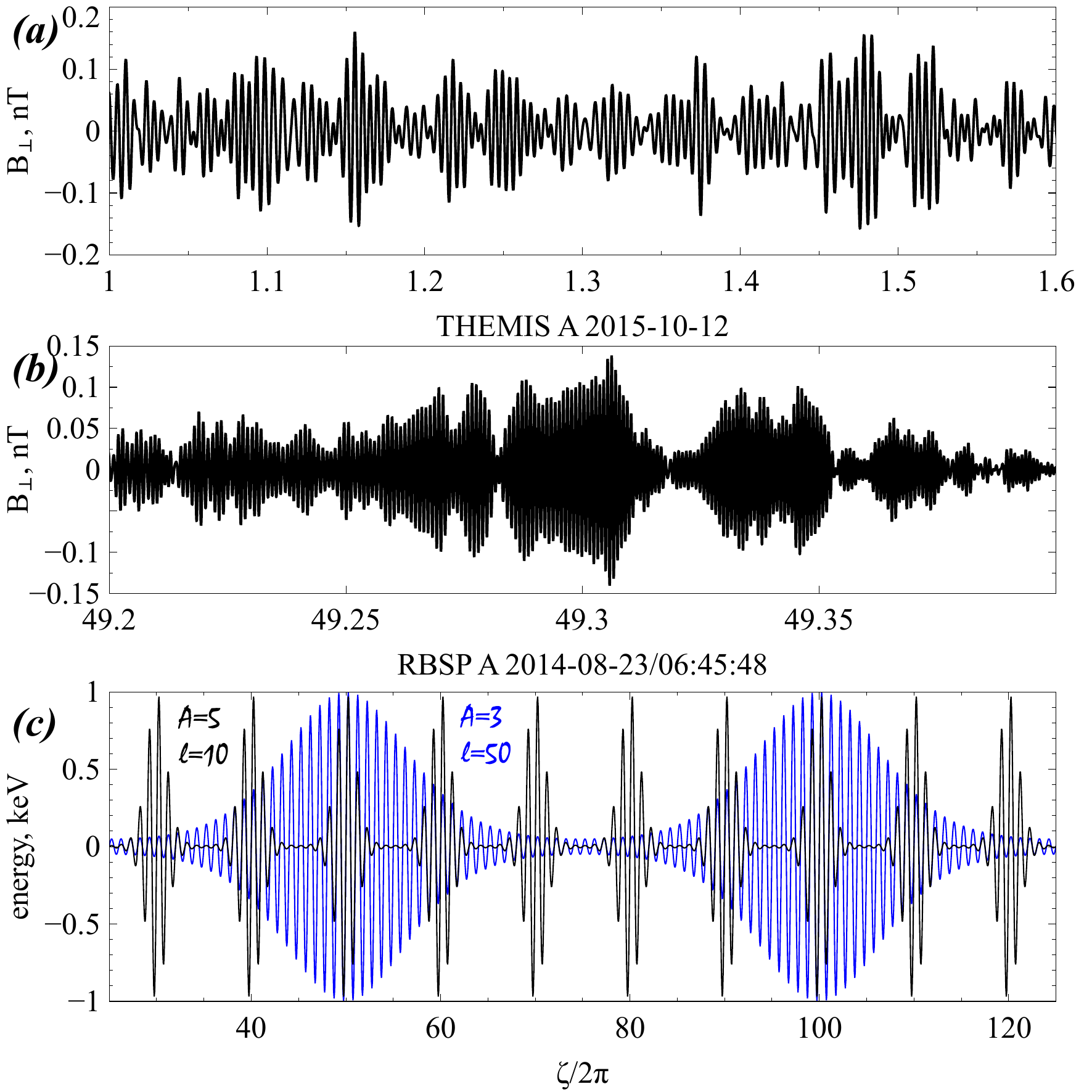}
\caption{Panel (a,b) shows four examples of whistler-mode wave-packets propagating with a small wave normal angle: measurements from THEMIS \cite{LeContel08} and Van Allen Probes \cite{Kletzing13} search-coil magnetometers ($L$-shell and $MLT$ are indicated within panels).Panel (c) shows two examples of model wave-packets.}
\label{fig21}
\end{figure}

The general theoretical results described by Eqs. (\ref{eq:kinetic1D}) and (\ref{eq:map_multi}) were obtained in the simplified case of a plane wave (such that the term $\sim B_w$ contains only $\sin\zeta$ in Hamiltonian (\ref{eq:hamiltonian})). However, in realistic plasma systems, such very long wave-packets (such that the wave packet envelope can be neglected) are rarely observed, and most of the whistler-mode waves present in the outer radiation belt are propagating in the form of short wave-packets (see examples in Fig. \ref{fig21}(a,b) and \citep{Zhang18:jgr:intensewaves,Zhang19:grl}). Typical wave-packets (or subpackets) comprise several (up to ten) wave periods and reach moderate peak amplitudes ($\B_w\approx 100-300$ pT), but the most intense ($\B_w>500$ T) wave-packets can be quite long and may include more than hundred wave periods. The formation of such long wave-packets is associated with the classical nonlinear mechanism of whistler-mode chorus wave generation \citep[see reviews in][]{Demekhov17,Omura13:AGU,Omura21:review,Tao20,Tao21}, and it is also the case for moderately long but intense subpackets \cite{HChen24}. On the other hand, most of the moderate intensity short wave-packets are likely generated due to wave superposition, also called wave beating \citep[see details in][]{Zhang20:grl:frequency,Nunn21,Mourenas22:jgr:ELFIN,An22:Tao}. Detailed comparison of wave generation models and spacecraft statistics of whistler-mode wave-packets confirms this scenario of separation of short and long wave-packets \cite{Zhang21:jgr:data&model}.  Note that basic empirical models of whistler-mode waves do not include information about wave-packets \cite{Meredith12,Agapitov13:jgr}, and this significantly complicates the investigation of nonlinear resonant effects \citep[see discussion in][]{Allanson24}. In this review, we utilize wave-packet statistics published in \cite{Zhang19:grl,Zhang20:grl:frequency}.

To model this effect for a monochromatic wave, we may adopt the scheme proposed in \cite{Chapman00:Matsoukis,Chapman01:Wykes} and rewrite the field term in Eq. (\ref{eq:hamiltonian}) as
\[
\B_{w}(\lambda)\sin\zeta \to \B_{w}(\lambda)\exp\left(-\a\sin^2\left(\phi/2\ell\right) \right)\sin\zeta
\]
where parameter $\a$ determines the {\it depth} of wave field modulation, $\ell$ is the wave-packet size in wave length (wave periods) units. Figure \ref{fig21}(c) shows examples of this wave-packet model.

An alternative and more sophisticated wave-packet model for whistler-mode chorus waves (with frequency drift) has been proposed in \citet{Furuya08} and actively used for investigation of the effect of wave field modulation on efficiency of electron resonant dynamics \citep{Tao12:GRL,Tao13,Gan20:grl}. This model includes the difference between group and phase wave velocities, and thus describes wave-packet modification during the propagation, because different waves contributing to each wave-packet propagate with different velocities. \citet{Kubota&Omura18, Hiraga&Omura20, Hsieh20} have investigated in detail this chorus wave-packet model and incorporated it into a general scheme of electron flux evaluation in systems with multiple nonlinear resonances based on the Green function approach \cite{Omura15}.

\begin{figure}
\centering
\includegraphics[width=1\textwidth]{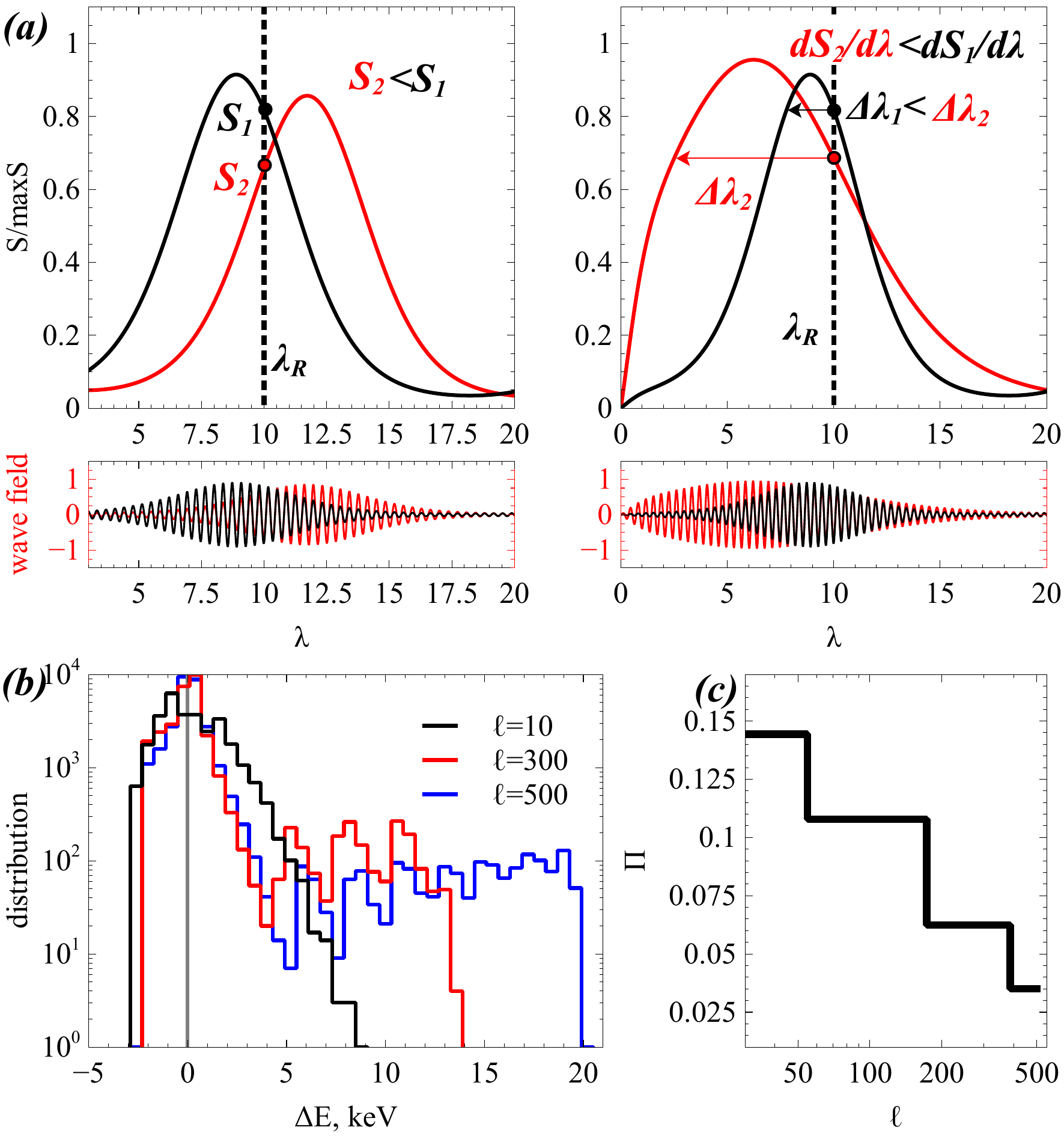}
\caption{Panel (a) shows a schematic view of electron resonant interactions with short wave-packets. The three main parameters are the dependencies of (1) the instantaneous wave amplitude (determining the efficiency of phase bunching) at the resonance and (2) the wave amplitude gradient (determining the phase trapping probability) on the initial electron location relative to the wave-packet generated at the equator, and (3) the wave-packet size (which determines the energy gain due to phase trapping, instead of the background magnetic field gradient in the case of a very long packet). Panel (b) shows the distribution of energy changes due to a single resonant interaction with long ($\ell=300$ and $\ell=500$) and short ($\ell=10$) wave-packets. Panel (c) shows the distribution of the trapping probability, $\Pi$, for a single resonant interaction, as a function of wave-packet length, $\ell$. For numerical simulations shown in panels (b,c) we use field-aligned whistler-mode waves with a peak amplitude of $500$ pT and electrons with an initial energy of $100$ keV and an equatorial pitch-angle of $30^\circ$; other system characteristics are the same as in Fig. \ref{fig02}.}
\label{fig22}
\end{figure}

Resonant interactions with such short wave-packets differ from the simplified picture of interaction with infinitely long packets in four main aspects:
\paragraph{Nonlinear phase bunching} is affected mostly by the wave-packet propagation in latitude. For an infinite wave-packet, the wave amplitude at the latitude of resonance (for fixed energy and pitch-angle) is determined by a fixed $B_w(\lambda)$ profile that does not depend on time. But for short wave-packets, the wave amplitude depends on the time delay between the particle and the wave-packet arriving at the latitude of resonance with that particle (see scheme in Fig. \ref{fig22}(a)). This significantly randomizes the effect of phase bunching and gives a broader energy distribution in comparison with results for the infinite plane wave. Figure \ref{fig22}(b) shows the probability distribution of energy changes due to a single resonant interaction with a very long wave-packet ($\ell=300$, $\ell=500$) and short wave-packets ($\ell=10$). In these simulations, the upper limit of the negative energy jump (the phase bunching effect) is the same, because it is determined by the peak wave amplitude, which is the same in all cases. However, for short wave-packets there is a broader distribution of negative energy changes, because particles can encounter various (randomized) instantaneous wave amplitudes at the resonance. These simulations have been set up in such a way that initial electron coordinates are uniformly distributed within the range corresponding to electrons arriving to the resonance when the wave-packet amplitude there exceeds $1/100$ of the peak value, i.e., we excluded from the simulations the particles arriving into the resonance between two successive wave-packets and which would not significantly interact with the waves.

\paragraph{The probability of nonlinear phase trapping} is also mostly affected by the strong wave field gradient at the wave-packet edges. For an infinite wave-packet, the trapping probability is determined by the latitudinal gradients of the wave field and background magnetic field, but for short wave-packets this probability is instead determined by the gradient of the wave-packet envelope \cite{Bortnik08}. In contrast to background latitudinal gradients, with typical spatial scales of $\sim LR_E \sim 10^4$km, the typical scale of the wave envelope gradient can shrink to only a few wavelengths, $\sim 10-100$km, which represents a huge increase of $d\area/d\gamma \propto d\B_w/d\lambda$ along the resonant trajectory. Figure \ref{fig22}(c) shows how the trapping probability depends on wave-packet size for a given wave-packet model, and increases for shorter packets. In particular, for a short wave-packet, electrons interacting resonantly with the packet leading edge have a larger probability of phase trapping. %Note, we exclude from this simulation all electron arriving into the resonance between two successive wave-packets and not resonating with waves.

\paragraph{Energy changes due to nonlinear phase trapping} are also strongly affected by the wave-packet size. For such short wave-packets, in principle phase trapping becomes possible at any latitude of resonance, without a strong influence of the background profiles of wave and magnetic fields. However, the resulting strong increase of the probability of trapping is compensated by the brevity of this trapping, because particles rapidly escape from the resonance (and from the trapping) at the edges of wave-packets. For the first cyclotron resonance, the resonant electrons and waves are propagating in opposite directions, corresponding to a maximum possible trapping time $\Delta t_{trap}\approx 2\pi \ell /k|v_g+v_R|$, where $v_g=\partial\omega/\partial k_\parallel$ denotes the wave group velocity and $v_R=(\omega-\Omega_{0}/\gamma)/k_\parallel<0$ the velocity of the particle in cyclotron resonance with the wave. Using a simple approximation $v_g\approx \omega/k_\parallel$ \citep[for parallel propagating waves, the typical ratio of $k_\parallel v_g/\omega$ is $1-2$, see][]{bookStix62}, we obtain $\Delta t_{trap}\approx 2\pi\ell/\Omega_{ce}$. Thus, a finite wave-packet size directly sets an upper limit on the electron acceleration. A possible exception has been considered in \cite{Hiraga&Omura20}, where the idea of {\it successive trapping} into series of wave-packets (wave-packet train) has been proposed. Such a mechanism allows to neglect the effect of particle detrapping at the edges of wave-packets, because such detrapped particles will soon be trapped again by the next packet. However, {\it successive trapping} requires that the wave phase remain coherent across the entire wave-packet train, without strong phase stochastization at the wave-packet edges, which limits its applicability \citep[see examples of observations of both coherent and non-coherent wave packet trains in][]{Zhang20:grl:phase}.

In Figure \ref{fig22}(b), wave-packets are assumed to be isolated, or separated by random jumps of wave frequency and phase as in most observations \citep{Zhang20:grl:frequency, Zhang20:grl:phase}. Compared with long packets, short wave-packets lead to a significantly broader distribution of small positive energy changes at $\Delta E=0-5$ keV through phase trapping, due to a randomization of the duration of trapping and of the related energy change, while the maximum energy changes are strongly reduced (from 20 keV to 8 keV). Together with the randomization of phase bunching effects for negative energy changes, short wave-packets therefore lead to a more symmetrical distribution of energy changes compared with long packets.

The periodical structures present in the distribution of energy changes due to trapping for long packets ($\ell=300$ and $\ell=500$) is due to the discretization of the trapping time: trapped particles can spend in acceleration regime time intervals divisible to trapping periods $\sim 2\pi/\Omega_{tr}$. As a result, resonant particles may makes $\sim \N_{trap}$ times rotations during their trapping, with $\N_{trap}=\Delta t_{trap}\Omega_{tr}/2\pi$. Changes of this number by $\pm 1$ due to uncertainties of trapping/de-trapping time result in the formation of several peaks in the $\Delta E$ distribution \citep[see details in][]{Vainchtein17}.

\paragraph{Nonlinear anomalous phase trapping effects} are also affected, because a very long packet corresponds to one long phase trapping interval, whereas short wave-packets result in multiple trapping occurrences, more probable but much more short-lived. For medium to high pitch-angles, this leads to a modification of the probability distribution of energy changes (see an example in Fig. \ref{fig22}(b)), whereas for small pitch-angles this may prevent an effective particle transport away from the loss-cone, i.e., it can reduce the efficiency of anomalous trapping \citep[see Appendix in][]{Mourenas22:jgr:ELFIN}. Figure \ref{fig23} illustrates this idea: for long wave-packets a significant fraction of low pitch-angle particles are transported by anomalous trapping to higher pitch-angles \citep[see also][]{Kitahara&Katoh19}, whereas for short wave-packets the resonant interactions become less regular and anomalous trapping competes with phase bunching, leading to a global transport of initially low equatorial pitch-angle electrons to both smaller and higher pitch-angles, in roughly similar quantities (note that only the first cyclotron resonance is considered here), corresponding to a symmetrization of pitch-angle changes.

\begin{figure}
\centering
\includegraphics[width=1\textwidth]{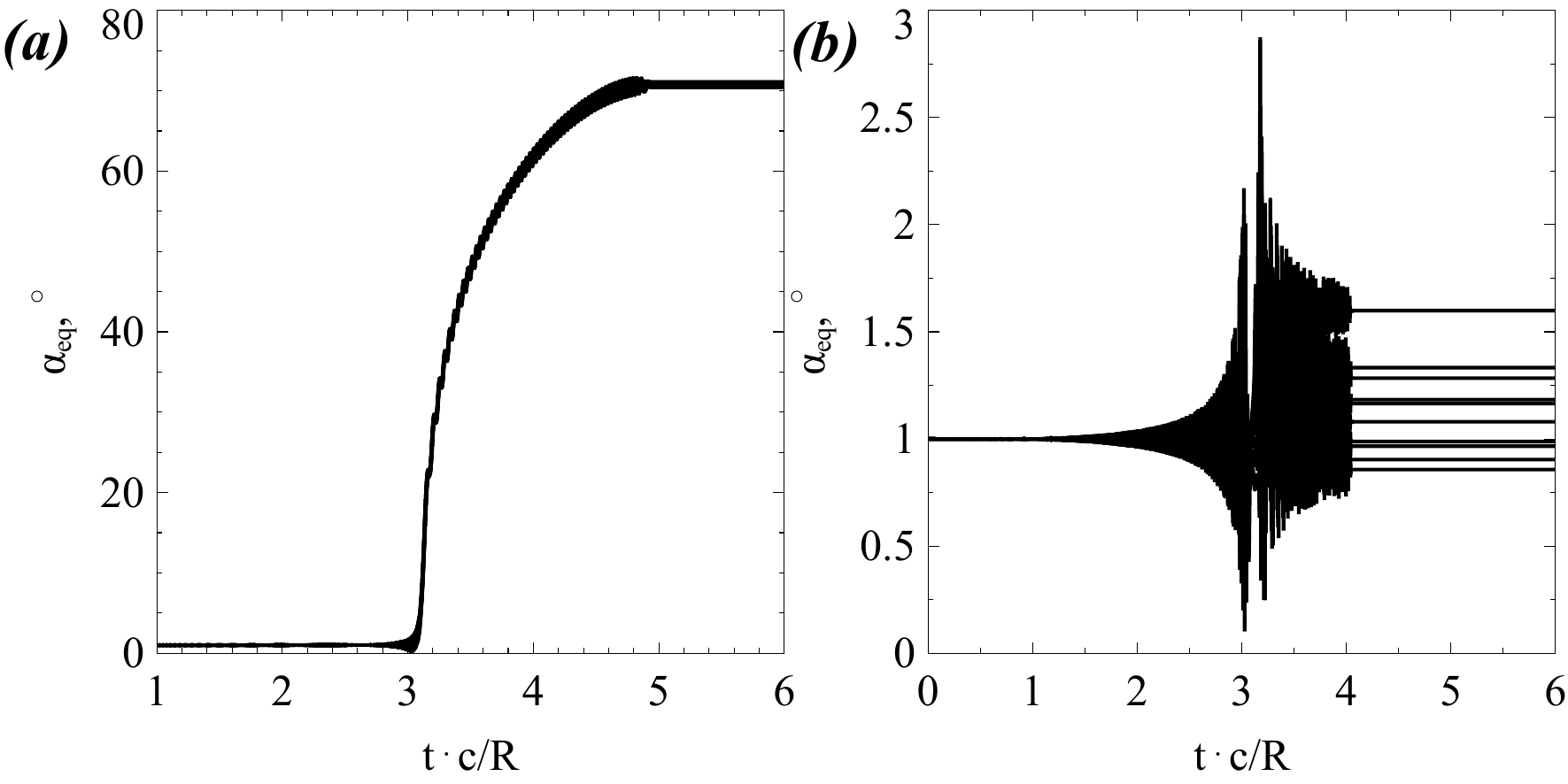}
\caption{Dynamics of electron equatorial pitch-angles due to a single resonant interaction with a plane wave, corresponding to an infinitely long ($\ell\to\infty$) wave-packet (a) and with a short ($\ell=10$) wave-packet (b). The system parameters are the same as in Fig. \ref{fig22}.}
\label{fig23}
\end{figure}

\subsection{Kinetic equation}
\begin{figure}
\centering
\includegraphics[width=1\textwidth]{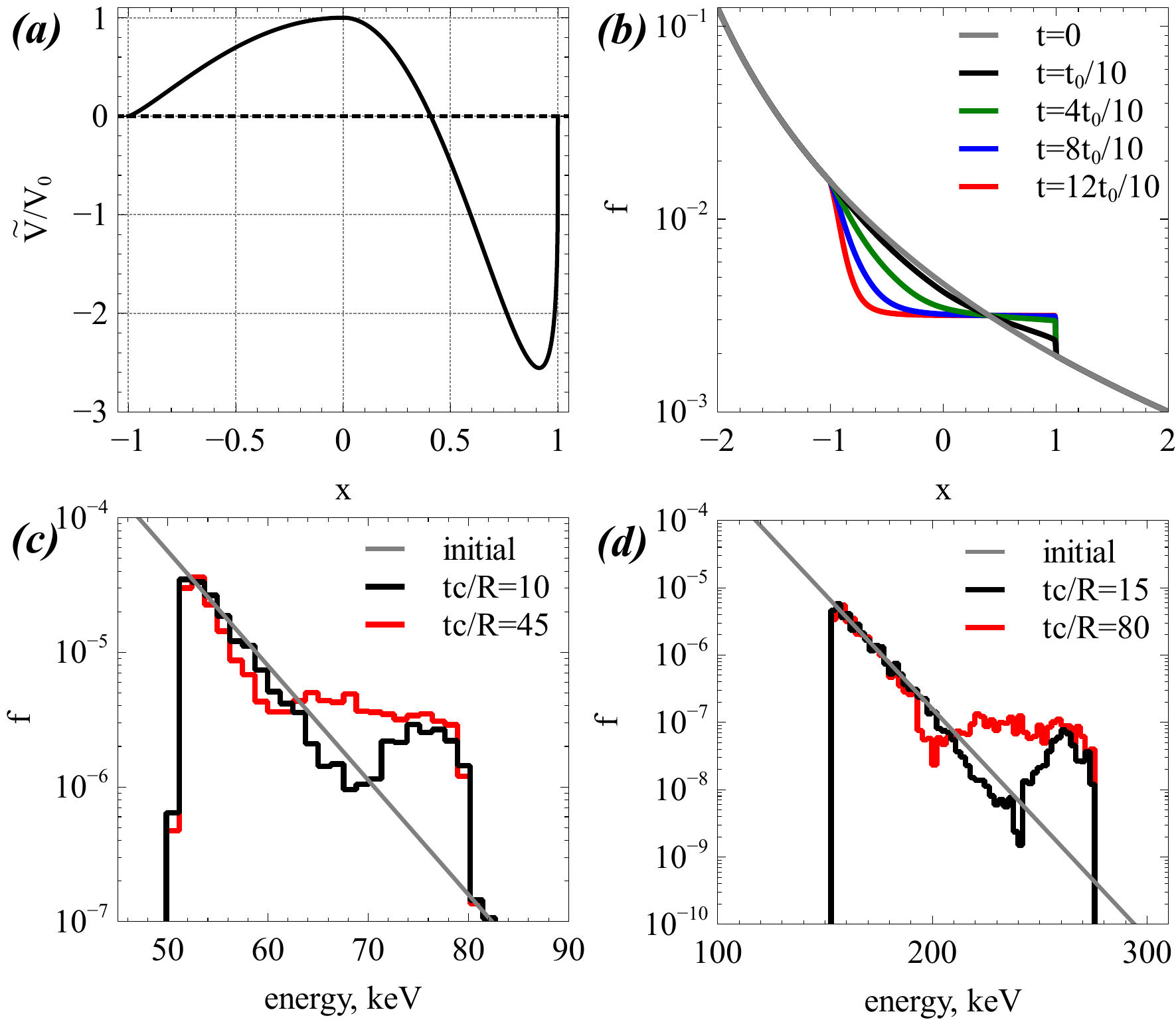}
\caption{Panel (a) shows the profile of $\tilde V(x)/V_0$. Panel (b) shows the solution to Eq. (\ref{eq:kinetic1D_small}) at several times $t$, with $t_0=V_0\cdot R/c$. Panels (c, d) show two examples of numerical solutions, using test particle tracing. System parameters are the same as in Fig. \ref{fig02}, but we use a wave amplitude $\B_w=100$ pT and a wave packet size $\ell=15$. Two values of the constant $h$ from Eq. (\ref{eq:h_const}) are used: $h=1.1$ (c) and $h=1.3$ (d).}
\label{fig24}
\end{figure}

Let us consider the kinetic equation (\ref{eq:kinetic1D}) under the short wave-packet approximation. The effects of phase trapping at arbitrary latitude (arbitrary $\gamma$ or $I$) and the variability of the phase bunching amplitudes cannot be straightforwardly included into this kinetic equation (we shall describe these effects using the mapping technique in Sect.~\ref{sec:scalings}). Thus, we limit our consideration to the effects in a small energy (or $I$) range with $V(\gamma) \propto \area(\gamma) \ne 0$. The maximum energy change due to trapping is equal to $mc^2(\gamma_{\max}-\gamma_{\min})$, where $\gamma_{\min,\max}$ are boundary values of the $\area(\gamma)$ profile, i.e., $\area(\gamma_{\min,\max})=0$. Therefore, for a small range $[\gamma_{\min},\gamma_{\max}]$ we can expand the nonlocal term of Eq. (\ref{eq:kinetic1D}) as:
\[
f\left(\gamma'\right) \approx f\left(\gamma\right) - \frac{{\partial f}}{{\partial \gamma }} \cdot \left( {\gamma  - \gamma'} \right) \approx f - 2\frac{{\partial f}}{{\partial \gamma }}\left( {\gamma  - \gamma _0 } \right)
\]
where $\gamma_0$ is the energy where $\area(\gamma)$ reaches its peak value. Substituting this expansion into Eq. (\ref{eq:kinetic1D}), we obtain
\begin{eqnarray}
 \frac{{\partial f}}{{\partial t}} = \frac{\partial }{{\partial \gamma }}\left( {D_{\gamma \gamma } \frac{{\partial f}}{{\partial \gamma }}} \right) - \tilde V\frac{{\partial f}}{{\partial \gamma }},\quad \tilde V = V_\gamma   + 2\frac{{dV_\gamma  }}{{d\gamma }}\left( {\gamma  - \gamma _0 } \right)\Theta  \nonumber\\
 \label{eq:kinetic1D_small}\\
 \Theta  = 1,\quad \gamma  > \gamma _0 ,\quad \& \quad \Theta  = 0,\quad \gamma  < \gamma _0  \nonumber
 \end{eqnarray}
Using the variable $x=(\gamma-\gamma_0)/\left(\gamma_{\max}-\gamma_{\min}\right)$ with $\gamma_0=\left(\gamma_{\max}+\gamma_{\min}\right)/2$ and $\area$ from Eq. (\ref{eq:area_small}), we can write
\[
V_\gamma   =  - V_0 \left( {1 - x^2 } \right)^{5/4} ,\quad \tilde V =  - V_0 \left( {1 - x^2 } \right)^{1/4} \left( {1 - x^2 \left( {5\Theta  + 1} \right)} \right)
\]
Figure \ref{fig24}(a) shows the profile of $\tilde V(x)$: $\tilde V$ is positive for $x<0$ ($\gamma<\gamma_0$, $\Theta=0$) and negative for $x>1/\sqrt{6}$ ($\gamma>0$, $\Theta=1$). Therefore, electrons with $x >1/\sqrt{6}$ and $x<1/\sqrt{6}$ are drifting in opposite directions, corresponding to the opposite effects of trapping-induced acceleration and deceleration by nonlinear scattering.

If we omit the diffusion term, $\sim D_{\gamma\gamma}$, Eq. (\ref{eq:kinetic1D_small}) represents a simple drift equation
\[
\frac{{\partial f}}{{\partial t}} =  - \tilde V\left( x \right)\frac{{\partial f}}{{\partial x}},
\]
and for an initial condition $f_{t=0}(x)$ this equation has an analytical solution: $f$ is a constant along trajectories $t-\int{dx/\tilde{V}(x)}=const$. The function $\int{dx/\tilde{V}(x)}\to \pm\infty$ at $x=1/\sqrt{6}$ where $\tilde{V}(x)=0$, implying that the particles should drift in energy space away from $x=1/\sqrt{6}$. Indeed, the solution of the drift equation displayed in Figure \ref{fig23}(b) demonstrates the formation of a plateau in energy space around $x=1/\sqrt{6}$. This effect is essentially the same as what occurs during the well-known quasi-linear evolution of a particle distribution function: resonant wave-particle interactions result in a flattening of the phase space density profile within the range of resonant energies \citep[e.g.,][]{Galeev&Sagdeev79, Vedenov&Ryutov75}. However, in contrast with quasi-linear diffusion, this peculiar type of nonlinear interaction is much faster and corresponds to drifts $\sim \tilde{V} \propto \left(\B_w/B_0\right)^{1/2}$. Figures \ref{fig23}(c,d) confirm this theoretical result of plateau formation within a narrow energy range for particles interacting resonantly with short wave-packets. These figures show the results of a numerical integration of Hamiltonian equations of motion for a large particle ensemble interacting with a monochromatic wave-packet (so that Eq. (\ref{eq:h_const}) is well satisfied and we may consider a 1D evolution of the particle distribution function).

Note that the location $x=1/\sqrt{6}$ of the plateau formation shown in Figure \ref{fig23}(b) is largely determined by the system parameters, while for an ensemble of wave-packets (even if such packets are well separated in space and time) these plateaus will be formed around different $x$ (at different energies). Therefore, the global evolution of the electron distribution function can ultimately lead to a reduction of phase space density gradients over a wide energy range \citep[see Section~\ref{sec:mapping_short} and][for details of such evolution]{Artemyev21:pre}.

\subsection{Limit of nonlinear diffusion}\label{sec:nldiffusion}
\begin{figure}
\centering
\includegraphics[width=1\textwidth]{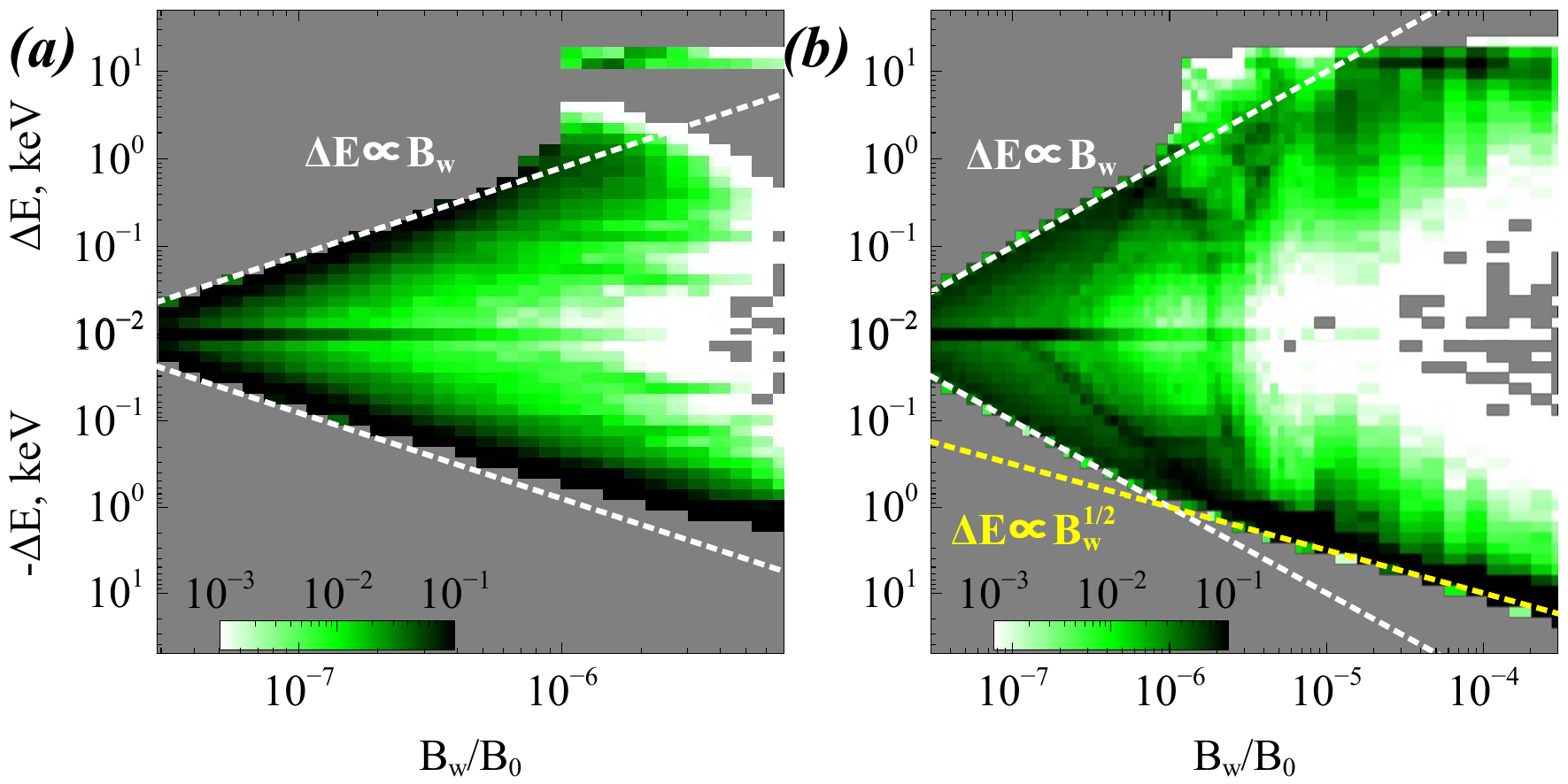}
\caption{
Distributions of energy changes $\Delta E$ for different $B_{eff}/B_0$ with $\a=0$ (a) and $\a=1$ and $\ell=20$ (b) waves, with $\Delta E \propto B_w$ fits (white dashed lines). System parameters are $E_0 = 100$ keV, $\alpha_{eq} = 60^{\circ}$, $L\text{-shell} = 6$, whistler-mode waves with a frequency equal to $0.35$ of the electron cyclotron frequency at the equator, and a constant plasma frequency equal to $10$ of the electron cyclotron frequency at the equator. For each $B_{eff}$, we use $10^4$ trajectories to evaluate the $\Delta E$ distribution, each particle resonating with the wave only once, and $\Delta E$ being the energy change for a single resonance. Initial particle and wave phases are random and thus in the modulated amplitude case the actual wave amplitude is different for different particles having the same energy/pitch-angle. See details in \cite{Frantsuzov23:jpp}.}
\label{fig25}
\end{figure}

Shorter wave-packets correspond to smaller energy changes due to trapping and a higher trapping probability (due to the high probability for electrons to resonate with wave-packet edges having a strong $d\area/d\gamma \propto dB_w/d\lambda$ gradient). Therefore, as discussed earlier, the decrease of the wave-packet size should lead to a more symmetrical $\Delta E$ distribution (e.g., see Figs. \ref{fig22}(b) and \ref{fig25}, and the works from \citet{Tao13,Zhang20:grl:phase, An22:Tao,Gan22,Frantsuzov23:jpp}). The natural limit is the fully symmetrical $\Delta E$ distribution with $\langle \Delta E \rangle \sim 0$ and an equal amount of trapped and phase bunched particles. This regime of wave-particle resonant interactions is equivalent to diffusion, but the diffusion rate may not anymore linearly depend on wave intensity, contrary to the case of quasi-linear diffusion, where we have $\langle (\Delta E)^2 \rangle \propto \B_w^2$ \cite{Kennel&Engelmann66}. Let us examine this regime of {\it nonlinear diffusion} in more detail. We start with test particle simulations showing the $\Delta E$ distributions for different wave amplitudes $B_w$ and wave-packet sizes. Figure \ref{fig25} shows $\Delta E$ distributions obtained from a numerical integration of Hamiltonian equations (\ref{eq:hamiltonian}) with
\[
B_w  = B_{w,peak} \frac{{1 - \exp \left( { - \a\sin ^2 \left( {\phi /2\ell } \right)} \right)}}{{1 - \exp \left( { - \a} \right)}}
\]
where $B_{w,peak}$ is the peak wave amplitude, $\ell$ defines the number of wave oscillations (periods) within one wave packet, and $\a$ controls the {\it depth} of amplitude modulations. We also introduce the effective wave amplitude, which corresponds to the wave intensity averaged over the wave-packet size:
\[
   \B_{w} = \sqrt{\left\langle B_w^2 \right\rangle}_{\phi\in[0,2\pi l]}= B_{m} \frac{\sqrt{1 - 2 I_0\left( \a/2 \right) e^{ -\a/2 } + I_0\left( h \right) e^{ -\a}}}{1 - e^{-\a}}
\]
where $I_n(z)$ is the modified Bessel function of the first kind.

Figure \ref{fig25}(a) shows the detailed distribution of individual (separate) electron energy changes $\Delta E$ for long wave packets. For small wave amplitudes, $\B_w/B_0<5\cdot 10^{-4}$, the distribution is symmetric relative to $\Delta E=0$ with a dispersion scaling as $\langle(\Delta E)^2\rangle \propto (\B_w/B_0)^2$. This is the regime of purely diffusive electron scattering. As the wave amplitude increases, the $\Delta E$ distribution shows the formation of a population of phase trapped particles with $\Delta E\sim 10$ keV (well separated from the main distribution) and of a large population of phase bunched particles (increase of the probability for $\Delta E<0$ in comparison with the probability of $\Delta E>0$; the distribution boundary scales with wave amplitude as $\propto \sqrt{\B_w/B_0})$. This is the standard picture of nonlinear resonant interactions with plane waves or very long wave-packets, when two distinct populations of phase trapped and phase bunched particles are formed, well separated in the $\Delta E$ space.

Figure \ref{fig25}(b) demonstrates the effect of including a wave-packet modulation, corresponding to electron interaction with a short wave-packet. The main difference with Fig. \ref{fig25}(a) is the absence of a distinct asymmetry in the $\Delta E$ distribution for large wave amplitudes, $\B_w/B_0<5\cdot 10^{-4}$. Although some asymmetry of $\Delta E>0$ vs. $\Delta E<0$ is present, the $\Delta E$ distribution is much more symmetric when a short wave-packet modulation is included, with a larger population of $\Delta E>0$. Such symmetrization occurs because the wave amplitude modulation, or limited size, of a short wave-packet, leads to a randomization of the magnitudes and occurrences of nonlinear effects, such that phase trapping provides smaller and more random, but also more probable energy increases, while phase bunching provides more random energy decreases \citep[see also][]{Tao13,Zhang20:grl:phase,An22:Tao}. As a result, the main effect of a wave-packet modulation (i.e., of a limited wave-packet size) is a symmetrization of the $\Delta E$ distribution, such that wave-particle interactions become more diffusive-live, even in the limit of high wave amplitudes. Note that the dispersion of the $\Delta E$ distribution for large amplitudes scales with $\B_w$ as $\langle(\Delta E)^2\rangle^{1/2} \propto \sqrt{\B_w/B_0}$, i.e., the diffusion by intense modulated wave-packets does differ from quasi-linear diffusion.

\begin{figure}
\centering
\includegraphics[width=0.75\textwidth]{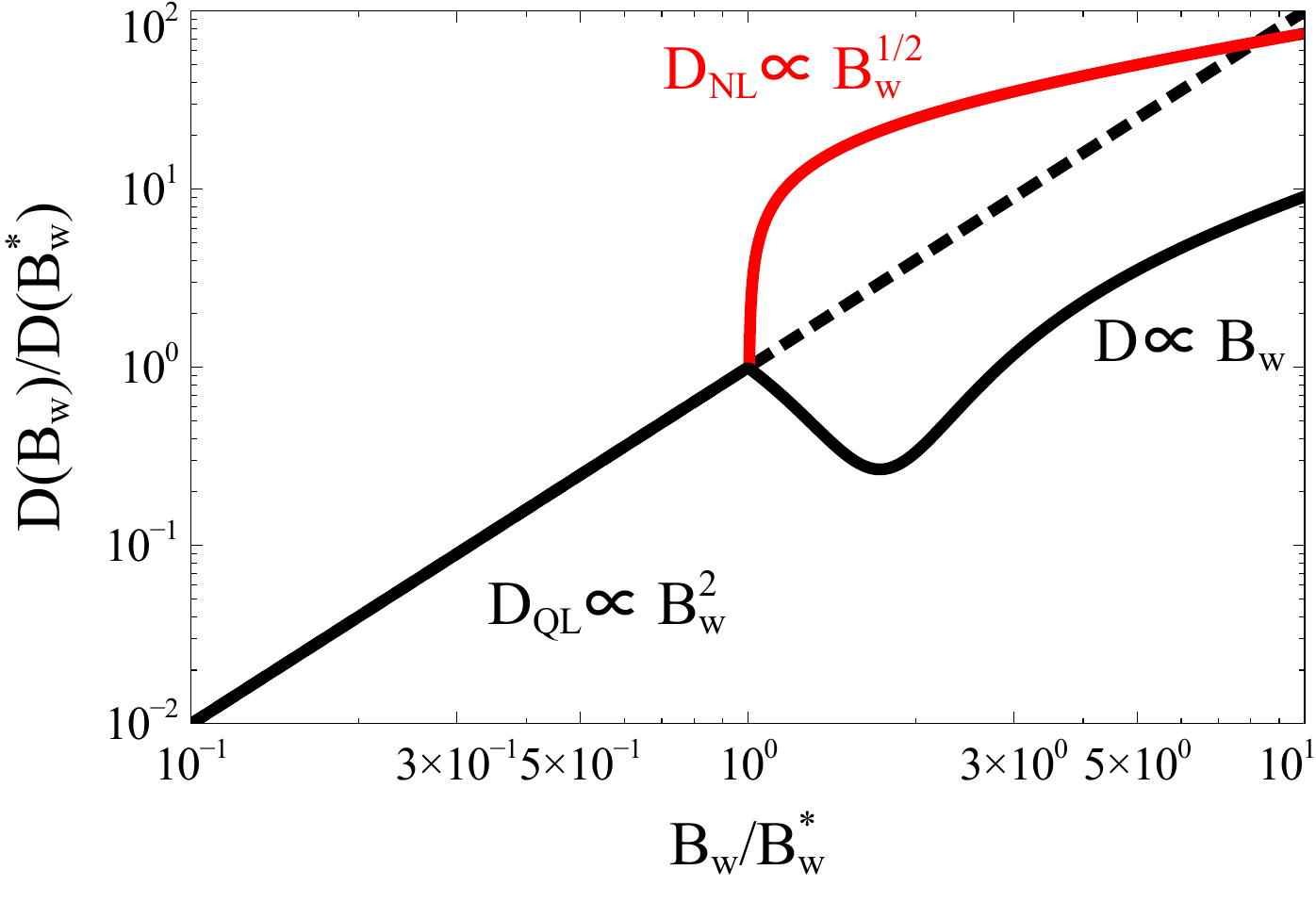}
\caption{Schematic view of diffusion rate dependence on wave intensity. Intensity $\B_w^*$ corresponds to the threshold for nonlinear resonant interactions, $a=1$. See details in \cite{Frantsuzov23:jpp}. }
\label{fig26}
\end{figure}

Let us briefly consider the limit of extremely short wave-packets, corresponding to fully incoherent waves. In such a case,  the $\Delta E$ distribution can be characterized by its variance alone, and the only system characteristic is the diffusion coefficient. Note that this is only an ideal limiting case: it should not be confused with the case of usual short wave-packets that lead to both energy diffusion and distinct energy drifts due to phase trapping and bunching. This ideal limiting case is interesting, though, because we can derive the corresponding characteristic diffusion coefficient analytically, using Eq. (\ref{eq:DeltaI}), as the resonant scattering is then a purely local process (in contrast with phase trapping that depends on the wave-packet shape determining the $\area(\gamma)\ne 0$ energy range). For such local processes, the wave-packet shape can be simply taken into account by introducing an effective wave amplitude parameter $\B_{w}$. Equation $m_ec^2\Delta\gamma=\omega\Delta I$ and (\ref{eq:DeltaI}) provide the energy change due to a single resonant interaction for an arbitrary wave intensity (exceeding the quasi-linear threshold) and, therefore, can be used to evaluate the variance of energy changes:
\[
\left( {\Delta \gamma } \right)^2  = \left( {\frac{{\omega \sqrt {2M\left| {\rm A} \right|} }}{{m_e c^2 }}} \right)^2 F^2 \left( {\xi ,a} \right)
\]
Figure \ref{fig09}(b) shows $\left\langle {F^2} \right\rangle _\xi$ and its mean value, $\left\langle {F} \right\rangle _\xi$, evaluated as a function of parameter $a\propto \B_{w}$. The variance has a peak at $a=1$, the boundary value between the weak wave approximation (no phase trapping and bunching for $a<1$) and the intense wave approximation ($a>1$). The diffusion coefficient for such intense waves takes the form
\begin{equation}
\left\langle {\left( {\Delta \gamma } \right)^2 } \right\rangle _\xi   = \left( {\frac{{\omega \sqrt {2M\left| {\rm A} \right|} }}{{m_e c^2 }}} \right)^2 \left\langle {F^2 \left( {\xi ,a} \right)} \right\rangle _\xi \label{eq:diffusion_strong}
\end{equation}
and for very large wave amplitudes, $a\gg 1$, this coefficient has an asymptotic form \cite{Frantsuzov23:jpp}
\[
\left\langle {\left( {\Delta \gamma } \right)^2 } \right\rangle _\xi   \approx \left( {\frac{{\omega \sqrt {2M\left| {\rm A} \right|} }}{{m_e c^2 }}} \right)^2 \frac{{32a}}{{\pi ^2 }} = \frac{{64}}{{\pi ^2 }}\left( {\frac{\omega }{{m_e c^2 }}} \right)^2 M{\rm B} \propto \B_{w}
\]
Therefore, the diffusion rate corresponding to intense monochromatic waves in the ideal limit of extremely short packets increases with wave intensity weaker than expected based on quasi-linear theory, where $D\propto \B_w^2$. Figure \ref{fig26} illustrates this effect: the diffusion rate scales with $\B_w$ as $D\propto \B_w^2$ up to the threshold for nonlinear resonant interactions, $\B_w^*$, and then significantly decreases relative the $D\propto \B_w^2$ expectation. This effect has been noted in numerical verifications of the quasi-linear diffusion equations, which demonstrated that the diffusion rate is lower than the expected quasi-linear level for fully incoherent waves of high intensity \cite{Tao12,Gan22}. These results underline that even extremely short, strongly modulated wave-packets will not lead to a purely quasi-linear diffusion of electrons for a sufficiently high wave intensity: in such a case, the wave intensity cannot be used anymore as a simple scaling factor \cite[i.e., the procedure of wave intensity averaging over spatial and temporal domains may affect the estimation of the diffusion rate; see also][for discussion of a similar effect of temporal/spatial variability of diffusion rates]{Watt17:D_variability,Watt21:D_variability}.

A natural question arising from the results in Fig. \ref{fig26} is whether quasi-linear diffusion models systematically overestimate diffusion rates during geomagnetically active periods, characterized by enhanced wave intensity. Previous simulations of radiation belt dynamics during geomagnetic storms rather suggest that quasi-linear diffusion underestimates, or estimates reasonably well, the rates of electron acceleration and losses \cite{Thorne13:nature,Li14:storm, Glauert14, Glauert18,Allison&Shprits20}. Therefore, there should be some balance between, on one hand, the reduction of the electron pure diffusion rate at high wave amplitude due to the strong modulation of intense wave-packets \cite{Frantsuzov23:jpp} and, on the other hand, the coexistence of a faster, inherently nonlinear and advective electron transport provided by the main population of short and moderately intense waves-packets \cite{Mourenas18:jgr} and an even faster nonlinear electron acceleration due to a much smaller but finite population of intense long wave-packets \cite{Hiraga&Omura20}. Over the long run, such electron interactions with various wave-packets should result in a quasi-diffusive transport roughly similar to, or somewhat faster than, diffusion by the global time-averaged wave intensity, whereas over shorter time scales, the finite population of intense and not-too-short wave-packets can occasionally provide a much faster nonlinear resonant electron acceleration \citep{Agapitov15:grl:acceleration, Foster17}. In the next sections, we discuss how the net effect of intense wave-packets can be estimated in terms of diffusion.

\section{Mapping technique for short wave-packets}\label{sec:mapping_short}
The mapping technique is a more flexible approach than the kinetic equation (\ref{eq:kinetic1D}), and this allows to make several generalizations for the inclusion of short wave packets. However, one of the main limitations for the description of short wave-packet systems, the absence of an analytical model for $\area(\gamma)$, cannot be overcome anyway. Thus, we should either use some synthetically generated $\area(\gamma)$ in mapping equations (\ref{eq:map}) or resort to a numerical integration of electron trajectories to determine the $\area(\gamma)$ profiles.
\subsection{Simplified mapping and typical scalings}\label{sec:scalings}
We start our consideration with the limit of very short wave-packets, when $\area(\gamma)$ can be approximated by Eq. (\ref{eq:area_small}), but the dynamics of wave-packets make it possible for electrons to interact resonantly with different wave amplitudes (and different $dB_w/d\lambda$ gradients) at any latitude. Note that the smallest wave-packet size (the shortest timescale of phase trapping) has been considered in Appendix C, while in the present section, we only consider short wave-packets with a size exceeding this minimum threshold size.

First, it is important to recall that nonlinear resonant interactions lead to significant energy drifts or jumps, and that such energy drifts and jumps cannot be accurately (fully) described within a diffusive framework, which intrinsically assumes very small and random individual energy changes, with zero mean value. Therefore, we focus here on a more general characteristic of the dynamic evolution of the electron distribution, common to both the quasi-linear and nonlinear regimes: the characteristic timescale of relaxation of the electron distribution toward its asymptotic steady-state. This asymptotic steady-state was indeed shown in Section \ref{sec:asymptote} to be essentially the same state for both regimes. As will be demonstrated below, this common characteristic parameter will provide a way to approximately incorporate nonlinear effects in the diffusive framework.

Let us examine the characteristic timescale of the evolution of the electron distribution in a system with nonlinear wave-particle resonant interactions. Numerical and analytical solutions of Eq. (\ref{eq:kinetic1D}) have shown that a plateau will ultimately be formed in the distribution function of electrons within the energy range of nonlinear resonant interactions, $\area\ne0$. This implies that the electron distribution will tend toward a solution identical to the asymptotic solution of the Fokker-Planck diffusion equation. Therefore, we may assume that such evolution can {\it on average} be approximated by a diffusive transport, albeit with an effective nonlinear diffusion coefficient $D_{NL}$ different from the quasi-linear coefficient $D_{QL} \propto \B_w^2$. Such effective diffusion cannot describe all the peculiarities of nonlinear wave-particle interactions (e.g., the formation of transient electron populations at high-energies or the drift due to phase bunching), but it should provide a reasonable description of both the asymptotic (plateau) state of the electron distribution and of the characteristic timescale for reaching this asymptotic state. This effective nonlinear diffusion coefficient $D_{NL}$ should depend on wave intensity $\B_w^2$ and wave-packet size $\ell$, and for very small $\ell$ this coefficient should tend toward the scaling $D_{NL}\propto \B_w$ derived in Sect.~\ref{sec:nldiffusion}.

\subsubsection{Effective nonlinear diffusion}\label{sec:diffusion}
\begin{figure}
\centering
\includegraphics[width=1\textwidth]{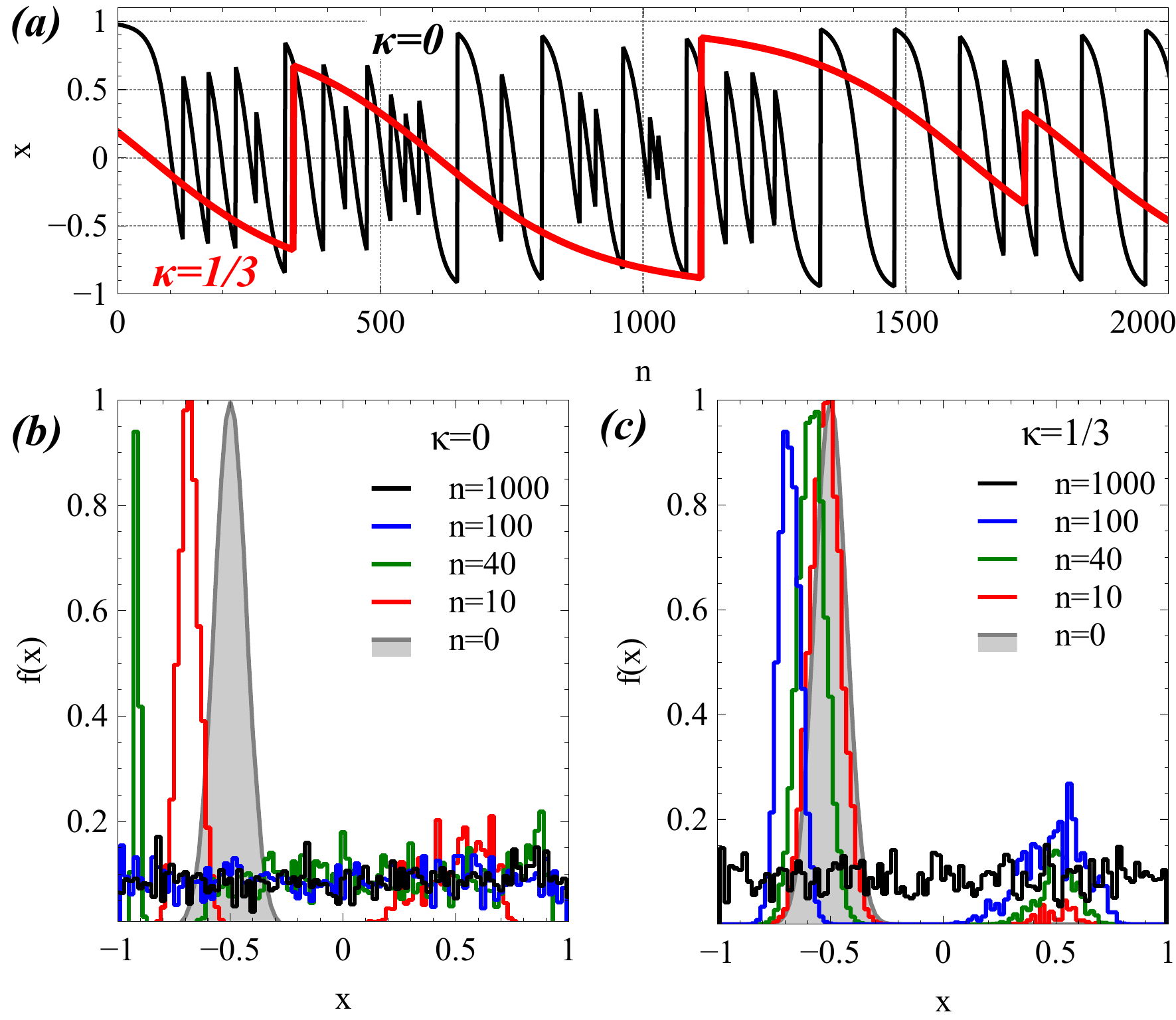}
\caption{Panel (a) shows two examples of $x_n$ trajectories described by map (\ref{eq:map_simple}) with $\area$ given by Eq. (\ref{eq:mapping_small_diffusion}). Panels (b,c) show evolution of  $f(x)$ distribution for two $\kappa$ values. Results are shown for $\B_w/B_0=10^{-3}$.}
\label{fig27}
\end{figure}

To develop the procedure of $D_{NL}$ derivation, let us start with the simple form of $\area(x)$ given by Eq. (\ref{eq:area_small}), with a multiplication factor $\delta_0^{5/4}$ (see Appendix C). We introduce the parameter $\kappa$ through the equation $\delta_0=\left(\B_w/B_0\right)^{4\kappa/5}$. This parameter $\kappa$ is related to the maximum number of trapped particle periods, $N_{trap}$, during resonant interaction with a wave-packet, and a larger $\kappa$ corresponds to a smaller $N_{trap}$ (see Appendix C, and here below). We rewrite Eq. (\ref{eq:area_small}) as
\begin{equation}
\area = \left(\B_w/B_0\right) ^{1/2 + \kappa } \left( {1 - x{}^2} \right)^{5/4}
\label{eq:mapping_small_diffusion}
\end{equation}
Figure \ref{fig27} shows examples of trajectories given by this mapping and examples of the evolution of the particle distribution function $f(x)$. For $\kappa=0$, the trajectory is very similar to the trajectory displayed in Fig. \ref{fig16}(a), but for $\kappa=1/3$ we have smaller $\area \propto (\B_w/B_0)^{1/2+\kappa}=(\B_w/B_0)^{5/6} \ll (\B_w/B_0)^{1/2}$, which results in smaller $x$ changes due to bunching (longer intervals of $x$ decrease) and a smaller probability of trapping (rarer jumps with $x$ increase). Despite these differences, the evolution of the particle distribution $f(x)$ is quite similar, but it takes more time for $\kappa=1/3$. Figure \ref{fig27}(b,c) shows that after $100$ resonant interactions ($1000$ for $\kappa=1/3$) the initially localized $f(x)$ is transformed into a uniform distribution, in agreement with the asymptotic solution derived in Sect. \ref{sec:asymptote} (see also Fig. \ref{fig24}(b)). Therefore, $\kappa$ regulates the time of $f(x)$ evolution, but does not change the main features of this evolution and its asymptotic solution.

\begin{figure}
\centering
\includegraphics[width=1\textwidth]{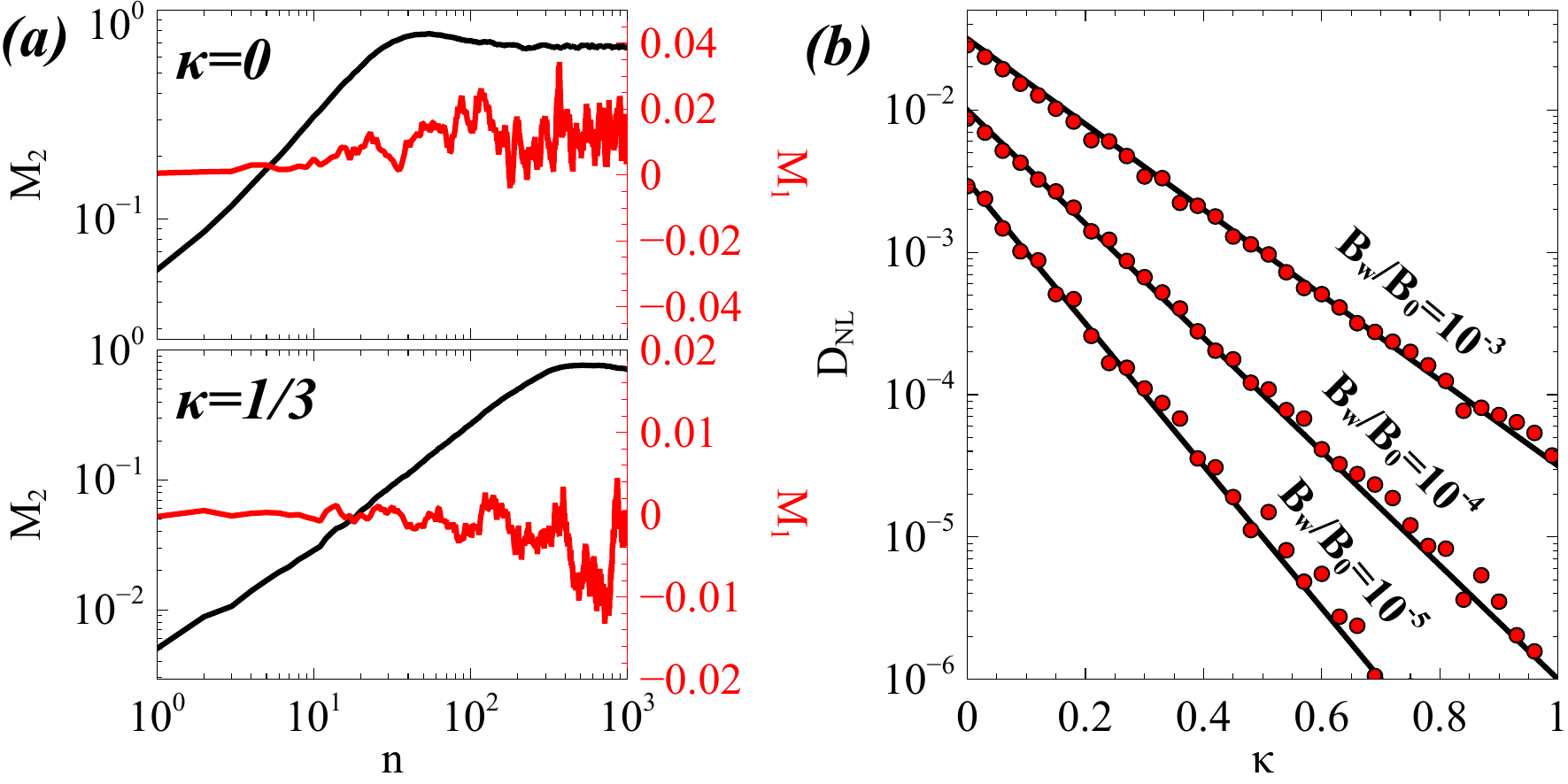}
\caption{Panel (a) shows $\M_{1,2}$ profiles for two $\kappa$ values and $\B_w/B_0=10^{-3}$. Panel (b) shows $D_{NL}$ dependence on $\kappa$; fitting $D_{NL}\propto (\B_w/B_0)^{1/2+\kappa}$ is shown by black lines.}
\label{fig28}
\end{figure}

To characterize the timescale of the evolution of $f(x)$ as a function of $\kappa$, we numerically integrate a large ensemble of trajectories described by the map (\ref{eq:mapping_small_diffusion}), for various values of $\kappa$. The initial $x_{0,i}$ values for the $i=0…N$ trajectories are uniformly distributed within the $[-1,1]$ range, and we calculate two characteristics:
\[
\M_1(n)=N^{-1}\sum_{i=0..N}\left(x_{n,i}-x_{0,i}\right)\;\;\; \M_2(n)= N^{-1}\sum_{i=0..N}\left(x_{n,i}-x_{0,i}\right)^2-\M_1^2(n)
\]
where $x_{0,i}$ are initial $x_i$ values. Figure \ref{fig28}(a) shows  $\M_1$ and $\M_2$ profiles for two $\kappa$ values: $\M_1$ is around zero and $\M_2$ first increases with the number of resonant interactions, and then saturates. This saturation means a full mixing of particles within the $\area(x)\ne0$ domain (i.e., the formation of a plateau in the corresponding $f(x)$ distribution, without any $df/dx$ gradient). Before the saturation, the growing fragment of $\M_2$ can be fitted as
\[
\M_2(n)=\Q n^q
\]
and if $q\approx 1$, we may interpret the proportionality coefficient $\Q$ as a diffusion rate $D_{NL}$, because this coefficient describes the rate of increase of particle variance $d\M_2/dn$ during particle mixing. Note that, instead of being a local diffusion rate (that should depend on $x$), the coefficient $D_{NL}$ determined from the $\M_2(n)$ profile is a global system characteristics that is {\it averaged} over the entire $x$ range of $\area(x)\ne 0$. Because the particle dynamics include very large $x$ jumps (due to phase trapping; see Fig. \ref{fig27}(a)), this dynamics cannot be described locally by a diffusion, but the overall particle mixing within the entire $x$ range stills follows a diffusive-like behavior. Therefore, $D_{NL}$ may be considered as a global characteristic of nonlinear wave-particle interactions, corresponding to the inverse characteristic timescale of evolution of the electron distribution, and it can be compared with average quasi-linear diffusion rates. Figure \ref{fig28}(b) shows the $D_{NL}$ scaling with $\kappa$ and $\left(\B_w/B_0\right)$: $D_{NL}\propto \left(\B_w/B_0\right)^{1/2+\kappa}$, i.e., it is proportional to $\area$ magnitude.

To explain the scaling $D_{NL}\propto \left(\B_w/B_0\right)^{1/2+\kappa}$, let us consider an ensemble of $N$ electrons with the same initial $x$ (i.e., with the same initial energy and $h$). The distribution of $\Delta x$ changes consists of two well separated populations: the most representative population contains almost all particles, $\sim N$, and have a finite $\langle x \rangle \propto \area \propto (\B_w/B_0)^{1/2+\kappa}$; this is the phase bunched population. A much smaller electron population contains only a tiny fraction $(\B_w/B_0)^{1/2+\kappa}$ of all resonant particles, $\sim N\Pi \sim N (B_w/B_0)^{1/2+\kappa}$ where $\Pi=d\area/dx\propto \area$. This population is characterized by a finite $\langle \Delta x \rangle \sim O(\B_w/B_0)$. Such a large $x$ change is due to phase trapping. The dispersion of the entire distribution of $\Delta x$ is about
\begin{eqnarray*}
 \left\langle {\left( {\Delta x } \right)^2 } \right\rangle  \approx \left\langle {\Delta x } \right\rangle _{bunching}^2 \frac{{N_{bunching} }}{N} + \left\langle {\Delta x } \right\rangle _{trapping}^2 \frac{{N_{trapping} }}{N} \\
 \sim\left( {\frac{{\B_w }}{{B_0 }}} \right)^{1+2\kappa} \cdot O\left( 1 \right) + O\left( 1 \right) \cdot \left( {\frac{{\B_w }}{{B_0 }}} \right)^{1/2+\kappa} \propto\left( {\frac{{\B_w }}{{B_0 }}} \right)^{1/2}
\end{eqnarray*}

The parameter $\kappa$ (or equivalently, $\delta_0$) can be expressed through a more physical variable -- the number of oscillations of phase trapped particles around the resonance, $N_{trap}$. For a small $\area$ in Eq. (\ref{eq:mapping_small_diffusion}), the period of such oscillations can be written as $N_{trap}\sim \left(\B_w/B_0\right) ^{3\kappa /5 - 1/2}$ (see Appendix C). We use the normalized number of trapping periods $\N_{trap}=N_{trap}/N_{trap}^\infty$, where $N_{trap}^\infty\sim O(1)/\Omega_{tr}\sim \left(\B_w/B_0\right)^{-1/2}$ is the number of periods in long wave-packets, where trapping and detrapping are determined by the latitudinal gradients of the background magnetic field and of the wave amplitude. Therefore, $N_{trap}/N_{trap}^\infty\sim\left(\B_w/B_0\right) ^{3\kappa /5}$ and $\left(\B_w/B_0\right) ^{\kappa}\sim \N_{trap}^{5/3}$. The effective diffusion rate $D_{NL}$ can be then written as
\[
D_{NL} \sim\left(\B_w/B_0\right) ^{1/2 + \kappa } \sim\left(\B_w/B_0\right) ^{1/2} \N_{trap}^{5/3}
\]
For $\N_{trap}\to 1$, in the limit of long wave-packets, the effective nonlinear diffusion rate is $D_{NL}\sim \left(\B_w/B_0\right)^{1/2}\propto \B_w^{1/2}$. The limit of intense but short wave-packets with $D_{NL}\sim \left(\B_w/B_0\right) \propto \B_w$ can be reached for $\N_{trap}\sim \left(\B_w/B_0\right) ^{3/10}$ (or $N_{trap}\sim \left(\B_w/B_0\right) ^{-1/5}$). It is worth emphasizing that the short wave-packet approximation (or, in another words, the approximation of a symmetrical distribution of $\Delta x$ changes; see Fig. \ref{fig25}) provides a full analytical description of the transition from the classical quasi-linear diffusion rate $D_{QL}\propto \B_w^2$ to the strongly nonlinear effective diffusion rate $D_{NL}\propto \B_w$ (such that $D_{NL}>D_{QL}$), including all the numerical factors quantifying the diffusion rates. Note also that the effective nonlinear diffusion rate $D_{NL}\propto \B_w$ produced by intense short wave-packets (which represent the overwhelming majority of the observed chorus wave-packets in the Earth's outer radiation belt) is larger than $D_{QL}$ because it includes a finite, important contribution from nonlinear interactions which hasten the evolution of the electron distribution, contrary to the ideal case of extremely short packets, briefly investigated in Section \ref{sec:nldiffusion}, where phase trapping and bunching strictly compensate each other, leaving only a purely diffusive transport at a rate lower than $D_{QL}\propto \B_w^2$ even in the high $\B_w$ range.

The effective nonlinear diffusion rate for long wave-packets, $D_{NL}\propto \B_w^{1/2}$ has a very large magnitude (see numerical test particle simulation results in Section \ref{sec:nl&ql}), allowing $D_{NL}\propto \B_w^{1/2}$ to be much higher than $D_{QL}\propto \B_w^{2}$ over a wide range of $\B_w$ values. In a real system containing a large distribution of wave-packets of various sizes and peak amplitudes, the properties of this distribution will determine the actual magnitude of the effective nonlinear diffusion rate $D_{NL}$. However, because the scaling of the larger $D_{NL}$ with $B_w$ in the case of (rare) long packets is slower ($D_{NL}\propto \B_w^{1/2}$) than for the weaker $D_{NL}$ due to short packets ($D_{NL}\propto \B_w$), which is itself increasing more slowly with $B_w$ than the even weaker quasi-linear diffusion rate $D_{QL}\propto \B_w^{2}$, it is no surprise that the total effective nonlinear diffusion rate may be close to the quasi-linear rate $D_{QL}\propto \B_w^{2}$ extrapolated into the high $\B_w$ range. This may explain why quasi-linear simulations often manage to reproduce the observed electron flux dynamics in the radiation belt \cite{Thorne13:nature,Li14:storm, Glauert18, Mourenas14:fluxes,Ma18} even when the measured wave intensity exceeds the threshold for nonlinear wave-particle interactions \cite{Zhang18:jgr:intensewaves,Zhang19:grl}.

\subsubsection{Diffusion by multiple short wave-packets} \label{sec:diffusion_waves}
The mapping (\ref{eq:mapping_small_diffusion}) can successfully describe electron resonant interactions with a short wave-packet, but this interaction is assumed to take place within the entire energy range where $\area\ne 0$. This is a simplified case. In a more realistic situation, we deal with multiple short wave-packets. At a given time, each of these wave-packets covers only a small range of magnetic latitudes and, thus, will interact resonantly with electrons such that their energy is comprised within only a small portion of all the resonant energies at these latitudes. We consider a bounce period as an elementary time-step between electron resonant interactions that change their energy and pitch-angle. After each time-step, wave-packets will have significantly propagated, and each electron resonant interaction will occur with a new distribution of wave-packets along the magnetic field line. Therefore, we should include into the model (1) the probability for each electron to meet one wave-packet at the resonant latitude, (2) the corresponding limited range of resonant energies for each wave-packet. This limited energy range of actually resonant electrons represents a fraction $\chi<1$ of the total energy range of particles potentially reaching resonance with a wave packet over the full length of the magnetic field line.

Electron trapping by a short packet leads to a smaller energy change than trapping by an ideal {\it infinitely long} wave packet, because the electron is released from trapping faster, corresponding to a reduction of its energy change by a factor $\sim\chi$. If short wave packets are rare (not appearing in close succession) and occur only approximately once every bounce period, this corresponds to an additional reduction of the occurrence rate of resonant interaction by a factor $\sim\chi<1$, as compared with both the case of a close succession of short packets and the ideal case of {\it infinitely long} packets assumed in the preceding subsection.

\begin{figure}
\centering
\includegraphics[width=1\textwidth]{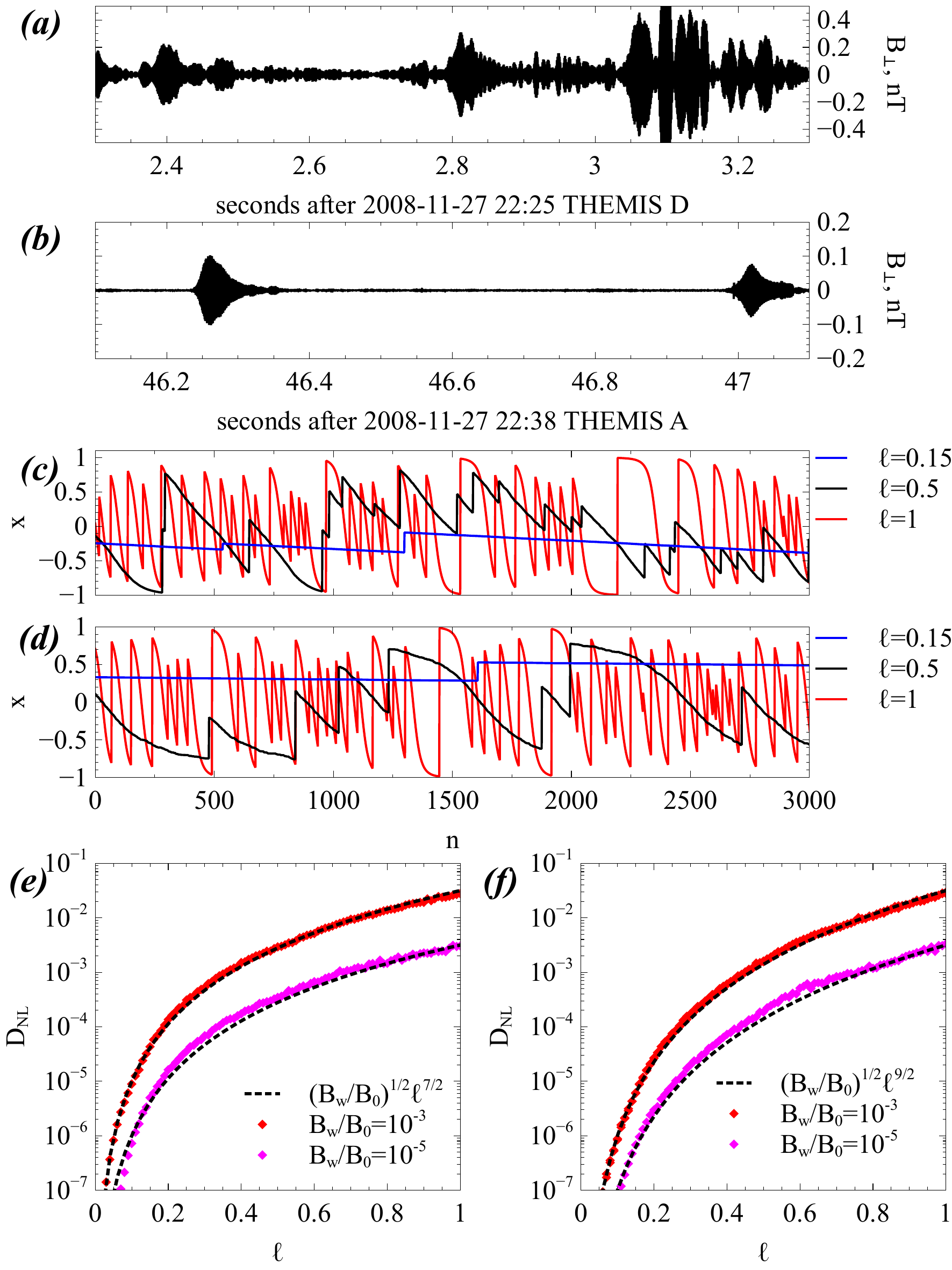}
\caption{(a,b) show observational examples of dense and rarefied filling of a magnetic field field line by short wave-packets. Data are collected by THEMIS search-coil magnetometer \cite{LeContel08}. Details of these wave events can be found in \cite{Zhang20:grl:phase}. Panels (c,d) show examples of trajectories $x(n)$ for the two types of models of mapping (\ref{eq:mapping_small_diffusion_multiple}). Panel (e) shows the corresponding scaling of $D_{NL}=d\M_2/dn$ with $\chi$ for the two models of mapping.}
\label{fig29}
\end{figure}

Accordingly, let us consider a realistic situation where the magnitude of $\area\sim \left(\B_w/B_0\right)^{1/2}$ remains the same, but the range of nonlinear resonant interactions shrinks. There are two kinds of systems with an total resonant range $x\in[-1,1]$, but where $\area$ is not equal to zero only for $x\in[-\ell, \ell]$, with $\chi=\sqrt{\delta_0}$. In systems of the first kind, we center $\area$ around a $\bar{x}_n$ randomly generated at each map iteration around $x_n$, i.e., for each iteration there is a finite change of $x_n$. This type of mapping mimics electron resonant interactions with a set of short wave-packets that would fill the entire magnetic field lines, but which cannot trap electrons for a long time because of the limited packet duration. Figure \ref{fig29}(a) shows observational examples of such a dense filling of the field-line by short wave packets.

In the system of the second kind, we center $\area$ around a randomly generated $\bar{x}_n$, but we do not control the position of $\bar{x}_n$ relative to $x_n$ and some map iterations can occur without any change of $x_n$ because $x_n$ is outside the $\area\ne0$ range. This type of mapping mimics electron resonant interaction with rare short wave-packets propagating with a significant time separation. It corresponds to a situation where, during each bounce period, only one wave packet is present and many electrons reach the latitude of cyclotron resonance without encountering this intense packet there. Figure \ref{fig29}(b) shows observational example of such rarefied filling of a magnetic field-line by short wave packets.

For both systems, the mapping (\ref{eq:mapping_small_diffusion}) can be rewritten as
\begin{equation}
x_{n + 1}  = x_n  + \left\{ {\begin{array}{*{20}c}
   {\bar x_n  - 2x_n ,} & {\xi  \in \left[ {0,\Pi } \right]}  \\
   { - \area_n \left( {x_n } \right) ,} & {\xi  \in \left( {\Pi ,1} \right]}  \\
\end{array}} \right.
\label{eq:mapping_small_diffusion_multiple}
\end{equation}
and
\begin{eqnarray}
 \area_n  &=& \left(\B_w/B_0\right) ^{1/2} \left( {\chi ^2  - \left( {x - \bar x_n } \right)^2 } \right)^{5/4} \nonumber \\
 \Pi  &=& \frac{{d\area_n}}{{dx}} =  - \frac{5}{2}\left( {x - \bar x_n } \right)\left(\B_w/B_0\right) ^{1/2} \left( {\chi ^2  - \left( {x - \bar x_n } \right)^2 } \right)^{1/4}  \nonumber
 \end{eqnarray}
where $\bar{x}_n=x_n+\chi\cdot R$ for the system of the first kind, and $\bar{x}_n=-1+\chi+2(1-\chi)\cdot R$ for the system of the second kind; here $R$ is a random number with a uniform distribution within $[-1,1]$ (note that for $\bar{x}_n=y_n+\chi\cdot R$ we also control that $\area\ne0$ range does not cross $x\pm1$ boundaries). An important property of both systems is that for small $\chi$, trapping is possible for any $x_n$ values, whereas for $\chi=1$ (the initial map given by Eq. (\ref{eq:mapping_small_diffusion})) trapping is possible only for $x_n<0$. Figure \ref{fig29}(c,d) shows a set of sample trajectories for each of the two kinds of systems. The rate of change of $x$ is going down as $\chi$ decreases, and this effect is stronger for the system of the second kind  (Fig. \ref{fig29}(d)).

For these two kinds of systems, we set the range of $\chi$ and for each $\chi$ value we calculate $\M_{2}(n)$. Then we fit the growing fragment of $\M_2(n)$ by $D_{NL}n$. Fig. \ref{fig29}(e) shows the corresponding $D_{NL}$ dependence on $\chi$. For both systems, $D_{NL}$ scales with $\chi$ as $D_{NL}\sim\left(\B_w/B_0\right)^{1/2}\chi^\eta$. But $\eta\approx 7/2$ for the system of the first kind, where resonances occur at each iteration, whereas $\eta\approx 9/2$ for the system of the second kind, where resonance occurrence is decreased by a factor $\chi\ll1$. Taking into account that $\chi=\sqrt{\delta_0}$, we can write $\N_{trap}\sim \chi^{3/2}$. Therefore, for the scaling of $D_{NL}$ as a function of $\N_{trap}$ we have $D_{NL}\sim \left(\B_w/B_0\right)^{1/2}\chi^\eta\sim \left(\B_w/B_0\right)^{1/2}\N_{trap}^{2\eta/3}$ and $D_{NL}\sim\left(\B_w/B_0\right) ^{1/2} \N_{trap}^{2\eta/3}$.
For $\eta=5/2$, this scaling allows to recover the scaling of the ideal case of electron interactions with {\it infinitely long} wave packets/waves present all the time along magnetic field lines: $D_{NL}\sim\left(\B_w/B_0\right) ^{1/2} \N_{trap}^{5/3}$. In the more realistic situation of short wave packets, two different kinds of systems are possible, corresponding to $\eta=7/2$ and $\eta=9/2$. For the system of the first kind, we get $D_{NL}\sim\left(\B_w/B_0\right) ^{1/2} \N_{trap}^{7/3}$. For the system of the second kind, we have $D_{NL}\sim\left(\B_w/B_0\right) ^{1/2} \N_{trap}^{3}$. The parameter $\N_{trap}$ depends on wave characteristics (dispersion, wave-packet size, and wave magnitude) and resonant electron characteristics. Although simplified scalings of this parameter with wave and electron characteristics can be derived analytically (see Figs. \ref{fig28} and \ref{fig29} and \citep{Artemyev21:pre}), a more accurate (likely numerical) approach is needed to complete the quantification of the diffusion regimes for different wave amplitudes and wave coherence. Beside a fully numerical approach through test particle simulations, there are two promising directions for solving the problem of the dynamics of a large ensemble of resonant electrons. The first direction requires a significant modification of the mapping technique, with inclusion of realistic distributions of $\B_w^2$ and $\ell$. We discuss this approach in Section \ref{sec:synthetic}. The second approach requires incorporating a mixture of quasi-linear and nonlinear resonant effects, weighted by the $\mP(\B_w,\ell)$ distribution, into the Fokker-Planck diffusion equation; we discuss this second approach in Section \ref{sec:nl&ql}.

\subsection{Synthetic map}\label{sec:synthetic}
\begin{figure}
\centering
\includegraphics[width=1\textwidth]{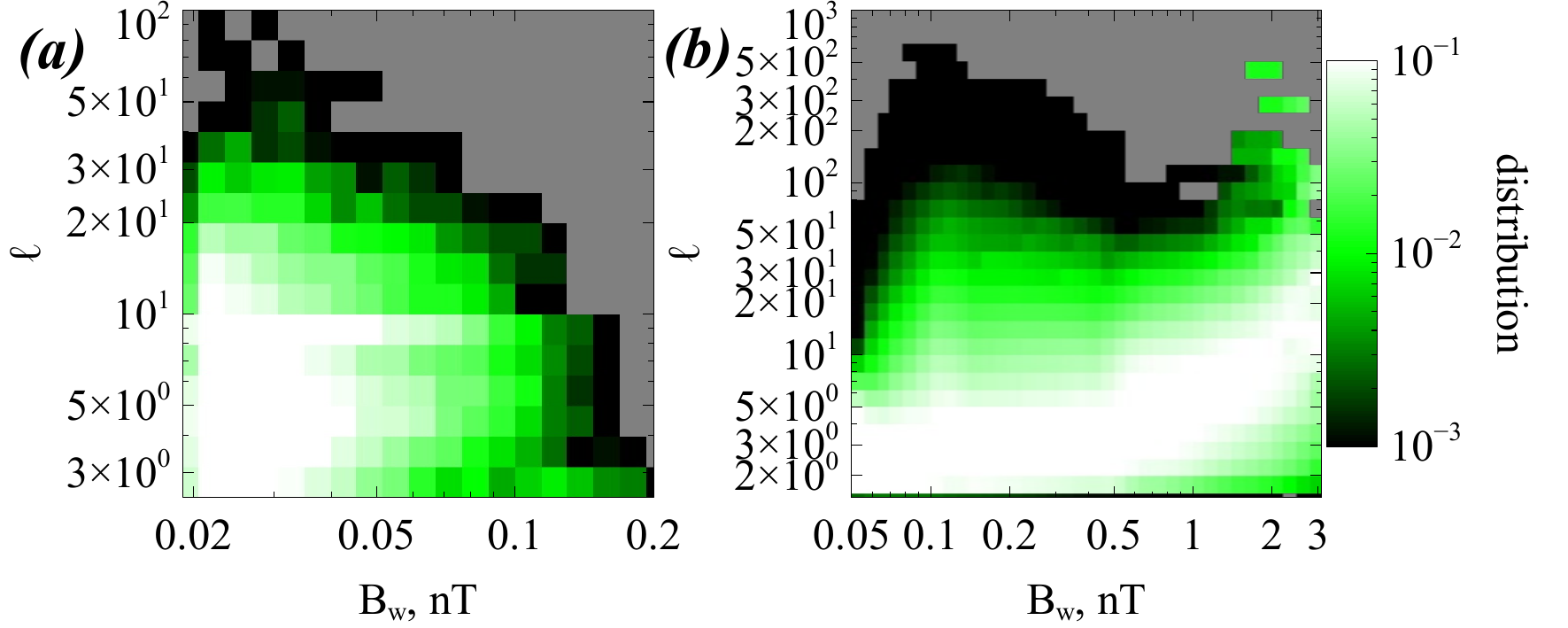}
\caption{Distributions of wave-packets amplitudes and sizes for very oblique (a) and field-aligned (b) whistler-mode waves ($L$-shell$\in[6,10]$ for very oblique waes, THEMIS statistics; $L$-shell$\in[4,6]$ for parallel waves, Van Allen Probe statistics \citep[see details of dataset in ][]{Zhang19:grl}). All levels of geomagnetic activity are included. }
\label{fig30}
\end{figure}

The mapping (\ref{eq:mapping_small_diffusion_multiple}) is a simplified version of the realistic map describing electron resonant interaction with multiple different wave-packets. Although this simplified version does reproduce many peculiarities of such interactions, we may still generalize it to incorporate the observed distribution of wave characteristics, $\mP(B_w,\ell)$, as in examples shown in Figure \ref{fig30}(b) for field-aligned whistler-mode waves. This distribution shows the prevalence of short packets, but also demonstrates the existence of very long packets with $\ell>100$. Note that the method of wave-packet determination (i.e., the criterion used for identifying the wave-packet edges) can significantly affect the shape of the $\mP(B_w,\ell)$ distribution \citep[see results in][]{Zhang19:grl,Zhang21:jgr:data&model}. Wave-packets for very oblique whistler-mode waves are generally shorter (compare panels (a) and (b) of Fig. \ref{fig30}) because their comparatively lower magnetic fields (corresponding to very high electric field amplitudes) are often closer to the noise level.  The dimension of wave characteristics can be further increased by including wave frequency, $\mP(\B_w,\ell)\to\mP\left(\B_w,\ell,\omega/\Omega_0\right)$. Each {\it bin} of this distribution corresponds to a specific map, but in contrast to the $\ell \to\infty$ limit where such a map can be evaluated analytically (see Section \ref{sec:mapping}), for an arbitrary $\ell$ there is no analytical model of wave-particle resonant interactions. Therefore, this approach should rely on test-particle simulations, and the results of such simulations should be either incorporated into the Green function approach \cite{Hsieh20,Hsieh22}, or somehow used for an improved mapping.

\begin{figure}
\centering
\includegraphics[width=1\textwidth]{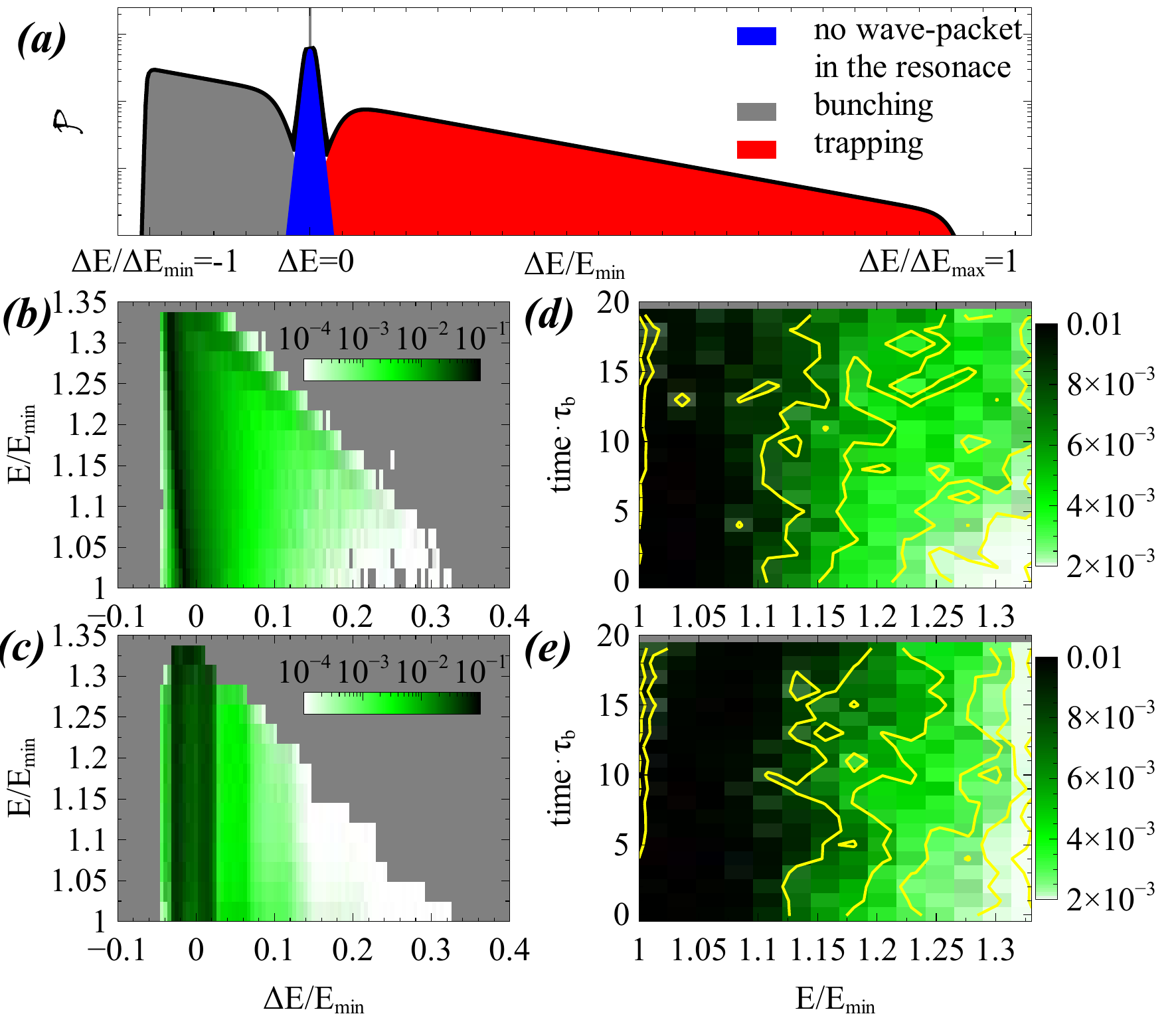}
\caption{Panel (a) provdies a schematic view of the construction of the $\hat m$ mapping using the numerically evaluated distribution of energy changes. Note that we normalize electron energy to the minimum energy, $E_{\min}$, of the range where the nonlinear resonant interactions are possible. Panels (b, c) show the numerically determined 2D probability distribution $\mP(E/E_{\min},\Delta E/E_{\min})$, as fitted by a sum of $\hat m$ maps. Panels (d, e) show the evolution of the electron distribution function obtained directly from test particle simulations and, indirectly, using the mapping technique with the synthetic map shown in panel (c). Details of this demonstration of the usefulness of the synthetic map can be found in \cite{Tonoian23:pop}. }
\label{fig31}
\end{figure}

The idea of incorporating such wave-packet distributions into the mapping technique has been proposed in \cite{Tonoian23:pop,Shi23:pop}, and can be named the {\it synthetic map} approach. This approach consists in the construction of a combined map
\begin{equation}
\hat M = \sum\limits_{\B_w ,\ell ,\omega } {P\left( {\B_w ,\ell ,\omega } \right)\hat m}
 \label{eq:map_sum}
\end{equation}
where $\hat m$ is the mapping described by Eq. (\ref{eq:mapping_small_diffusion_multiple}) with the
\[\area_{\B_w ,\ell ,\omega}=C_{\B_w ,\ell ,\omega}\left(\B_w/B_0\right)^{1/2}\left(\chi^2-(x_n-x)^2\right),\] where $=C_{\B_w ,\ell ,\omega}$, $\chi$, and the probability distribution of $\mP(x_n)$ have to be determined through test particle simulations. Such simulations (for fixed  $\B_w$ ,$\ell$ ,$\omega/\Omega_0$) provides the probability distribution function of particle energy changes, and the main elements of $\area_{\B_w ,\ell ,\omega}$ model can be derived from this distribution -- see schematic view in Fig. \ref{fig31}(a). The maximum energy change due to trapping determines the parameter $\chi$, which should be proportional to the wave-packet size parameter $\ell$ (see Fig. \ref{fig22}(b)). The magnitude of negative energy changes due to phase bunching determines the parameter $C_{\B_w ,\ell ,\omega}$, which denotes the normalized magnitude of the area $\area$. The relative number of test particles such that their energy does not change in the first resonant interaction, due to an absence of wave-packet at the latitude of resonance, determines the distribution $\mP(x_n)$ \citep[see details in][]{Tonoian23:pop}. Therefore, even a quite roughly reproduced (due to limited statistics from test particle simulations) probability distribution of electron energy changes may allow to reproduce the main mapping characteristics and to construct a synthetic map. Figure \ref{fig31}(c) shows an example of such a synthetic map, constructed to reproduce the numerically obtained distribution of electron energy changes from Fig. \ref{fig31}(c). Although the synthetic map is simplified and omits many details of this distribution, the application of this map for the simulation of the dynamics of an ensemble of electrons shows very similar results to what can be obtained directly from test particle simulations (e.g., compare panels (d) and (e) in Fig. \ref{fig31}). The general dynamics of the electron energy distribution for this simulation shows an increase of energetic particle population due to trapping, but because we deal with an ensemble of short wave-packets, the formation of the plateau in the distribution function takes a much longer time than for simulations with long wave-packets.

The method of construction of a synthetic map resembles a general decomposition technique where the results of multiple different processes (bunching and trapping by wave-packets having different amplitudes and sizes) is expressed as a linear sum of elementary processes, each of them being described by a simple model (e.g., the mapping model (\ref{eq:mapping_small_diffusion_multiple})). Such a decomposition requires a minimization of the difference between $\mP(\Delta E)$ derived numerically and obtained from Eq. (\ref{eq:map_sum}). Numerical schemes for this minimization are not yet well developed (\cite{Tonoian23:pop} used a quite simplified form of determination of the parameters of Eq. (\ref{eq:mapping_small_diffusion_multiple}) shown in Fig. \ref{fig31}) and, therefore, the synthetic map approach has not been thoroughly investigated yet. However, this approach looks quite promising for complicated (realistic) systems, and it would be worth generalizing and developing it in the future.

\subsection{SDE versus mapping}\label{sec:sde&mapping}
The mapping approach utilizes the probability distribution function of electron resonant energy changes, $\mP(\Delta E)$, and thus, it is quite similar to the more widespread stochastic differential equation (SDE) approach, which is mostly used for simulations of the purely diffusive regime of wave-particle interactions. The basic idea of SDE is that, instead of directly solving the Fokker-Planck equation, one can write and solve the corresponding Ito stochastic differential equations of {\it quasi-particle} trajectories  \cite{Tao08:stochastic,Zheng14:stochastic}:
\begin{equation}\label{eq:Ito_vec}
    \boldsymbol{p}(t+\Delta t)= \boldsymbol{p}(t) + \boldsymbol{\mu}( \boldsymbol{p}(t),t)\Delta t+ \boldsymbol{\sigma}(\boldsymbol{p}(t),t)\boldsymbol{dW_t}
\end{equation}
where $\Delta t$ is the time step over which we calculate the change of $\boldsymbol{p}=(p_\parallel, p_\perp)$, $\boldsymbol{\mu}(\boldsymbol{p}(t),t)$ is a $2$-dimension vector of the drift coefficient, $\boldsymbol{\sigma}(\boldsymbol{p}(t),t)$ is a $2\times 2$-dimension matrix related to the diffusion coefficients written in such a way that $\hat D_{pp}=\frac{1}{2}\boldsymbol{\sigma} \boldsymbol{\sigma}^T$, $\hat D_{pp}$ is a $2\times 2$ matrix of diffusion coefficients\cite{bookLyons&Williams,bookSchulz&anzerotti74,Albert18:jastp}, $\boldsymbol{W_t}$ is a $N$-dimension standard Wiener process; $\boldsymbol{dW_t}=\sqrt{\Delta t}\boldsymbol{N}$, where $\boldsymbol{N}$ is a vector of standard normal random values, $N_i\sim N(0,1)$.

The term {\it quasi-particles} means that we do not directly integrate the equations of motion, but treat the change of $\boldsymbol{p}$ as a stochastic process and approximate it by the equation (\ref{eq:Ito_vec}). As a result, two {\it quasi-particles} having equal initial conditions $\boldsymbol{p}_0$ may have different trajectories $\boldsymbol{p}(t)$ (in the numerical integration of the equation (\ref{eq:Ito_vec}), one can fix the seed of the pseudo random number generator to preserve the sequence of random numbers and make the results repeatable). We can examine electron distributions in the energy space, and thus the Ito equation (\ref{eq:Ito_vec}) can be rewritten in the following form
\begin{equation}\label{eq:Ito_E}
    E(t+\Delta t)=E(t)+\mu_E\left(E(t)\right)\Delta t+\sqrt{2D_{EE}\left(E(t)\right)}dW_t
\end{equation}
This equation describes energy evolution for a fixed $h$ given by Eq. (\ref{eq:h_const}), i.e., for a monochromatic wave we may reduce the energy, pitch-angle evolution to the energy evolution only and calculate pitch-angle changes from $h$ conservation. In the more general case of a wave spectrum, there are three diffusion rates (energy, pitch-angle, and mixed energy-pitch-angle) and, thus, we would need to solve a system of equations for energy and pitch-angle evolution \citep[see detail in][]{Tao08:stochastic}. Note also that if we rewrite the Fokker-Planck equation in terms of energy, additional coefficients of variable transformation from velocity(momentum) to energy and pitch-angle (Lam\'e coefficients) should be added \cite[e.g.,][]{Glauert&Horne05}, but the Ito equation will still have the same form. Equation (\ref{eq:Ito_E}) is quite similar to the mapping equations (compare with Eqs. (\ref{eq:map_simple})), but instead of a more complicated probability distribution function containing phase trapped and bunched populations, this equation utilizes a Gaussian distribution of random energy jumps.

 \begin{figure}
\centering
\includegraphics[width=1\textwidth]{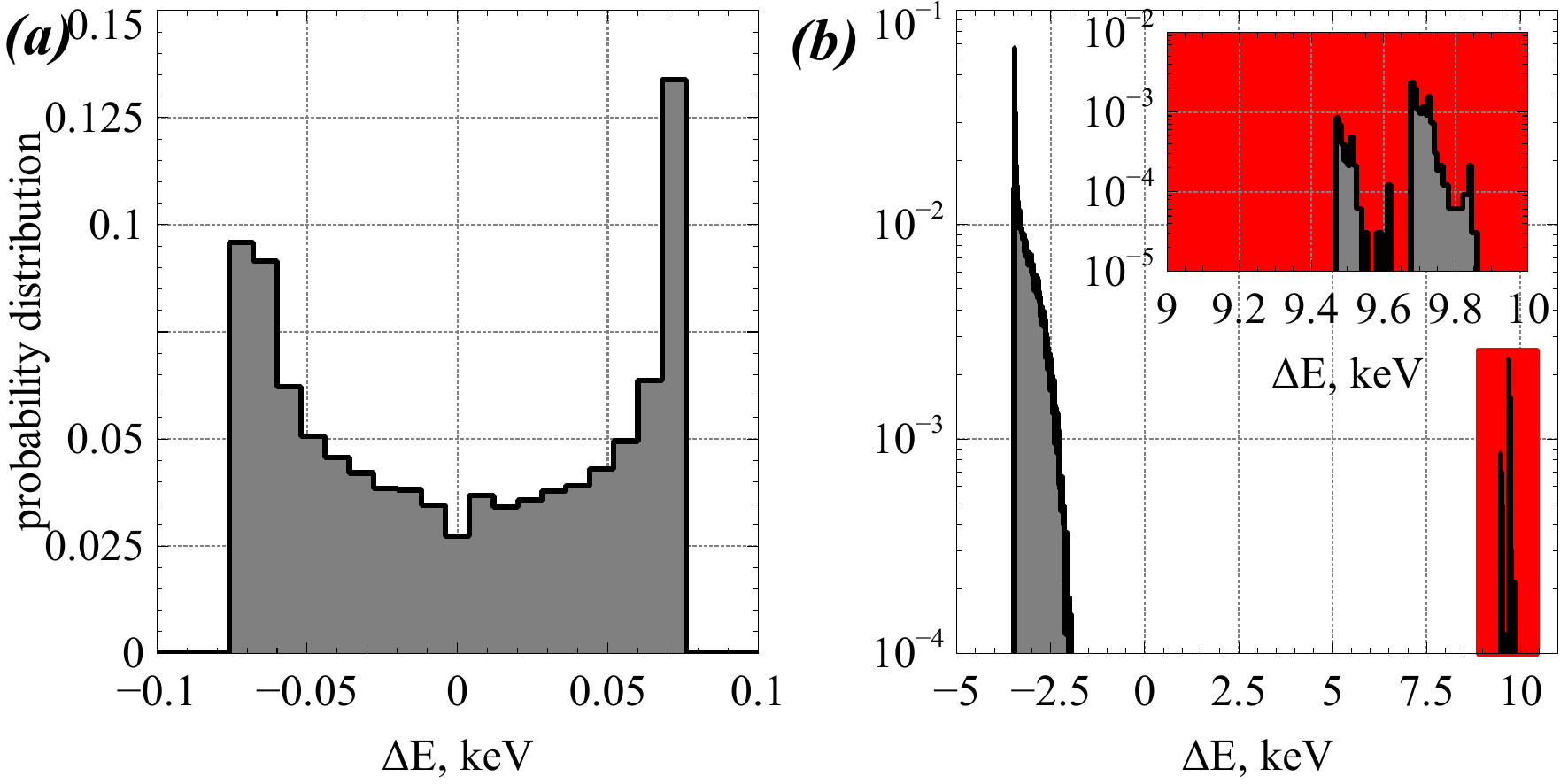}
\caption{Two examples of $\mP(\Delta E)$ distributions for low-amplitude waves with $\B_w/B_0=10^{-4}$(a) and intense waves with $\B_w/B_0=10^{-3}$(b). In panel (b) the red insert zooms in on $\mP(\Delta E)$ for trapped particles. System parameters are: $\omega/\Omega_0(0)=0.3$, $L=5$, $h=1.5$, field-aligned waves \citep[see details in][]{Lukin24:jgr}.}
\label{fig32}
\end{figure}

\begin{figure}
\centering
\includegraphics[width=1\textwidth]{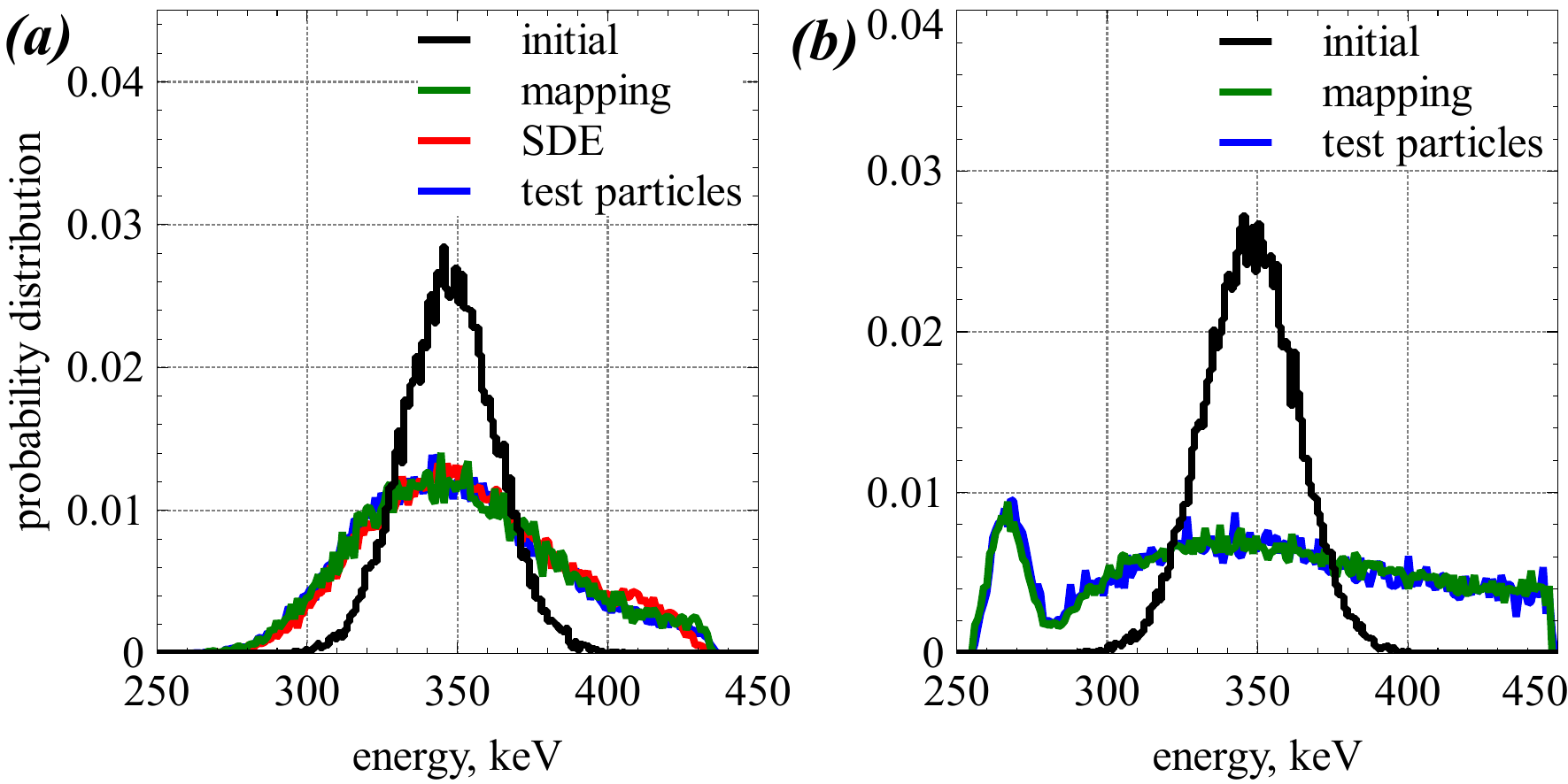}
\caption{Two examples of evolution of electron distribution function, evaluated using different techniques, for low-amplitude waves with $\B_w/B_0=10^{-4}$(a) and for intense waves with $\B_w/B_0=10^{-3}$(b). System parameters are given in the caption of Fig. \ref{fig33}, the typical bounce period of electrons is $\tau_b\approx 0.7$s, and results are shown after $1100$ s of real time \citep[see details in][]{Lukin24:jgr}.  }
\label{fig33}
\end{figure}

In this Section, we shall compare three methods (test particles, SDE, and mapping) to simulate the evolution of the electron distribution function, and these three methods are expected to give the same results for small wave amplitudes, when wave-particle resonant interactions are quasi-linear. The test particles approach is expected to be more precise, because it does not rely on a constructed $\Delta E$ distribution and it is based on the full set of equations of motion. The main advantage of SDE and mapping techniques is their computational efficiency in long-term simulations. Therefore, one can accept a lower accuracy of SDE and mapping, provided that they still describe the main features of the evolution of the electron distribution and that their results statistically reproduce the results from direct test particle simulations. Figure \ref{fig33}(a) shows the evolution of the electron distribution function for a small wave amplitude ($\left(\B_w/B_0\right)=10^{-4}$): we show the initial distribution and the distributions obtained by three methods after $\sim 100$ bounce periods (see figure caption for details). Without nonlinear resonant interactions, both SDE and mapping technique show results consistent with the test particle simulation. In this case, the evolution is diffusive (as expected from quasi-linear theory) and shows a spread of the initially localized electron phase space density peak. The difference between SDE and the test particle simulation, most clearly seen around $E\sim 420$ keV, is due to an overestimation of the diffusion coefficients. We evaluate diffusion coefficients as half of the variance of $\Delta E$ distributions, and thus we assume that $\Delta E$ distributions are symmetric relative to their mean value. However, even in the case of low-amplitude waves (see, e.g., Figure \ref{fig32}(a)) this assumption may not work, leading to an overestimation of the diffusion rate. The mapping technique does not require any assumptions about $\Delta E$ distributions and, therefore, it performs better even in the case of low-amplitude waves.

Figure \ref{fig33}(a) shows that for small wave amplitudes, all three methods, SDE, mapping, and test particle simulations, provide the same evolution of the electron distribution function. This results underlines that the mapping technique is equivalent to SDE approach for the quasi-linear regime. Moreover, for such amplitudes the details of the $\mP(\Delta E)$ distribution are not important: despite the fact that the distributions $\mP(\Delta E)$ from Fig. \ref{fig32}(a) are not Gaussian, their dispersion well describes the electron dynamics due to a very large number of resonant interactions. This is a consequence of the {\it Central limit theorem}, which states that the mean value of weakly correlated variables is normally distributed as long as the sample size is large enough, whatever the probability distribution of these variables \cite{book:Elliott80,book:Dudley14}. Therefore, we should not care too much about describing the fine details of the probability distributions of energy changes, but rather focus on their main characteristics ($\Delta E$ range, mean value, etc.).

Figure \ref{fig33} (b) shows the evolution of the electron distribution function for a large wave amplitude ($\left(\B_w/B_0\right)=10^{-3}$): we show the initial distribution and the distributions obtained by test particle simulations and by the mapping technique after $\sim 100$ bounce periods (see figure caption for details). For such intense waves, the SDE approach becomes fully inapplicable, but we can still compare the results from test particle simulations and from the mapping technique. The mapping technique accounts for nonlinear resonant interactions (e.g., phase trapping) and describes well the evolution of electron distribution. After several wave-particle resonant interactions, the main electron population propagates to lower energies due to the phase bunching, while a small population becomes trapped by waves and gains energy. During the drift of the main population toward smaller energy, the probability of particle trapping increases (see Figure \ref{fig32}(b)) and more particles become trapped and accelerated. Accelerated particles are transported to latitudes of resonance where phase trapping is not possible anymore, and thus, these particles start losing their energy due to phase bunching. Around the time when the main population (at the initial peak of electron phase space density) reaches the left boundary of the allowed energies, the processes of phase bunching and phase trapping statistically compensate each other. This results in the formation of a plateau in the distribution function; this moment is shown in Fig. \ref{fig33}(b). Such an evolution of the electron distribution is consistent with theoretical predictions for the system with multiple nonlinear resonances (see Section \ref{sec:asymptote} and \cite{Artemyev19:pd}).

\section{Incorporation of nonlinear resonant interactions into radiation belt models }\label{sec:nl&ql}
The currently existing radiation belt models can be formally separated into two classes: test particle simulations in the global electromagnetic fields provided by MHD models, and solvers of the Fokker-Planck diffusion equations \cite[see reviews in][]{Shprits08:JASTP_local,Shprits08:JASTP_transport}. Let us discuss the possible approaches for an incorporation of nonlinear resonant interactions into such existing models.

We start with test-particle simulation models, which build on advances in high-resolution global MHD simulations \cite{Elkington04, Hudson99,Hudson12:simulation}. These models reproduce well many important details of energetic electron injections into the inner magnetosphere by fast plasma flows \cite{Sorathia20}, losses across the realistic dynamical magnetopause \cite{Hudson14,Sorathia17}, and radial transport by ultra-low-frequency (MHD) waves \cite{Hudson15,Sorathia18}. A quite obvious deficiency of such models is the absence of any wave-particle interaction effects corresponding to electron scattering by whistler-mode (and any other non-MHD mode) waves. This issue may be resolved if test electron trajectories can be evaluated within the SDE approach, i.e., wave-particle interactions can be included as perturbations of electron dynamics for prescribed quasi-linear diffusion coefficients \cite{Chan23:frontiers}. Although this combined approach (test particle trajectories in MHD field and SDE for wave-particle interaction effects) is still under construction, we can suggest that it can be further generalized. The inclusion of nonlinear resonant effects can be performed by changing the probability distribution of electron energy/pitch-angle changes from the Gaussian one (describing diffusion) to a more complicated form including phase bunching and phase trapping effects. Section~\ref{sec:sde&mapping} describe such a generalization, which formally consists in replacing the SDE approach by the mapping technique \cite[see also][]{Artemyev22:jgr:DF&ELFIN}.

\begin{figure}
\centering
\includegraphics[width=1\textwidth]{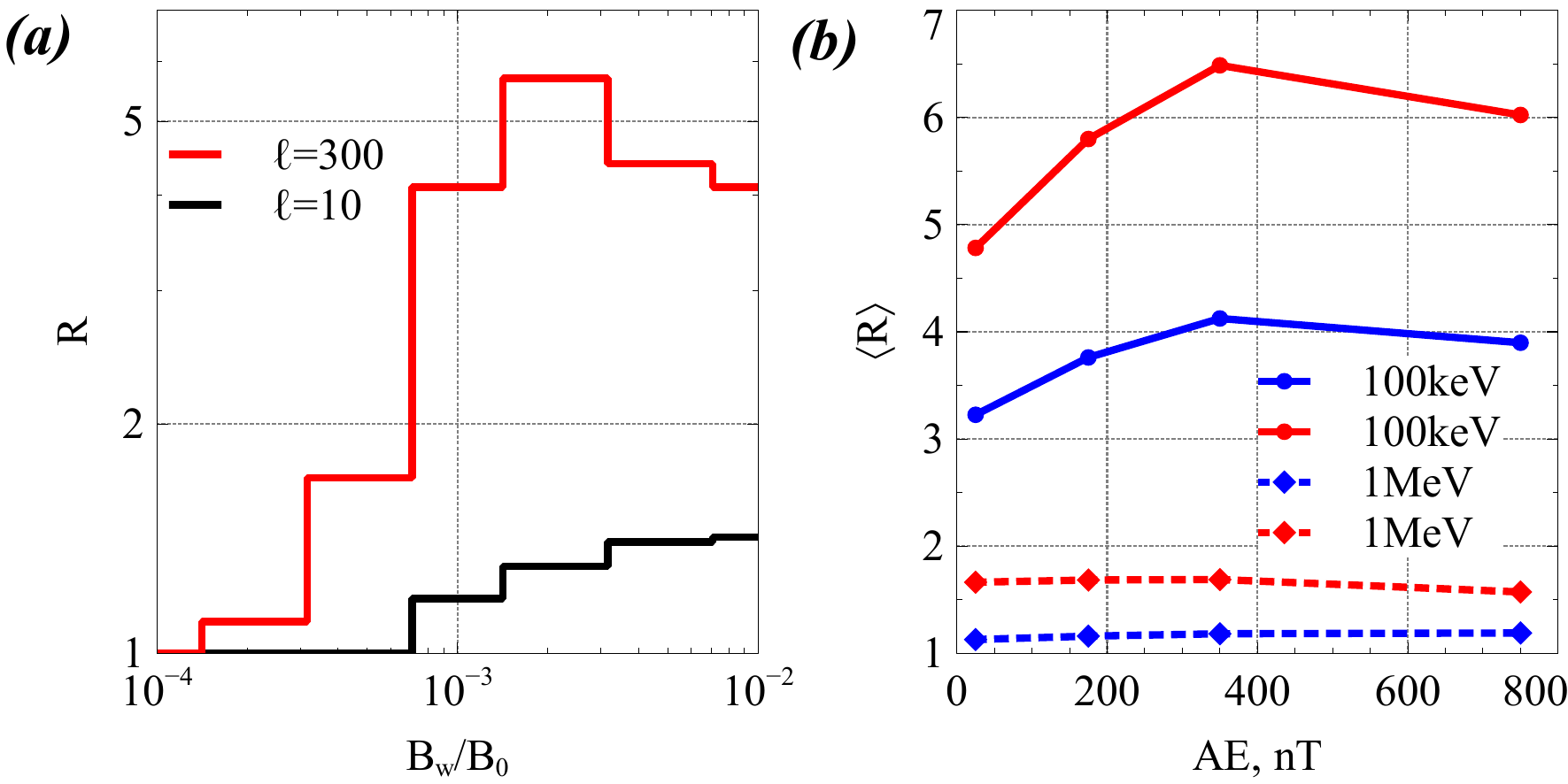}
\caption{(a) $R$ as a function of $\B_w$ in the case of short ($\ell=10$) and long ($\ell=300$) wave-packets. (b) $\langle R \rangle$ coefficient for two typical energies and two datasets of whistler-mode waves-packets (red and blue) \citep[see details in][]{Artemyev22:jgr:NL&QL}.
\label{fig34}}
\end{figure}

Finally, let us discuss the inclusion of nonlinear effects into existing radiation belt models based on Fokker-Planck diffusion equations. The numerical scheme of these models cannot be easily generalized for solving equations with integral operators, and it looks more reasonable to include nonlinear effects in the form of a renormalization of diffusion rates. Sections~\ref{sec:diffusion}, \ref{sec:diffusion_waves} show that over the long term, electron interactions with intense waves indeed result in an effective nonlinear diffusion, with a nonlocal diffusion rate, $\tilde{D}_{NL}$, which represents the rate of electron mixing in the energy, pitch-angle domain of $\area\ne 0$. Such effective diffusion rates can be calculated for realistic wave-packet distributions in two regimes: for realistically high wave amplitudes, and for very small wave amplitudes, $\B_{w,small}$. The second regime will mimic quasi-linear diffusion and provide an estimate of the corresponding nonlocal mixing rate, $\tilde{D}_{QL}$. This rate can be further renormalized to the actual wave amplitude, $\tilde{D^*}_{QL}=\tilde{D}_{QL}\cdot\left(\B_w/\B_{w,small}\right)^2$. The ratio of these rates can be considered as a normalized factor $R=\tilde{D}_{NL}/\tilde{D^*}_{QL}$ indicating the relative contribution of nonlinear effects. Figure \ref{fig34}(a) shows examples of $R$ dependencies on wave amplitude $\B_w$, in the case of short ($\ell=10$) and long ($\ell=300$) wave packets. For sufficiently small $\B_w$, there is no contribution from nonlinear effects, and for both $\ell$ values we have $R\approx 1$. As the wave amplitude $\B_w$ increases, the contribution of nonlinear effects increases, providing a larger $R$. This $R$ increase is stronger for longer wave-packets, due to a more important contribution of phase trapping into $\tilde{D}_{NL}$.

The coefficients $\tilde{D}_{NL}$ and $\tilde{D^*}_{QL}$ can be derived for different wave amplitudes and wave-packet sizes, and then averaged over the $\mP(\B_w,\beta)$ distribution. Such averaging will provide the weighted factor $\langle R \rangle$ that should depend on the geomagnetic activity level, since $\mP$ varies with geomagnetic activity (see Fig. \ref{fig30}). This factor can be determined for some energy, pitch-angle domains (roughly corresponding to $\area\ne0$ domains for different constants $h$ given by Eq. (\ref{eq:h_const})). Figure \ref{fig34}(b) shows examples of $\langle R \rangle$ evaluated for several electron energies such that $h$ remains constant. The factor $\langle R \rangle$ weakly increases with geomagnetic activity, but it varies significantly with energy. Depending on the specific procedure of $\tilde{D}_{NL}$ averaging over the distribution of wave-packet sizes and amplitudes, $\langle R \rangle$ can vary from $1.5$ up to $6$ for $100$ keV, whereas for $1$ MeV $\langle R \rangle$ can vary between $1.1$ and $1.5$ \citep[see details in][]{Artemyev22:jgr:NL&QL}. By definition, $\langle R \rangle$ is determined from the statistics of $I_x$ variations. Therefore, this factor is equally applicable to pitch-angle and energy diffusion, i.e., the derived $\langle R \rangle$ will increase the rate of electron acceleration and precipitation.

Figure \ref{fig34} shows $\langle R \rangle$ for intermediate equatorial pitch-angles at $\sim 100$ keV and for small pitch-angles at $\sim 1$ MeV. Simulations for a multiple pitch-angle, energy domains provide a set of $\langle R \rangle$ values that can be fitted by a function of energy and pitch-angle. The coefficients of such fitting depend on properties of the $\mP(\B_w,\beta)$ distribution. In particular, the relative contribution of long wave-packets in realistic situations is largely determined by the definition of the wave-packet edge, i.e., by the chosen method of determination of the wave-packet length in statistical observations \cite{Artemyev22:jgr:NL&QL, Zhang18:jgr:intensewaves, Zhang19:grl, Zhang20:grl:frequency}.

For instance, if the wave-packet length is calculated between two consecutive minima of $B_w$ below $\sim50$ pT but above $10$ pT \cite{Artemyev22:jgr:NL&QL, Zhang20:grl:frequency}, $\mP(\B_w,\beta)$ contains a significant fraction of long wave-packets, and nonlinear effects (phase trapping) significantly amplify electron mixing in the phase space. Then, we can use
\[
\left\langle R \right\rangle  \approx 1 + A \cdot \frac{{1 + \sin \alpha _{eq} }}{{1 + \left( {E/E_0} \right)}}
\]
where $E_0\approx 50$ keV and $A\approx7$, so that $\left\langle R \right\rangle \approx 3$ for $\alpha_{eq}=0$ and $\left\langle R \right\rangle \approx 5.6$ for $\alpha_{eq}=\pi/2$ for $100$ keV, whereas $\left\langle R \right\rangle \approx 1.3$ for $\alpha_{eq}=0$ and $\left\langle R \right\rangle \approx 1.7$ for $\alpha_{eq}=\pi/2$ for $1$ MeV.

If the fraction of long wave-packets is reduced in $\mP(\B_w,\beta)$, nonlinear effects are also reduced. For example, if we now assume that when the wave amplitude reaches one half of the packet peak amplitude, the wave phase fluctuations \cite{Santolik14:rbsp,Hanzelka20} destroy nonlinear resonances, then the wave-packet size distribution will contain only a small population of long packets \cite{Artemyev22:jgr:NL&QL, Zhang19:grl}. Under such an assumption, we get $A\approx3.5$, so that $\left\langle R \right\rangle \approx 2.2$ for $\alpha_{eq}=0$ and $\left\langle R \right\rangle \approx 3.3$ for $\alpha_{eq}=\pi/2$ for $100$ keV, whereas $\left\langle R \right\rangle \approx 1.17$ for $\alpha_{eq}=0$ and $\left\langle R \right\rangle \approx 1.3$ for $\alpha_{eq}=\pi/2$ for $1$ MeV.

\begin{figure}
\includegraphics[width=1\textwidth]{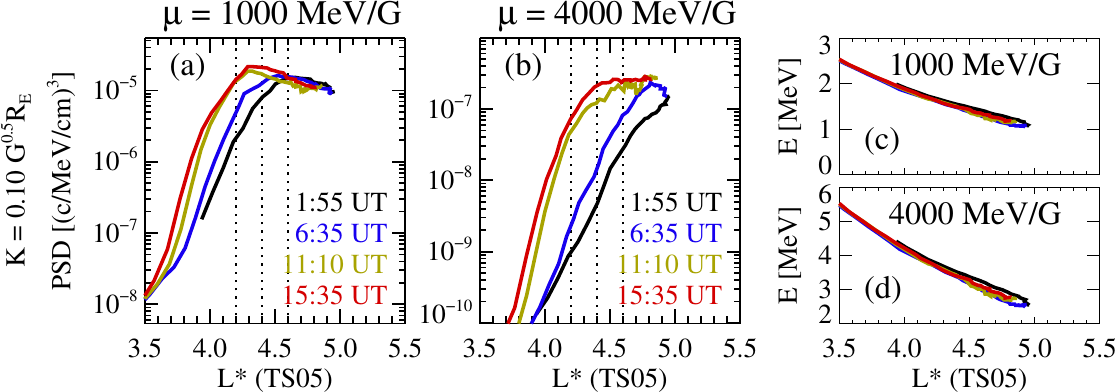}
  \caption{(a, b) Evolution of the electron phase space density for two magnetic moment values with a fixed second adiabatic invariant, as derived from Van Allen Probe A measurements on 01 September 2019, and (c, d) the corresponding electron energy as a function of $L^*$. Colors represent the times when Van Allen Probe A was at $L^{*}=4.2$. The phase space density is derived with the TS05 \citep{Tsyganenko&Sitnov05} magnetic field model.}
\label{fig35}
\end{figure}

To demonstrate the applicability of the proposed approach, which consists in the introduction of a renormalization factor $\langle R \rangle$, we consider a case study during the 31 August 2019 - 01 September 2019 storm \citep[see details of this event in][]{Hudson21,Nasi22}. Figure \ref{fig35} shows examples of electron phase space density (PSD) derived from measurements of Van Allen Probe A MagEIS (for $\mu = 1000$ MeV/G) and REPT (for $\mu = 4000$ MeV/G) instruments \citep{Claudepierre21,Blake13, Baker13}. The fast, localized (in $L^*$ evaluated with the TS05 \citep{Tsyganenko&Sitnov05} magnetic field model) increase in phase space density indicates an effective, local acceleration via wave-particle resonant interactions \citep{Chen07:NatPh,Thorne13:nature}. As an example, we analyze the electron acceleration at $L^{*}=4.2$, where the phase space density increased $\sim$20-fold from 6:35 UT to 11:10 UT on 01 September (Figure \ref{fig35}b).

\begin{figure}
\includegraphics[width=1\textwidth]{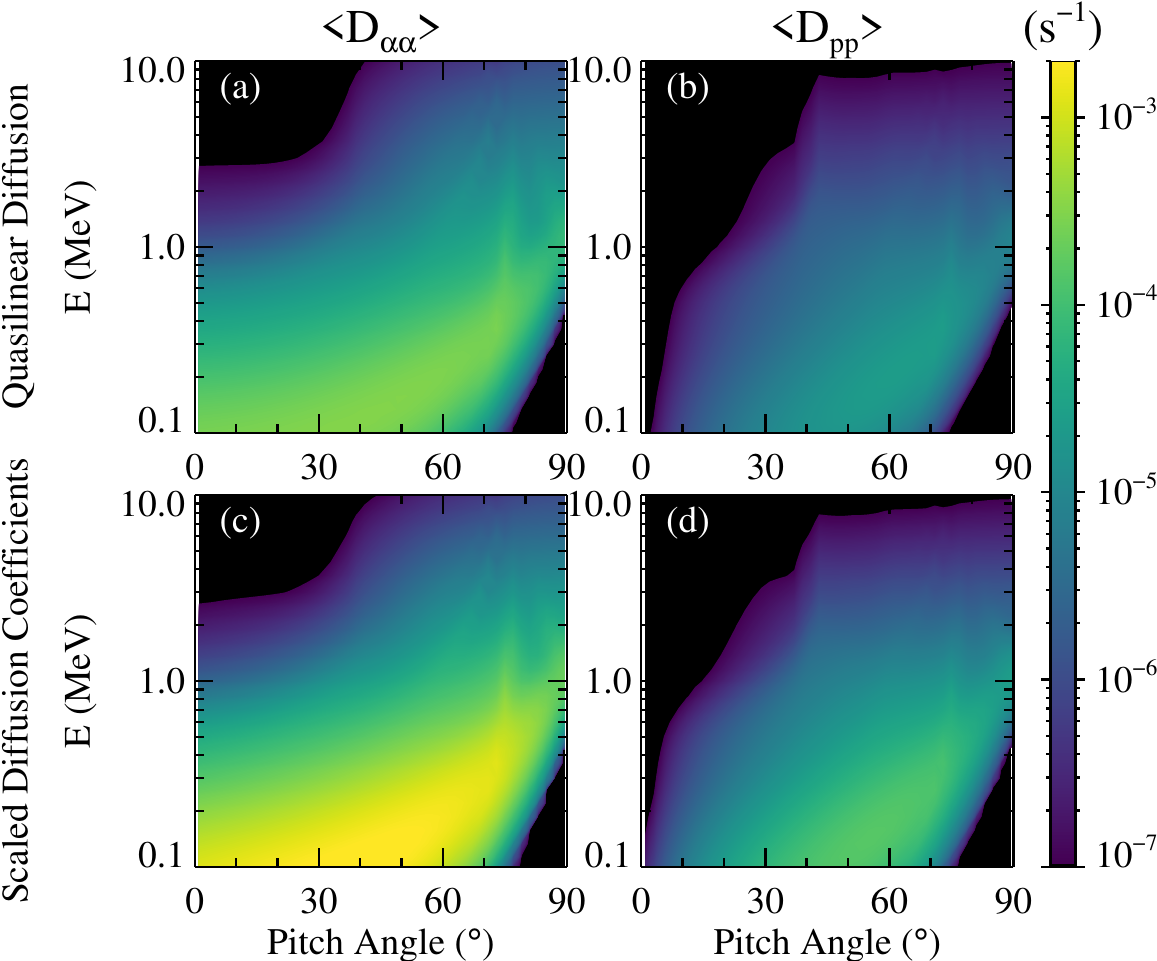}
  \caption{Time, MLT and bounce-averaged pitch-angle (left) and momentum (right) diffusion rates for electron acceleration at $L^{*}=4.2$ observed from 6:35 UT to 11:10 UT on 01 September 2019 \cite[see equations of quasi-linear diffusion rates in][]{Glauert&Horne05}; (a, b) quasi-linear diffusion rates $D_{QL}$ (top), and (c, d) rescaled rates $\langle R \rangle D_{QL}$ (bottom) based on the multiplication factor $\langle R \rangle >1$ obtained for the wave model in Fig. \ref{fig34}.
  }\label{fig36}
\end{figure}

During this event, the characteristics of quasi-parallel whistler-modes wave and cold plasma are available from measurements of Van Allen Probe A ($MLT\in[9,15]$) and ERG ($MLT\in[3,9]$) \citep[see details of missions in][]{Mauk13,Miyoshi17:arase}. We use background magnetic field measurements from the Magnetic Field Experiment \cite{Matsuoka18:ERG_MGF} at $8$-second resolution, and high resolution electric and magnetic field wave spectra from the Onboard Frequency Analyzer \citep[see][]{Matsuda18} onboard on the Plasma Wave Experiment \cite{Kasahara18:ERG_PWE}. We also utilize measurements from the Electric and Magnetic Field Instrument Suite and Integrated Science Waves instrument \citep{Kletzing13} providing all components of the wave electric and magnetic fields. To exclude measurements within the plasmasphere, we use the electron plasma density inferred from the upper hybrid resonance frequency line in the $10-400$ kHz range for Van Allen Probes \citep{Kurth15}  and in the 10 kHz - 10 MHz range for ERG \citep{Kumamoto18}. To obtain the full MLT coverage of waves and total electron density, we use measured wave intensities to scale an empirical wave model \cite{Meredith12}, whereas ambient plasma measurements are used to scale global empirical density models \cite{Sheeley01,Denton06}. Measurements by Van Allen Probe A and ERG are combined at $L^{*} \sim 4.2$, binned and interpolated in time to an 1-h resolution, and then we calculate diffusion rates at each $1$-h time step (see Fig. \ref{fig36}).

\begin{figure}
\includegraphics[width=1\textwidth]{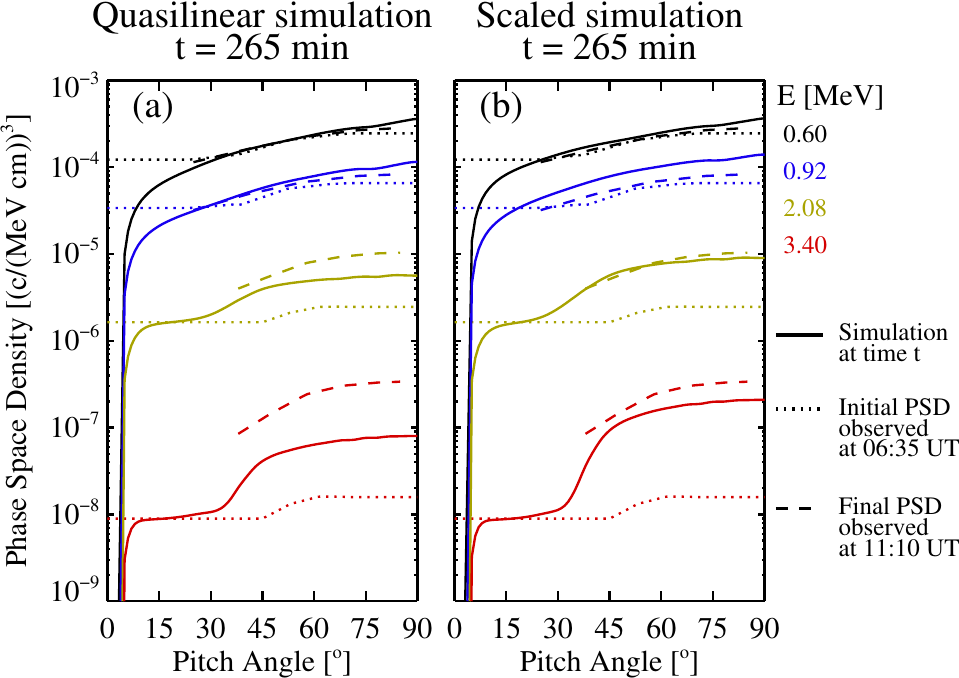}
  \caption{Evolution of pitch-angle distributions of electron PSD at different energies  at $L^* = 4.2$: solid lines show simulation results after 265 min in the simulation, dotted lines show the PSD observed at 06:35 UT (used as the initial condition), and dashed lines show the PSD observed at 11:10 UT (used as the final state to verify the model results). (a) Simulation results using quasi-linear diffusion rates, and (b) simulation results obtained with the rescaled (using a renormalization factor $\langle R \rangle >1$) diffusion rates (see Fig. \ref{fig36}).}
\label{fig37}
\end{figure}

We use classical quasi-linear diffusion coefficients, without or with the nonlinear renormalization factor $\langle R \rangle>1$, to solve the bounce-averaged Fokker-Planck equation \citep{Lyons72,Glauert&Horne05} using University of California, Los Angeles (UCLA) full diffusion code \citep[see][]{Ni08,Ni11,Ma15,Ma18}. The wave normal angle distributions and latitude ranges at different MLT sectors are from \citep{Thorne13:nature}.  After the diffusion coefficients are calculated, we perform 2D Fokker Planck simulation \cite{Ma12} to examine the electron phase space density evolution due to chorus waves. Figure \ref{fig37} compares the observed electron flux dynamics (dashed) and the simulation results (solid). The first simulation, which includes only pure quasi-linear diffusion (with $\langle R \rangle=1$), clearly underestimates the electron acceleration rate above $1$ MeV, where the simulation fails to reproduce the observed rapid $\sim5$-hour electron flux increase (Figure \ref{fig37}a). The second simulation, with renormalized diffusion rates (with $\langle R \rangle>1$), leads to a much better agreement between the modelled increase of relativistic electron fluxes and actual spacecraft observations (Figure \ref{fig37}b). This benchmark comparison therefore confirms the validity of our approach for incorporating the effects of nonlinear resonant interactions into classical Fokker-Planck simulations of energetic electron dynamics. It also underlines the importance of such generalized radiation belt models for an accurate description of electron dynamics in the presence of intense wave-packets \citep[see][for another example of the application of the same approach]{Kondrashov24:NL&QL}.

\section{Discussion and Conclusions}\label{sec:discussion}

\subsection{Effects beyond the classical resonance model}
For simplicity, we may define wave-particle resonant interactions as processes satisfying the cyclotron resonance condition with $O(1)$ accuracy in the $\B_w/B_0$ parameter:
\begin{equation}
\frac{{p_\parallel }}{{m_e }}k_\parallel  - \gamma \omega  + \Omega _0  = 0 \label{eq:resonance_condition}
\end{equation}
Then, we can discuss several effects going beyond this definition, calling them {\it nonresonant}, although some of them can still be called nonlinear resonant effects. Many of these effects have been investigated in details for electromagnetic ion cyclotron waves \cite{Grach&Demekhov18:I,Grach&Demekhov20,Grach22:elfin,Bortnik22:emic} and for magnetosonic waves \cite{Bortnik&Thorne10,Bortnik15}, whereas the importance of these effects for whistler-mode waves is not always obvious. However, to offer a comprehensive picture of nonlinear interactions, we discuss below all these effects, as applied to electron resonant interaction with whistler-mode waves.

\paragraph{Anomalous phase trapping}
One example of such effects has already been considered in Section \ref{sec:nl}, where we discussed wave-particle interactions for small pitch-angle electrons, when the term $\propto (\B_w/B)/\sqrt{I_x}$ should be included into the equation of phase variation
\begin{equation}
\frac{{p_\parallel }}{{m_e }}k_\parallel  - \gamma \omega  + \Omega _0
 = \sqrt {\frac{{2\Omega _0 }}{{Im_e c^2 }}} \frac{{eB_w }}{k}\cos \zeta
\label{eq:resonance_condition_with_theta}
\end{equation}
Such a generalization results in a new type of resonant dynamics, where electrons appear to be phase trapped by default; the so-called {\it autoresonance} effect \cite{Fajans&Friedland01, Friedland09, Neishtadt13, Neishtadt75, Sinclair72:MNRAS}. This effect is often called {\it anomalous trapping} \cite{Kitahara&Katoh19,Gan20:grl} and consists in nearly $100$\% of trapping probability for small pitch-angle electrons. So high trapping probability significantly exceeds estimates given in Section~\ref{sec:nl} for $\Pi\sim \sqrt{\B_w/B_0}$. This discrepancy is due to the fact that the scaling $\Pi\sim \sqrt{\B_w/B_0}$ has been derived for the classical resonance conditions given by Eq. (\ref{eq:resonance_condition}) and the corresponding specific hierarchy of timescales of $(\zeta, I)$ and $(s, p)$ variations. However, an accurate description of a system with a significant wave amplitude impact on the variation of the phase $\zeta$ requires using different approaches and the evaluation of a new Hamiltonian system (see details in Appendix B and \cite{Albert21,Artemyev21:pop}).

%\begin{figure}
%\centering
%\includegraphics[width=1\textwidth]{figs/fig36pdf}
%\caption{\todo{Panel (a) illustrates the anomalous trapping effect. Panel (b) illustrates the positive phase bunching effect. Panel (c) illustrates the resonant broadening effect. Panel (d) illustrates the nonresonant scattering effects.} }
%\label{fig38}
%\end{figure}

\paragraph{Positive phase bunching}
Another effect that cannot be explained within the classical framework of resonant interactions is the positive phase bunching, which corresponds to the situation when resonant electrons gain energy during phase bunching, instead of losing energy. To explain the positive bunching effect, we shall consider the dimensionality of initial particle coordinates in velocity space. With the introduction of the magnetic moment as an adiabatic invariant, the 3D electron velocity vector can be represented by energy, local pitch-angle, and gyrophase, whereas the magnetic moment conservation allows projecting the local pitch-angle to the equatorial one. Both energy and equatorial pitch-angle are integrals of motion and will change only at resonance with the waves. The electron gyrophase is the fast oscillating variable, and introducing the magnetic moment excludes (averages out) this variable from the equations \cite{Cary&Brizard09}. However, the energy and pitch-angle changes at the resonance depend on the gyrophase, and this dependence is usually treated as a dependence on some random variable. Therefore, for fixed initial electron energy and equatorial pitch-angle, the particle ensemble for a single resonant interaction will be characterized by some probability distribution of energy, pitch-angle changes (see examples in Section \ref{sec:mapping} and \citep{Bortnik08,Tao12:GRL,Allanson19,Frantsuzov23:jpp}). The conventional approach is to determine the main characteristics of such distributions through phase averaging. However, such a phase averaging approach can result in a loss of information concerning the particular resonant particles that do not follow one of the two main trends, i.e., which do not drift with the main phase bunched population or do not experience energy, pitch-angle jumps with the phase trapped population. Test particle simulations show a population of electrons experiencing a positive phase bunching: such resonant electrons do not follow either of the two main trends, i.e., they do not lose energy (as phase bunched electrons do) nor gain a large amount of energy (as phase trapped electrons do) \cite{Albert21,Gan22}. Theoretical interpretation of this positive bunching effect has been proposed in \cite{Albert22:phase_bunching}, where the distribution of energy change as a function of the phase in the resonance (or initial gyrophase) has been considered. This distribution indeed includes a small population of bunched particles with the energy gain. Although a contribution of this population into the net electron energy change is not large \cite{Vargas23:pop}, this effect may be important for short-scale precipitations (a couple of resonant interactions).

\paragraph{Resonance broadening}
The finite resonance width effect (also known as {\it resonance broadening}) is most important for large-amplitude waves, such that the particle momentum resonance range is proportional to the square root of the wave amplitude \cite{Karney78,Palmadesso72}. This effect effectively extends the pitch-angle, energy range of electrons affected by resonant interactions with intense waves. An important consequence of this effect for whistler-mode waves is that it provides a potential solution (at least in the presence of very intense waves) to the so-called {\it $90^\circ$ problem}: the problem of absence of resonant scattering for very high equatorial pitch-angle electrons \citep[see discussion in][]{Allanson22,Cai23:broadening}. This is a problem of quasi-linear theory excluding scattering of $\alpha_{eq}\approx 90^\circ$ particles, and thus leaving in the numerical simulations a phase space density peak at large pitch-angles. In test particle simulations such problem disappears \citep{Camporeale&Zimbardo15,Camporeale15:grl}, likely due to the resonance broadening effect. The basic theoretical description of the resonance broadening has been proposed in \citep{Karimabadi&Menyuk91,Karimabadi92}, and then adopted for description of diffusion coefficients for whistler-mode waves \cite{Cai23:broadening} and electromagnetic ion cyclotron waves \cite{Tonoian22:pop}.

\paragraph{Scattering by short wave-packets}
For the sake of simplicity, the resonance condition given by Eq. (\ref{eq:resonance_condition}) is used in many studies for the main frequency characterizing the highest-amplitude plane wave component of the observed wave-packets. However, electron resonant interactions can also be affected by the other frequencies (and corresponding wave numbers) of the wave-packet envelop, even if these waves are usually much less intense than the main wave. This effect is stronger when the wave-packet is shorter, because the temporal/spatial scales of short wave-packets can become comparable to the main wave characteristics. Importantly, the wave-packet envelop may be formed by whistler-mode waves: either by the beating of two waves with close frequencies \citep[see][]{Zhang20:grl:frequency,Nunn21} or by oscillations of currents of electrons nonlinearly interacting with waves around the wave generation region \citep[see][and references therein]{Zhang21:jgr:data&model}. However, the envelope may also be formed by external effects, e.g. by a modulation of the whistler-mode wave source region by ultra-low-frequency waves \citep[e.g.,][]{Xia16:ulf&chorus,Xia20:ulf&vlf}, or by rapid wave damping or diffraction. Therefore, the contribution of the wave-packet envelop may not always be taken into account only via a simple broadening of the wave spectrum, especially in the case when the main wave interacts with electrons nonlinearly.

The short size of wave-packets can contribute to a widening of the energy range of electrons affected by the waves compared with the case of the main wave at peak power alone, an effect which was mostly investigated for electromagnetic ion cyclotron waves \cite{An22:prl,Grach&Demekhov23:theory,Hanzelka23:emic}, where this effect can account for the observed sub-relativistic electron precipitation by such waves, well below the minimum resonant energy with the main wave \citep[see discussions in][]{Angelopoulos23:ssr, Shi24:emics, An24:jgr:ELFIN}. So far, there is no investigations of short wave-packet effect for electron scattering by whistler-mode waves \citep[although numerical simulations may include such effect; see ][]{Camporeale&Zimbardo15,Camporeale15:grl}.

\paragraph{Fractional resonances}
The resonant condition given by Eq. (\ref{eq:resonance_condition}) is written for unperturbed electron trajectory. his equation does not include the wave amplitude. Observed waves, however, can be sufficiently intense for modifying electron trajectories and change the resonance condition, but still not strong enough to result in nonlinear electron dynamics for these new conditions. A good example of this type of effects is the fractional resonance \citep{Lewak&Chen69,Smirnov&FrankKamenestkii68}, recently reevaluated in details for electromagnetic ion cyclotron waves \citep{Hanzelka23:emic}. The basic idea of fractional resonance for whistler-mode waves have been discussed in Hamiltonian formalism in \cite{Fu15:radiation_belts}. In the equation of phase (\ref{eq:A10}) we should take into account a perturbation of electron gyrorotation by wave field: $\psi=\psi_0+Q\sin(\psi_0+\phi_\parallel)$ where $Q\propto \B_w/B_0$ and $\dot\psi_0=\Omega_0/\gamma$. Thus, after couple Jacobi--Anger expansions instead of sum $\sim\sum_{\Nr}\sin\left(\phi_\parallel- \Nr\psi\right)$ we obtain
\begin{eqnarray*}
&\sim&\sum\limits_{\Nr ,n_l } {h_0^{^{\left( {\Nr ,n_l } \right)} } \sin \left( {\phi _\parallel n_l  - \left( {\Nr  - n_l } \right)\psi _0 } \right)}  + \sum\limits_{\Nr ,n_l } {h_{ + 1}^{^{\left( {\Nr ,n_l } \right)} } \sin \left( {\phi _\parallel n_l  - \left( {\Nr  - n_l  + 1} \right)\psi _0 } \right)}  \\
 &+ &\sum\limits_{\Nr ,n_l } {h_{ + 2}^{^{\left( {\Nr ,n_l } \right)} } \sin \left( {\phi _\parallel n_l  - \left( {\Nr  - n_l  + 2} \right)\psi _0 } \right)}
 \end{eqnarray*}
with a following resonant condition:
\[
\dot \phi _\parallel  - n_q \dot \psi _0  = k_\parallel \frac{{p_\parallel }}{{\gamma m_e }} - \omega  - n_q \frac{{\Omega _0 }}{\gamma } = 0
\]
where $n_q=-1+(\Nr+1)/n_l,\; -1+(\Nr+2)/n_l,\; -1+\Nr/n_l$, and $N_l\ne 0$. Therefore, besides the classical resonances with $n_q=\Nr$ (for $n_l=1$), we will have fractional resonances, e.g., $n_q=-1/2$ (for $n_l=2$ and $\Nr=-1$). Such fractional resonances can be considered within the approach proposed in \citep{Terasawa&Matsukiyo12,Hanzelka23:emic}. Note that a fractional resonance requires both a high wave amplitude and an oblique propagation of the wave (see in Eq. (\ref{eq:A10}) term with the gyrophase has an amplitude proportional to $k_\perp$). The main importance of this effect is associated with the scattering of electrons at {\it sub-cyclotron resonant energies}, i.e., the fractional resonance can scatter electrons with energies insufficiently large for classical cyclotron resonant scattering. Such {\it sub-cyclotron scattering} may be important for electron interaction with electromagnetic ion cyclotron waves \citep[because the minimum cyclotron resonant energy for these waves is quite high, see][]{Summers&Thorne03}, but it seem to be less important for whistler-mode waves.

\paragraph{Transient scattering}
The last effect that we would like to discuss in this section is the transient scattering, when the resonance condition in Eq. (\ref{eq:resonance_condition}) can be satisfied only within a limited spatial domain, smaller than the wavelength \cite{Bortnik&Thorne10,Bortnik15}. The energy variation in the resonance is a periodic function of phase gain (see Section~\ref{sec:nl} and Fig. \ref{fig09}(a)), and in the nonresonant situation this integral is equal to zero. However, if the wave field varies within one period (one wavelength), this integral becomes finite even outside the resonance. This is some analog of the {\it short wave-packet} effect, but this transient scattering has been mainly considered for the extremely low frequency branch of whistler-mode waves, fast magnetosonic waves \cite{Bortnik&Thorne10}, which are indeed strongly confined near the equator \cite{Mourenas13:JGR:magnetosonic}. For chorus and hiss modes (propagating quasi field-aligned with a frequency exceeding the lower-hybrid frequency) such a spatial localization might occasionnally be possible, but only in the case of wave scattering and trapping by small scale density fluctuations that would lead to a significant variation of wave characteristics within a typical scale of one wavelength \cite{Streltsov12,Streltsov&Bengtson20,Hanzelka&Santolik22}.

\subsection{Stability of nonlinear resonant interactions}
In this review, we have discussed the importance of wave coherence for nonlinear wave-particle resonant interaction: for less coherent waves with a more modulated wave field (shorter wave-packets) and phase jumps at the wave-packet edges, nonlinear phase trapping becomes quite ineffective. The asymptotic limit of such very ineffective phase trapping is the nonlinear diffusion described in Section~\ref{sec:nldiffusion}. However, even in the case of long wave-packets with potentially strong electron acceleration due to phase trapping, some additional low-coherency waves or non-resonant waves may interrupt this trapping and thereby strongly reduce the acceleration efficiency \cite{Brinca78, Brinca80}. The same effect of trapping destruction may be provided by resonant sideband waves \cite{Nunn86}. Let us consider the corresponding {\it stability} of nonlinear resonant interactions for two most extreme variants: (1) the destruction of phase trapping by the nonresonant noise-like waves, (2) the reduction of the efficiency of nonlinear resonant interactions in case of a broadband wave spectrum.

\paragraph{Trapping destruction}
The effect of trapping destruction by non-resonant waves involves the idea that the phase trapped electron motion can be affected by high-frequency fluctuations \cite{Artemyev11:pre}. To describe this effect, let us introduce the distribution function of trapped particles $f_{tr}$. Trapping is characterized by a periodical motion in the $(\zeta, P_\zeta)$ plane (see Fig. \ref{fig07}(b)), and the area surrounded by the trapped electron trajectory, $I_\zeta=(2\pi)^{-1}\oint{P_\zeta d\zeta}$ is conserved during the slow evolution of the coefficients of the Hamiltonian (\ref{eq:hamiltonian_zeta}). These coefficients depend on the coordinate and momentum $(s,p)$ and vary with time along the trapped trajectory. Therefore, the trapped particles can be characterized by $I_\zeta$: when $I_\zeta=\area/2\pi$, particles either get trapped (if $\area$ grows) or escape from the trapping (if $\area$ decreases). However, non-resonant particle scattering by magnetic field fluctuations with a time-scale smaller than the trapping period $\sim 1/\Omega_{tr}$ can interrupt the conservation of $I_\zeta$. This results in an escape of the particle from the resonance much earlier than in the case of a conserved $I_\zeta$. This process of destruction of the $I_\zeta$ conservation is described by the diffusion equation written for $f_{tr}(I_\zeta)$ \cite{Artemyev15:pop:stability}:
\[
\frac{{\partial f_{tr} }}{{\partial t}} = \frac{\partial }{{\partial I_\zeta  }}\left( {Q\left( t \right)I_\zeta  \frac{{\partial f_{tr} }}{{\partial I_\zeta  }}} \right)
\]
where $t$ is the time along the trapped trajectory, and $Q$, which depends on $(s,p)$, can be rewritten as a function of time along the trapped trajectory. The diffusion coefficient $Q$ is proportional to the power density of nonresonant fluctuation, $Q\propto \B_{w,nonres}^2 \nu_{nonres}$ where $\nu_{nonres}$ is the typical frequency of such fluctuations.

The above diffusion equation shows how $f_{tr}(I_\zeta)$ spreads over the $I_\zeta$ space and reaches the boundary $I_\zeta=\area/2\pi$, where particles escape from trapping and join transient particles. The slow diffusive character of the destruction of $I_\zeta$ conservation suggests that this effect should become important only for long wave-packets, because in such a case the particles are supposed to spend a sufficient time within each phase trapping. Therefore, the most important application of this trapping limitation mechanism is the destruction of the {\it turning acceleration}, which is due to a particularly prolonged electron acceleration through protracted phase trapping \cite{Omura07,Summers&Omura07}.

\paragraph{Nonlinear resonant interaction in a broadband wave spectrum}
There are two competitive processes for systems with a broadband wave ensemble. Firth, a broadband wave ensemble includes multiple resonant waves, and each of these waves may perturb electron trajectories and destroy/interrupt resonant trapping. Second, even during a quarter of the bounce period (the time scale of a single resonant interaction with a monochromatic wave), electrons can experience multiple interactions with waves constituting the broadband ensemble, and thus the overall electron energy, pitch-angle changes can be large despite trapping destruction. This situation is roughly similar to multiple trapping by a wave-packet train \citep{Hiraga&Omura20}, except that trapping duration is very short. The effects of multiple resonant interactions with broadband intense waves has been numerically investigated in \cite{Gan22}, where the electron pitch-angle, energy change due to multiple resonances has been called {\it successive resonant acceleration}. It can be shown that the maximum pitch-angle, energy change via this {\it successive resonant acceleration} is bounded by values of pitch-angle, energy change due to phase trapping by a monochromatic wave along a magnetic field line \citep{Gan22}. Overall, these changes result in an electron diffusion rate smaller than quasi-linear rates for fully destroyed nonlinear resonances. Therefore, similarly to the scaling shown in Fig. \ref{fig26}, numerical simulations demonstrate that electron scattering by intense broadband waves is less effective than the quasi-linear diffusion estimates for the same high wave intensity \citep[see also, e.g.,][]{Tao11:GRL,Tao12,Allanson20,Allanson21}. However, nonlinear drifts may then need to be taken into account, similarly to the case of short intense wave-packets.

\subsection{Feedback to waves}
This review focuses on dynamics of energetic electrons, which fluxes are usually low enough to neglect the electron feedback to waves. However, the energy range of electrons responsible for whistler-mode wave generation can cover $10-100$keV particles \citep{Fu14:radiation_belts}, formally associated with energetic component of electron spectrum. A realistic feedback of such particles to waves scattering and accelerating them is quite important, because this feedback can limit the efficiency of acceleration and determine the wave characteristics. The most straightforward approach for accounting of electron feedback to waves is the self-consistent simulations \citep[see][]{Katoh&Omura04,Katoh&Omura07:acceleration,Katoh08:angeo,Tao14} where electrons are traced as individual particles \citep[see alternatives in][]{Nunn05,Nunn21,Demekhov11,Demekhov&Trakhtengerts08,Tao21}. This approach is mostly applicable to a small-scale and short simulation setups focused on the wave generation problem \citep[e.g.,][]{Katoh&Omura16,Katoh&Omura13,Demekhov17}, whereas the simulation of long-term dynamics of energetic electron populations requires too large numerical resources within such approach.

The approach alternative to the full numerical simulations consists in theoretical determination of impact of energetic electrons on wave growth/damping processes controlling wave amplitude variation \citep[][]{Shklyar11:angeo,Shklyar17,Omura08,Omura13:AGU,Omura21:review,Tao20}. This approach reproduces many important characteristics of whistler-mode waves \citep[see details in][]{Cully11,Tao12:sweeprate,Mourenas15} and principally can provide self-consistent variation of wave amplitude \citep[see examples in][]{Shklyar21,Shklyar&Luzhkovskiy23,Luzhkovskiy&Shklyar23}. The main advantage of such theoretical consideration is that it can be incorporated into evaluation of main characteristics of particle acceleration models based on mapping technique or the Green function approach. In perspective, this merging of particle feedback and acceleration models should provide more realistic results for wave-particle nonlinear interactions in the radiation belts and should naturally limit nonlinear effects.

\section{Conclusions}
This review aimed to provide a detailed overview of the effects of nonlinear resonant interactions of radiation belt electrons with intense whistler-mode waves. We focused on the idea that such interactions should change the timescales of electron acceleration and losses in comparison with quasi-linear diffusion, and then investigated different approaches for modeling this change. The most basic approach described here, the generalized Fokker-Planck equation, accounted for all effects of bunching and phase trapping for the evolution of the electron distribution, and, most importantly, allowed us to show that the asymptotic solutions for nonlinear interactions and for diffusion are identical. Although this is a simplified approach and many realistic generalizations would require significant modifications, the existence of this asymptotic solution remains a key for further constructions of more advanced techniques for the simulation of nonlinear wave-particle interactions. One of the most promising technique, including many realistic properties of whistler-mode waves, is the mapping technique, which allowed us to consider the effects of a wave ensemble as well as the effects of short wave-packets. This technique, carefully applied to observed wave and electron distributions, may model electron flux dynamics with a sufficient accuracy to be verified by comparisons with spacecraft observations (see examples in Appendix E). Two alternative techniques, the {\it Green function approach} and the {\it probabilistic approach}, may include many realistic wave details \citep[see][]{Omura15,Kubota&Omura18,Hiraga&Omura20,Hsieh22,Hsieh&Omura17,Hsieh&Omura23}, but they still require verification against spacecraft observations. Besides the mapping technique, we discussed a possible approach for the incorporation of the effects of nonlinear resonant interactions in the existing global numerical Fokker-Planck models of the radiation belts. Although this last approach takes wave-particle interactions into account only in a very simplified, somewhat integral way, it is particularly promising, because it directly brings nonlinear physics into already well developed state of the art simulations.

\begin{acknowledgements}
This review incorporates several results obtained by Alexander Lukin, Victor Frantsuzov, and David Tonoian \cite{Lukin24:jgr,Frantsuzov23:jpp,Tonoian23:pop}. We are deeply grateful to these researchers for providing datasets and for helpful discussion of their results.

The review summarizes Authors' current understanding of nonlinear resonant interactions and theoretical approaches for evaluation and description of these interactions. An invaluable contribution to this understanding has been provided by multiple fruitful discussions and join research projects with leading specialists of radiation belt physics and nonlinear dynamical systems, far from complete list of whom includes Adnane Osmane, Andrei Demekhov, David Shklyar, David Nunn, Li Li, Longzhi Gan, Lunjin Chen, Jay Albert, Miroslav Hanzelka, Oliver Allanson, Veronica Grach, Xin An, Xin Tao, Xuzhi Zhou, Yikai Hsieh, Yoshiharu Omura.

The work of O.V.A. was supported by NASA contracts 80NSSC20K0218, 80NSSC22K0433, 80NSSC22K0522, 80NSSC20K0697, and 80NSSC21K1770. Q.M. would like to acknowledge the NASA grants 80NSSC20K0196 and 80NSSC24K0572 and NSF grant AGS-2225445. J.B. gratefully acknowledges support from NASA 80NSSC22K1637, NASA/HTMS 80NSSC20K1270 subcontract through Boston University, NSF/GEM award 2225613 subcontract to UCLA through the University of Iowa, and NSF/GEM award AGS‐ 2025706. The work of V.V.K. was supported by CNES {\it Solar Orbiter} and {\it Parker Solar Probe} projects, and by NASA contract 80NSSC21K1770. The work of A.V.A., X.-J.Z., and D.L.V. was supported by NASA contracts 80NSSC24K0138, 80NSSC23K0100,  80NSSC23K0108, 80NSSC22K0522, 80NSSC23K1038, 80NSSC24K0561, 80NSSC19K0844.

\end{acknowledgements}
\section*{Conflict of Interest}

The authors have no relevant financial or non-financial interests to disclose.

\section*{Appendix A}\label{Appendix:A}
This Appendix is devoted to derivation of the main Hamiltonian equations for wave-particle resonant interaction in two systems (two types of resonances): cyclotron  resonance with field-aligned whistler-mode waves and Landau resonance with highly-oblique (quasi-electrostatic) whistler-mode waves. The general approach for the derivation of these Hamiltonians repeats results presented in \citep{Albert93,Vainchtein18:jgr,Artemyev18:jpp}: we start with writing wave field through the vector and scalar potentials, then we introduce the magnetic moment, and an adiabatic invariant resulted from averaging of cyclotron rotations, and finally we consider a general Hamiltonian system around the resonance. We provide and expand results on the separation of Hamiltonians describing slow electron bounce motion and fast resonant phase oscillations. The latter Hamiltonian is the classical Hamiltonian of the pendulum with torque \cite{bookAKN06}. We derive coefficients of this Hamiltonian for two resonances, and then describe a very important property of this Hamiltonian leading to the equation of particle energy change due to the nonlinear phase bunching \cite{Neishtadt75}.

\subsection{Wave equations}
The cold plasma dispersion relation \citep{bookStix62} provides the relationship between electric and magnetic field components of whistler mode waves. Let $B_y=B_{w0}\Re(e^{i\phi})=B_{w0}\cos\phi$ where $\phi$ is the wave phase, then \cite{Tao&Bortnik10}
\begin{eqnarray}
 \frac{{B_x }}{{\B_{w} }} &=&  - \varepsilon _2 \frac{{N^2 \sin ^2 \theta  - \varepsilon _3 }}{{\varepsilon _3 \left( {\varepsilon _1  - N^2 } \right)}}\sin \phi ,\quad \frac{{B_z }}{{\B_{w} }} = \tan \theta \varepsilon _2 \frac{{N^2 \sin ^2 \theta  - \varepsilon _3 }}{{\varepsilon _3 \left( {\varepsilon _1  - N^2 } \right)}}\sin \phi,\nonumber \\
 \frac{{E_x }}{{\B_{w} }} &=&  - \frac{{N^2 \sin ^2 \theta  - \varepsilon _3 }}{{\varepsilon _3 N\cos \theta }}\cos \phi ,\quad \frac{{E_y }}{{\B_{w} }} = \frac{{\varepsilon _2 }}{{\varepsilon _1  - N^2 }}\frac{{N^2 \sin ^2 \theta  - \varepsilon _3 }}{{\varepsilon _3 N\cos \theta }}\sin \phi \nonumber \\
 \quad \frac{{E_z }}{{\B_{w} }} &=&  - \frac{{N\sin \theta }}{{\varepsilon _3 }}\cos \phi  \label{eq:A01}
\end{eqnarray}
Functions $\varepsilon_1$, $\varepsilon_2$, and $\varepsilon_3$ are components of the cold plasma dielectric tensor:
\begin{equation}
\varepsilon _1  = 1 - \frac{{\Omega _{pe}^2 }}{{\omega ^2  - \Omega _{0}^2 }},\quad \varepsilon _2  = \frac{{\Omega _{0} }}{\omega }\frac{{\Omega _{pe}^2 }}{{\omega ^2  - \Omega _{0}^2 }},\quad \varepsilon _3  = 1 - \frac{{\Omega _{pe}^2 }}{{\omega ^2 }}
\label{eq:A02}
\end{equation}
where $\omega$ is the wave frequency, $k_\perp$ and $k_\parallel$ are two components of the wave vector lying in the $(x,z)$ plane ($k_\perp=k\sin\theta$, $k_\parallel=k\cos\theta$), $\Omega_{0}>0$ is the electron gyrofrequency in the background magnetic field directed along the $z$ axis, $\Omega_{pe}$ the plasma frequency. Wave frequency and wave vector satisfy the wave dispersion relation $N=N(\omega,\theta)$ where $N=kc/\omega$ is the wave refractive index \citep{bookStix62}. Equations (\ref{eq:A01}) lead to the following expressions for the total wave magnetic and electric fields:
\begin{eqnarray}
 \bar\B_w^2  &=& \frac{{1 }}{{2\pi }}\int\limits_0^{2\pi } {\left( {B_x^2  + B_y^2  + B_z^2 } \right)d\phi }  = \frac{1}{2}\B_{w}^2 \left( {C_2 C_1  + 1} \right) \nonumber\\
 \bar\E_w^2  &=& \frac{{1}}{{2\pi }}\int\limits_0^{2\pi } {\left( {E_x^2  + E_y^2  + E_z^2 } \right)d\phi }  = \frac{1}{2}\frac{{\B_{w}^2 }}{{N^2 }}\left( {C_1 \left( {1 + C_2 } \right) + \left( {N^2 \varepsilon _3^{ - 1} \sin \theta } \right)^2 } \right) \nonumber\\
 C_1 & =& \left( {\frac{{N^2 \varepsilon _3^{ - 1} \sin ^2 \theta  - 1}}{{\cos \theta }}} \right)^2 ,\quad C_2  = \left( {\frac{{\varepsilon _2 }}{{\varepsilon _1  - N^2 }}} \right)^2
\label{eq:A03}
\end{eqnarray}
For parallel propagating waves ($\theta=0$), we have $C_1=1$, $N^2=\varepsilon_1-\varepsilon_2$, $C_2=1$, and $\bar\B_w=\B_{w}$, $\bar\E_w=\B_{w}/N$.

We use Coulomb gauge to write field components through components of vector potential ${\bf A}$ and scalar potential $\varphi$:
\begin{eqnarray}
A_y  &=& \frac{{\B_{w} }}{k}\sqrt {C_2 C_1 } \cos \phi ,\quad A_x  = \frac{{\B_{w} }}{k}\cos \theta \sin \phi ,\quad  \label{eq:A04}\\
 A_z  &=&  - \frac{{\B_{w} }}{k}\sin \theta \sin \phi ,\quad \varphi  = \frac{{\B_{w} }}{{kN}}\left( {N^2 \varepsilon _3^{ - 1}  - 1} \right)\tan \theta \sin \phi  \nonumber
\end{eqnarray}
We introduce $A_w=\B_{w}/k$, $C=\sqrt{C_1C_2}$, $C_3=(N^2\varepsilon_3^{-1}-1)/N\cos\theta$ and write $A_x=A_w\cos\theta\sin\phi$, $A_y=A_wC\cos\phi$, $A_z=-A_w\sin\theta\sin\phi$, $\varphi=A_wC_3\sin\theta\sin\phi$ where $\B_{w}=\bar\B_{w}\sqrt{2/(C^2+1)}$. Highly oblique waves, with $\theta$ close to the resonance cone angle $\theta_r$ \citep{bookStix62}, are characterised by a significant variation of vector and scalar potential amplitudes, whereas these amplitudes are almost constant for parallel propagating waves.

\subsection{Electron dynamics}
We start with the Hamiltonian of an electron (of charge $-e$ and mass $m_e$) moving in the background magnetic field (described by the $y$-component of the vector potential $A_{0}=-xB_0(z)$ \citep{Bell84}) and the wave electromagnetic field (given by Eqs.(\ref{eq:A04})):
\begin{equation}
H = \sqrt {m_e^2 c^4  + c^2 \left( {{\bf p} + \frac{e}{c}{\bf A}(x,z,t) + \frac{e}{c}A_{0} (x,z){\bf e}_y } \right)^2 }  - e\varphi (x,z,t)
\label{eq:A05}
\end{equation}
where ${\bf p}$ is the particle generalized momentum. Note that $(x,y,z)$ are Cartesian coordinates, i.e., we do not take into account the magnetic field line curvature of the dipole field \citep[see a more general consideration in][]{Shklyar09:review}. Although we use a $B_0(z)$ function simulating the distribution of dipole field strength along a magnetic field line, the same consideration is valid for arbitrary ambient magnetic fields with a sufficiently large spatial scale of inhomogeneity.

The Hamiltonian (\ref{eq:A05}) does not depend on $y$ and the conjugated momentum is constant, $p_y=const$. We consider particles motion around field-line with $x=0$, and thus $p_y$ can be taken equal to zero. There are two important assumptions here. First, we assume that the spatial scale of $B_0(z)$ variation is much larger than the wavelength $\sim (\partial \phi/\partial x)^{-1},\, (\partial \phi/\partial z)^{-1}$. Therefore, the period of particle oscillations in the plane $(z, p_z)$ is much longer than the wave period $\phi$ (wave frequency $-\partial \phi/\partial t$ is about electron gyrofrequency $eB_0/m_ec$ and much larger than the inverse of the time-scale of $(z,p_z)$ oscillations). Second, we assume a small wave amplitude $eA_0/m_ec^2\ll 1$ (for typical conditions in the Earth radiation belts, this limit is satisfied since it means that the wave magnetic field amplitude must be much smaller than $10$ nT) and expand the Hamiltonian (\ref{eq:A05}) as:
\begin{eqnarray}
H &=& H_0  + \frac{{ce}}{{H_0 }}{\bf A}\left( {{\bf p} + \frac{e}{c}A_{0} {\bf e}_y } \right) - e\varphi  \label{eq:A06}
\\
 H_0  &=& \sqrt {m_e^2 c^4  + c^2 p_z^2  + c^2 p_x^2  + c^2 \left( {m_ex\Omega _0 (z)} \right)^2 } \nonumber
\end{eqnarray}
where $\Omega_0=eB_0(z)/m_ec$. Considering the unperturbed system $H=H_0$, we introduce the adiabatic invariant (magnetic moment):
\begin{equation}
I_x  = \frac{1}{{2\pi }}\oint {p_x dx}  = \frac{{H_0^2  - m_e^2 c^4  - c^2 p_z^2 }}{{2m_e c^2 \Omega _0 }}
\label{eq:A07}
\end{equation}
The corresponding coordinates $(x,p_x)$ can be rewritten as: $p_x=-\sqrt{2I_x\Omega_0m_e }\sin\psi$ and $x=\sqrt{2I_x/\Omega_0m_e}\cos\psi$ where $I_x, \psi$ are conjugated variables (i.e., the transformation $(x,p_x)$ to $(\psi, I_x)$ is canonical \citep{bookLL:mech}). To introduce these variables we use the generating function $W_0=\int p_xdx + \tilde{p}_{z}z$ where
\begin{equation}
\tilde{p}_{z}  = p_{z}-\frac{{\partial W_0}}{{\partial z}}=p_z-I_x\frac{1}{{2}}\frac{{\partial\ln \Omega_0}}{{\partial z}}\sin2\psi
\label{eq:A07a}
\end{equation}
and $z$ does not change. Therefore, the Hamiltonian (\ref{eq:A06}) can be rewritten in new variables:
\begin{eqnarray}
H &=& m_e c^2 \sqrt {1 + \frac{{1}}{{m_e^2 c^2 }}\left(\tilde{p}_{z}+ I_x\frac{1}{{2}}\frac{{\partial\ln \Omega_0}}{{\partial z}}\sin2\psi\right)^2 + \frac{{\rho ^2 \Omega _0^2 }}{{c^2 }}}  + H_1,  \label{eq:A08a}\\
 H_1  &=& \frac{1}{{\gamma _0 }}\left( {\frac{{\rho \Omega _0 }}{c}\left( {eA_y \cos \psi  - eA_x \sin \psi } \right) + \frac{{eA_z }}{{m_e c}}\left(\tilde{p}_{z}+ I_x\frac{1}{{2}}\frac{{\partial\ln \Omega_0}}{{\partial z}}\sin2\psi \right) } \right) - e\varphi  \nonumber
\end{eqnarray}
where $\rho=\sqrt{2I_x/m_e\Omega_0}$. There is a fast oscillating term $\sim \sin2\psi$.  This term results in variation of adiabatic invariant $I_x$ ($\dot I_x=-\partial H/\partial \psi$). The amplitude of this term is $\sim \partial\ln\Omega_0/\partial z$, i.e. $I_x$ is conserved with the accuracy $\sim \partial\ln\Omega_0/\partial z$. In absence of wave perturbations (i.e., in absence of external forces depending on wave phase $\phi$) we can always introduce the improved adiabatic invariant $J_x$, which is conserved with the accuracy $\sim (\partial\ln\Omega_0/\partial z)^2$ \citep{bookAKN06}. In the system with wave depending on phase $\phi$, the variation of $I_x$ is determined by resonance of wave phase $\Nr\psi+\phi$, $\Nr=0,\pm1,\pm2,...$. For resonant particles with $d(\Nr\psi+\phi)/dt\approx 0$ the phase $2\psi$ is a fast oscillating nonresonant phase, and, after expansion of $\gamma_{new}$ over $\partial\ln\Omega_0/\partial z$, the corresponding term $\sim\sin2\psi$ can be omitted as a fast oscillating term with mean zero value. Therefore, we can rewrite Hamiltonian (\ref{eq:A08a}) as
\begin{eqnarray}
H &=& m_e c^2 \gamma  + H_1,\; \gamma  = \sqrt {1 + \frac{{p_{z}^2}}{{m_e^2 c^2 }} + \frac{{\rho ^2 \Omega _0^2 }}{{c^2 }}}  \label{eq:A08}\\
 H_1  &=& \frac{1}{{\gamma}}\left( {\frac{{\rho \Omega _0 }}{c}\left( {eA_y \cos \psi  - eA_x \sin \psi } \right) + \frac{{eA_z }}{{m_e c}}p_z } \right) - e\varphi  \nonumber
\end{eqnarray}
were $\tilde{p}_{z}\approx p_z$.

Note that the typical spatial scale of electron dynamics in the strong magnetic field, of the order of the electron gyroradius $\rho_e$, is much smaller than the magnetic field curvature radius, $\R_c=R_EL/3$ for a dipole field. This allows to omit effects of electron curvature drifts, $\sim \rho_e/\R_c$, and consider electron motion along a magnetic field line as a straight trajectory, what makes $z$ coordinate and conjugated momentum $p_z$ equal to the field-aligned coordinate $s$ and the corresponding momentum $p_\parallel$. Therefore, we now can replace $(z,p_z)$ by $(s,p_\parallel)$, underlying that we deal with the field-aligned coordinate and momentum of electrons. Systems with a finite $\rho_e/\R_c$ effect include the so-called curvature scattering \citep[e.g.,][]{Birmingham84,BZ89,Artemyev13:angeo:scattering,Brizard23} and for such systems the $(z,p_z)\to(s,p_\parallel)$ transition requires application of a quite sophisticated procedure, either based on non-canonical \citep[e.g.,][ and references therein]{Cary&Brizard09} or canonical \citep[e.g.,][and references therein]{Neishtadt&Artemyev20} guiding center theory. Some examples of such transformations for systems with nonlinear wave-particle resonant interactions can be found in \cite{Vainchtein09,Neishtadt11:ppcf,Artemyev22:jgr:DF&ELFIN}, whereas the application of such transformations for charged particle diffusion by waves was described in details by \cite{Brizard&Chan22}.

We then substitute Eqs. (\ref{eq:A04}) into the $H_1$ expression to get:
\begin{equation}
H_1  = \frac{{eA_w }}{{\gamma }}\frac{{\rho \Omega _0 }}{c}\left( {C\cos \phi \cos \psi  - \cos \theta \sin \phi \sin \psi } \right) - eA_w \left( {\frac{{p_\parallel }}{{\gamma m_e c}} + C_3 } \right)\sin \theta \sin \phi \label{eq:A09}
\end{equation}
where wave phase $\phi$ can be written as
\begin{equation}
\phi  = \phi _\parallel   + k_\perp \rho \cos \psi, \; \phi _\parallel  = \int\limits_{}^s {k_\parallel (\tilde s)d\tilde s} - \omega t
\label{eq:A10}
\end{equation}
Using the Jacobi--Anger expansion, we rewrite Eq. (\ref{eq:A09}) as
\begin{eqnarray}
H_1  &=& eA_w \sum\limits_{\Nr =  - \infty }^\infty  {h^{(\Nr)} } \left( {I_x ,p_\parallel ,s} \right)\sin \left( {\phi _\parallel   + \Nr\frac{\pi }{2} - \Nr\psi } \right) \label{eq:A11}\\
 h^{(\Nr)}  &=& \frac{{\rho \Omega _0 }}{{2c\gamma }}\left( {\left( {C - \cos \theta } \right)J_{\Nr - 1} \left( {k_\perp \rho } \right) - \left( {C + \cos \theta } \right)J_{\Nr + 1} \left( {k_\perp \rho } \right)} \right)\nonumber\\ &-& \left( {\frac{{p_\parallel }}{{\gamma m_e c}} + C_3 } \right)\sin \theta J_{\Nr} \left( {k_x \rho } \right) \nonumber
 \end{eqnarray}
where $J_{\Nr}$ is the Bessel function. We consider waves with insufficiently strong amplitudes to allow an overlap of resonances, i.e., such that each term in the sum of resonances can be considered independently \citep[i.e., the resonance width $\sim\sqrt{|eA_wh^{(\Nr)}/m_e|}$ is much smaller than the smallest distance between two resonances $\sim \Omega_0/k\cos\theta$, see the corresponding discussion in][]{Shklyar81}. For one particular $\Nr$, Eq. (\ref{eq:A08}) takes the form
\[
H^{(\Nr)} = m_e c^2 \gamma  + eA_w h^{(\Nr)} \sin \left( {\phi _\parallel   + \Nr\frac{\pi }{2} - \Nr\psi } \right)
\]
and we rewrite this Hamiltonian as
\begin{equation}
H^{(\Nr)} = m_e c^2 \gamma  + eA_w h^{(\Nr)} \cos \left( {\phi _\parallel   + (\Nr+1)\frac{\pi }{2} - \Nr\psi } \right)
\label{eq:A12}
\end{equation}

Numerically solving the Hamiltonian equations for $(s,p_\parallel)$, $(\psi, I_x)$ variables for the Hamiltonian (\ref{eq:A12}) allows to describe the resonant interaction of charged particles with a whistler mode wave. In the next subsection, we investigate the Hamiltonian (\ref{eq:A12}) to determine the main characteristics of this resonant interaction.

\subsection{Wave-particle resonance}
First, we would like to consider a conservative system, with conserved energy, whereas the Hamiltonian (\ref{eq:A12}) depends on time through phase $\phi_\parallel$. To exclude this temporal dependence, we introduce new conjugated variables $(\zeta_{\Nr}, I)$ and $(s, p)$ with $\zeta_{\Nr}=\phi_\parallel -(\Nr+1)\psi+\Nr\pi/2$. This variable change excludes phase $\psi$ from the Hamiltonian and, thus, the new momentum $\tilde{I_x}$ is constant -- i.e., instead of two pairs of conjugated variables $(s, p_\parallel)$, $(\psi, I_x)$ we introduce $(s, p)$, $(\zeta_{\Nr}, I)$. To this aim, we use the generating function $W=(\phi_\parallel +\Nr\psi+\Nr\pi/2)I+sp+\psi \tilde{I}_x$ corresponding to:
\begin{equation}
\zeta _{\Nr}  = \phi _\parallel   + \Nr\frac{\pi }{2} - \Nr\psi ,\quad p_\parallel = k_\parallel I + p ,\quad I_x  =  - \Nr I + \tilde I_x
\label{eq:A13}
\end{equation}
Therefore, the new Hamiltonian $\h=H^{(\Nr)}+\partial W/\partial t$ takes the form
\begin{eqnarray}
\h &=&  - \omega I + m_e c^2 \gamma  + eA_w h^{(\Nr)} \cos \zeta _n  \label{eq:A14}\\
 \gamma  &=& \sqrt {1 + \frac{{\left( {k_\parallel I + p } \right)^2 }}{{m_e^2 c^2 }} + \frac{{2\Omega _0 }}{{m_e c^2 }}\left( {\tilde I_x  - \Nr I} \right)}  \nonumber
\end{eqnarray}
where $h^{(\Nr)}= h^{(\Nr)}(\tilde{I}_x-\Nr I, k_\parallel I+p,s)$. Hamiltonian (\ref{eq:A14}) does not depend on $\psi$, and thus $\tilde{I}_x$ is the constant. For $\Nr=0$ we have $I_x=\tilde{I}_x=const$, i.e. $\tilde{I}_x$ is the initial magnetic moment that is conserved for Landau resonance. For $\Nr\ne0$ we have $I_x=-\Nr I+\tilde{I}_x$ and we can set $\tilde{I}_x=0$ by choosing the initial $I$ value (far from the resonance). For example, for the first cyclotron resonance with $\Nr=-1$ we can set initial $I$ equals to the initial magnetic moment $I_x$, and thus $\tilde{I}_x=0$.

The resonance condition $\dot\zeta_{\Nr}=0$ can be written through Hamiltonian equations:
\begin{equation}
\dot \zeta _{\Nr}  = \frac{{\partial \h}}{{\partial I}} =  - \omega  + \frac{{k_\parallel \left( {k_\parallel I + p } \right) - m_e \Omega _0 \Nr}}{{m_e \gamma }} = 0\label{eq:A15}
\end{equation}
where we omit perturbations $\sim eA_w\ll m_ec^2$. The solution of Eq. (\ref{eq:A15}) provides the resonant value $I_R$ of momentum $I$:
\begin{equation}
\frac{{k_\parallel I_R }}{{m_e c}} =  - \frac{{p }}{{m_e c}} + \frac{{\Nr\Omega _0 }}{{\omega N_\parallel  }} + \frac{1}{{\sqrt {N_\parallel ^2  - 1} }}\sqrt {1 + \beta _ \bot ^2  - \frac{{\Nr^2\Omega _0^2 }}{{\omega^2 N_\parallel ^2 }} + 2\frac{{\Nr\Omega_0 }}{{\omega N_\parallel  }}\frac{{p }}{{m_e c}}} \label{eq:A16}
\end{equation}
where $N_\parallel=k_\parallel c/\omega$, and $\beta_\bot=\sqrt{2\tilde{I}_x\Omega_0/m_ec^2}$. Particles with $I=I_R$ are in resonance with the wave. Therefore, to consider such resonant particles we expand Hamiltonian (\ref{eq:A04}) around $I_R$:
\begin{eqnarray}
\h &=& \Lambda  + \frac{1}{2M}\left( {I - I_R } \right)^2  + eA_w h_R^{(n)} \cos \zeta _n  \nonumber\\
 \Lambda  &=&  - \omega I_R  + m_e c^2 \gamma _R  = \frac{{m_e c^2 }}{{N_\parallel ^2 }}\left( {\frac{{N_\parallel  p }}{{m_e c}} - \frac{\Nr\Omega _0}{\omega}  + \left( {N_\parallel ^2  - 1} \right)\gamma _R } \right) \nonumber\\
 \gamma_R  &=& \frac{{N_\parallel  }}{{\sqrt {N_\parallel ^2  - 1} }}\sqrt {1 + \beta _ \bot ^2  - \frac{{\Nr^2\Omega _0^2 }}{{\omega^2N_\parallel ^2 }} + 2\frac{{n\Omega _0 }}{{\omega N_\parallel  }}\frac{{p }}{{m_e c}}}  \nonumber\\
 \frac{1}{M} &=& \left. {\frac{{\partial ^2 \gamma }}{{\partial I^2 }}} \right|_{I = I_R }  = \left( {N_\parallel ^2  - 1} \right)\frac{{\omega ^2 }}{{m_e c^2 \gamma _R }} \label{eq:A17}
\end{eqnarray}
where $h_R^{(\Nr)}=h^{(\Nr)}(nI_R+\tilde{I}_x, k_\parallel I_R+p,s)$. We introduce variables $P_\zeta=I-I_R$, $\tilde p$, $\tilde s$ through the generating function $W=(I-I_R)\zeta_{\Nr}+p\tilde s$ (we keep the old notation for $\zeta_n$, because this transformation does not change the phase):
\begin{equation}
P_\zeta = I - I_R ,\quad \tilde{p} = p  - \frac{{\partial I_R }}{{\partial s}}\zeta _{\Nr} ,\quad \tilde{s} = s + \frac{{\partial I_R }}{{\partial p }}\zeta _{\Nr}
\label{eq:A18}
\end{equation}
The new Hamiltonian $\h=\Lambda(\tilde{s}, \tilde{p})+m_ec^2P_\zeta^2/2M(\tilde{s}, \tilde{p})+eA_0h_R^{(\Nr)}(\tilde{s}, \tilde{p})\sin\zeta_{\Nr}$ contains the $\Lambda$ function depending on $(\tilde{s}, \tilde{p})$. Terms $\sim \partial I_R/\partial s$, $\partial I_R/\partial p$ in Eq. (\ref{eq:A18}) are much smaller than $(s,p)$ terms due to the condition $\partial/\partial s \ll k_\parallel$. Therefore, we can expand the function $\Lambda$:
\begin{eqnarray}
\Lambda \left( {s,p } \right) &=& \Lambda \left( {s - \frac{{\partial I_R }}{{\partial \tilde{p} }}\zeta _{\Nr} ,p + \frac{{\partial I_R }}{{\partial \tilde{s}}}\zeta _{\Nr}} \right)  \label{eq:A19}\\
&=& \Lambda \left( {s,p} \right) + \left( {\frac{{\partial I_R }}{{\partial \tilde{p} }}\frac{{\partial \Lambda }}{{\partial \tilde{s}}} - \frac{{\partial I_R }}{{\partial \tilde{s}}}\frac{{\partial \Lambda }}{{\partial \tilde{p}}}} \right)\zeta _{\Nr}  = \Lambda  - {\rm A}\zeta _{\Nr} \nonumber
 \end{eqnarray}
where ${\rm A}=\{\Lambda, I_R\}_{\tilde{s},\tilde{p}}\approx \{\Lambda, I_R \}_{s,p}$. Therefore, the new Hamiltonian $\h$ can be split into two Hamiltonians $\h=\Lambda+\h_\zeta$, where $\Lambda=\Lambda(s,p)\sim m_ec^2$ describes the slow evolution of $(s,p)$ variables, while $\h_\zeta\sim eA_0\ll m_ec^2$ depends on these variables as on parameters and describes the fast variations of $(\zeta_{\Nr}, P_\zeta)$:
\begin{eqnarray}
 \h_\zeta &= &\frac{1}{2M}P_{\zeta}^2  + {\rm A}\zeta _n  + {\rm B} \cos \zeta _{\Nr}, \;\;\; {\rm B}= eA_w h_R^{(\Nr)}  \nonumber\\
\label{eq:A20}\\
 {\rm A} &=& -\frac{{m_e c^2 \D N_\parallel ^2}}{{N_\parallel ^2  - 1}}\frac{1}{{\gamma _R }}\left( {\frac{{\left( {\gamma _R  + \Nr\Omega_0/\omega } \right)^2 }}{{N_\parallel ^2 }}\frac{{\partial \ln N_\parallel  }}{{\partial \ln \Omega _0 }} - \frac{{\Nr\Omega _0 }}{{\omega N_\parallel  }}\frac{p}{{m_e c}} - \frac{1}{2}\beta _ \bot ^2 } \right) \nonumber
 \end{eqnarray}
where $\D=c(\partial \ln \Omega_0/\partial s)/N_\parallel\omega\ll 1$ is a dimensionless factor determining the scale ratio of the inhomogeneity scale $\sim (\partial/\partial s)^{-1}$ and wavelength scale $\sim 1/k_\parallel$.

\subsection{Parallel propagating waves: cyclotron resonance}
Let us consider nonlinear resonant electron interaction with parallel propagating whistler-mode waves (i.e., with $k_\perp=0$ and $C_1=C_2=C=1$ in Eq. (\ref{eq:A11})). Therefore, $\Nr=-1$ and the Hamiltonian $H_1$ takes the form
\begin{equation}
H_1  =  eA_w \frac{{\rho \Omega _0 }}{{c\gamma}}\cos \left( {\phi _\parallel   + \psi} \right)
\label{eq:A31}
\end{equation}
and there are no other resonances (the corresponding Bessel functions in Eq. (\ref{eq:A11}) are equal to zero). We consider a magnetic field model with $\Omega_0(s/R)$, where $R$ is the spatial scale of the field gradient. The coefficient ${\rm A}$ from Eq. (\ref{eq:A20}) takes the form:
\begin{eqnarray}
 {\rm A} &=& -\frac{{m_e c^2 \D}}{{\gamma _R }}\frac{{N_\parallel^2 }}{{N_\parallel^2  - 1}}\left( {\left(\frac{p_{\parallel,R}}{{m_e c}}\right)^2\frac{{\partial \ln N_\parallel }}{{\partial \ln \Omega _0 }} + \frac{{\Omega _0 }}{{\omega N_\parallel }}\frac{p}{{m_e c}} - \frac{1}{2}\beta _ \bot ^2 } \right)\nonumber \\
 &=&-\frac{{m_e c^2 \D}}{{\gamma _R }}\frac{{N_\parallel^2 }}{{N_\parallel^2  - 1}}\left( {\left( {\frac{{p_{\parallel,R} }}{{m_e c}}} \right)^2 \frac{{\partial \ln N_\parallel }}{{\partial \ln \Omega _0 }} + \frac{{\Omega _0 }}{{\omega N_\parallel }}\frac{{p_{\parallel,R} }}{{m_e c}} - \frac{{\Omega _0 \left( {I_R + \tilde I_x } \right)}}{{m_e c^2 }}} \right) \label{eq:AA}  \\
 &=&-\frac{{m_e c^2 \D}}{{\gamma _R }}\frac{{N_\parallel^2 }}{{N_\parallel^2  - 1}}\left( {\left( {\frac{{p_{\parallel,R} }}{{m_e c}}} \right)^2 \frac{{\partial \ln N_\parallel }}{{\partial \ln \Omega _0 }} + \frac{{\Omega _0 }}{{\omega N_\parallel }}\frac{{p_{\parallel,R} }}{{m_e c}} - \frac{{\Omega _0 I_{x,R} }}{{m_e c^2 }}} \right) \nonumber
\end{eqnarray}
where $p_{\parallel,R}/m_ec=(\gamma_R-\Nr\Omega_0/\omega)/N_\parallel=(\gamma_R\omega-\Omega_0)/k_\parallel c$ is the normalized resonant momentum, and $I_{x,R}$ is the magnetic moment in the resonance. For $I_{x,R}$ we can use $I_R$ expression, because constant $\tilde{I}_{x}$ can be set to zero for $\Nr\ne 0$. We aslo can use definition of $\gamma$ in the resonance, $\gamma_R=\sqrt{1+\left(p_{\parallel,R}/m_ec\right)^2+2I_{x,R}\Omega_0/m_ec^2}$, and rewrite Eq. (\ref{eq:AA}) as
\[
   {\rm A} = -\frac{{m_e c^2 \D}}{{\gamma _R }}\frac{{N_\parallel^2 }}{{N_\parallel^2  - 1}} \left( {\left( {\frac{{p_{\parallel,R} }}{{m_e c}}} \right)^2 \left(\frac{{\partial \ln N_\parallel }}{{\partial \ln \Omega _0 }}+\frac{1}{2}\right) + \frac{{\Omega _0 }}{{\omega N_\parallel }}\frac{{p_{\parallel,R} }}{{m_e c}} + \frac{1}{2}\left(1-\gamma_R^2\right)} \right)
\]

Using $\zeta=\zeta_{-1}$, we can rewrite general equation (\ref{eq:A20}) for the first cyclotron resonance as:
\begin{eqnarray}
 \h_\zeta   &=& \frac{1}{{2M}}P_\zeta ^2  + {\rm A}\zeta  + {\rm B}\cos \zeta  \label{eq:A:Hzeta_cyclotron}  \\
 \frac{1}{M} &=& \frac{{N_\parallel^2  - 1}}{{\gamma _R }}\frac{{\omega ^2 }}{{m_e^2 c^4 }},\quad {\rm B} = eA_0 \sqrt {\frac{{2I_{R} \Omega _0 }}{{m_e c^2 }}}  \nonumber
 \end{eqnarray}

%We also use the dipole magnetic latitude $\lambda$ instead of $s$: $ds/d\lambda=R\sqrt{1+3\sin^2\lambda}\cos\lambda$ and $\Omega_0=\Omega_0(0) \sqrt{1+3\sin^2\lambda}/\cos^6\lambda $.

Note that for the simplified dispersion relation valid for parallel propagating waves \citep{bookStix62}, we can write:
\begin{equation}
N_\parallel   = \frac{{\Omega _{pe} }}{\omega }\left( {\frac{\Omega_0}{\omega}  - 1} \right)^{ - 1/2} ,\quad \frac{{\partial \ln N_\parallel  }}{{\partial \ln \Omega _0 }} + \frac{1}{2}=  - \frac{1}{2}\frac{\Omega_0}{\Omega_0-\omega} + \frac{1}{2} =  - \frac{1}{2}\frac{\omega }{{\Omega _0  - \omega }}
\label{eq:A32}
\end{equation}
where $\Omega_{pe}=const$ is the plasma frequency.

\subsection{Strongly Oblique waves: Landau resonance}
In this section, we consider electrons in Landau resonance $\Nr=0$ with strongly oblique whistler mode waves propagating with $\theta$ within one degree from the resonant cone angle $\theta_r={\rm arccos}(\omega/\Omega_0)$ \citep{bookStix62}. Therefore, the Hamiltonian $H_1$ from Eq. (\ref{eq:A11}) takes the form
\begin{eqnarray}
 H_1  &=&  + eA_0 h^{(0)} \cos \phi _\parallel  \label{eq:A34} \\
 h^{(0)}  &=&  - \frac{{\rho \Omega _0 }}{{c\gamma }}CJ_1 \left( {k_\perp \rho } \right) - \left( {\frac{{p_\parallel }}{{\gamma m_e c}} + C_3 } \right)\sin \theta J_0 \left( {k_\perp \rho } \right) \nonumber
 \end{eqnarray}
where $p_\parallel=k_\parallel I_R+p_{\parallel,R}=\gamma_R/N_\parallel$. %As in the previous section, we consider a magnetic field model with $\Omega=\Omega(0) \sqrt{1+3\sin^2\lambda}/\cos^6\lambda $ and $ds/d\lambda=R\sqrt{1+3\sin^2\lambda}\cos\lambda$, where $R$ is the spatial scale of magnetic field inhomogeneity.

Note that using a simplified dispersion relation valid for obliquely propagating waves in a high-density plasma \citep{bookStix62}, we write:
\begin{equation}
N_\parallel   = \frac{{\Omega _{pe} \cos \theta }}{{\sqrt {\Omega_0\omega\cos \theta  - \omega^2} }},\quad \frac{{\partial \ln N_\parallel  }}{{\partial \ln \Omega _0 }} =  - \frac{1}{2}\frac{{\Omega_0 \cos \theta }}{{\Omega_0 \cos \theta  - \omega}}
\label{eq:A35}
\end{equation}
where $\Omega_{pe}=const$ is the plasma frequency.

%Using $\omega_{\Nr}=0$, we rewrite $\varepsilon=\beta_\bot^2-2\gamma_{init}\omega_n$ as:
%\begin{equation}
%\varepsilon  = \beta _ \bot ^2  = \frac{{\Omega _0 }}{\omega }\frac{{2\omega \tilde I_x }}{{m_e c^2 }} = \varpi \varepsilon _0
%\label{eq:A36}
%\end{equation}
%where $\varepsilon_0$ is a constant.

\subsection{Properties of pendulum equation}
Hamiltonian (\ref{eq:A14}) is conservative, i.e. $\h=const$. Therefore, the energy change $m_ec^2\Delta\gamma$ due to resonant wave-particle interaction is approximately equal to $\omega \Delta I$. For resonant scattering (crossing of the resonance), $\Delta I$ can be written as
\begin{eqnarray*}
 \Delta I &=& \int\limits_{ - \infty }^{ + \infty } {\dot Idt}  = \int\limits_{ - \infty }^{ + \infty } {\frac{{\partial \h}}{{\partial \zeta }}dt}  = eA_w h^{\left( {n_r } \right)} \int\limits_{ - \infty }^{ + \infty } {\sin \zeta dt}  = 2eA_w h^{\left( {n_r } \right)} \int\limits_{ - \infty }^{\zeta _R } {\frac{{\sin \zeta d\zeta }}{{\dot \zeta }}}  \\
  &=& 2eA_w h^{\left( {n_r } \right)} M\int\limits_{ - \infty }^{\zeta _R } {\frac{{\sin \zeta d\zeta }}{{P_\zeta  }}}  = eA_w h^{\left( {n_r } \right)} \sqrt {\frac{{2M}}{{\rm A}}} \int\limits_{ - \infty }^{\zeta _R } {\frac{{\sin \zeta d\zeta }}{{\sqrt {h_\zeta   - \zeta  - a\cos \zeta } }}}
 \end{eqnarray*}
where $a=|{\rm B}/{\rm A}|$, $h_\zeta=\h_\zeta/{\rm A}$, $\zeta_R$ is $\zeta$ value at the resonance. This equation gives $\Delta I$ as a function of the energy $h_\zeta$, but the precise $h_\zeta$ value depends on initial electron gyrophase and can be considered as a random number. Therefore, we should consider $\langle \Delta I \rangle_{h_\zeta}$. An important property of the $\h_\zeta$ system is that the integral
\[
F\left( {a,h_\zeta  } \right) = \int\limits_{ - \infty }^{\zeta _R } {\frac{{\sin \zeta d\zeta }}{{\sqrt {h_\zeta   - \zeta  - a\cos \zeta } }}}
\]
averaged over $h_\zeta$ gives \cite{Neishtadt75}
\[
\left\langle F \right\rangle _{h_\zeta  }  = - \frac{1}{\pi }\int\limits_{\zeta _ -  }^{\zeta _ +  } {\sqrt {\zeta _ +   - \zeta  + a\left( {\cos \zeta _ +   - \cos \zeta } \right)} d\zeta }
\]

Let us re-derive this equation. We introduce
\[
\tilde h_\zeta   = \left\{ {\begin{array}{*{20}c}
   {0,} & {\zeta  < \zeta _ -  }  \\
   {\zeta  + a\cos \zeta ,} & {\zeta  > \zeta _ -  }  \\
\end{array}} \right.
\]
and $h_-=\zeta_-+a\cos\zeta_-$, where $\zeta_-$ is the solution of $1-a\sin\zeta=0$. Then we rewrite the integral from
\[
\left\langle {F} \right\rangle  = \frac{1}{{2\pi }}\int\limits_0^{2\pi } {F\left( {a,h_\zeta  } \right)dh_\zeta  }  = \frac{1}{{2\pi }}\int\limits_{ - \infty }^{\zeta _R } {d\zeta \int\limits_0^{2\pi } {\frac{{a\sin \zeta dh_\zeta  }}{{\sqrt {h_\zeta   - \zeta  - a\cos \zeta } }}} }
\]
as a sum of two terms
\[
\int\limits_{ - \infty }^{\zeta _R } { \int\limits_0^{2\pi } {\frac{{a\sin \zeta dh_\zeta d\zeta  }}{{\sqrt {h_\zeta   - \zeta  - a\cos \zeta } }}} }  = \int\limits_{ - \infty }^{\zeta _ -   + 2\pi } { \int\limits_{\tilde h_\zeta   + h_ -  }^{2\pi  + h_ -  } {\frac{{a\sin \zeta dh_\zeta d\zeta  }}{{\sqrt {h_\zeta   - \zeta  - a\cos \zeta } }}} }  - \int\limits_{\zeta _ -  }^{\zeta _ +  } { \int\limits_{\tilde h_\zeta   + h_ -  }^{h_ -  } {\frac{{a\sin \zeta dh_\zeta d\zeta }}{{\sqrt {h_\zeta   - \zeta  - a\cos \zeta } }}} }
\]
where $\zeta_+$ is the solution of $\zeta_-+a\cos\zeta_0-\zeta-a\cos\zeta=0$. The first integral can be written as
\[
\begin{array}{l}
 \int\limits_{ - \infty }^{\zeta _ -   + 2\pi } {\int\limits_{\tilde h_\zeta   + h_ -  }^{2\pi  + h_ -  } {g^{ - 1} \left( {h_\zeta  } \right)a\sin \zeta d\zeta dh_\zeta  } }  = 2\int\limits_{ - \infty }^{\zeta _ -   + 2\pi } {a\sin \zeta \left( {g\left( {2\pi  + h_ -  } \right) - g\left( {\tilde h_\zeta   + h_ -  } \right)} \right)d\zeta }  \\
  = 2\int\limits_{ - \infty }^{\zeta _ -  } {a\sin \zeta \left( {g\left( {2\pi  + h_ -  } \right) - g\left( {h_ -  } \right)} \right)d\zeta }  + 2\int\limits_{ - \infty }^{\zeta _ -   + 2\pi } {a\sin \zeta g\left( {2\pi  + h_ -  } \right)d\zeta }  \\
  = 2\mathop {\lim }\limits_{N \to \infty } \int\limits_{\zeta _ -   - 2\pi N}^{\zeta _ -  } {a\sin \zeta \left( {g\left( {2\pi  + h_ -  } \right) - g\left( {h_ -  } \right)} \right)d\zeta }  + 2\int\limits_{ - \infty }^{\zeta _ -   + 2\pi } {a\sin \zeta g\left( {2\pi  + h_ -  } \right)d\zeta }  \\
  =  - 2\int\limits_{\zeta _ -   - 2\pi }^{\zeta _ -  } {a\sin \zeta g\left( {2\pi  + h_ -  } \right)d\zeta }  + 2\int\limits_{ - \infty }^{\zeta _ -   + 2\pi } {a\sin \zeta g\left( {2\pi  + h_ -  } \right)d\zeta }  = 0 \\
 \end{array}
\]
where $g\left( {h_\zeta  } \right) = \sqrt {h_\zeta   - \zeta  - a\cos \zeta } $. The second integral has the form
\[
\begin{array}{l}
 \int\limits_{\zeta _ -  }^{\zeta _ +  } {\int\limits_{\tilde h_\zeta   + h_ -  }^{h_ -  } {g^{ - 1} \left( {h_\zeta  } \right)a\sin \zeta d\zeta dh_\zeta  } }  = 2\int\limits_{\zeta _ -  }^{\zeta _ +  } {g\left( {h_ -  } \right)a\sin \zeta d\zeta }  \\
  =  - 2\int\limits_{\zeta _ -  }^{\zeta _ +  } {g\left( {h_ -  } \right)\left( {1 - a\sin \zeta } \right)d\zeta }  + 2\int\limits_{\zeta _ -  }^{\zeta _ +  } {g\left( {h_ -  } \right)d\zeta }  = 2\int\limits_{\zeta _ -  }^{\zeta _ +  } {g\left( {h_ -  } \right)d\zeta }  \\
 \end{array}
\]
Therefore, for $\left\langle {F} \right\rangle$ we get
\begin{equation}
\left\langle {F} \right\rangle_{h_{\zeta}}  =  - \frac{1}{\pi }\int\limits_{\zeta _ -  }^{\zeta _ +  } {\sqrt {h_ -   - \zeta  - a\cos \zeta } d\zeta }
\label{eq:Scattering}
\end{equation}

\section*{Appendix B}\label{Appendix:B}
This Appendix is devoted to the problem of whistler-mode wave resonance with small pitch-angle electrons \citep{Lundin&Shkliar77,Inan78}. Nonlinear resonance in such a case can be very different from the classical version: the primary effect of small pitch-angles is that the probability of phase trapping increases significantly and can reach $100$\% \cite{Kitahara&Katoh19}. This is so-called {\it auto-resonance} \cite{Fajans&Friedland01, Friedland09, Neishtadt13, Neishtadt75, Sinclair72:MNRAS}, and we provide a basic theoretical description of this phenomenon for electrons resonating with field-aligned whistler-mode waves.

In Hamiltonian (\ref{eq:hamiltonian_const}) the equation for phase $\zeta$ can be written as
\[
\dot \zeta  = \frac{{\partial \h_I }}{{\partial I}} =  - \omega  + \frac{{k_\parallel \left( {p + k_\parallel I} \right)}}{{\gamma m_e }} + \frac{{\Omega _0 }}{\gamma } - \frac{{\partial eU_w }}{{\partial I}}\cos \zeta
\]
where $I$ is equivalent to the magnetic moment $I_x$, and thus small pitch-angle electrons have small $I$. For field-aligned waves $U_w\propto\sqrt{I}$ and $\partial U_w/\partial I = U_w/2I$. For sufficiently small $I$ the term
\[
\frac{{eU_w }}{{2I}}\cos \zeta\propto I^{-1/2}
\]
will be larger than the main resonance term
\[
\frac{{k_\parallel  \left( {p + k_\parallel  I} \right)}}{{m_e\gamma }} - \omega  + \frac{{\Omega _0 }}{\gamma }
\]
and this should change the applicability of expansion around resonance (small $I-I_R$) that we used to derive Hamiltonian (\ref{eq:hamiltonian_slow}). Indeed, numerical simulations show that resonant interactions in small-$I$systems differs significantly from the theoretical predictions based on analysis of these Hamiltonians \cite{Kitahara&Katoh19, Grach&Demekhov18:II, Grach&Demekhov20}. In this Appendix we describe an approach for analysis of small-$I$ systems (see details in \cite{Albert21,Albert22:phase_bunching,Artemyev21:pop}).

Let us start with Hamiltonian (\ref{eq:hamiltonian_const}) for $\Nr=-1$ (when constant $\tilde{I_x}=0$ and $U_w$ is given by Eq. (\ref{eq:A31})). We expand this Hamiltonian around small $I$:
\begin{eqnarray}
 \h_I &=& m_ec^2 \gamma _0  + I\left( {\frac{{k_\parallel  p}}{{m_e\gamma _0 }} + \frac{{\Omega _0 }}{{\gamma _0 }} - \omega } \right) + \frac{1}{{2M_0}} I^2  - \sqrt {\frac{{2I\Omega _0 }}{{m_ec^2 }}} \frac{{eA_w }}{{\gamma _0 }}\cos \zeta \nonumber \\
\label{eq:Csmall01}\\
 & =& \Lambda  + \frac{1}{{2M_0}}K^2 \left( {I - I_R } \right)^2  - \sqrt {\frac{{2I\Omega _0 }}{{m_ec^2 }}} \frac{{eA_w }}{{\gamma _0 }}\cos \zeta  \nonumber
 \end{eqnarray}
where
\begin{eqnarray*}
 \frac{1}{M_0}  &=& \frac{{k_\parallel ^2 }}{{m_e\gamma _0^3 }}\left( {1 - 2\frac{p}{{m_ec^2 }}\frac{{\Omega _0 }}{{k_\parallel  }} - \left( {\frac{{\Omega _0 }}{{k_\parallel  c^2 }}} \right)^2 } \right),\quad \gamma _0  = \sqrt {1 + \left( {\frac{p}{{m_ec}}} \right)^2 }  \\
 \Lambda  &=& m_ec^2 \gamma _0  - \frac{1}{{2M_0}} I_R^2 ,\quad I_R  = M_0\left( {\omega  - \frac{{\Omega _0 }}{{\gamma _0 }} - \frac{{k_\parallel  p}}{{m_e\gamma _0 }}} \right) \\
\end{eqnarray*}
Note that $I_R$ is the resonant momentum, because $\partial\h_I/\partial I \propto I-I_R$ and the resonant condition is $\partial\h_I/\partial I=0$.

Hamiltonian equations for $I$ and $\zeta$ can be written as
\begin{eqnarray}
 \dot I &=&  - \frac{\partial\h_I}{\partial \zeta} =  - \sqrt {\frac{{2I\Omega _0 }}{{m_ec^2 }}} \frac{{eA_w }}{{\gamma _0 }}\sin \zeta  \nonumber\\
\label{eq:Csmall02}\\
 \dot \zeta  &=& \frac{\partial\h_I}{\partial I} = \frac{1}{{2M_0}} \left( {I - I_R } \right) - \frac{1}{{2I}}\sqrt {\frac{{2I\Omega _0 }}{{m_ec^2 }}} \frac{{eA_w }}{{\gamma _0 }}\cos \zeta  \nonumber
 \end{eqnarray}
and $\zeta$ changes much faster than $(s,p,I)$. Thus, for frozen $(s,p)$ we can rewrite Eqs. (\ref{eq:Csmall02}) as
\begin{equation}
\frac{\partial Y}{\partial \tau} =  - \sqrt {2Y} \sin \zeta ,\quad \frac{\partial \zeta}{\partial \tau}  = Y - Y_R  - \frac{u}{{\sqrt {2Y} }}\cos \zeta
\label{eq:Csmall03}
\end{equation}
where
\[
\tau  = \Omega _0 t \cdot \left( {\frac{{eA_w }}{{m_ec^2 }}} \right)^{2/3} \left( {\frac{{Kc}}{{\Omega _0 }}} \right)^2 ,\quad Y = \frac{{\Omega _0 I}}{{m_ec^2 }}\left( {\frac{{eA_w }}{{m_ec^2 }}} \right)^{2/3} ,\quad u = \frac{1}{{\gamma _0 }}\left( {\frac{{\Omega _0 }}{{Kc}}} \right)^2
\]
and we introduce $K=\sqrt{m_e/M_0}$, which has the dimensionality of a wave number.

Equations (\ref{eq:Csmall02}) does not contain a small parameter $\propto eA_w/m_ec^2$, i.e. if $\Omega_0I/m_ec^2$ is of the order of $(eA_w/m_ec^2)$, time-scales of $\zeta$ and $I$ variations become comparable. This makes this system different from the one described by Hamiltonian (\ref{eq:hamiltonian_zeta}), where $(\zeta, P_\zeta)$ are fast variables, but $(s,p,I)$ are slow variables.

Let us write the Hamiltonian for $(\zeta,Y)$, i.e., such that Eqs. (\ref{eq:Csmall03}) will be Hamilton's equations:
\begin{equation}
H_Y  = \frac{1}{2}\left( {Y - Y_R } \right)^2  - \sqrt {2Y} u\cos \zeta
\label{eq:Csmall04}
\end{equation}
To describe dynamics for Hamiltonian (\ref{eq:Csmall05}) we will introduce new variables $(q,P)$, such that
\[
q =  - \sqrt {2Y} \cos \zeta ,\quad P = \sqrt {2Y} \sin \zeta ,\quad \frac{{\partial q}}{{\partial \zeta }}\frac{{\partial P}}{{\partial Y}} - \frac{{\partial q}}{{\partial Y}}\frac{{\partial P}}{{\partial \zeta }} = 1
\]
The latter equation shows that $(q,P)$ are new canonical variables, whereas a new Hamiltonian is
\begin{equation}
\mF = \frac{1}{2}\left( {\frac{{P^2  + q^2 }}{2} - Y_R } \right)^2  + uq
\label{eq:Csmall05}
\end{equation}
Hamiltonian (\ref{eq:Csmall05}) describes {\it second type} resonant systems and has been investigated in \cite{Neishtadt75,Henrard&Lemaitre83,Sinclair72:MNRAS}.

\begin{figure*}
\centering
\includegraphics[width=1\textwidth]{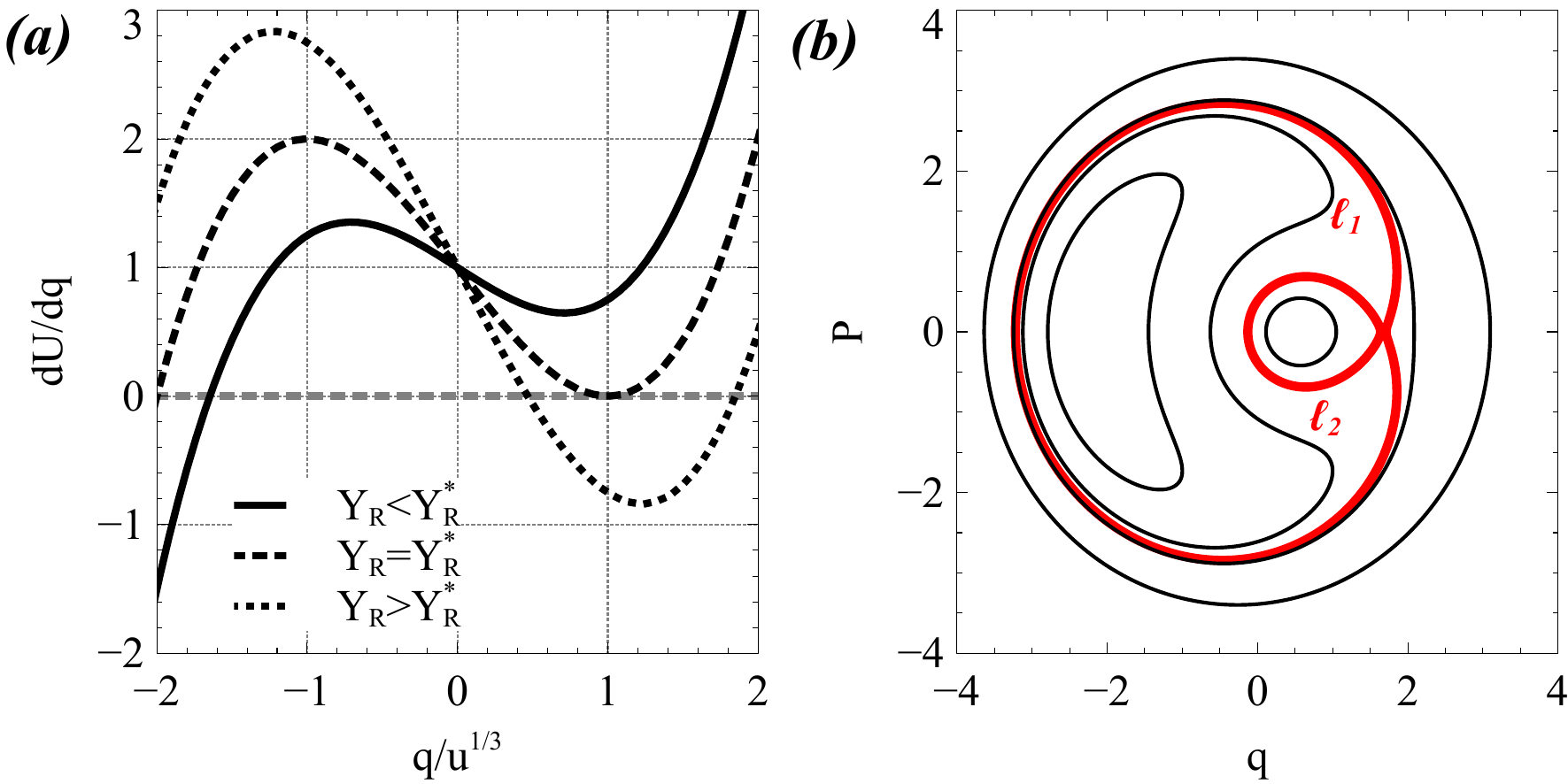}
\caption{(a) Profiles of $dU/dq=\tilde{q}^3/2-(3/2)\tilde{q}(Y_{R}/Y_{R}^*)+1$ for different $Y_{R}$; $Y_{R}^*=(3/2)u^{2/3}$ and $\tilde{q}=q/u^{1/3}$. (b) Phase portraits of Hamiltonian (\ref{eq:Csmall05}) with $u=2$ for $Y_R>Y_{R}^*$. Bold red curve shows the separatrices $\ell_{1,2}$.\label{fig:C03}}
\end{figure*}

Let us consider a profile of Hamiltonian (\ref{eq:Csmall05}) on the axis $P=0$: $U=\mF_{P=0}=(1/2)(q^2/2-Y_{R})^2+uq$. The equation determining  extrema of function $U(q)$  is $dU/dq=(1/2)q^3-Y_{R}q+u=0$, which can be rewritten as  $(1/2)\tilde{q}^3-(3/2)\tilde{q}(Y_{R}/Y_{R}^*)+1=0$ with $\tilde{q}=q/u^{1/3}$ and $Y_R^*=(3/2)u^{2/3}$. Figure \ref{fig:C03}(a) shows that for $Y_{R}<Y_{R}^*$ there is only one extremum and for $Y_{R}>Y_{R}^*$ there are three extrema of $U(q)$. Therefore, the phase portraits of Hamiltonian (\ref{eq:Csmall05}) have two types shown in Fig. \ref{fig:C03}(b): for $Y_{R}<Y_{R}^*$ there is only one O-point in the phase plane and phase trajectories rotate around this point, whereas for $Y_R>Y_R^*$ there are two O-points and one X-point (saddle point), and two separatrices $\ell_{1,2}$ demarcate the phase portrait onto three domains. Ratio $Y_R/Y_R^*$ can be written as
\[
 \frac{{Y_R }}{{Y_R^* }} =\frac{2}{3}\frac{{\kappa ^{2/3} k^2 c^2 }}{{\gamma _0^{4/3} \Omega _{0}^2 }}\frac{{I_R \Omega _{0} }}{{m_e c^2 \left( {B_w /B} \right)^{2/3} }}
 \]
where
\[
\kappa  = 1 - \frac{{\Omega _{0}^2 }}{{k^2 c^2 }} - 2\frac{{\Omega _{0} }}{{kc}}\frac{p}{{m_e c}}
\]
and at the resonance $I_R=I_x$, $\kappa=\gamma _0^2 \left( {1 - \left( {\omega /kc} \right)^2 } \right)$.

For constant $u$, the $Y_R$ system with Hamiltonian (\ref{eq:Csmall05}) is integrable, whereas for slowly changing $u$, $Y_R$ (slowly changing $(s,p)$) we can introduce an adiabatic invariant $I_{P}=(2\pi)^{-1}\oint{Pdq}$ (because all phase trajectories in the portrait shown in Fig. \ref{fig:C03}(b) are closed). In absence of separatrix (for $Y_{R}<Y_{R}^*$), $I_P$ would be conserved with the exponential accuracy $\sim \exp\left(-(\B_w/B_0)^{-1/3}\right)$ where $(\B_w/B_0)^{1/3}\ll 1$ separates time-scales of $u$, $Y_R$ change ($s$, $p$ change) and $P$, $q$ change. For conserved $I_P$ the system becomes integrable and $Y$ well before the resonance is equal to $Y$ well after the resonance. Note that $I_P=(2\pi)^{-1}\oint{Pdq}=2Y(2\pi)^{-1}\oint{\cos^2\zeta d\zeta}=Y$ far from $\ell_{1,2}$ or in absence of $\ell_{1,2}$.

Figure \ref{fig12} shows the relation between phase portraits in Fig. \ref{fig:C03}(b) and Fig. \ref{fig07}(b). Hamiltonian $\h_\zeta$ from Eq. (\ref{eq:hamiltonian_zeta}) describes the charged particle motion from the domain $G_{1}$ to the domain $G_{2}$, and this motion may include a temporary particle trapping into the domain $G_{0}$. The trapping probability is typically quite small and most of particles will move from  $G_{1}$ to $G_{2}$ via a single resonance $\dot\zeta \propto I-I_R=0$ crossing. However, if initial particle $I_x$ ($I$) is small, we should consider Hamiltonian $\mF$ to describe particle motion. Small $I$ means small $Y$, and thus small area $I_P$ in the phase portrait of Hamiltonian $\mF$. Thus, particles will small $I$ (small $I_P$) appear within the $G_{0}$ domain in the phase portrait, whereas particles with large $I_P$ will move along the orbits outside the $\ell_1$ separatrix (see Fig. \ref{fig:C03}(b)), within the $G_{1}$ domain. This predicts that most of particles with small $I$ experience the phase trapping.

Particles  having $I$ sufficiently large to start motion in $G_{1}$ domain, but sufficiently small to be described by $\mF$ Hamiltonian, can experience the phase trapping (transition from $G_{1}$ to $G_{0}$) or phase bunching (scattering; transition from $G_{1}$ to $G_{2}$). The latter transition happens when crossing the separatrix $\ell_1$, which will result in $I_P$ jump, $\Delta I_P=\Delta \area/2\pi$, where $\Delta \area$ is the difference of areas of $G_{1}$ and $G_{2}$ domains. This $I_P$ jump is directly related to the jump of $I$ (and $I_x$)
\begin{equation}
\Delta I_x  = \Delta I  = \frac{{m c^2 \Delta \area}}{{2\pi \Omega _{0} }}\left( {\frac{{eA_w }}{{m_ec^2}}} \right)^{2/3}
\label{eq:Csmall06}
\end{equation}
This scaling $\Delta I_x \propto A_w^{2/3} \propto \B_w^{2/3}$ differs from the phase bunching scaling $\Delta I_x \propto \B_w^{1/2}$ derived in Section~\ref{sec:nl}. Therefore, two main features of small $I$ (small pitch-angle) systems are: (a) a large probability of trapping, when electrons are already within the trapping region when the separatrix appears, (b) a different scaling of momentum (energy) jump due to phase bunching.

\section*{Appendix C}
\begin{figure}
\centering
\includegraphics[width=1\textwidth]{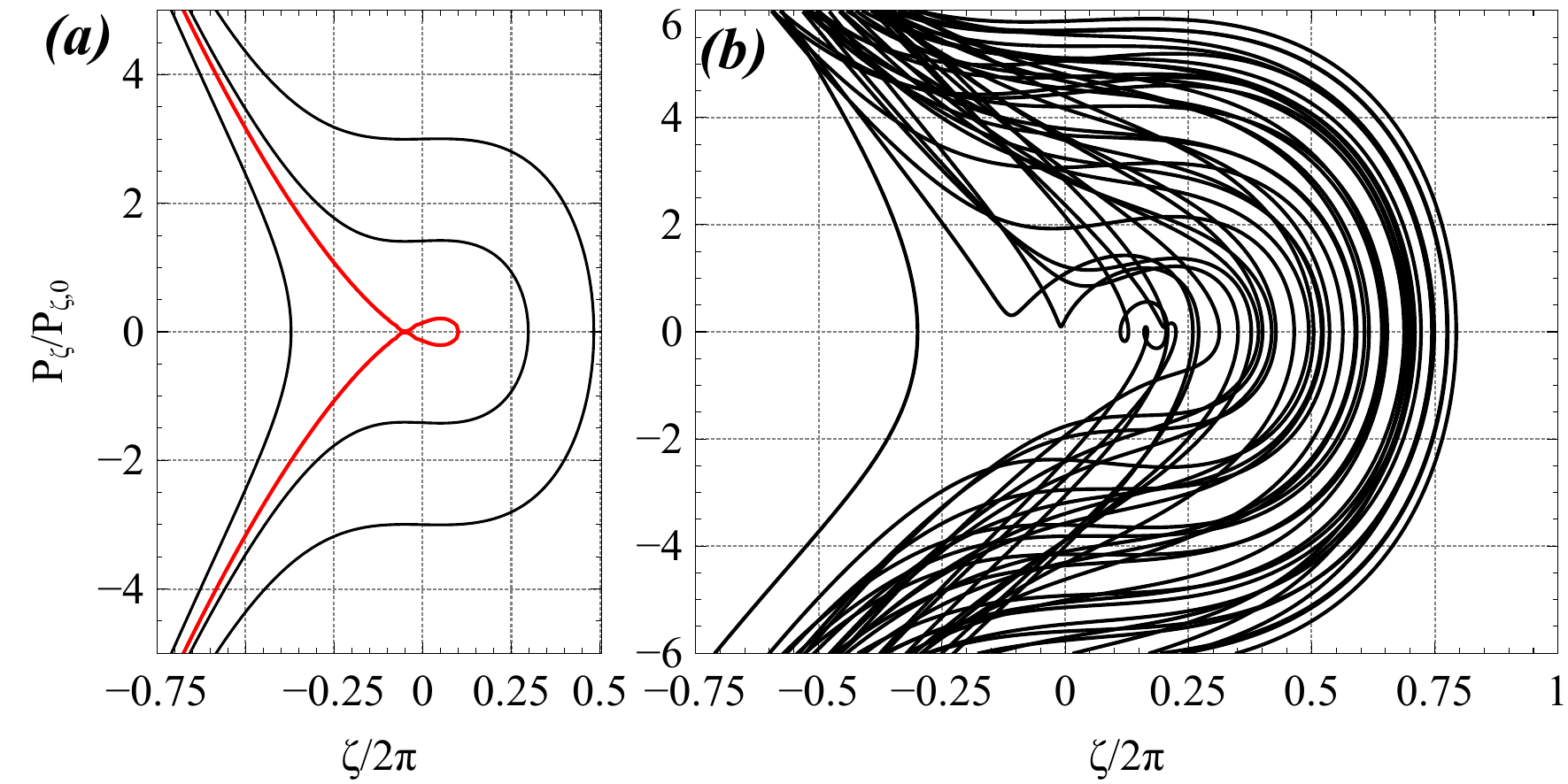}
\caption{
\label{figC} The phase portrait of Hamiltonian (\ref{eq:C14}) for $r/u$ around $1$ (a). Set of trajectories obtained by numerical integration of Eq. (\ref{eq:C18}) with $\delta_0=\varepsilon^{2/3}$.}
\end{figure}

In this Appendix we describe the limit of a small area $\area$ system, i.e., such systems where parameter $a=|{\rm B}/{\rm A}|$ from Hamiltonian (\ref{eq:hamiltonian_zeta}) is about one. To make nonlinear resonant interaction possible, we assume that $|{\rm A}|$ becomes smaller than $|{\rm B}|$, but to keep $\area$ small the inequality $a>1$ holds only within a short range of magnetic latitudes, i.e., over a short range of resonant energies, $\gamma_R$. Then, the $\zeta$ range of the area filled with closed trajectories (where $\zeta \in[\zeta_-,\zeta_+]$ in Fig. \ref{fig07}(b), $\zeta_-$ is the saddle point, and $\zeta_+$ is the coordinate of separatrix crossing $P_\zeta=0$, right from the $\zeta_-$) is small and Hamiltonian (\ref{eq:hamiltonian_zeta}) can be expanded as:
\begin{equation}
     \h_\zeta = \frac{1}{2M}P_\zeta^2  + {\rm B} \cdot \left( {\frac{1}{a}\zeta  + \cos \zeta } \right)   \approx \frac{1}{2M}P_\zeta^2  + {\rm B} \cdot \left( {\left( {\frac{1}{a} - 1} \right)\zeta  + \frac{1}{6}\zeta ^3 } \right) \label{eq:C14}
\end{equation}
where $1/a$ slowly changes along the trajectory from $>1$ (no closed trajectories in the phase portrait) to $\min 1/a<1$ (maximum area filled by closed trajectories), and then to $>1$ again. This expansion is around $\zeta = pi/2$, because $\zeta_{\pm} \to \pi/2$ with $a\to 1$ accordingly to equation $\sin\zeta=1/a$. Figure \ref{figC}(a) shows the phase portrait of Hamiltonian (\ref{eq:C14}) for such $1/a<1$. There are two small system parameters: ${\rm B}\propto \D \propto k^{-1}(\partial\ln\Omega_0/\partial) \ll 1$ and ${\rm A} \propto \B_w/B_0 \ll 1$. When these two parameters are of the same order of magnitude, $a=|{\rm B}/{\rm A}| \sim 1$. In this Appendix we introduce $\varepsilon \ll 1$ such that ${\rm B}\propto \varepsilon$ and ${\rm A}\propto \varepsilon$, so that $\tilde{\rm A}={\rm A}/\varepsilon$ is of the order of one. To model the system evolution we can write $1/a=1-\delta_0+(\varepsilon t)^2$ where the small parameter $\delta_0>0$ determines how far $1/a$ is from $1$ and $(\varepsilon t)^2$ models the slow evolution of $1/a$ along the trajectory (i.e., we change here the slow coordinate $s$ to a slow time):
\begin{equation}
\h_\zeta \approx \frac{1}{2M}P_\zeta^2  - \varepsilon \tilde{{\rm A}} \cdot \left( {\left( {\delta _0  - \left( {\varepsilon t} \right)^2 } \right)\zeta  - \frac{1}{6}\zeta ^3 } \right)
\label{eq:C15}
\end{equation}
and $M$, $\tilde{\rm A}$ can be considered as constant along a short interval of $\varepsilon t \in [-\sqrt{\delta_0}, \sqrt{\delta_0}]$. The equation for $\area$ for Hamiltonian (\ref{eq:hamiltonian_zeta}) can be written as
\begin{eqnarray}
 \area &=& \sqrt {8\varepsilon\tilde{\rm A} M} \int\limits_{\zeta _ +  }^{\zeta  - } {\left( {\left( {\delta _0  - \left( {\varepsilon t} \right)^2 } \right)\left( {\zeta   - \zeta _ + } \right) - \frac{{\zeta ^3  - \zeta _ + ^3 } }{6}} \right)^{1/2}d} \zeta \nonumber \\
  &=& \sqrt {8\varepsilon\tilde{\rm A} M}  \delta _0^{5/4}\left( {1 - \frac{{t^2 }}{{t_0^2 }}} \right)^{5/4}  \int\limits_{\bar \zeta _ +  }^{\bar \zeta  - } {\left( {\left( {\bar \zeta   - \bar \zeta _ + } \right) - \frac{{\bar \zeta^3  - \bar \zeta _ +  ^3 }}{6}} \right)^{1/2}d} \bar \zeta  \nonumber \\
  &=& \sqrt {8\varepsilon\tilde{\rm A} M}  \delta _0^{5/4} \left( {1 - \left( {t/t_0 } \right)^2 } \right)^{5/4}\cdot\frac{{12 \cdot 2^{3/4} }}{5}
 \label{eq:C16}
 \end{eqnarray}
where $\bar\zeta_+=-\sqrt{2}$, $\bar\zeta_-=\sqrt{8}$, $t_0=\sqrt{\delta_0}/\varepsilon$ and $t/t_0$ is equivalent to $\gamma_R$, the energy at resonance, because different slow time values imply here different values of slow $s$, i.e., different values of the resonant $s_R$ related to $\gamma_R$ through Eq. (\ref{eq:h_const}). Equation (\ref{eq:C16}) demonstrates that the $\delta_0$ parameter controls the magnitude of $\area$, and effects of nonlinear interactions should disappear when $\delta_0\to 0$. In the nonlinear regime, there are well separated populations of trapped particles (a small number of particles gaining a large energy) and phase bunched particles (a large number of particles losing energy). In contrast, in the diffusive regime the numbers of particles gaining and losing energy are (approximately) equal. Therefore, there is a threshold $\delta_0$ value (or $\area$ value) such that for $\delta_0$ below this threshold, we cannot separate trapping and nonlinear scattering. Let us derive an expression for this threshold $\delta_0$ (or $\area$) value.

Hamiltonian equations for Hamiltonian (\ref{eq:C15}) can be combined to get the second order equation for $\zeta$:
\begin{equation}
\frac{{d^2 \zeta }}{{d\tilde t^2 }} =   \varepsilon  \cdot \left( {\delta _0  - \left( {\varepsilon \tilde t} \right)^2  - \frac{1}{2}\zeta ^2 } \right)
\label{eq:C17}
\end{equation}
where $\tilde{t}=t\sqrt{\tilde{\rm A} M}$ and $\tilde{\rm A}$, $ M$ are constants of the order of one (not dependent on $\varepsilon$). Equation (\ref{eq:C17}) can be rewritten in normalized variables $\tau=\tilde{t}\varepsilon^{1/2}\delta_0^{1/4}$, $\xi=\zeta/\sqrt{\delta_0}$
\begin{equation}
\frac{{d^2 \xi }}{{d\tau ^2 }} =   \left( {1 - \frac{\varepsilon }{{\delta _0^{3/2} }}\tau ^2 } \right) - \frac{1}{2}\xi ^2
\label{eq:C18}
\end{equation}
Equation (\ref{eq:C18}) shows that for $\delta_0=\varepsilon^{2/3}$ there is no separation of time-scales, i.e. the equation describing electron motion around the resonance does not contain a slow time. Thus, both trapped and nonlinearly scattered particles should stay in the resonance approximately the same time, and there is no separation between these two types of trajectories. Figure \ref{figC}(b) confirms this conclusion: we solve equation Eq. (\ref{eq:C18}) with $\delta_0=\varepsilon^{2/3}$ for set of trajectories and plot these trajectories in the phase plane $(\xi,d\xi/d\tau)$. There are still some trajectories similar to trapped trajectories of the original system, but particles on these trajectories do not fulfil a single oscillation across the resonance $d\xi/d\tau=0$, i.e., we cannot separate particles on trapped and scattered trajectories.

The parameter $\delta_0$ controls the effective range of energy change due to trapping, and should be related to the number of complete rotations around the resonance of trapped particles, $N_{trap}$. Thus, we can derive the $\delta_0(N_{trap})$ dependence. We define $N_{trap}$ as the maximum number of periods of a trapped particle's rotation in the phase space (see phase portrait in Fig. \ref{figC}(b)). This parameter is quite universal: it can be determined for any particular wave-field model (see, e.g., \cite{Zhang18:jgr:intensewaves} for a discussion of $N_{trap}$ values typical for whistler-mode waves observed in the Earth's radiation belts). The trapped period for Hamiltonian (\ref{eq:C15}) can be written as
\begin{equation}
T_{trap} \left( {\varepsilon t} \right) = 2\int\limits_{\zeta _ -  }^{\zeta _{_ +  } } {\frac{{d\zeta }}{{\dot \zeta }}}  = \frac{C}{{\varepsilon ^{1/2} \left( {\delta _0  - \varepsilon ^2 t^2 } \right)^{1/4} }}
\label{eq:C20}
\end{equation}
where constant $C\sim O(1)$ is determined by the distance from the separatrix in the phase portrait. The maximum time of trapped particle motion is of the order of $\sim2\delta_0^{1/2}/\varepsilon$. Therefore we can write for $N_{trap}$
\begin{equation}
N_{trap}  = \frac{2}{\varepsilon }\int\limits_0^{\sqrt {\delta _0 } } {\frac{{d\varepsilon t}}{{T_{trap} \left( {\varepsilon t} \right)}}}  = \frac{{\delta _0^{3/4} }}{{\sqrt \varepsilon  }}\tilde C\sim\varepsilon ^{3\kappa /5 - 1/2} \label{eq:C21}
\end{equation}
where constant $\tilde{C}\sim O(1)$. Equation (\ref{eq:C21}) shows that $N_{trap}\sim \varepsilon^{3\kappa/5-1/2}$. For the threshold value $\kappa=5/6$, we get $N_{trap}\sim O(1)$, i.e., for $\kappa=5/6$ ($\delta_0\sim \varepsilon^{2/3}$) the number of trapping periods does not depend on $\varepsilon$ and there is no separation between trapped and scattered particles anymore.

\section*{Appendix D}
In this Appendix we consider an important property of resonant systems described by the Hamiltonian $\h_I=\h_0\left(I,s,p\right)+\varepsilon  \h_1\left(I,\zeta,s,p\right)$ with $\varepsilon=const\ll 1$: the gain of phase $\zeta$ between two successive resonances is a random value. The rigorous proof of this %$\Delta \xi$
property can be found in \citep{Gao23:RCD23}, whereas a simplified version of this proof is provided below \citep[this derivation is somewhat similar to  the derivation provided in][]{Neishtadt05}. Examples with $\h_0=-\omega I+\gamma(s,p,I)$, $\varepsilon \h_1=-eU_w(s,I)\cos\zeta$, and  $\varepsilon\sim \B_w/B_0$ are provided by Eq. (\ref{eq:hamiltonian_const}), but within this Appendix we do not use explicit forms for $H_0$, $H_1$. In this Hamiltonian $(s, \varepsilon^{-1} p)$, $(\zeta, I)$ are pairs of conjugate variables (hence $(\zeta, I)$ are fast variables, $(q, p)$ are slow variables), and $\h_1$ is periodic in $\zeta$. Momentum $I$ is an adiabatic invariant: $\dot I=-\varepsilon\partial \h_1/\partial \zeta$, and  $I$ is constant in the averaged over $\zeta$  system. There is no explicit dependence on time, and thus $\h_I=h={\rm const}$. The resonance condition is determined by the equation $\partial \h_0/\partial I=0$. Solving this equation for $I$ gives the equation $I=I_R(s,p)$ of the surface of resonance. Denote $\Lambda(s,p)=\h_0(I_R(s,p),s,p)$.  The Hamiltonian can be expanded around the resonance surface similarly to Eq. (\ref{eq:hamiltonian_zeta}). For variables $\zeta$ and $P_{\zeta}=I-I_R$ we get Hamiltonian
\begin{equation*}
\h_\zeta   = \frac{1}{{2M}}P_\zeta ^2  + {\rm A}\zeta  + \h_1
 \label{eq:hamiltonian_zeta_A}
\end{equation*}
where
\begin{equation*}
M^{-1} = \left. {\frac{{\partial ^2 \h_0 }}{{\partial I^2 }}} \right|_{I = I_R }  \approx {\rm const}, \quad {\rm A} =  \left\{ {\Lambda, I_R } \right\} \approx {\rm const} .
\end{equation*}

\begin{figure}
\includegraphics[width=1\textwidth]{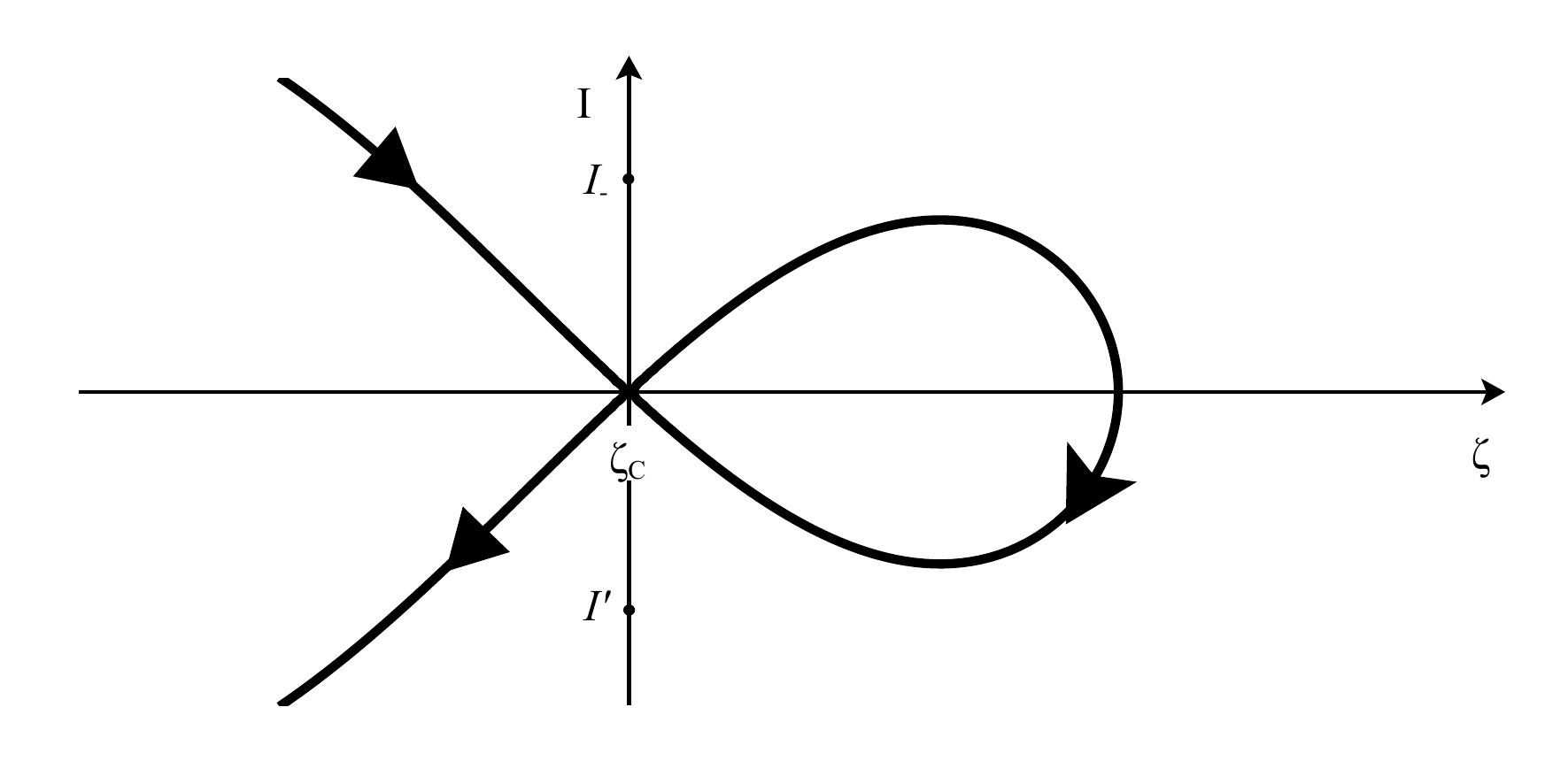}
\caption{Schematic of phase portrait. \label{figA} }
\end{figure}

We assume that the phase portrait of the Hamiltonian $\h_\zeta$ looks like one shown in Fig. \ref{figA} (we put $I$ instead of $P_\zeta$ onto the vertical axis there).

We introduce the improved adiabatic invariant $J$ with the variable transformation $(I,\zeta, p,s)\mapsto (J,\theta, P,Q)$ such that the new Hamiltonian is $\h=\h_0(J, P, Q)+\varepsilon\bar \h_1(J, P,Q)$,  where $\bar \h_1$ is the average of $\h_1$ over $\zeta$ (in the leading approximation).

Far from the resonance $\theta$ changes with the frequency
\begin{equation}
\dot \theta  = \frac{{\partial \h_0 }}{{\partial J}} + \varepsilon \frac{{\partial \bar \h_1 }}{{\partial J}}
\label{eq01A}
\end{equation}
with $J={\rm const}$, and
\begin{equation}
\dot Q = \varepsilon \frac{{\partial \h_0 }}{{\partial P}} + \varepsilon ^2 \frac{{\partial \bar \h_1 }}{{\partial P}},\quad \dot P =  - \varepsilon \frac{{\partial \h_0 }}{{\partial Q}} - \varepsilon ^2 \frac{{\partial \bar \h_1 }}{{\partial Q}}
\label{eq02A}
\end{equation}

We introduce $\omega_0(J,P,Q)=\partial \h_0/\partial J$, $\omega_1(J,P,Q)=\partial \bar H_1/\partial J$, and consider a large number $N\gg 1$ of rounds of $\zeta$ from $t=t_0$ (when phase point is far from the resonance and moves towards the resonance) to $t=t_N$;  the last round is sufficiently far from the resonance and $\theta \approx \zeta$ in the leading approximation. The last round ends at at $\zeta=\zeta_{c N}  =\zeta_{c}\ {\rm mod}\ 2\pi$;
 (see Fig. \ref{figA} for the definition of $\zeta_c$). Then
\begin{equation}
\zeta _{c}  = \zeta _0  + \int\limits_{t_0 }^{t_N } {\left( {\omega _0  + \varepsilon \omega _1 } \right)dt} \quad {\rm mod}\ 2\pi.
\label{eq04A}
\end{equation}
 We introduce $t_*$ as  the time of crossing  the resonance, i.e. $\omega_0(J, P, Q)|_{t=t_*}=0$, and rewrite Eq. (\ref{eq04A}):
\begin{equation}
\zeta _{c} = \zeta _0  + \int\limits_{t_0 }^{t_* } {\left( {\omega _0  + \varepsilon \omega _1 } \right)dt}  - \int\limits_{t_N }^{t_* } {\left( {\omega _0  + \varepsilon \omega _1 } \right)dt} \quad {\rm mod}\ 2\pi
\label{eq05A}
\end{equation}
Because $t_*-t_N\ll 1/\varepsilon$ we can use $\dot Q=\varepsilon \partial \h_0/\partial P$, $\dot P=-\varepsilon \partial \h_0/\partial P$ and $Q\approx q$, $P\approx p$ in the last integral in Eq. (\ref{eq05A}). We also assume that $\zeta_{c}\approx \zeta_{c*}=\zeta_c|_{t=t_*}$. Thus, we can replace $\zeta_{c}$ with $\zeta_{c*}$ in (\ref {eq05A}).

To describe system dynamics for $t\in [t_N,t_*]$ we use the expansion of the Hamiltonian around the resonance: $H=\Lambda+\Upsilon $ and
\begin{equation}
\Upsilon  = \frac{1}{2M}\left( {I - I_R } \right)^2  + \varepsilon \h_1 .\;\;
\label{eq06A}
\end{equation}
The Hamiltonian in new variables $(J,\theta, P, Q)$ can be expanded as
\begin{equation}
\h = \Lambda  + \frac{1}{2M}\left( {J - I_R } \right)^2  + \varepsilon \bar H_1
\label{eq07A}
\end{equation}
We introduce $\e=(J-I_R)^2/2M$ and write
\begin{eqnarray}
\dot {\e} &=&- \varepsilon M^{-1}\left( {J - I_R } \right){\rm A},\quad {\rm A} =   \left\{ {\Lambda, I_R } \right\} \approx {\rm const} \nonumber\\
 \omega _0  &=& \frac{{\partial \h}}{{\partial J}} = M^{-1}\left( {J - I_R } \right) \label{eq08A}
\end{eqnarray}
Using $dt= d\e/(d\e/dt)$ we rewrite integral
\begin{equation}
\int\limits_{t_N }^{t_* } {\left( {\omega _0  + \varepsilon \omega _1 } \right)dt}  \approx -\frac{1}{\varepsilon }\int\limits_{ \e_N }^0 {\frac{{\omega _0 d \e}}{{g\left( {J - I_R } \right){\rm A} }}}  =- \frac{1}{\varepsilon }\int\limits_{ \e_N }^0 {\frac{{d \e}}{{\rm A}}}  =   \frac{{ \e_N }}{{\varepsilon {\rm A} }}
\label{eq09A}
\end{equation}
where $\e_N$ is the value $\e$ along the trajectory at $t=t_N$,    and we omit $\varepsilon \omega_1$ because $t_*-t_N\ll 1/\varepsilon$. Using $ \e+\varepsilon \bar \h_1=\Upsilon $, we write
\begin{equation}
 \e_N  = \Upsilon _N  - \varepsilon \bar \h_1  \approx \Upsilon _N  - \varepsilon \bar \h_{1*}
\label{eq10A}
\end{equation}
where $\Upsilon _N$ is the value $\Upsilon $ along the trajectory at $t=t_N$, and  $\bar \h_{1*}$ is the resonant value of $\bar \h_1$. Substituting Eqs. (\ref{eq09A}, \ref{eq10A}) to Eq. (\ref{eq05A}), we obtain
\begin{equation}
\zeta _{c*}  = \zeta _0  + \int\limits_{t_0 }^{t_* } {\left( {\omega _0  + \varepsilon \omega _1 } \right)dt}  - \frac{{\Upsilon _N  - \varepsilon \bar \h_{1*} }}{{\varepsilon  {\rm A} }} \quad {\rm mod}\ 2\pi
\label{eq11A}
\end{equation}
or
\begin{equation}
\frac{{\Upsilon _N }}{{2\pi \varepsilon {\rm A} }} = \frac{{\bar \h_{1*} }}{{2\pi {\rm A}}} + \frac{1}{{2\pi }}\left( {\zeta _0  - \zeta _{c*}  + \int\limits_{t_0 }^{t_* } {\left( {\omega _0  + \varepsilon \omega _1 } \right)dt} } \right)\,\,\, \bmod 1
\label{eq12A}
\end{equation}
We define $\Upsilon _{last}$ as the value of  $\Upsilon $ along the trajectory  at the last crossing of the line  $\zeta=\zeta_c$ before crossing the resonance. Thus, $\Upsilon _{last}=\Upsilon _N \ {\rm mod}\ 2\pi \varepsilon {\rm A}$, because the change of $\Upsilon $ for  one round of  $\zeta$  equals  $2\pi\varepsilon {\rm A}$. We introduce $\xi=(\Upsilon _{last}-\varepsilon \h_{1c*})/(2\pi\varepsilon {\rm A})$,  where $\h_{1c*}$ is value  of  $\h_{1}$ at $\zeta=\zeta_c$, $t=t_*$  and write
\begin{equation}
\xi  = {\mathop{\rm Frac}\nolimits} \left( \frac{{\bar \h_{1*}  - \h_{1c*} }}{{2\pi {\rm A}}} + \frac{{\zeta _0  - \zeta _{c*}}}{{2\pi }}  + \frac{1}{2\pi\varepsilon } \int\limits_{\tau _0 }^{\tau _* } \left( \omega _0  + \varepsilon \omega _1  \right)d\tau \right)
\label{eq13A}
\end{equation}
Here $\tau=\varepsilon t$, $\tau_*=\varepsilon t_*$.
Note that $\xi$ can be written as
\begin{equation}
\xi  = \frac{{\Upsilon _{last}  - \varepsilon H_{1c*} }}{{2\pi \varepsilon r}} = \frac{{\Upsilon _{last}  + \varepsilon {\rm A}\zeta _{c*}  - \left( {\varepsilon H_{1c*}  + \varepsilon {\rm A}\zeta _{c*} } \right)}}{{2\pi \varepsilon {\rm A}}}
= \frac {\E_{last}  - \E_{c*}}{2\pi\varepsilon {\rm A}}
\label{eq14A}
\end{equation}
where
\begin{equation}
\E = \frac{1}{2M}\left( {I - I_R } \right)^2  + \varepsilon {\rm A}\zeta  + \varepsilon \h_1,
\label{eq15A}
\end{equation}
and $\E_{c*}$ is the value of $\E$ at $\zeta=\zeta_c$, $t=t_*$, $\E_{last}$ is the value of $\E$ at the last crossing of the line $\zeta=\zeta_c$ before crossing  the resonance.

Let us use Eq. (\ref{eq13A}) to consider two successive crossings of the same resonance in the same direction. During the period  of a slow motion a particle crosses the same resonance twice in opposite directions.  We consider  crossing the resonance in direction from positive to negative value of frequency, as in Fig.\ref{figA}, assuming that  for crossing in the opposite direction we have  $H_1\equiv 0$.    Then there is only one resonant interaction for one period of slow motion, i.e. two successive resonance crossings are separated by one slow period. Far from the resonance the improved adiabatic invariant $J$ can be considered as a  constant.  Denote $\tau_{-}$ and $\tau_{+}$ slow time moments of the resonance crossings ($\tau=\varepsilon t$). Let   $\xi_{\pm}$ be values of the variable $\xi$ corresponding to these two crossings. We  are looking for a relation between  $\xi_+$ and $\xi_-$. Due to periodicity of the slow motion, values of $\zeta_{c*}$, $\h_{1c*}$, $\bar \h_{1*}$, ${\rm A}$ are the same at $\tau=\tau_{+}$ and $\tau= \tau_{-}$.

We consider value $\xi_-$ for the first of the resonance crossings, and the corresponding value $\E=\E_{last-}$. At $\tau =\tau_-$ the phase point is on the line $\zeta=\zeta_{c*}$ with $I>I_R$ at the position indicated by the symbol $I_-$ in Fig. \ref{figA}. We assume that this phase point crosses the resonance without trapping.
Thus, at some $\tau=\tau'$ it arrives again to the line  $\zeta=\zeta_{c}= \zeta_{c*}'$ with the value $\E=\E'$ and $I<I_R$. The phase point position is indicated by the symbol $I'$ in Fig. \ref{figA}.
We denote $\xi'=  ({\E'  - \E_{c*}})/({2\pi\varepsilon {\rm A}})$.
%We have $ \E'\approx \E_{last-},\,  \xi'\approx \xi_-  $.

At some moment of the slow time $\tau_0\in (\tau_{-}, \tau_{+})$ the phase point is far from the resonance and has $\zeta=\zeta_0$. Then Eq. (\ref{eq13A}) gives
\begin{equation}
\xi_+  = {\mathop{\rm Frac}\nolimits} \left( \frac{{\bar \h_{1*}  - \h_{1c*} }}{{2\pi {\rm A}}} + \frac{{\zeta _0  - \zeta _{c*}}}{{2\pi }}  + \frac{1}{2\pi\varepsilon } \int\limits_{\tau _0 }^{\tau _{+} } \left( \omega _0  + \varepsilon \omega _1  \right)d\tau \right)
\label{eq13A+}
\end{equation}
Similarly, considering the backward motion on the time interval from $\tau_0$ to $\tau'$ we get
\begin{equation}
\xi' = {\mathop{\rm Frac}\nolimits} \left( {\frac{{\bar \h_{1*}  - \h_{1c*} }}{{2\pi {\rm A}}} +\frac{{\zeta _0  - \zeta _{c*}}}{{2\pi }}  -\frac{1}{2\pi\varepsilon} \int\limits_{ \tau'}^{\tau _0} {{{\left( {\omega _0  + \varepsilon \omega _1 } \right)}}d\tau }  } \right)
\label{eq20A-}
\end{equation}
In this expression we can replace in the leading approximation $\xi'$ with $\xi_{-}$ and $\tau'$ with $\tau_{-}$ (note that
\[
\frac{1}{2\pi\varepsilon} \int\limits_{ \tau_{-}}^{\tau'} \left( {\omega _0  + \varepsilon \omega _1 } \right)d\tau
\]
is small, because $\omega _0$ vanishes on the resonant surface).
Thus for the value
$\Delta\xi=\xi_+ -\xi_-$
we get
\begin{equation}
 \Delta\xi=\xi_+ -\xi_-=  \frac{1}{2\pi\varepsilon}\int\limits_{\tau _{-} }^{\tau _{+} } {{\left( {\omega _0  + \varepsilon \omega _1 } \right)}}{d\tau } \ {\rm mod}\ 1\phantom{***}
 \label{gain}
\end{equation}
 and this describes the phase $\zeta$ gain between two resonance crossings (between moments $\tau_{-}$ and $\tau_{+}$) normalized on $2\pi$.  In the main text $\bar \h_1=0$ and thus $ \omega_1= \partial \bar H_1/\partial J=0$.

\section*{Appendix E}
In this Appendix, we demonstrate the applicability of the mapping technique for two types of events observed in the outer radiation belts. Both events are associated with energetic electron precipitations via scattering by whistler-mode waves. Such scattering corresponds to a decrease of the electron equatorial pitch-angle $\alpha_{eq}$ and a change of electron energy. Field-aligned chorus waves interact with electrons through the first cyclotron resonance. The nonlinear regime of this resonant interaction includes phase trapping (with $\alpha_{eq}$ and energy increase, see, e.g., \cite{Omura15}) and phase bunching (with $\alpha_{eq}$ and energy decrease, see, e.g., \cite{Albert00,Vainchtein18:jgr}). The competition between phase trapping and phase bunching \cite{Istomin73:electrostatic, Solovev&Shkliar86, Shklyar11:angeo, Artemyev16:pop:letter} controls the net electron transport into the loss-cone, but all precipitating electrons lose energy, because precipitation is provided only by phase bunching. Thus, both linear and non-linear electron interaction with chorus via the cyclotron resonance is associated with electron deceleration. Such effective scattering and deceleration of electrons by field-aligned waves is modeled with the mapping technique in Subsection Appendix E.2.

Although field-aligned chorus waves are the most intense whistler-mode wave population in the radiation belt \cite{Li11,Agapitov13:jgr}, there is also a significant population of very oblique whistler-mode waves (almost electrostatic mode, see \cite{Artemyev16:ssr}) propagating near the resonance cone angle \cite{Agapitov13:jgr,Agapitov18:jgr,Li16:statistics}. These oblique waves can interact resonantly with electrons through Landau resonance \cite[e.g.,][]{Shklyar09:review,Nunn&Omura15} and can accelerate electrons  efficiently via phase trapping \cite{Artemyev12:pop:nondiffusion,Agapitov14:jgr:acceleration,Hsieh&Omura17,Hsieh20}. The corresponding Landau resonant energies are much lower than cyclotron resonant energies with quasi-parallel waves, allowing very oblique waves to resonate with $100$ eV - $10$ keV electrons \cite{Mourenas14,Artemyev15:natcom,Artemyev16:ssr} which can almost never be scattered by field-aligned lower-band chorus waves \cite{Li10chorus}. Moreover, in contrast to the cyclotron resonance associated with an energy decrease of precipitating electrons, the Landau resonant trapping is associated with energy increase (see Figs. \ref{fig06} and \ref{fig20}). Thus, precipitating electrons (those with decreasing $\alpha_{eq}$) are simultaneously accelerated, corresponding to an increase of low-$\alpha_{eq}$ energetic electron flux \cite{Agapitov15:grl:acceleration}. Since very oblique whistler-mode waves can be quite intense \cite{Cully08,Cattell08,Cattell11:Wilson,Agapitov14:jgr:acceleration}, these waves may in principle interact nonlinearly with electrons and efficiently trap them through Landau resonance \cite{Agapitov15:grl:acceleration,Mourenas16}. Such trapping results in intense, short-lived precipitation of $\lesssim 200$ keV electrons \cite{Zhang22:natcom,Artemyev22:jgr:Landau&ELFIN}. An important feature of this precipitation mechanism is the {\it loss-cone overfilling}, when precipitating fluxes at low altitudes appear to be larger than quasi-trapped fluxes just outside of the loss-cone \cite{Zhang22:natcom}. In Subsection Appendix E.1 we provide a simulation of this effect using the mapping technique.

The loss-cone size $\alpha_{LC}$ (the maximum pitch-angle range of precipitating electrons) in the outer radiation belt around the magnetic equator is only a few degrees, which makes it challenging for near-equatorial spacecraft to directly measure such electron losses \cite{Kasahara18:nature}. Spacecraft at low altitudes (where magnetic field is large and $\alpha_{LC}$ reaches $60-70^\circ$), however, can easily measure electron fluxes within the loss cone. Conjugate spacecraft measurements of near-equatorial waves responsible for electron scattering into the loss cone and low-altitude electron fluxes within it, are optimal for testing theoretical models of electron scattering and precipitation by waves. \cite{Li13:POES}. Thus, we use such a combination of observations from the equatorial THEMIS-E spacecraft \cite{Angelopoulos08:ssr} and the low-altitude ($\sim 400-450$ km) ELFIN spacecraft \cite{Angelopoulos20:elfin}.

The two identical CubeSats of the Electron Losses and Fields Investigation (ELFIN) mission were launched on September 15th, 2018, with onboard energetic particle detectors of ions (EPD-I) and electrons (EPD-E) \cite{Angelopoulos20:elfin}. Electrons are measured in an energy range from 50 keV to $\sim 6$ MeV with an energy resolution ($\Delta E/E$) less than 40 \% and a time resolution ($T_{spin}$) of $\sim 2.85$ s \cite{Angelopoulos23:ssr,Tsai24:review}. The high resolution of the pitch-angle of the EPD-E ($\Delta \alpha \simeq 22.5 ^{\circ}$) enables it to distinguish trapped electrons (outside the local bounce loss cone) from precipitation electrons (within the local bounce loss cone) \cite[see details]{Zhang22:microbursts,Angelopoulos23:ssr}. The ratio of the precipitating electron flux and the trapped electron flux, $j_{prec}/j_{trap}$, indicates the efficiency of equatorial electron scattering by waves \cite{Mourenas21:jgr:ELFIN,Zhang22:natcom,Shen22:jgr:WISP,Capannolo23:elfin} or of the magnetic field line curvature process \cite{Wilkins23,Zou24:elfin}. In this Appendix, we consider two ELFIN events with electron scattering by whistler-mode waves \citep[additional examples are provided in][]{Chen22:microbursts,Tsai22,Zhang23:jgr:ELFIN&scales,Gan23:grl_elfin,Kang24:elfin}.

\subsection*{Appendix E.1: Nonlinear Landau resonance: loss-cone overfilling effect}
\begin{figure}
\includegraphics[width=1\textwidth]{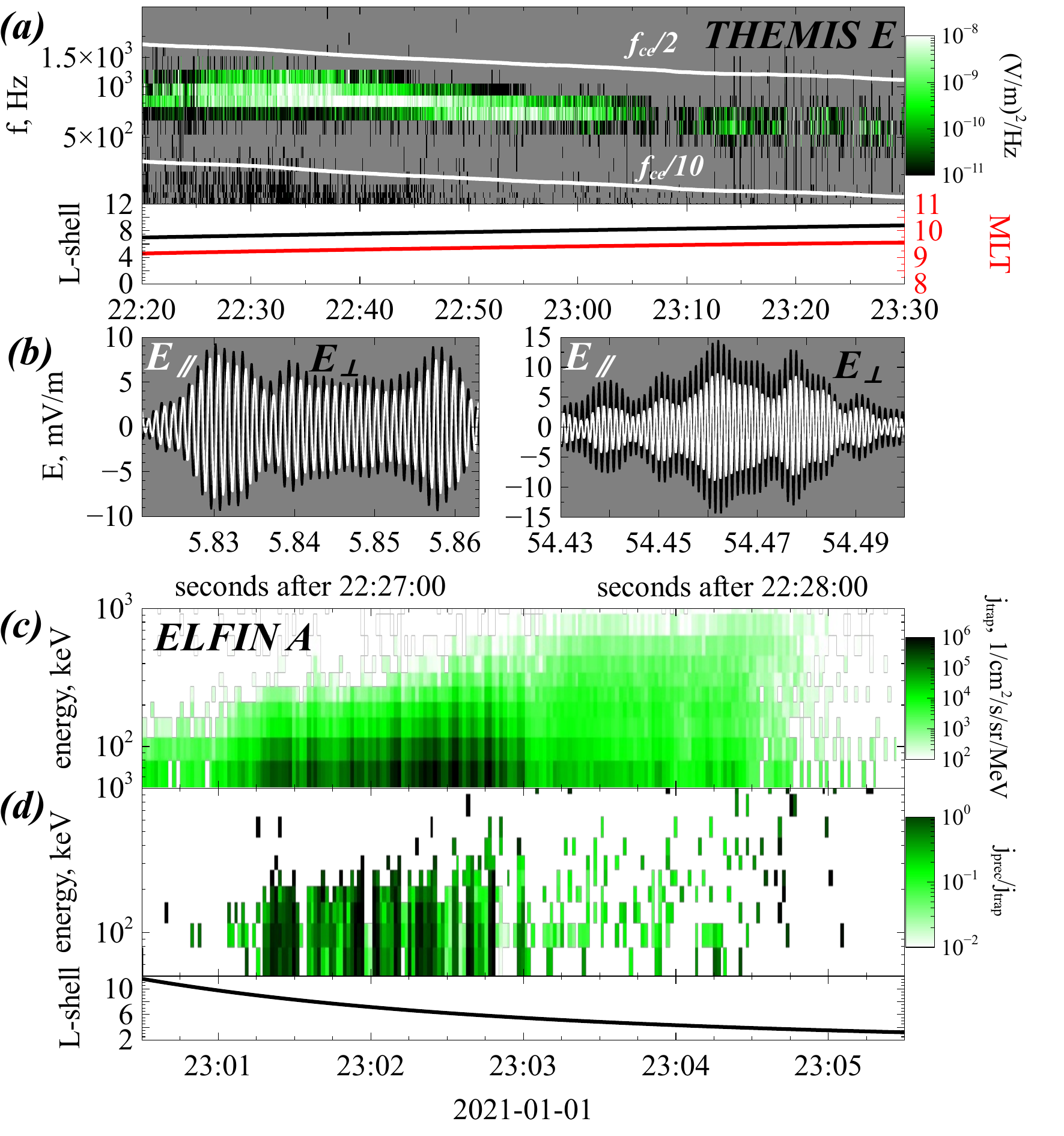}
\caption{Overview of the event with the effect of nonlinear Landau resonance: near-equatorial THEMIS observations of whistler-mode electric field (a, b), low altitude ELFIN observations of trapped fluxes (c) and precipitating-to-trapped flux ratio (d).
\label{figE1} }
\end{figure}

Figure \ref{figE1}(a) is an overview of THEMIS E measurements on 2021-01-01 from 22:20 to 23:30 UT in the dayside inner magnetosphere, when the spacecraft moved outward from $L\sim7$ to $L\sim9$ ($L$ is evaluated with the \cite{Tsyganenko89} magnetic field model). There is strong whistler-mode wave activity within the frequency range $f/f_{ce}\in[0.2,0.4]$, where the electron cyclotron frequency $f_{ce}$ is calculated from in-situ measurements from the fluxgate magnetometer \cite{Auster08:THEMIS} and the wave spectrum is obtained from the search coil \cite{LeContel08, Cully08:ssr}. Using plasma density estimates from the spacecraft potential \cite{Nishimura13:density}, we can estimate the plasma frequency to electron gyrofrequency ratio, $f_{pe}/f_{ce}\sim 3$. Figure Figure \ref{figE1}(b) shows several whistler-mode wave packets with a large parallel electric field component, suggesting the presence of very oblique whistler-mode waves \cite{Artemyev16:ssr}. Using $f_{pe}/f_{ce}$ and 3D electric fields, we estimate the wave normal angle $\theta$ for these wave-packets: $\theta\in[60,70]$ (see details of $\theta$ estimation method in \cite{Ni11,Agapitov14:jgr:acceleration}), with a Gendrin angle $\theta_g={\rm acos}(2f/f_{ce})\approx45^\circ$ and a resonance cone angle $\theta_r={\rm acos}(f/f_{ce})\approx 69^\circ$. Thus, the observed waves indeed propagate obliquely with $\theta\in[\theta_g,\theta_r]$, i.e., in the quasi-electrostatic mode \cite{Agapitov13:jgr,Li16:statistics}, and can interact with electrons through the Landau resonance \cite{Shklyar09:review,Artemyev16:ssr}. An interesting and important feature of the observed whistler-mode wave activity is the transient nature of wave bursts, which may be due to quasi-periodic wave generation modulated by compressional ultra-low-frequency waves \cite{Li11:modulation1,Li11:modulation2,Xia16:ulf&chorus,Zhang19:jgr:modulation}. Indeed, measurements of precipitating electrons by the Precipitating Electron and Ion Spectrometer onboard two DMSP satellites \cite{Hardy84:DMSP,Rich85:DMSP} show clear spatially/temporally localized enhancements of $\sim 5$ keV electron precipitation, consistent with the presence of a transient (quasi-periodic) equatorial generation of very oblique whistler-mode waves \cite[see details in][]{Artemyev22:jgr:Landau&ELFIN}.

There are a couple of orbits of ELFIN A and B within the same time interval and $L$-shell, MLT range \cite[see][for THEMIS, DMSP, and ELFIN orbits as a function of MLT and $L$-shell]{Artemyev22:jgr:Landau&ELFIN}. Figure \ref{figE1}(c, d) shows that within the $L$-shell range corresponding to the outer radiation belt, ELFIN observed transient, strong precipitation of $<200$ keV electrons. The timescale of these precipitation bursts is about (or even less than) ELFIN's half-spin-period, 1.5s. Therefore, we infer  that these  precipitation events are probably driven by electron scattering by transient whistler-mode waves which, due to their amplitude modulation by ULF waves, do not maintain the same spatial distribution at the equator over time intervals longer than $\sim1-2$ minutes. Most of the precipitation bursts reach a ratio $\sim1$ (and even exceed it) at energies $<200$ keV, i.e., they represent very strong precipitation with an entirely filled loss-cone, roughly corresponding to the so-called strong diffusion limit \cite{Kennel69} (or the loss-cone overfilling \cite{Zhang22:natcom}).

To model electron precipitation by very oblique whistler-mode waves, we use the mapping technique as described by Eqs. (\ref{eq:map}). To evaluate the trapping probability, $\Pi$, and energy/pitch-angle changes due to trapping and bunching, we specify several properties of the wave model. We use a simplified model that fits the observed $\B_w(\lambda)$ distribution obtained from satellite statistics by \cite{Agapitov18:jgr}:
\begin{equation}
\B_w  = \B_{w,eq}  \cdot \frac{1}{2}\left( {1 + \tanh \left( {\lambda /\delta \lambda _1 } \right)} \right) \cdot \exp \left( { - \lambda ^2 /\delta \lambda _2^2 } \right)
\label{eq02:E}
\end{equation}
where $\delta\lambda_1$ determines the spatial extent of the wave generation region, and $\delta\lambda_2$ determines the latitude range of wave propagation before damping at middle latitudes. This model describes waves in the $\lambda>0$ hemisphere, and we assume a symmetric wave field distribution relative to the equator, i.e., $\B_{w}(\lambda)=\B_{w}(-\lambda)$.
Wave frequency can be assumed to be fixed along the wave propagation (or slowly varying in time, if we deal with chorus waves), whereas the wave number profile $k(\lambda)$ can be obtained from the cold plasma dispersion \cite{bookStix62} for a given wave normal angle distribution $\theta(\lambda)$ at this particular frequency. For very oblique whistler-mode waves, we use a $\theta=\theta_r(\lambda)-\Delta\theta$ model, with $\theta_r={\rm arccos}(\omega/\Omega_{ce})$ being the resonant cone angle and $\Delta\theta=const$ the model parameter specifying the wave normal angle deviation from $\theta_r$. This wave normal angle model is based on previous observations of very oblique whistler-mode wave propagation around the resonance cone angle \cite{Agapitov13:jgr,Li16:statistics}. The cold plasma dispersion relation does not work for $\Delta\theta\to 0$, where the thermal electron contribution to the wave dispersion properties becomes crucial \cite{Sazhin&Horne90}. Theoretical estimates \cite{Mourenas14,Artemyev16:ssr} and spacecraft observations \cite{Ma17} suggest that the minimum $\Delta\theta$ can be determined from the limitation of the wave refractive index $N=kc/\omega$ to values $N<N_{hot}\approx 100-300$ for typical thermal electron energies in the inner magnetosphere.

During Landau resonance ($\omega\gamma-k_\parallel c\sqrt{\gamma^2-1}\cos\alpha=0$), the electron propagates in the same direction as the wave, in contrast to cyclotron resonance characterized by opposite propagations of wave and electron. For a fixed wave frequency at Landau resonance, the resonance curve $(\omega/\Omega_{ce,eq})(\gamma^2-1)\sin^2\alpha_{eq}/2=const$ (see Fig. \ref{fig06}) is given by the equation of the constant magnetic moment (we use the normalized moment $I_x=(\omega/\Omega_{ce,eq})(\gamma^2-1)\sin^2\alpha_{eq}/2$). Combining this equation with the resonance condition, we obtain the equation for dependence of the Landau resonant energy on the magnetic latitude:
\begin{equation}
\gamma _R  = \frac{1}{{\sqrt {1 - \left( {\omega /ck_\parallel  } \right)^2 } }}\sqrt {1 + 2I_x \Omega _{ce}}
\end{equation}
\label{eq03:E}

\begin{figure}
\centering
\includegraphics[width=0.9\textwidth]{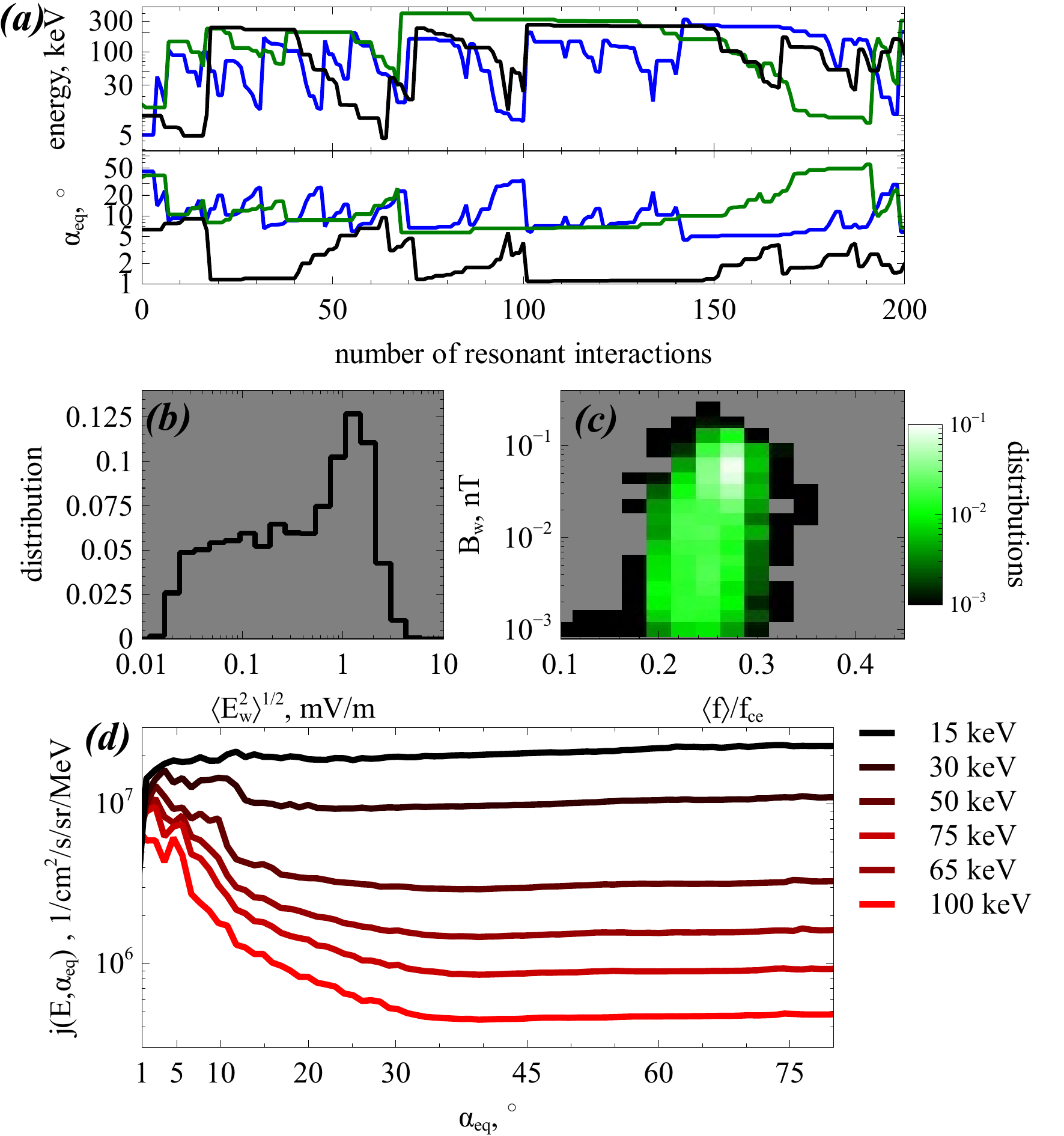}
\caption{Panels (a) shows energy and equatorial pitch-angle for three trajectories evaluated with the mapping technique (\ref{eq:map_simple}). Panel (b) shows the distribution of electric field amplitudes derived from the wave frequency spectra ($fff$ data) shown in Fig. \ref{figE1}. Panel (c) shows the distribution of wave amplitudes and frequencies for the event in Fig. \ref{figE1}. Wave magnetic field amplitudes are converted from the wave electric field amplitudes using cold plasma dispersion relation \cite{Tao&Bortnik10}. Panel (d) shows pitch-angle distributions for different energies after $\sim 10R/c\sim2$s of interactions (sufficient to make at least one bounce period for all considered particles). The initial (equatorial) electron phase space density is obtained by fitting THEMIS E observations. We use a $K=3$ factor for this simulation, with the corresponding probability of wave-particle interactions $1/K^2$. \label{figE2}}
\end{figure}

The wave intensity increases with latitudes $dB_w/d\lambda>0$ around the equator \cite{Agapitov18:jgr,Omura08}. This condition suggests that electrons may be trapped by waves. Being trapped at small resonant latitudes, electrons are transported by the wave to higher latitudes (with a decrease of their pitch-angle) and escape from the resonance at a latitude where $dB_0/d\lambda$ becomes sufficiently strong (for a precise description of the trapping and de-trapping conditions, see, Section \ref{sec:resonance} and e.g., \cite{Artemyev12:pop:nondiffusion, Artemyev13:pop}). Such an electron transport is associated with an energy increase, e.g., a $5$ keV equatorial electron can gain $100$ keV before reaching resonant latitudes $\sim 30^\circ$ and finishing within the loss-cone \cite{Artemyev22:jgr:Landau&ELFIN}. Therefore, the trapping moves lower energy/larger pitch-angle electrons toward the larger energy/lower pitch-angle region in phase space (see schematic in Fig. \ref{fig06}). The Landau resonance scattering moves electrons along the same resonance curve $I_x=const$, but in the direction opposite to the trapping motion. Therefore, Landau resonance results in electron drifts toward smaller energy/larger pitch-angle (the phase bunching effect) and more rare jumps to larger energy/smaller pitch-angle (the phase trapping effect). Figure \ref{figE2}(a) shows three examples of such electron dynamics evaluated with the mapping technique (\ref{eq:map}) for typical wave characteristics (see also Fig. \ref{fig20}).

Near-equatorial spacecraft measurements of whistler-mode waves provide distributions of wave characteristics, e.g., the average wave intensity for various frequency ranges, $L$-shells, $MLT$, and $\lambda$ \cite{Meredith01, Meredith12,Agapitov13:jgr,Agapitov15:jgr}. Such time-averaged intensities can be directly applied to the evaluation of quasi-linear diffusion rates \cite[e.g.,][]{Horne13:jgr,Agapitov18:jgr,Ma18}, but they are not suitable for the evaluation of nonlinear wave-particle interactions. Instead of time-averaged intensities, the models of nonlinear resonant wave-particle interactions require a knowledge of the distributions (occurrence rates) of intense waves that can interact with electrons nonlinearly \cite{Zhang18:jgr:intensewaves,Zhang19:grl}. Therefore, we use THEMIS E measurements during the event in Fig. \ref{figE1} to obtain the distribution of wave amplitudes $\E_w$, as inferred from the wave frequency spectrum $\hat E_w^2(f)$: $\E_w^2=\int_{f_{ce}/20}^{f_{ce}/2}\hat E_{w}^2(f)df$. As shown in Figure \ref{figE2}(b), most waves have amplitudes $<3$ mV/m. However, $E_w$ here is inferred from the 1s-averaged spectrum \cite{Cully08:ssr} and the actual magnitude of individual wave-packets is often significantly larger (see Fig. \ref{figE1}). To account for such an underestimation of $E_w$ due to time averaging, we introduce a multiplicative factor $K$, i.e., we assume that electrons interact with waves of amplitudes $K\times E_w$. To keep the same average intensity as without the $K$ factor, we assume that electrons resonate with waves only a fraction ($1/K^2$) of the simulation time, whereas during the remaining fraction ($1-1/K^2$) of the time there is no resonant interaction.

The wave-particle resonant interaction is not determined solely by $E_w$, but also by the wave frequency, wave normal angle, and the $\E_w(\lambda)$ profile along the magnetic field line. The wave frequency can be determined from the wave frequency spectrum: $\langle f \rangle =\int_{f_{ce}/20}^{f_{ce}/2}f\cdot\hat E_{w}^2(f)df/\E_w^2$ (with $f_{ce}=\Omega_{0}/2\pi$ and $f=\omega/2\pi$). Figure \ref{figE2}(c) shows the $\P(B_w, \langle f \rangle)$ distribution for the event in Fig. \ref{figE1}, with $B_w$ recalculated from $E_w$ using the cold plasma dispersion \cite[see examples in, e.g.,][]{Ni11:plasma_sheet,Agapitov14:jgr:acceleration}. Using this $\P(\B_w, \langle f \rangle)$ distribution, we can rewrite the mapping in Eqs. (\ref{eq:map}) as a two-step equation. During each half of the bounce period, we generate a random number to select a $(\B_w, \langle f \rangle)$ bin from the $\P$ distribution. According to this number, we choose the wave amplitude $B_w$ and wave frequency $\langle f \rangle/f_{ce}$ to calculate three main characteristics of wave-particle interactions: $\Delta\gamma_{trap}$, $\Delta\gamma_{scat}$, and $\Pi$. Then, we make one iteration for energy and pitch-angle using Eq. (\ref{eq:map_multi}). Thus, the observed distribution of wave characteristics, $\P(\B_w, \langle f \rangle)$, determines  $\Delta\gamma_{trap}$, $\Delta\gamma_{scat}$, and $\Pi$ for each resonant interaction. Figure \ref{figE2}(a) shows several examples of electron trajectories calculated with the observed $\P$ distribution.

Using the mapping technique with the observed distribution $\P(\B_w, \langle f \rangle)$, we integrate a large ensemble of electron trajectories having an initial (energy, pitch-angle) distribution measured by THEMIS E around equator. As we are interested in precipitation patterns, we restrict the integration time to $\sim 1$ bounce period of $100$ keV electrons.  All electrons with $\alpha_{eq}<2^\circ$ are considered to be precipitating (as appropriate at $L<8$). Figure \ref{figE2}(d) shows several pitch-angle distributions for different energies after wave-particle resonant interactions. There is a clear peak of small $\alpha_{eq}$ electron fluxes, and this peak is most pronounced for energies higher than $\sim30$ keV. The formation of such a field-aligned (and partially precipitating) electron population is produced by the electron trapping acceleration in the Landau resonance. Despite the small probability of electron Landau trapping $\Pi$ \cite[see, e.g.,][]{Artemyev12:pop:nondiffusion,Artemyev14:pop}, this acceleration mechanism is quite effective, because waves trap particles with initially smaller energies ($\sim$ a few keV) and accelerate them up to $\sim30-100$ keV. Therefore, the large difference of initial phase space densities in the energy of trapping and release compensates the small probability of such trapping ($\Pi$) and allows for the formation of a very substantial field-aligned electron population at $\sim30-100$ keV.

The time-scale of such electron acceleration and precipitation is about a bounce period, that is, the time-scale of electron transport in the Landau trapping. This represents an insufficiently long time for nonlinear wave-particle interactions to establish a fine balance between trapping and phase bunching (nonlinear scattering) and, thus, electron transport toward the loss-cone (due to trapping) may not be fully compensated by electron transport to higher pitch-angles (due to bunching). After a sufficiently long wave-particle interaction, such a fine balance will establish a new distribution function without strong gradients along the resonance curve $I_x=const$ (see Section \ref{sec:asymptote} and  \cite{Artemyev19:pd}). Thus, electron precipitation associated with the Landau trapping acceleration should be provided by bursts of very oblique whistler-mode waves, and they can share properties of microbursts. Therefore, intense (microburst) precipitation events can indicate an effective electron acceleration and an increase of $\sim 50-200$ keV electron fluxes at small pitch-angles, rather than being a signal of electron flux depletion. The accelerated electrons ending up outside the loss-cone, however, have already small pitch-angles, and can be later scattered into the loss-cone by much weaker field-aligned whistler-mode waves \cite[see discussions of such double-resonance mechanism of electron losses due to Landau trapping and cyclotron scattering in][]{Mourenas16,Ma16:tds,Hsieh22}.

\subsection*{Appendix E.2: Nonlinear cyclotron resonance: electron precipitation from the injection region}

In this Subsection we consider the mapping modeling of electron nonlinear resonant interaction with field-aligned whistler-mode waves. The Subsection structure repeats Appendix E.1: we combine the mapping technique from Section \ref{sec:mapping}, near-equatorial THEMIS E measurements, and low-altitude ELFIN measurements. We examine the event with plasma sheet injection observed by THEMIS E \citep[see details in][]{Artemyev22:jgr:DF&ELFIN}. We use magnetic field measurements by the fluxgate magnetometer \cite{Auster08:THEMIS} with $1/3$ s resolution, and magnetic field fluctuations in the $10-4000$ Hz frequency range measured by the search-coil \citep{LeContel08}. We use both waveforms measured in the wave burst mode and wave spectra evaluated on board at $1$ s time resolution \citep{Cully08:ssr}.

\begin{figure}
\includegraphics[width=1\textwidth]{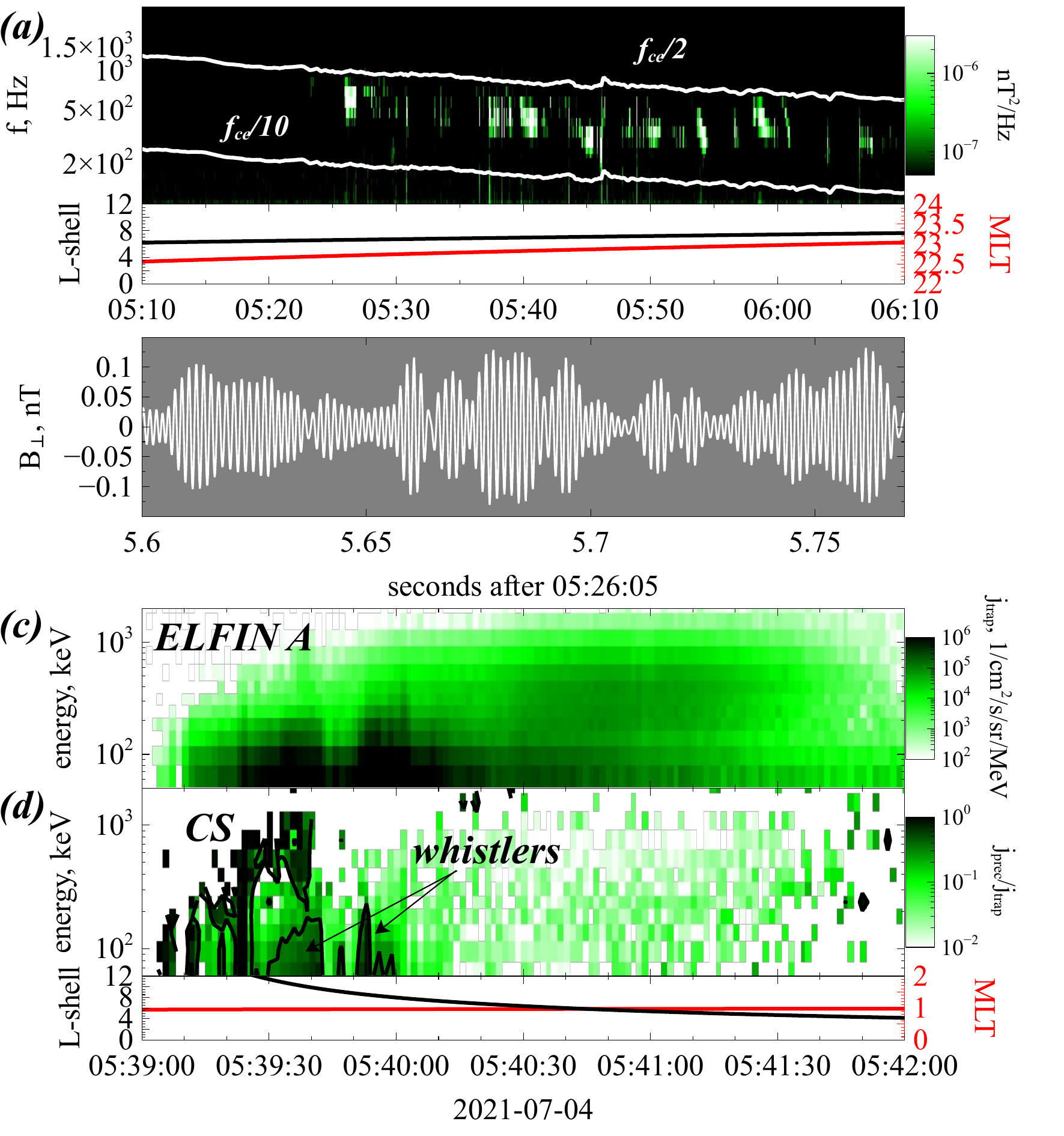}
%\caption{\todo{Overview of the event with the effect if nonlinear cyclotron resonance: near-equatorial THEMIS observations of whistler-mode electric field (a, b), low altitude ELFIN observations of trapped fluxes (c) and precipitating-to-trapped (d, e) flux ratio, distribution of wave intensity and wave frequency (f).  }}
\caption{Overview of the event with the effect of nonlinear cyclotron resonance: near-equatorial THEMIS observations of whistler-mode magnetic field (a, b), low altitude ELFIN observations of trapped fluxes (c) and precipitating-to-trapped flux ratio (d).  The precipitation burst due to electron scattering by whistler-mode waves is embedded into the dispersive precipitation structure (higher energies of precipitating electrons closer to the Earth) due to electron scattering by the magnetotail current sheet (CS) \citep[see details in][]{Artemyev23:ELFIN&dispersion,Wilkins23}.
\label{figE3} }
\end{figure}

\begin{figure}
\centering
\includegraphics[width=1\textwidth]{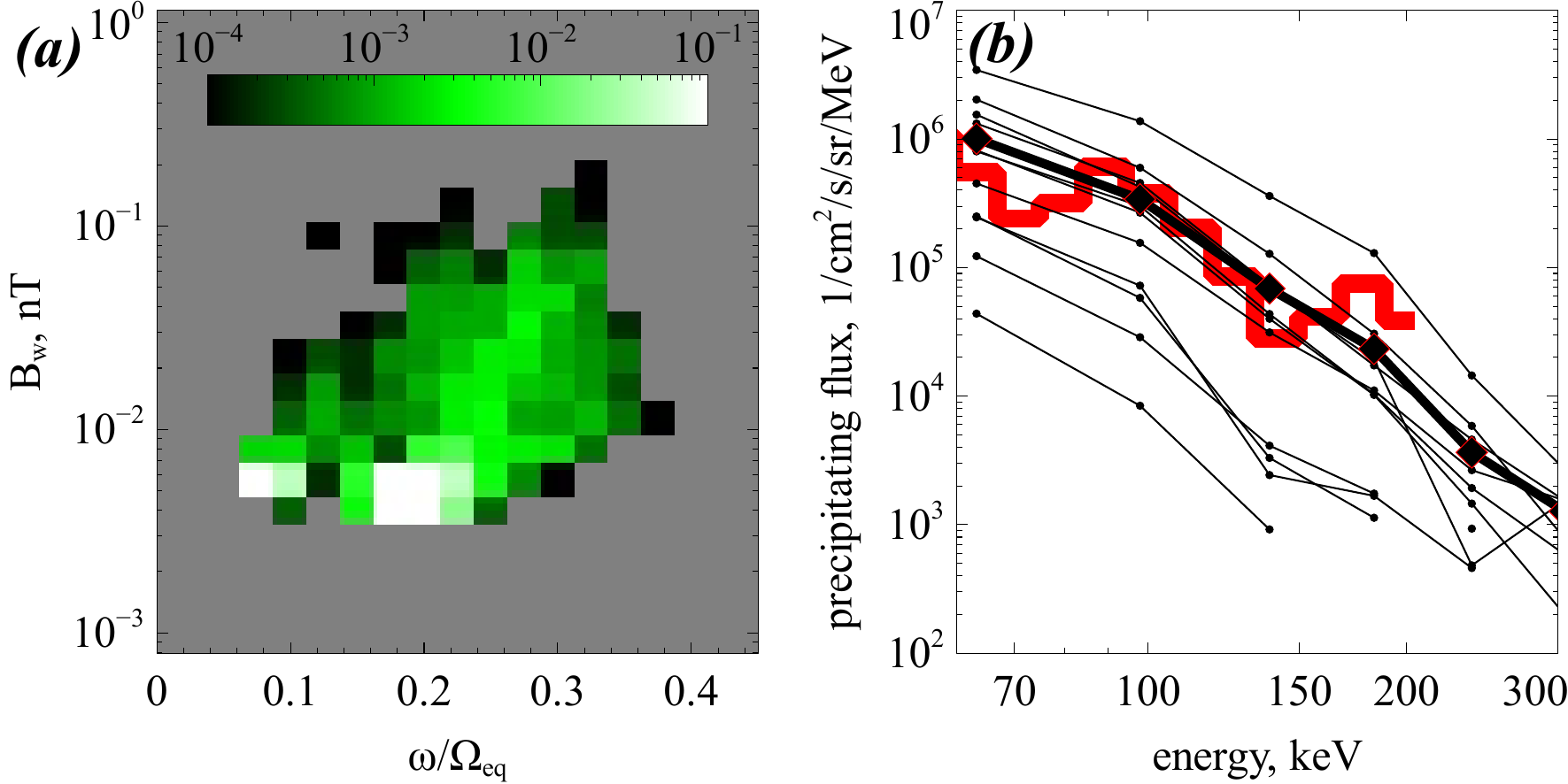}
\caption{Panels (a) shows the distribution of wave amplitudes and frequencies for the event in Fig. \ref{figE3}.  Panel (b) shows energy spectra from ELFIN A for the time interval associated with enhanced electron precipitations. The red curve shows the model result. \label{figE4}}
\end{figure}

Figure \ref{figE3} shows observations of the plasma injection, associated with sporadic and transient bursts of whistler-mode waves (see panel (a)). The wave spectrum $\B_w^2(f)$ with $f=\omega/2\pi$ shows intense bursts around $f\in[0.1,0.4]f_{ce}$ (with $f_{ce}=\Omega_{0}/2\pi$and $f=\omega/2\pi$), in the typical frequency range of whistlers captured around dipolarizing flux bundles \citep{LeContel09,Breuillard16,Zhang18:whistlers&injections,Grigorenko20:whistlers}. For each $1$ s spectrum we calculate the root-mean-square time-averaged wave amplitude $\B_w=\left(\int_{f_{ce}/10}^{f_{ce}/2}\hat B_{w}^2df\right)^{1/2}$ and mean wave frequency $f_m=\int_{f_{ce}/10}^{f_{ce}/2}\hat B_{w}^2fdf/\B_w^2$. These wave characteristics are input parameters for the mapping technique. $\B_w^2$ is the time-averaged wave intensity calculated from wave spectra. The amplitude of individual wave-packets can be significantly larger than $\B_w$. This difference is not important for the quasi-linear diffusion describing wave-particle resonant interaction by diffusion rates that depend only on the time-averaged $\B_w$ \cite{bookSchulz&anzerotti74,bookLyons&Williams}. For nonlinear resonant interactions, however, the electron dynamics are determined by peak wave amplitudes and the occurrence rate of intense waves. Therefore, we need to recalculate the $\B_w$ distribution derived from the wave spectra integration to estimate the actual occurrence rate of intense waves. This is done by using simultaneous measurements of whistler wave packets (see Figure \ref{figE3}(b)). Typical wave packet amplitudes during this event are $\sim 100$ pT, i.e., a factor $K\approx 20$ larger than the average $\B_w\sim 5$pT and a factor $K\approx 5$ larger than the level $\B_w\sim 20$pT of the most intense wave population (see the $\P(\B_w,f/f_{ce})$ distribution in Fig. \ref{figE4}(a)). Accordingly, we multiply the time-averaged $\B_w$ by $K=5$. To keep the same time-averaged wave intensity $\B_w^2$ we assume that such intense wave packets are observed only a fraction $1/K^2$ of the time, whereas during the rest $1-1/K^2$ of the time there are no whistlers. This may be a simplistic approach, but nonetheless it captures the essential features of the wave packet for the purpose of nonlinear interactions. Thus, during each half of the bounce period, there is a probability $1/K^2$ that an electron resonates with one of the waves from the $\P$ distribution, and a probability $1-1/K^2$ that there is no resonant interaction and thus electron characteristics remain unchanged. To set the initial electron distribution function we use THEMIS E electron measurements \cite{McFadden08:THEMIS,Angelopoulos08:sst}.

Electron scattering by whistlers may significantly decrease electron pitch-angles and move particles to the loss-cone. An accurate estimate of the loss-cone size for the near-Earth magnetotail (injection region) is almost impossible, because empirical magnetic field models \citep{Tsyganenko95,Tsyganenko&Sitnov07,Andreeva&Tsyganenko19}, with only few exceptions \citep{Stephens19,Sitnov21,Tsyganenko21}, do not include the effects of dipolarization associated with plasma injection and strong variations of the equatorial magnetic field (such variations may significantly change the loss-cone size and precipitating electron fluxes, see, e.g. \citep{Eshetu18}). Typical loss-cone angle $\alpha_{LC}$ estimates give $\leq 2^\circ$ in the magnetotail (see Fig. 3(d) in  \cite{Zhang15:ECH}), and this value can be larger closer to the Earth. For a simulation of precipitating electron fluxes, we use $\alpha_{LC}=3^\circ$. This value could overestimate the precipitating flux magnitude, but there is almost no energy change due to the wave-particle resonant interaction at pitch-angles $<3$ degrees. Therefore, the energy spectrum of precipitating electrons derived from our simulation should reproduce the expected spectrum of electrons precipitating from the dipolarizing flux bundle region. To check this, we compare model results with observations from the low-altitude, polar-orbiting CubeSats ELFIN \cite{Angelopoulos20:elfin} that provide conjugate measurements for the event in Fig. \ref{figE3}. Figure \ref{figE3} shows an overview of ELFIN A trapped and precipitating fluxes: pattern of strong $j_{trap}$ increase at the inner edge of the plasma sheet (05:39:00-05:39:45 for ELFIN A) is also associated with a high ratio $j_{loss}/j_{trap}\sim 1$ showing strong energy dispersion: larger energies with  $j_{loss}/j_{trap}\sim 1$ are closer to the Earth). This pattern most likely corresponds to the electron isotropic boundary \cite{Imhof79,Yahnin97,Sergeev12:IB,Wilkins23} , i.e., energetic electron scattering in the current sheet due to the magnetic field line curvature \cite{Birmingham84,BZ89,Young02, Lukin21:pop:df}. The curvature scattering efficiency depends on energy and on the equatorial magnetic field curvature radius and intensity. These dependencies are responsible for the observed energy dispersion \cite{Sergeev18:grl,Dubyagin21}. Closer to the Earth, ELFIN shows both a secondary burst of $j_{loss}/j_{trap}\sim 1$ and an increase of $j_{trap}$ (around 05:39:55), and this second precipitation pattern is not due to electron curvature scattering, because the energies of precipitating electrons are lower than energies of the pattern associated with the isotropic boundary observed farther away from the Earth. For ELFIN A this second precipitation pattern is observed around its conjunction to the equatorial injection region observed by THEMIS. We thus attribute this pattern to electron scattering from the injection region. Figure \ref{figE4}(b) shows spectra of precipitating electrons. Their average spectral shape is similar to the precipitating electron fluxes derived from the simulation driven by THEMIS equatorial measurements of electron fluxes and whistlers (compare red and grey curves). Therefore, our simulation confirms that the mapping technique can reproduce electron resonant interactions with field-aligned whistler-mode waves.

%\subsection{Combination of Nonlinear Cyclotron and Landau resonances}

%\bibliographystyle{apalike}      % mathematics and physical sciences
\bibliographystyle{agsm}
%\bibliography{full,addon,addon2}   % name your BibTeX data base

%\begin{thebibliography}{}
% etc
%\end{thebibliography}

\end{document}